%% file: ClassicThesis.tex
\RequirePackage{lineno}
\documentclass[ twoside,openright,titlepage,numbers=noenddot,headinclude,
                footinclude=true,cleardoublepage=empty,abstractoff, 
                BCOR=5mm,paper=a4,fontsize=11pt,
                ngerman,american,%
                ]{scrreprt}

\newcommand*\patchAmsMathEnvironmentForLineno[1]{%
  \expandafter\let\csname old#1\expandafter\endcsname\csname #1\endcsname
  \expandafter\let\csname oldend#1\expandafter\endcsname\csname end#1\endcsname
  \renewenvironment{#1}%
     {\linenomath\csname old#1\endcsname}%
     {\csname oldend#1\endcsname\endlinenomath}}%
\newcommand*\patchBothAmsMathEnvironmentsForLineno[1]{%
  \patchAmsMathEnvironmentForLineno{#1}%
  \patchAmsMathEnvironmentForLineno{#1*}}%
\AtBeginDocument{%
\patchBothAmsMathEnvironmentsForLineno{equation}%
\patchBothAmsMathEnvironmentsForLineno{align}%
\patchBothAmsMathEnvironmentsForLineno{flalign}%
\patchBothAmsMathEnvironmentsForLineno{alignat}%
\patchBothAmsMathEnvironmentsForLineno{gather}%
\patchBothAmsMathEnvironmentsForLineno{multline}%
}
\newenvironment{changemargin}[2]{%
\begin{list}{}{%
\setlength{\topsep}{0pt}%
\setlength{\leftmargin}{#1}%
\setlength{\rightmargin}{#2}%
\setlength{\listparindent}{\parindent}%
\setlength{\itemindent}{\parindent}%
\setlength{\parsep}{\parskip}%
}%
\item[]}{\end{list}}

\input{classicthesis-config}
\usepackage{textcomp}
\usepackage{amsmath}

\usepackage{epstopdf}
\usepackage{svg}
\usepackage{color}
\usepackage{transparent}
\usepackage{etoolbox}
\preto\subequations{\ifhmode\unskip\fi}
\graphicspath{{gfx/chapter1/}{gfx/chapter2/}{gfx/chapter3/}{gfx/chapter4/}}
\usepackage{cancel}
\usepackage{oldgerm}
\usepackage{soul}
\usepackage[strict]{changepage}
\usepackage[flushleft]{threeparttable}
\definecolor{LightMidnightBlue}{rgb}{0.86275,0.91765,0.97255}
\sethlcolor{LightMidnightBlue}
\begin{document}
\frenchspacing
\raggedbottom
\selectlanguage{american} 
\pagenumbering{roman}
\pagestyle{plain}
\include{FrontBackmatter/DirtyTitlepage}
\include{FrontBackmatter/Titlepage}
\include{FrontBackmatter/Titleback}
\cleardoublepage\include{FrontBackmatter/Abstract}
\cleardoublepage\include{FrontBackmatter/Overview}
\cleardoublepage\include{FrontBackmatter/Acknowledgments}

\cleardoublepage\include{FrontBackmatter/Contents}
\pagenumbering{arabic}
\cleardoublepage

\include{Chapters/Chapter01}

\include{Chapters/Chapter02}

\include{Chapters/Chapter03}

\include{Chapters/Chapter04}
\include{Chapters/Chapter05}


\cleardoublepage\include{FrontBackmatter/Bibliography}
\include{FrontBackmatter/Colophon}
\end{document}

%% file: classicthesis-config.tex
\PassOptionsToPackage{eulerchapternumbers,listings, 
				 pdfspacing,
				 subfig,beramono,parts,dottedtoc}{classicthesis}										

\usepackage{ifthen}
\newboolean{enable-backrefs} 
\setboolean{enable-backrefs}{true} 

\newcommand{\myTitle}{Inference from modelling the chemodynamical evolution of the Milky Way disc\xspace}

\newcommand{\myName}{Jan Rybizki\xspace}

\newcommand{\myFaculty}{Astronomisches Rechen-Institut\xspace}

\newcommand{\myUni}{Ruprecht-Karls-Universit\"at Heidelberg\xspace}

\newcommand{\myVersion}{version 0.1\xspace}

\usepackage[binary-units=true]{siunitx} 

\usepackage[autolanguage]{numprint}
\usepackage{expl3}
\newcounter{dummy} 
\providecommand{\mLyX}{L\kern-.1667em\lower.25em\hbox{Y}\kern-.125emX\@}

\PassOptionsToPackage{utf8x}{inputenc}	
 \usepackage{inputenc}				

\PassOptionsToPackage{ngerman,american}{babel}   
 \usepackage{babel}					

\PassOptionsToPackage{square,authoryear}{natbib}
 \usepackage{natbib}				

\PassOptionsToPackage{fleqn}{amsmath}		
 \usepackage{amsmath}

\PassOptionsToPackage{LGR,T1}{fontenc} 
	\usepackage{fontenc}     
\usepackage{textcomp} 
\usepackage{scrhack} 
\usepackage{xspace} 
\usepackage{mparhack} 
\usepackage{fixltx2e} 
\usepackage{relsize}
\PassOptionsToPackage{printonlyused,smaller,withpage}{acronym}
	\usepackage{acronym} 
	
\newcommand{\textgreek}[1]{\begingroup\fontencoding{LGR}\selectfont#1\endgroup}

\usepackage{tabularx} 
\usepackage[format=plain,font=footnotesize,labelfont={bf,sf},textfont=sf,tableposition=top]{caption}
\usepackage{subfig}  
\usepackage{floatrow}

\usepackage[wide,ragged]{sidecap}

\usepackage{listings} 
\PassOptionsToPackage{pdftex,hyperfootnotes=false,plainpages=false}{hyperref} 
	\usepackage{hyperref}  
\PassOptionsToPackage{pdftex}{graphicx}
	\usepackage{graphicx} 

\newcommand{\backrefnotcitedstring}{\relax}
\newcommand{\backrefcitedsinglestring}[1]{(Cited on page~#1.)}
\newcommand{\backrefcitedmultistring}[1]{(Cited on pages~#1.)}
\ifthenelse{\boolean{enable-backrefs}}%
{%
		\PassOptionsToPackage{hyperpageref}{backref}
		\usepackage{backref} 
		   \renewcommand*{\backref}[1]{}  
		   \renewcommand*{\backrefalt}[4]{
		      \ifcase #1 %
		         \backrefnotcitedstring%
		      \or%
		         \backrefcitedsinglestring{#2}%
		      \else%
		         \backrefcitedmultistring{#2}%
		      \fi}%
}{\relax}    

\hypersetup{%
    colorlinks=true, linktocpage=true, pdfstartpage=1, pdfstartview=FitV,%
    breaklinks=true, pdfpagemode=UseNone, pageanchor=true, pdfpagemode=UseOutlines,%
    plainpages=false, bookmarksnumbered, bookmarksopen=true, bookmarksopenlevel=1,%
    hypertexnames=true, pdfhighlight=/O,
    urlcolor=MidnightBlue, linkcolor=MidnightBlue, citecolor=MidnightBlue, 
    pdftitle={\myTitle},%
    pdfauthor={\textcopyright\ \myName, \myUni, \myFaculty},%
    pdfsubject={},%
    pdfkeywords={},%
    pdfcreator={pdfLaTeX},%
    pdfproducer={LaTeX with hyperref and classicthesis}%
}   

\makeatletter
\@ifpackageloaded{babel}%
    {%
       \addto\extrasamerican{%
				}%
       \addto\extrasngerman{%
				}%
    }{\relax}
\makeatother

\usepackage[]{geometry}

\usepackage{classicthesis} 

\usepackage{pdfpages}

\usepackage{enumitem,pifont}

\setlist[itemize]{itemsep=0pt,topsep=6pt}
\setlist[itemize,1]{label={\color{MidnightBlue}\Pifont{pzd}{\char70}}} %

\usepackage{longtable,tabulary}

\usepackage{textcomp}
\usepackage{wasysym}
\usepackage{units}

\usepackage[osf]{libertine}

\usepackage{float}
\floatstyle{plaintop}
\restylefloat{table}

\makeatletter

\def\ltabulary{%
\def\endfirsthead{\\}%
\def\endhead{\\}%
\def\endfoot{\\}%
\def\endlastfoot{\\}%
\def\tabulary{%
  \def\TY@final{%
\def\endfirsthead{\LT@end@hd@ft\LT@firsthead}%
\def\endhead{\LT@end@hd@ft\LT@head}%
\def\endfoot{\LT@end@hd@ft\LT@foot}%
\def\endlastfoot{\LT@end@hd@ft\LT@lastfoot}%
\longtable}%
  \let\endTY@final\endlongtable
  \TY@tabular}%
\dimen@\columnwidth
\advance\dimen@-\LTleft
\advance\dimen@-\LTright
\tabulary\dimen@}

\makeatother

\usepackage{amssymb}
\usepackage[eulergreek,EULERGREEK]{sansmath}

\DeclareMathVersion{sanssym}
\SetSymbolFont{operators}{sanssym}{OT1}{iwona}{m}{n}
\SetSymbolFont{letters}{sanssym}{OML}{iwona}{m}{it}
\SetSymbolFont{symbols}{sanssym}{OMS}{iwona}{m}{n}
\SetMathAlphabet{\mathit}{sanssym}{OT1}{iwona}{m}{sl}
\SetMathAlphabet{\mathbf}{sanssym}{OT1}{iwona}{bx}{n}
\SetMathAlphabet{\mathtt}{sanssym}{OT1}{iwona}{m}{n}
\SetSymbolFont{largesymbols}{sanssym}{OMX}{iwona}{m}{n}

\DeclareCaptionFont{sansmath}{\mathversion{sanssym}\color{MidnightBlue}}

\usepackage[figuresright]{rotating}

\makeatletter
\renewcommand{\p@subfigure}{}
\makeatother

\usepackage{pdflscape}

\usepackage{verbatimbox}

\usepackage{textcomp}
\usepackage{upgreek}

\definecolor{color3}{RGB}{60,60,60}
\definecolor{color1}{RGB}{0,60,80} 
\definecolor{color2}{RGB}{70,130,140} 

\usepackage[nostamp]{draftwatermark}

\SetWatermarkColor{MidnightBlue!25}
\SetWatermarkFontSize{.8cm}
 \SetWatermarkScale{1}
 \SetWatermarkAngle{0}
 \SetWatermarkText{%
\parbox{15cm}{
     \begin{center}
\vspace{25cm}
\sffamily DRAFT\ \ -\ \ NOT FOR DISTRIBUTION\\
\vspace{-1.5cm}
	\end{center}
}
  }

  \usepackage{tikz}

  \makeatletter
  \newcommand*{\cleartoleftpage}{%
  	\clearpage
  	\if@twoside
  	\ifodd\c@page
  	\hbox{}\newpage
  	\if@twocolumn
  	\hbox{}\newpage
  	\fi
  	\fi
  	\fi
  }
  \makeatother

%% file: FrontBackmatter/DirtyTitlepage.tex

\begin{titlepage}
\setlength{\marginparwidth}{0pt}%
\setlength{\marginparsep}{0pt}%
\setlength{\marginparpush}{0pt}%
\setlength{\oddsidemargin}{26pt}
\begin{center}

    \begingroup
        \color{MidnightBlue}\spacedallcaps{\myTitle}
    \endgroup\medskip\\
\spacedlowsmallcaps{\myName} \\ \medskip 

\vfill
  
\begin{figure}[h]

\includegraphics[width=\textwidth]{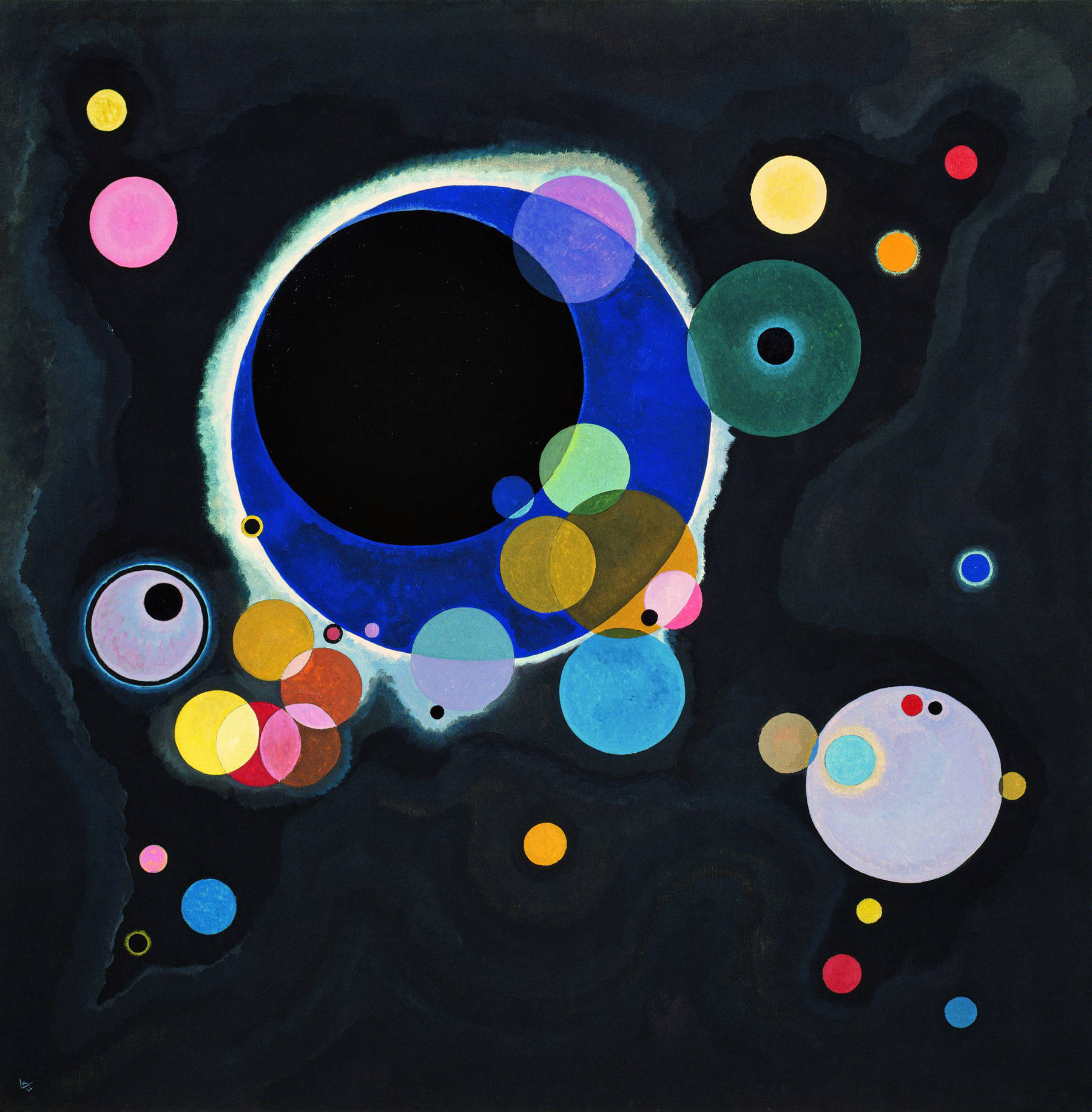} \\ \medskip

\end{figure}
\vfill
\end{center} 
\end{titlepage}

\thispagestyle{empty}
\hfill
\vfill
\noindent Printed in October 2015\\
Cover picture:  {\textit Several Circles}, Wassily Kandinsky (1926)
\setlength{\marginparpush}{6pt}%
\setlength{\oddsidemargin}{3pt}
\setlength{\marginparwidth}{85pt}%
\setlength{\marginparsep}{10pt}%

%% file: FrontBackmatter/Titlepage.tex
\begin{titlepage}
 \begin{changemargin}{-0cm}{-1.75cm}   
     \large
     
    \begin{center}
     
     Dissertation\hspace{0.4cm} \medskip
     
     submitted to the \medskip
     
     Combined Faculties of Natural Sciences and Mathematics \medskip
     
     of the Ruperto-Carola-University of Heidelberg, Germany \medskip
     
     for the degree of \medskip
     
     Doctor of Natural Sciences
     
     \vfill
     
     Put forward by\medskip
     
     Jan Rybizki \medskip
     
     born in: R\"udersdorf \medskip
     
     Oral examination: December 8th, 2015
         
    \end{center}
  \end{changemargin}    
\end{titlepage}  

%% file: FrontBackmatter/Titleback.tex
\thispagestyle{empty}

\begin{changemargin}{-1.75cm}{-0cm}
  
\begin{center}
\large
    \begingroup
         \color{MidnightBlue}\spacedallcaps{\myTitle}
    \endgroup

\vfill

\large
Referees: \hspace{8.9em}		Prof. Dr. Andreas Just \newline
\hphantom{Referees: }\hspace{6em}	Prof. Dr. Norbert Christlieb
\end{center}
\normalsize

\end{changemargin} 

%% file: FrontBackmatter/Abstract.tex

\begingroup
\let\clearpage\relax
\let\cleardoublepage\relax
\let\cleardoublepage\relax

\pdfbookmark[1]{Zusammenfassung}{Zusammenfassung}
\chapter*{Zusammenfassung - R\"uckschl\"usse aus der Modellierung der chemodynamischen Entwicklung der Milchstra\ss enscheibe}
In der vorliegenden Arbeit werden die anf\"angliche Massenfunktion (IMF) von Feldsternen und Parameter zur chemischen Anreicherung der Milchstra\ss e, unter Verwendung von Bayesscher Statistik und Modellrechnungen, hergeleitet.\\
Ausgehend von einem lokalen Milchstra\ss enmodell \citep{JJ} werden, f\"ur verschiedene IMF-Parameter, Sterne synthetisiert, die dann mit den entsprechenden \ac{hip} Beobachtungen verglichen werden. Die abgelei-tete IMF ist in dem Bereich von 0.5 bis 8 Sonnenmassen gegeben durch ein Potenzgesetz mit einer Steigung von -1.49\,$\pm$\,0.08 f\"ur Sterne mit einer geringeren Masse als 1.39\,$\pm$\,0.05\,M$_\odot$ und einer Steigung von -3.02\,$\pm$\,0.06 f\"ur Sterne mit Massen dar\"uber.\\
Im zweiten Teil dieser Arbeit wird die IMF f\"ur Sterne mit Massen schwerer als 6\,M$_\odot$ be-stimmt. Dazu wurde die Software \emph{Chempy} entwickelt, mit der man die chemische Anrei-cherung der Galaktischen Scheibe simulieren und auch die Auswahl von beobachteten Sternen nach Ort und Sternenklasse reproduzieren kann. Unter Ber\"ucksichtigung des systematischen Effekts der unterschiedlichen in der Literatur verf\"ugbaren stellaren Anreicherungs-tabellen ergibt sich eine IMF-Steigung von -2.28\,$\pm$\,0.09 f\"ur Sterne mit Massen \"uber 6\,M$_\odot$.\\
Dies zeigt, dass die von der chemischen Entwicklung abgeleitete IMF, \"ahnlich wie die Ergebnissen aus hydrodynamischen Simulationen der Milchstra\ss e, f\"ur hohe Massen der \citet{Sa55} IMF entspricht. Aufgrund der geringen Anzahldichte von massereichen Sternen kann dies durch Sternz\"ahlungen nur schwer nachgewiesen werden.

\vfill
\acresetall
\pdfbookmark[1]{Abstract}{Abstract}
\chapter*{Abstract - \myTitle}
In this thesis, the field star \ac{imf} and chemical evolution parameters for the \ac{mw} are derived using a forward modelling technique in combination with Bayesian statistics.\\
Starting from a local \ac{mw} disc model \citep{JJ}, observations of stellar samples in the Solar Neighbourhood are synthesised and compared to the corresponding volume-complete observational samples of \ac{hip} stars. The resulting \ac{imf}, derived from observations in the range from 0.5 to 8\,M$_\odot$, is a two-slope broken power law with powers of -1.49\,$\pm$\,0.08 and -3.02\,$\pm$\,0.06 for the low-mass slope and the high-mass slope, respectively, with a break at 1.39\,$\pm$\,0.05\,M$_\odot$.\\
In order to constrain the \ac{imf} for stars more massive than 8\,M$_\odot$, a fast and flexible chemical enrichment code, \emph{Chempy}, was developed, which is also able to reproduce spatial and stellar population selections of observational samples. The inferred high-mass slope for stellar masses above 6\,M$_\odot$ is -2.28\,$\pm$\,0.09, accounting for the systematic effects of different yield sets from the literature.\\
This shows that constraints from chemical modelling, similarly to hydrodynamical simulations of the Galaxy, demand a \citet{Sa55} high-mass index. This is hard to recover from star count analysis given the rareness of high-mass stars.

\acresetall
\endgroup			

\vfill

%% file: FrontBackmatter/Overview.tex
\pdfbookmark[1]{Overview}{overview}
\chapter*{Overview and Publications}
In this thesis, the mass distribution of stars and the evolution of elements in the \ac{mw} are investigated. For this, physically motivated models are constructed, their input parameters are changed, and the outcome is compared to observational data of discriminative power. This forward modelling approach is combined with the concept of Bayesian inference. Central to this technique is the construction of a likelihood function in order to sample the posterior probability distribution, which is performed by means of \ac{mcmc} methods.\\
The thesis is organised in the following way. The first chapter is an introduction to the evolution of the \ac{mw}, with a focus on the processes governing the synthesis of the chemical elements.\\
The statistical methodology will be introduced in chapter \ref{ch:methods}. The first example of Bayesian model inference, where the true distance of a star will be inferred from a parallax measurement, is taken in great detail from \citet{Coryn2015}. This leads to the question of determining stellar number densities in a specific volume, which will be discussed in the second part of chapter\,\ref{ch:methods}. The ideas presented in section\,\ref{sec:prob_distances} have been published in \citet{Just2015} with a contribution from the author.\\
In chapter \ref{ch:imf}, the \ac{imf} parameters are inferred from Solar Neighbourhood stars, using a local \ac{mw} disk model. Most of this chapter has been published in \citet{Rybizki2015}.\\
In chapter\,\ref{ch:chemistry}, a chemical evolution model of the local \ac{mw} disc is presented and important model parameters, constraining the high-mass \ac{imf}, the infall and the \ac{sn1a}, are inferred from abundances of Solar Neighbourhood stars. Section\,\ref{sec:mapping} of this chapter contains results, which are submitted as conference proceedings \citep{Just}.\\
The thesis will be concluded with a summary and a concise outlook.
\acresetall

%% file: FrontBackmatter/Acknowledgments.tex
\pdfbookmark[1]{Acknowledgements}{acknowledgements}


\bigskip

\begingroup
\let\clearpage\relax
\let\cleardoublepage\relax
\let\cleardoublepage\relax
\chapter*{Acknowledgements}
First and foremost I would like to thank Prof. Andreas Just for his confidence in my work and his guidance. The generous financial support and the freedom to ponder ideas are very much appreciated as well as his positive supervision style.\\
I am much obliged to PD Coryn Bailer-Jones and Dr. Ewan Cameron for their help with statistical questions, which I came across frequently, and for nurturing my fascination of Bayesian statistics.\\
For most enjoyable lunch breaks and for partaking in my science and non-science related challenges I wish to express my deep gratitude to Dr. Robert Schmidt. His presence made the Astronomisches Rechen-Institut (ARI) feel more like a home for me.\\
I am very grateful to Prof. Norbert Christlieb for readily agreeing to referee this thesis. Furthermore I wish to thank Prof. Eva Grebel and Prof. Luca Amendola for willingly joining my examination committee.\\
I also want to mention the highly valued advice and helpful strategic planning from my thesis committee consisting of Prof. Eva Grebel, Dr. Glenn van de Ven and Prof. Andreas Just.\\
Special thanks go to PD Christian Fendt who coordinated the International Max Planck Research School (IMPRS) activities that formed a sociable community out of our IMPRS generation.\\
At ARI, I would like to thank Hendrik Heinl for caring about the social activities and for his support especially towards the end of my thesis writing.\\
For numerous entertaining lunch breaks I want to thank Dr. Markus Demleitner and Aksel Alpay who also helped me to master the computationally expensive MCMC simulations.\\
In the same respect Dr. Peter Schwekendiek and Sven Weimann always provided me with helpful computing advice and maintained the excellent {\textit Perseus} machine.\\
I want to thank the whole ARI administration, especially mentioning Martina Buchhaupt for thoroughly taking care of my administrative issues and Stefan Leitner for mocking me for my poor hand crafting skills.\\
I enjoyed the scientific and non-scientific exchange with Dr. Corrado Boeche with  whom I could lament about the intriguing complexity of selection functions.\\
Last but not least I would like to thank Alex Hygate, Carolin Wittmann, Clio Bertelli, PD Coryn Bailer-Jones, Dr. Corrado Boeche, Daniel Haydon, Dr. Frederik Sch\"onebeck, Dr. Gernot Burkhardt, Hendrik Heinl, Peter Zeidler, Reza Moetazedian, Dr. Robert Schmidt, Timo Hirscher and Tina Gier for proof reading this thesis and for making the ARI a sociable workspace.\\

\noindent My PhD project was funded by SFB 881 {\textit The Milky Way System}.\\
I would also like to acknowledge financial support from the Heidelberg Graduate School of Fundamental Physics and the IMPRS which allowed me to participate in several interesting summer schools and useful workshops giving me the opportunity to learn from the best in the field and be an active part of the international community.
\endgroup

%% file: FrontBackmatter/Contents.tex
\refstepcounter{dummy}
\pdfbookmark[1]{\contentsname}{tableofcontents}
\setcounter{tocdepth}{2} 
\setcounter{secnumdepth}{3} 
\manualmark
\markboth{\spacedlowsmallcaps{\contentsname}}{\spacedlowsmallcaps{\contentsname}}
\tableofcontents 
\automark[section]{chapter}
\renewcommand{\chaptermark}[1]{\markboth{\spacedlowsmallcaps{#1}}{\spacedlowsmallcaps{#1}}}
\renewcommand{\sectionmark}[1]{\markright{\thesection\enspace\spacedlowsmallcaps{#1}}}
\clearpage

\begingroup 
    \let\clearpage\relax
    \let\cleardoublepage\relax
    \let\cleardoublepage\relax
    \refstepcounter{dummy}
    \pdfbookmark[1]{\listfigurename}{lof}
    \listoffigures

	\newpage
    \refstepcounter{dummy}
    \pdfbookmark[1]{\listtablename}{lot}
    \listoftables
        
    \vspace*{8ex}
    

       
    \refstepcounter{dummy}
    \pdfbookmark[1]{Acronyms}{acronyms}
    \markboth{\spacedlowsmallcaps{Acronyms}}{\spacedlowsmallcaps{Acronyms}}
    \chapter*{Acronyms}
    \begin{acronym}[UML]
  		\acro{2mass}[$2$MASS]{\acroextra{The }Two Micron All Sky Survey\acroextra{ at IPAC \citep{Skrutskie2006}}}
  		\acro{abc}[ABC]{Approximate Bayesian Computation}
  		\acro{agb}[AGB]{Asymptotic Giant Branch\acroextra{ star (late evolutionary state of stars with masses $0.8\,\mathrm{M}_\odot<\mathrm{m}<8\,\mathrm{M}_\odot$)}} 		
  		\acro{agn}[AGN]{Active Galactic Nuclei\acroextra{ (a super massive black hole sitting in the center of a galaxy and accreting gas)}}
  		\acro{amr}[AMR]{Age Metallicity Relation}
 		\acro{apogee}[APOGEE]{APO Galactic Evolution Experiment \citep{Alam2015}} 
 		\acro{avr}[AVR]{Age Velocity-dispersion Relation}
  		\acro{bao}[BAO]{Baryonic Acoustic Oscillations}
  		\acro{bb}[BB]{Big Bang}
  		\acro{bd}[BD]{Brown Dwarf}		
  		\acro{besancon}[Besan\c{c}on model]{Besan\c{c}on Galaxy model\acroextra{ (first version from \citet{Robin2003} updated in \citet{Ro12} latest update in \citet{Cz14} and \citet{Robin2014})}}
  		\acro{besb}[Besan\c{c}on\,B]{default model B of \citet{Cz14}}
  		\acro{bf}[BF]{Bayes Factor\acroextra{ (used for Bayesian model comparison)}}
  		\acro{bh}[BH]{Black Hole}
  		\acro{cdm}[CDM]{Cold Dark Matter}
  		\acro{cem}[CEM]{Chemical Enrichment Model}  
  		\acro{chabrier}[Chabrier\,$03$]{ \citet{Ch03}}
 		\acro{cmb}[CMB]{Cosmic Microwave Background}  		
  		\acro{cmd}[CMD]{Colour-Magnitude Diagram}
  		\acro{cno}[CNO cycle]{Carbon-Nitrogen-Oxygen cycle\acroextra{ (main fusion reaction to form helium from hydrogen for stars above $1.3$\,M$_\odot$ and with an initial abundance of at least carbon)}}
  		\acro{cns}[CNS]{Catalogue of Nearby Stars\acroextra{ - version 5 (used here but not yet published) is an update of version 4 \citep{Ja97}}}
  		\acro{co}[CO]{Carbon-Oxygen\acroextra{ \acl{wd} (remnant of \acl{agb} stars)}}
  		\acro{corot}[COROT]{COnvection ROtation and planetary Transits \citep{Baglin1998}\acroextra{ (asteroseismic mission)}}
  		\acro{deltam}[$\Delta$mag]{$\Delta$magnitude \acroextra{(magnitude difference of stars in a binary system)}}
  		\acro{dm}[DM]{Dark Matter\acroextra{ (which comes in different flavours but we assume it to be 'cold')}}
  		\acro{dtd}[DTD]{Delay Time Distribution\acroextra{ (the time it takes a statistical ensemble of different \ac{sn1a} progenitors to go supernova)}}
  		\acro{fh06}[FH\,$06$]{\citet{Fl06}\acroextra{ disc mass model}} 
 		\acro{gaia}[Gaia]{Gaia \citep{Lindegren2008}\acroextra{ (astrometric satellite mission)}}
 		\acro{ges}[GES]{Gaia-ESO \citep{Gilmore2012}\acroextra{ (spectroscopic follow-up of Gaia)}}
		\acro{galah}[GALAH]{GALactic Archeology with HERMES \citep{Anguiano2014}\acroextra{ (spectroscopic survey)}}
 		\acro{galaxia}[Galaxia]{Galaxia \citep{Sh11}\acroextra{ (a tool to synthesise observations)}}  
  		\acro{gcs}[GCS]{Geneva Copenhagen Survey \citep{Nordstroem2004}}
		\acro{hb}[HB]{Horizontal Branch}
		\acro{hip}[Hipparcos]{Hipparcos \citep{Pe97}\acroextra{ satellite, also the derived astrometric catalogue. Revised parallaxes from \citet{le07}}}
		\acro{hn}[HN]{HyperNova \acroextra{ (Feedback process of stars with mass > 20\,M$_\odot$)}}
		\acro{hrd}[HRD]{Hertzsprung-Russel Diagram\acroextra{ (a plot of effective temperature vs. the luminosity of a star, which is the theoretical counterpart of the \acs{cmd})}}
  		\acro{ir}[IR]{infrared \acroextra{(electromagnetic waves with $1000\mu\text{m}>\lambda>0.7\mu\text{m}$)}}
        \acro{imf}[IMF]{\acroextra{Stellar }Initial Mass Function}
		\acro{ira}[IRA]{Instantaneous Recycling Approximation}
		\acro{ism}[ISM]{InterStellar Medium}
		\acro{jj}[JJ-model]{MW disc model \citep{JJ}}
		\acro{kepler}[Kepler]{Kepler \citep{Borucki1997}\acroextra{ (asteroseismic mission)}}
		\acro{ktg}[KTG\,$93$]{\citet{Kr93}}
		\acro{lcdm}[$\Lambda$CDM]{Lambda Cold Dark Matter\acroextra{ (cosmological standard model)}}
		\acro{lf}[LF]{Luminosity Function}
        \acro{mcmc}[MCMC]{Markov Chain Monte Carlo \acroextra{ (an iterative sampling technique)}}
  		\acro{ms}[MS]{Main Sequence}
        \acro{mw}[MW]{Milky Way}
		\acro{ngp}[NGP]{North Galactic Pole}
        \acro{nir}[NIR]{Near-InfraRed\acroextra{ (photons with a wavelength of $0.8\,\mu \mathrm{m}<\lambda<2.5\,\mu\mathrm{m}$)}}
        \acro{ns}[NS]{Neutron Star\acroextra{ (a type of stellar remnant)}}
        \acro{pp}[pp]{Proton-Proton\acroextra{ chain (main fusion reaction to form helium from hydrogen for stars below $1.3$\,M$_\odot$)}}
        \acro{ps1}[PS\,$1$]{Pan-STARRS\,$1$\acroextra{ (the Panoramic Survey Telescope \& Rapid Response System \citep{Hodapp2004})}}
        \acro{pdf}[PDF]{Probability Density Function}
        \acro{pdmf}[PDMF]{Present-Day stellar Mass Function}  		
        \acro{pn}[PN]{Planetary Nebula\acroextra{ (the end stage of an \ac{agb} star)}}
  		\acro{paper1}[paper\,I]{\citet{JJ}}
  		\acro{paper2}[paper\,II]{\citet{Ju11}}		
  		\acro{quasar}[quasar]{Quasi-stellar object\acroextra{ (the radiation of an active galactic nuclei from an early galaxy looking like stellar point sources with peculiar spectra)}}		
  		\acro{rave}[RAVE]{The Radial Velocity Experiment \citep{Steinmetz2006}\acroextra{ (spectroscopic survey)}}
  		\acro{rc}[RC]{Red-Clump\acroextra{ (overdensity in the giant branch of the \acs{cmd})}}
  		\acro{rgb}[RGB]{Red Giant Branch}
  		\acro{rprocess}[r-process]{Rapid-neutron-capture-process\acroextra{ element (produced in massive stars and in neutron star merger)}}
  		
  		\acro{sdss}[SDSS]{Sloan Digital Sky Survey \citep{York2000}}				
  		\acro{segue}[SEGUE]{Sloan Extension for Galactic Understanding and Exploration \citep{Yanny2009}\acroextra{ (low-resolution spectroscopic survey of $\approx240,000$\,stars)}}				
  		  		
  		\acro{sfr}[SFR]{Star Formation Rate}
  		\acro{sfh}[SFH]{Star Formation History}
		\acro{sgp}[SGP]{South Galactic Pole}
  		\acro{sprocess}[s-process]{Slow-neutron-capture-process\acroextra{ element (produced in \acl{agb} and also in massive stars)}}
 		\acro{sn1a}[SN\,Ia]{SuperNova of type\,Ia}
 		\acro{sn2}[SN\,II]{SuperNova of type\,II}
 		\acro{smbh}[SMBH]{Super Massive Black Hole\acroextra{ (sitting in the center of massive galaxies and sometimes being the engine of an \ac{agn})}}
 		\acro{ssp}[SSP]{Simple Stellar Population}
  		\acro{tycho2}[Tycho\,2]{Tycho\,2 catalogue \citep{Hog2000}}
  		\acro{vmag}[VMag]{Absolute V\,magnitude}
  		\acro{avmag}[Vmag]{Apparent V\,magnitude}
  		\acro{wd}[WD]{White Dwarf}
  		\acro{zams}[ZAMS]{Zero-Age Main Sequence\acroextra{ (star begins to burn hydrogen)}}   	    	
    \end{acronym}                     
\endgroup

\cleardoublepage

%% file: Chapters/Chapter01.tex
\ctparttext{

\begin{center}
    \emph{"Wenn man in dem unerme{\ss}lichen Raume, darinn alle Sonnen der Milchstrasse sich gebildet haben, einen Punkt annimmt, um welchen durch, ich weiss nicht was vor eine Ursache, die erste Bildung der Natur aus dem Chaos angefangen hat; so wird daselbst die gr\"o{\ss}te Masse, und ein K\"orper von der ungemeinsten Attraction, entstanden sein, der dadurch f\"ahig geworden, in einer ungeheuren Sph\"are um sich alle in der Bildung begriffene Systeme zu n\"othigen, sich gegen ihn, als ihren Mittelpunkt, zu senken, und um ihn ein gleiches System im Ganzen zu errichten, als derselbe elementarische Grundstoff, der die Planeten bildete, um die Sonne im Kleinen gemacht hat."} \\ \medskip
   	Immanuel Kant (1755)\\ Universal Natural History and Theory of Heaven (p. 102)
\end{center}
\bigskip
\bigskip
\begin{figure}[h]
\centering
\includegraphics[width=.6\textwidth]{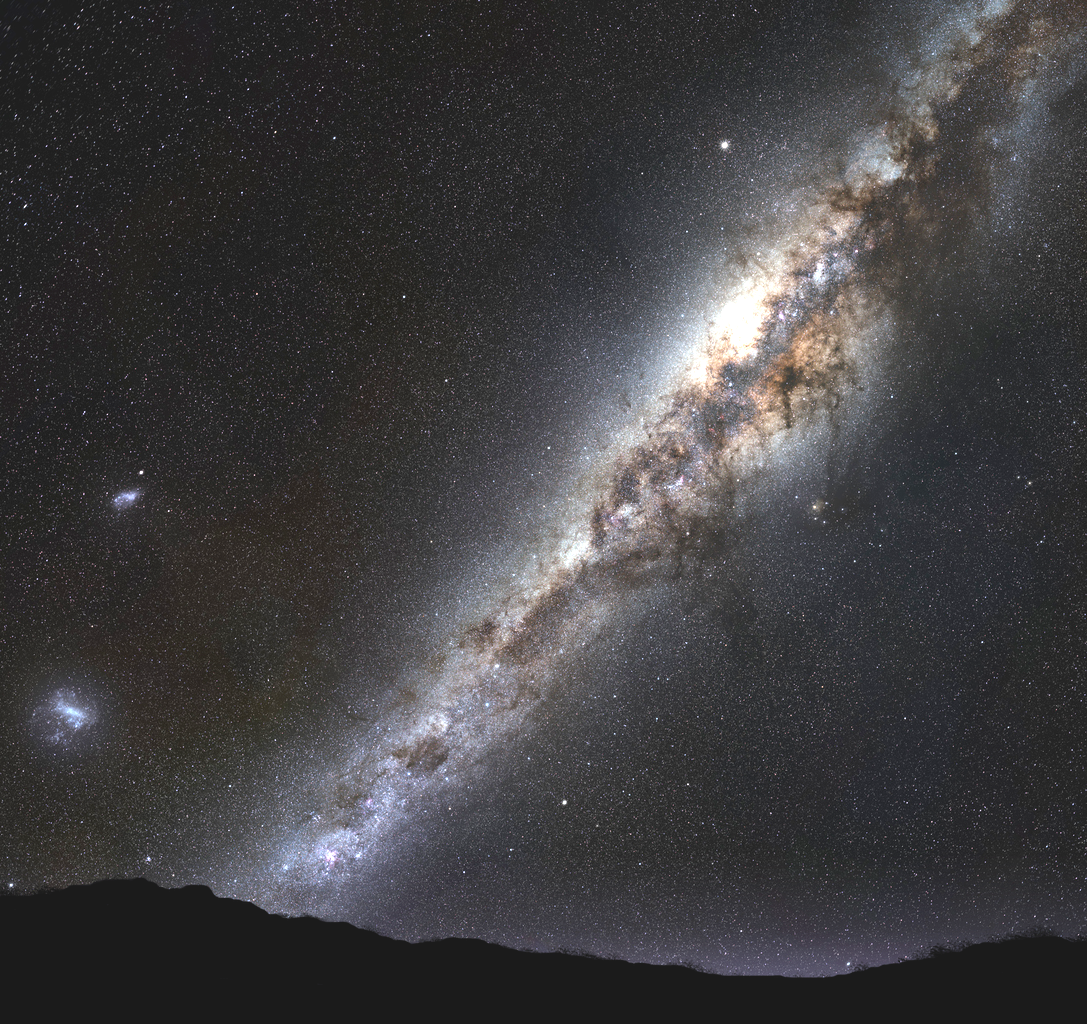}
\caption[Milky Way from Earth]{Milky Way from Earth, credit:\,ESO/S. Brunier}
\label{fig:nightsky}
\end{figure}
}

\part{Introduction}

\chapter{The Milky Way Galaxy}\label{ch:introduction}
\acresetall

\section{Etymology}
\begin{figure}[h]
  \centering
  \includegraphics[width=.45\linewidth]{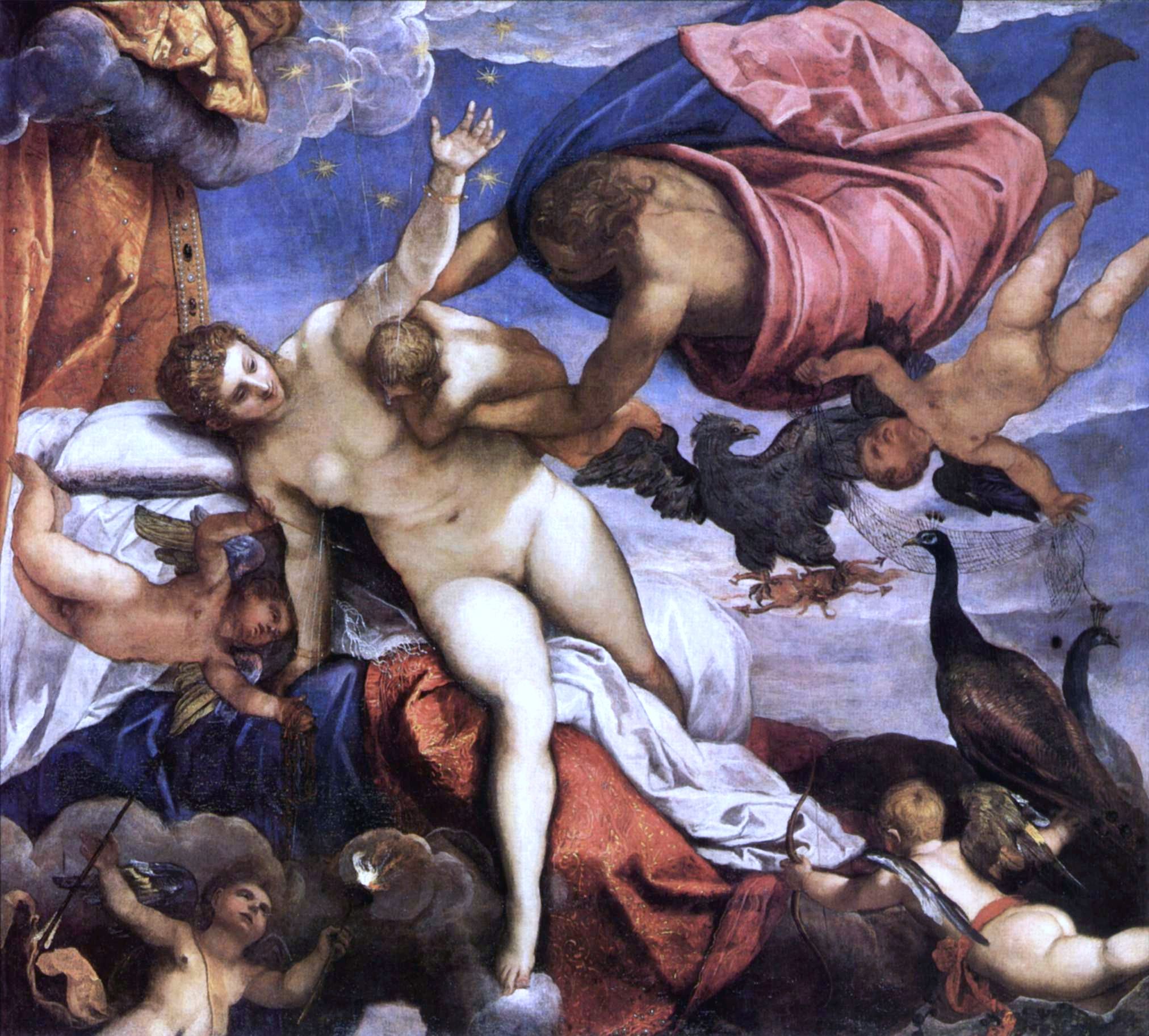}
  \captionof{figure}[The origin of the Milky Way]{\emph{The Origin of the Milky Way}, Jacopo Tintoretto (ca. $1575$ - $1580$)}
  \label{fig:origin}

\end{figure}

The ancient Greeks referred to the dim band visible across the sky on a clear night, depicted in figure\,\ref{fig:nightsky}, as \emph{galaxias kuklos} (\textgreek{γαλαξίας κύκλος}), meaning \emph{milky circle}. Today, the word \emph{Galaxy} is still a synonym for \emph{Milky Way}, which is a translation from the Latin \emph{via lactea}, itself coming from the Greek \emph{galaxias}. The Greek mythology explains the glowing appearance of the Milky Way with spilled milk from goddess Hera's breast, who pushed back Heracles, not being her own son, but from her husband Zeus and the mortal woman Alkmene. Zeus had put the baby to sleeping Hera's breast, in order for his child to gain divine powers as illustrated in Jacopo Tintoretto's oil painting on canvas, shown in figure\,\ref{fig:origin}.

\section{History}
Greek philosophers like Democritus already speculated about the nature of the \ac{mw}, proposing that it is composed of a huge number of distant stars. Actual proof had to wait until the invention of the Dutch telescope (from Greek \textgreek{τηλεσκόπος}, teleskopos "far-seeing"), and Galileo Galilei improving its magnifying power one year later in $1609$. Observing the \ac{mw}, he realised that it was composed of a huge number of individual stars, which was published in the pamphlet \emph{Sidereus Nuncius} in $1610$.\\
Immanuel Kant, more renown for his work as philosopher, speaks of the \ac{mw} as a gravitationally bound rotating system, in his treatise {\textit Universal Natural History and Theory of Heaven} from $1775$, building up on the work by Thomas Wright. From the regular elliptical shape of some 'nebulous' stars, Kant also derived that they should be distant \emph{galaxies}\marginpar{Galaxy is written capitalised if referring to the \ac{mw} and lowercase, when referring to external galaxies}
 on their own. Evidence for this correct assumption accumulated in the early $20$th century. 
 The period-luminosity relation for Cepheid variable stars was found by \citet{Leavitt1908} and was calibrated by \citet{Shapley1918}, using parallactic motions\footnote{Harley Shapley used the period-luminosity relation to measure the distribution of globular clusters within the \ac{mw} and rightly located the Solar system in the outer regions of our Galaxy. During the \emph{Great Debate} in 1920 he was an opponent of the idea of nebulae being galaxies like the \ac{mw}.}. This relation was used by \citet{Hubble1925}, who could determine the period of cepheids in the Andromeda nebula with the $2.5$\,meter Hooker telescope at the Mount Wilson Observatory, thus placing them well outside the sphere of the \ac{mw}. After presenting these findings, the idea of the \ac{mw} being a galaxy among others was generally accepted, although already \citet{Slipher1917} had inferred from the radial velocity measurements of nebulae that they should have extragalactic origin, and also \citet{Curtis1917} had come to the same conclusion from apparent magnitudes of novae associated with nebulae\marginpar{The term \emph{nebulae} is no longer referring to \emph{galaxies} but to interstellar gas or dust instead}.\\
 
\begin{figure}
  
  \centering
  \includegraphics[width=.6\linewidth]{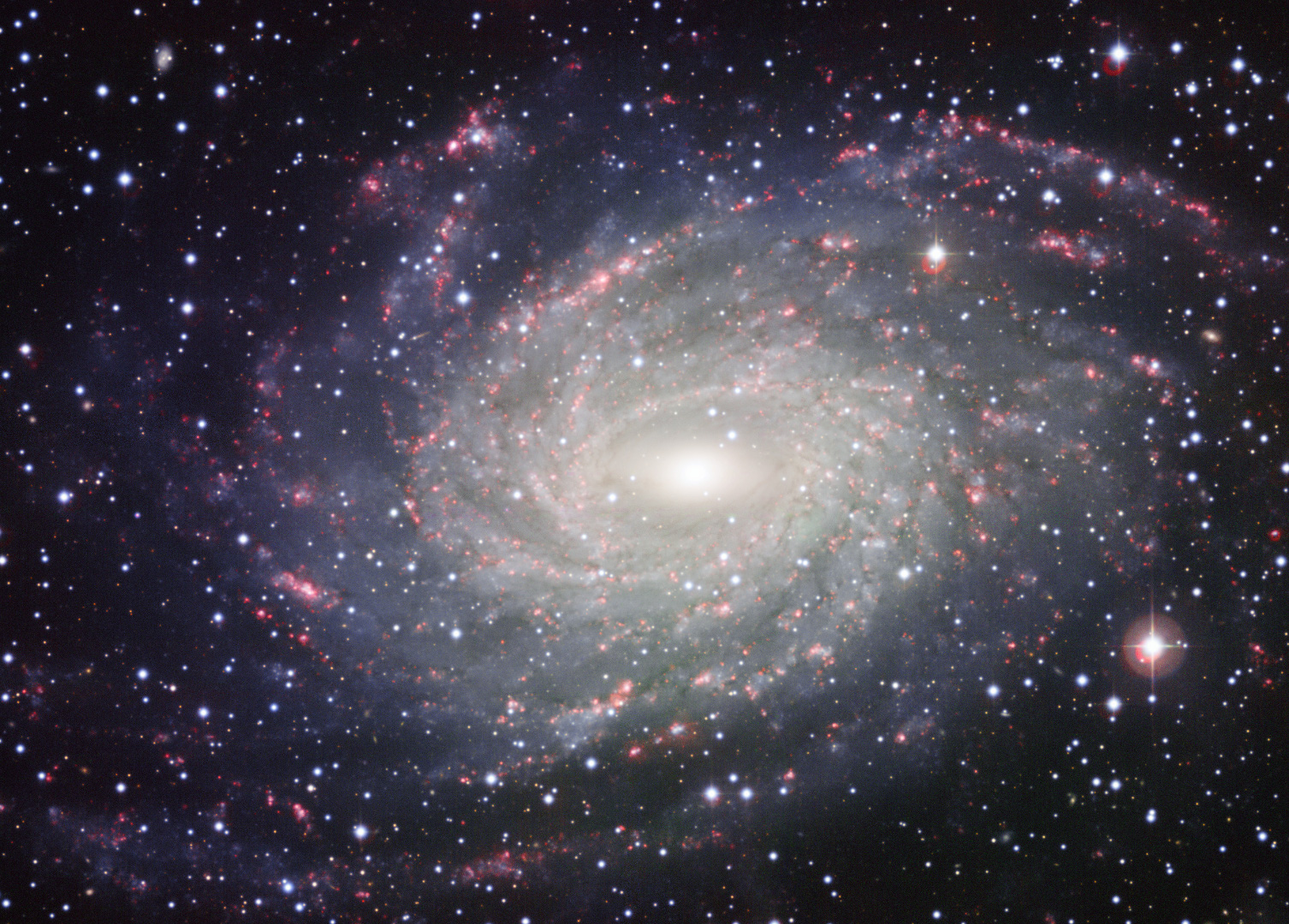}
  \captionof{figure}[Milky Way-like galaxy]{Milky Way-like galaxy NGC\,6744. Single stars on this image are not part of NGC\,6744, but foreground stars from our own Galaxy. Credit:\,ESO}
  \label{fig:spiral}

\end{figure}

 Nowadays, the \ac{mw} is believed to be a barred spiral galaxy, similar to the one depicted in figure\,\ref{fig:spiral}. Due to the fact that we can only observe the current evolutionary state of our Galaxy from its inside, the formation history and actual structure are not well constraint and still a matter of ongoing research, which this thesis would like to contribute to. \\
 In order to give an overview over the present-day knowledge of the \ac{mw}, we need to place it within the evolution of the Universe, before painting a more detailed picture of its substructure.\\
\section{Cosmology}
The basis for the development of models describing the evolution of the Universe was set by \citet{Einstein1915}, when he published his field equations and the theory of general relativity, describing gravitational processes comprehensively, e.g. by the idea of spacetime being curved by matter and energy. 
 In the following years \citet{Friedman1922,Lemaitre1927,Robertson1929,Walker1935} independently studied a set of homogeneous and isotropic models for the evolution of the Universe, which would expand (or contract), when applied to Einstein's field equations. \citet{Lemaitre1927} was the first to conclude that the increasing recessional velocity of observed galaxies with distance is due to the expansion of the Universe \citep{Livio2011}. With more data at hand, \citet{Hubble1929} found a linear relationship between the redshift\footnote{The light, being send out by an object moving away from us, is red-shifted due to the Doppler shift. This can be superimposed on the redshift caused by the expansion of the Universe. These two causes of redshift can be observationally distinguished \citep{Davis2004}.} of an object and its distance. Consequently, if going back in time the whole Universe must have started from a hot and dense state \citep{Lemaitre1931}, which is why these universe models were later termed \emph{Big Bang} cosmology.\\
 This theory was widely accepted when the predicted signature of radiation decoupling from matter\footnote{When the expanding Universe cools and its electrons and protons become bound in neutral hydrogen, radiation is not constantly scattered by electrons and can travel freely.} was serendipitously found by \citet{Penzias1965} and identified by \citet{Dicke1965} as the \ac{cmb}, with an almost perfect black body radiation. The anisotropies in the \ac{cmb} were indeed so small, that pure-\emph{baryonic}\marginpar{baryons are interacting with all fundamental forces, i.e. gravitationally, strongly, weakly and especially electromagnetically and are therefore visible, in contrast to dark matter, which only interacts gravitationally and maybe weakly} universe models could hardly explain the formation of galaxies \citep{Peebles1970,Sunyaev1970}. Already earlier \citet{Zwicky1933} had found, when studying the Coma galaxy cluster, that a large amount of dark matter was necessary to explain the observed internal velocity dispersion when applying the virial theorem. It was not until \citet{Rubin1970} rigorously analysed the galactic rotation curve of the Andromeda galaxy that dark matter became a topic of active research. \citet{Peebles1982} could reconcile the low \ac{cmb} anisotropies with the \ac{bb} cosmology assuming that a large mass fraction of the Universe is composed by dark matter. Dark matter was further constraint to be \emph{cold} by \citet{Davis1985}, whose simulations showed that the cosmological structure formation could only be explained if the velocity distribution of dark matter particles was already non-relativistic in the early Universe. The discovery of the predicted \ac{cmb} anisotropy with the COBE satellite \citep{Smoot1992} lent further strong evidence to dark matter.\\
 
 The expansion rate of \ac{bb} models is governed by the matter and radiation density, the curvature of space and the cosmological constant $\Lambda$, which was originally introduced by \citet{Einstein1917} to obtain a static Universe from his field equations. $\Lambda$ can be interpreted as the \emph{energy density of vacuum space}\marginpar{vacuum energy density is also called \emph{dark energy}} that exerts a negative pressure, thus driving space apart. The \emph{curvature} of the Universe seems to be \emph{zero}\footnote{A universe with zero curvature is called flat, meaning that space is Euclidean.}, so that only the matter and radiation density are counteracting the dark energy. Since space, and therefore the total vacuum energy is growing, but the matter content of the Universe stays the same, the expansion rate should be accelerated, which was confirmed by \ac{sn1a} observations of \citet{Riess1998} and \citet{Perlmutter1999} and was awarded with the Nobel Prize in 2011.\\
 
  Although on scales larger than $\sim100$\,M\emph{pc}\marginpar{$1$\,pc is approximately $3.26$\,light-years or $3,1\times10^{16}$\,m} the matter distribution in the Universe seems to be homogeneous and isotropic, on the scales of Galaxy clusters and smaller, considerable structure can be found, which must have originated from primordial density fluctuations, growing non-linearly through gravitation. 
  To set the initial conditions for this phenomenon and, more importantly, to account for the \ac{cmb} being nearly constant over the full sky and also the flatness of space, an elegant explanation was proposed and developed by \citet{Guth1981,Linde1982,Linde1983,Albrecht1982} and is referred to as \emph{inflation}. It was a short period, right after the \ac{bb}, in which the inflaton field dominated and inflated the microscopic Universe together with its quantum fluctuations, to macroscopic scales.\\
  To date, the described \ac{lcdm} model with inflation is the simplest cosmological model able to explain the bulk of observational evidence, among which the most spectacular is  probably the mapping of the small anisotropies of the \ac{cmb} onto the \ac{bao} of the galaxy distribution in the Universe \citep{Eisenstein2005a}. Together with the precise distance measurement of standard candles at high redshift \citep{Freedman2001}, all the parameters of this model can be well determined \citep{PlanckCollaboration2015}. Traces of the structure and evolution of the Universe are also found in weak lensing signals \citep{Mandelbaum2013} and in the \emph{Lyman-$\alpha$ forest}\marginpar{Lyman-$\alpha$ forest is a spectroscopic trace of interstellar hydrogen, seen in objects with high redshift} \citep{Borde2014}, giving us independent measurements to verify the \ac{lcdm} model.\\
  Since baryons form structure within the \ac{cdm} halos, physical feedback processes of baryonic matter, such as supernova explosions of stars or jets from \ac{agn} complicate the picture, because the inferred mass function of galaxies will not be an exact, downscaled representation of the \ac{cdm} halo mass function. Hydrodynamical simulations that are based on the \ac{lcdm} model, and include baryonic feedback processes reproduce the observable Universe in an increasingly accurate way. Still, deviations between observations and simulations exist, for example the \emph{missing satellite problem} \citep{Klypin1999}. It is not clear whether this is a problem in the cosmological model, in the prescription of physical feedback processes, or (but less likely) in the observations. For the missing satellite problem, it might be solved by decreasing the assumed virial mass of the \ac{mw} \citep{Wang2012} or by satellite halos, being stripped off their baryons \citep{Simon2007}. Based on these discrepancies, a few alternative theories have been put forward \citep{Kroupa2012,Famaey2012}, but generally these are not seriously challenging the overwhelming evidence for the \ac{lcdm} model, such that we assume this cosmology for the purpose of this thesis.\\
  
  How can such a diverse structure, like the \ac{mw}, differentiate out of such a simple model?
  The answer lies in the interplay of baryons with the gravitational force of the \ac{cdm} and hierarchical structure growth. The structure of today's Universe started from tiny fluctuations in the \ac{cdm} density distribution, which grew gravitationally and hierarchically (through mergers and accretion) to form \ac{cdm} halos on different scales, the biggest of which are just about able to withstand cosmic expansion. At the same time, baryons (able to cool faster than \ac{dm} due to radiative cooling) could settle deep into the potential wells of the \ac{dm} halos and thus form stars and galaxies. Since the gravitational collapse is anisotropic, tidal forces lead to the distribution of angular momentum, which can produce disc-like structures and will leave dynamical imprints. The feedback processes from stars and \acp{agn} will regulate star formation and the build-up of structure. Its effect depends on the baryonic mass available, on the mass of the surrounding \ac{dm} halo, and on the halo environment \citep{Dressler1980}.\\
  
  Before coming to the description of the present-day picture of the \ac{mw}, we will introduce the concepts of the matter cycle of the Universe, which will give us an understanding of the element synthesis thus providing a powerful tool to analyse the \ac{mw} and its evolution.
   
\section{Elemental synthesis and stellar evolution}
The production of chemical elements, of which the world around us consists, took almost $8$\,Gyr, when our Solar system formed. Imprints of the evolution of our Universe, as well as the formation history of our Galaxy, can be deduced from the elemental abundances. We will find that the nucleosynthesis is strongly entwined with the evolution of stars and the ending of their lives. Historically \citet{Weizsacker1937}, \citet{Bethe1939} and the seminal \citet{Burbidge1957} developed an understanding of nuclear fusion in stars. A good overview is given in \citet{Ryan2010}.\\
To trace the elemental synthesis, we need to go to the beginning of our Universe, right after the \ac{bb}. A good overview and the historical development of cosmological nucleosynthesis are given in \citet{Weinberg2008}, with \citet{Gamow1946}, \citet{Hayashi1950} and \citet{Alpher1953} laying the foundations.\\

\subsection{Primordial nucleosynthesis}
In the \ac{lcdm} framework, the first elements condensated out of the cooling plasma of the expanding Universe. One second after the \ac{bb}, the temperature of the Universe was below $10^{9}$\,K and protons could no longer be transformed into neutrons via weak interaction. Therefore the neutron to proton ratio dropped, which is one parameter determining the final elemental abundance. From ten seconds to about three minutes the temperature allowed neutrons and protons to fuse into deuterium and helium.
Because of the high binding energy of helium and no stable nuclei with mass number five and eight, only light elements up to a \emph{mass number}\marginpar{mass number of an element is the added number of neutrons and protons it consists of} of seven could form. In stars, usually the triple-$\alpha$ process would overcome this barrier, but the density was not high enough for three-body reactions and also the time scale too short. In the end, $76\,\%$ of the mass was in hydrogen and $24\,\%$ in Helium, with traces of deuterium, tritium, helium-3 and small amounts of beryllium- and lithium-isotopes. Left-overs of neutrons, that did not find a reaction partner, decayed away, together with beryllium and tritium, fixing the primordial elemental composition. Since element abundance measurements of very early intergalactic gas has become feasible from \ac{quasar} spectra \citep{Kirkman2003}, primordial element abundances are a further constraining observational test for \ac{lcdm}.\\
\subsection{First stars}
\label{sec:first_stars}
For a long period of time, the Universe expands and its plasma cools. $380,000$\,yr after the \ac{bb}, neutral hydrogen forms and the thermal radiation runs freely, for the first time, producing the \ac{cmb}. As already outlined in the previous section, the baryons settle as proto-galactic clouds in overdense \ac{dm} regions, which grow non-linearly from tiny anisotropies to a filamentary structure. Those proto-galactic gas clouds are immense and contract, so that the gas in their inner cores clumps, heats up and produces pressure to resist gravitational collapse. Radiative cooling is necessary, for further compression of the gaseous clumps. At the same time, angular momentum needs to be dissipated away, for the contraction to proceed, forming accretion discs, until a runaway collapse can form the first stars. They are supposed to have higher masses than present-day stars, since line cooling via hydrogen and helium is less efficient than \emph{metal}-\marginpar{metals are all elements heavier than helium} or dust-cooling, so that higher Jeans masses \citep{Jeans1902} are required. These stars are purely hypothetical\footnote{What we know of population\,III stars, comes from hydrodynamical simulations and from abundances of very old stars, that are assumed to be the second generation of stars.}, because yet no population\,III stars, as the first stars with no metals are called, have been observed. This is probably due to their high masses, resulting in very short (on the order of $3-5$\,Myr) lifetimes\footnote{If they fragmented, which might not have been possible during formation, low-mass population\,III stars could have survived until the present day.}. Indeed, they must have been extremely heavy, compared to present-day stars, with masses between $100$ and $1000$\,M$_\odot$ of pure hydrogen and helium and also with high rotational velocities, giving them special properties. Some of the first stars might have been so heavy during their formation, that they collapsed directly into a \ac{bh}, forming the seed for the first \ac{smbh}, powering the first \ac{quasar}s \citep{Larson2000}.\\
These heavy population\,III, stars would have fused elements up to the \emph{iron peak}\marginpar{iron-peak elements are formed during the last burning stage in nuclear statistical equilibrium, where nuclei with the highest binding energy per nucleus are favoured (Ti, V, Cr, Mn, Fe, Co, Ni, Cu, Zn)} in their inner cores, after which their energy source, delaying the gravitational pull, runs dry. This is the star's end stage, when it consists of successive layers of different burning products from hydrogen, helium at the surface over carbon, neon, oxygen, silicon up to iron in the core. With the fading pressure, provided by the fusion, the core contracts over the limit of electron degeneracy\footnote{In very densely packed electron gas, the Pauli exclusion principle prohibits the overlapping of electrons in the same quantum state.}, loosing energy via photodisintegration and electron-capture (also called neutronization) and then approaching \ac{ns} densities, which suddenly stops the contraction. Most of the infalling outer layers is rebounded and expelled into the \ac{ism} during this event called \ac{sn2} \citep[Chapter\,7.1]{Ryan2010}.\\
The interstellar gas is mainly enriched with \emph{$\alpha$-elements}\marginpar{$\alpha$-elements are all stable elements, that can be build up from $^4$He nuclei (C, O, Ne, Mg, Si, S, Ar, Ca)}, a bit of iron-peak elements, together with traces of heavier elements, which use the explosion energy and the abundant neutrons to produce \ac{rprocess} elements, from iron seed nuclei, in an endothermic nuclear reaction. The cores of the stars will end as stellar remnants, which can be either \ac{ns}s or \ac{bh}s.
\subsection{Reionization}
These first stars, with their intense radiation, are reionizing the \ac{ism} and ending, together with the igniting galactic machinery, the dark ages of the Universe. Since the gas is less dense, than during recombination, the opaqueness of the Universe decreases. The view onto the first galaxies is disclosed, and they are observable from a redshift of about $\mathrm{z}\approx7$ \citep{Mortlock2011}. The first stars are believed to have formed around $150-400$\,Myr, also depending on the strength of the overdensity in that region of the Universe. The abundance pattern of the oldest observed stars, could have come from only one single \ac{sn2} event, giving us insight on the properties of the first stars \citep{Frebel2005,Chiappini2011,Keller2014}.\\

With the \ac{ism} metallicity\marginpar{metallicity is the metal mass fraction, with the Sun having an approximate value of about $1.4\,\%$} increasing and the distribution of gas getting more inhomogeneous, the mass range of stars of the upcoming generations, goes down and is assumed to be between $0.08\,\mathrm{M}_\odot<\mathrm{m}<150\,\mathrm{M}_\odot$. The number distribution of stars being born, is called \ac{imf} and is a stochastic average, well described by a lognormal or broken power-law at small masses and a steep power-law at high masses, making massive stars comparatively rare. Depending on the overall mass and to a lesser degree also on the helium mass fraction and the metallicity, the stellar evolution can be quite different and we will trace it and the contribution towards the chemical evolution from small- to high-mass stars. Beside the mass, the helium fraction and the metallicity, also rotation, binarity and elemental abundances will have an effect on the evolution and the feedback of a star.
\subsection{Low-mass stars}
The lower mass limit, $0.08$\,M$_\odot$, for stars is determined by the heat and pressure that is generated in the core of a compact object, consisting mainly of hydrogen and helium, which needs to be high enough to start the \ac{pp}-chain. In objects smaller than about $0.08$\,M$_\odot$, called \ac{bd}, the temperature needed for fusion is not reached, because of the onset of electron degeneracy, prohibiting a further increase in temperature (for the same reasons the lower mass limit for helium burning in stars is $0.5$\,M$_\odot$). Up to about $0.8$\,M$_\odot$, stars remain on the \ac{ms}, meaning that they quiescently burn hydrogen in their core, for longer than a \emph{Hubble time}\marginpar{Hubble time is the inverse of the Hubble parameter, $\mathrm{H}_0=67.80\pm0.77\,\tfrac{\mathrm{km}}{\mathrm{s}}\mathrm{Mpc}^{-1}$ \citep{PlanckCollaboration2015}, which describes the expansion rate of the Universe. The inverse is the time, when a constant expansion would have started, which is a simple approximation for the age of the Universe, $\mathrm{H}_0^{-1}=14.4\,\mathrm{Gyr}$ and only slightly off the fiducial value of $\mathrm{t}_0\approx13.8\,\mathrm{Gyr}$}. From a matter-cycle point of view, these low-mass stars are a sink, binding material in them and not contributing by releasing newly formed elements. Stars above $0.3$\,M$_\odot$ will enter the \ac{rgb} phase, which will be described in the next subsection.\\
Most of the formed stars are actually low-mass stars. Since the radiation pressure from photons, released by nuclear fusion, equates the gravitational pull and this hydrostatic equilibrium is highly temperature- and therefore mass-dependent, low-mass stars, contrary to high-mass stars, lead quiescently dim and long lives. This leads to the effect, that when looking at a stellar population, most of its mass will be contained in low-mass stars and most of its light will be emitted by a few, very bright high-mass (and also evolved) stars.\\
The main reaction powering a star on the \ac{ms} changes with mass. Up to about $1.3$\,M$_\odot$ the \ac{pp} chain dominates. In more massive stars the \ac{cno}, using C, N and O as catalysts, becomes more efficient in turning hydrogen into helium (though population\,III stars would not have had this option). Independent of the initial composition of these three catalyst elements, the \ac{cno} will alter their abundances in the core, which will have an influence on the nucleosynthesis of intermediate-mass stars. 
\subsection{Intermediate-mass stars}
\subsubsection{\acl{agb}}
\begin{figure}[h]
\caption[Hertzsprung-Russel diagram]{Hertzsprung-Russel diagram with stellar effective surface temperature in Kelvin on the x-axis and luminosity in Solar luminosity on the y-axis. The evolutionary tracks of stars in the range of $1$\,M$_\odot$ to $8$\,M$_\odot$ and with a sub-Solar metallicity of $\text{Z}=0.01$ from \citet{Bressan2012} are shown, beginning with their \ac*{zams}. The end of the \ac*{ms} is marked with the colour-coded age of the corresponding track. This is where the contraction of the core heats the surface, until the ignition of the hydrogen burning shell cools it again, resulting in the zigzag for stars heavier than $2$\,M$_\odot$. Timescales of the different fusion processes are indicated for the $1$\,M$_\odot$ track. Different phases of the stellar evolution are shown for the $5$\,M$_\odot$ track. The area of the \ac*{rc} is indicated, as well as the position of \ac*{pn} and \ac*{wd}, which are outside of this diagram. In dashed grey the Hayashi limit is indicated and in dotted grey the stellar radii are indicated qualitatively. Main features of this figure are taken from \citet[Fig. 4.5, 5.5, 6.3]{Ryan2010}.}
\centering
\def\svgwidth{\columnwidth}
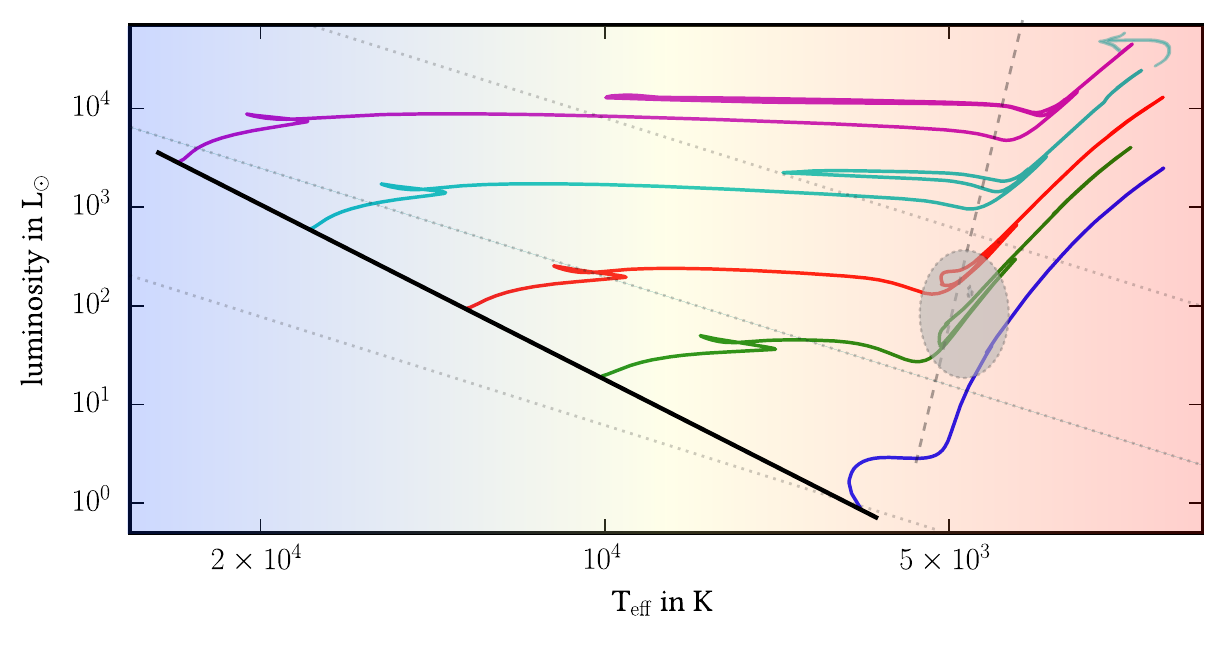
\label{fig:hrd}
\end{figure}

Stars with masses between $0.8\,\mathrm{M}_\odot$ and $8\,\mathrm{M}_\odot$ are heavy enough to use up the hydrogen fuel in their inner cores within a Hubble time. When around $10\,\%$ of the star's hydrogen has been turned into helium, it leaves the \ac{ms}. Since the core is full of helium, but not yet hot enough to burn it, hydrogen will fuse in a shell around the core, leading to an expansion of the whole star, though the luminosity stays the same. This results in the cooling of the photosphere, moving the star horizontally across the \ac{hrd}, as can be traced in figure\,\ref{fig:hrd}. Until it reaches the Hayashi-limit, where the increase of the outer convection zone\footnote{Eventually the convection can go so deep, that signatures of the hydrogen fusion, like decreased $^{12}$C/$^{13}$C and C/N ratios, as well as depleted lithium and beryllium abundances, can be exposed at the surface of the star. This effect is called first dredge-up and needs to be taken into account when interpreting stellar spectra.}, inhibits further cooling and marks the end of the subgiant branch and the beginning of the \ac{rgb}.\\
 The core keeps growing , increasing its temperature and connectedly the efficiency of the hydrogen burning shell, thereby increasing the luminosity and the radius of the star. For stars with around Solar mass the shell burning only ignites while already on the \ac{rgb} (see $1$\,M$_\odot$ stellar evolutionary track in figure\,\ref{fig:hrd}). Once the core reaches $10^8$\,K, a temperature high enough to ignite helium burning\footnote{To turn helium into carbon, via the triple-$\alpha$ process, and also helium and carbon into oxygen.}, the core will increase, lowering the energy output of the hydrogen burning shell, which leads to an overall contraction of the star. The position in the \ac{hrd} of the quiescent phase of helium burning is called \ac{rc}, for metal-rich stars (\emph{population\,I}\marginpar{population\,I\,\&\,II stars was termed by \citet{Baade1944}}), and \ac{hb} for metal-poor stars (\emph{population\,II}). Stars of mass lower than $2$\,M$_\odot$ will have a completely degenerate core, before starting helium burning, so that the fusion ignites in a degenerate gas, leading to the so called \emph{helium flash}, increasing the energy output of the core by several orders of magnitude. More massive stars will have a smoother transition into the helium burning phase and also move further blue-wards, as their cores are non-degenerate, which decreases their radius, resulting in a hotter surface.\\
Once helium is depleted in the core, a similar process to the first ascend of the \ac{rgb} will happen, now with two shells, one burning hydrogen and one burning helium. That again drives an expansion of the star, taking a similar route in the \ac{hrd}, only in a much shorter period of time and now called \ac{agb} phase. With the outer layer getting convective again, a second thermal dredge-up takes place (but only in stars heavier than $\approx3.5$\,M$_\odot$), increasing the helium and nitrogen abundances and decreasing the carbon and oxygen abundance in the photosphere of the star. Once the helium shell used up its fuel (increasing the \ac{co} core) and the hydrogen shell becoming the major energy source, the \ac{agb} star starts pulsing thermally. This happens, because the star switches consecutively from hydrogen to helium fusion, with the hydrogen shell producing new fuel for the helium shell to re-ignite. The back and forth of burning layers results in strong convective movements between the two shells and can also bring material to the outer layers, which is referred to as third dredge-ups. These can further increase the helium abundance and will also bring freshly synthesised carbon and \ac{sprocess} elements to the surface. \ac{sprocess} elements are produced during these pulses, when iron seeds \emph{grow} to heavier elements, via successive neutron-capture under moderate flux and $\beta$-decay. Neutrons are mainly provided from the $^{13}\mathrm{C}+\,^{4}\mathrm{He}\rightarrow \,^{16}\mathrm{O}+\mathrm{n}$ reaction, fuelled by the convective mixing. In \ac{agb} stars the main \ac{sprocess} produces heavy elements from strontium up to lead. There is also a \emph{weak} \ac{sprocess} occurring in massive stars, after their helium- and carbon-burning, producing elements beyond iron up to yttrium.\\  
The thermal pulses of the \ac{agb} phase become stronger each time, increasing the probability of moving material to the surface of the star. Due to the increased luminosity and molecules forming in the outer layers, strong stellar winds expel almost all of the envelope of the \ac{agb} star into the \ac{ism}, contributing to the chemical enrichment. The contracting \ac{co}-core becomes hotter and starts to ionize the surrounding material, upon reaching $10^4$\,K, which illuminates the \ac{pn}. The star moves blue-wards in the \ac{hrd}, following the post-\ac{agb} arrow in figure\,\ref{fig:hrd}, until it leaves the boundary of the figure, with the \ac{pn} position indicated at the bottom left. The electron-degenerate remnant has a typical mass range of $0.5$\,M$_\odot$ to $1.2$\,M$_\odot$, is called \ac{co}-\ac{wd} and will slowly radiate its thermal energy away. This is called \ac{wd} cooling sequence, which also exceeds the figure limits and is indicated in the bottom left.\\
The evolution timescales of stars are strongly mass-dependent, as can be seen from their \ac{ms}-lifetimes in figure\,\ref{fig:hrd}. The relative durations of subsequent burning phases are also written for the $1\,\mathrm{M}_\odot$ evolutionary track, which shows that a star spends most of its life on the \ac{ms}. Stars with $8\,\mathrm{M}_\odot$ have a \ac{ms} lifetime of about $40$\,Myr and stars of $0.8\,\mathrm{M}_\odot$ have one of $13$\,Gyr, which distributes the feedback from intermediate-mass stars over a long time. 
\subsubsection{\acl{sn1a}}
\label{sec:sn1a}
Another very important mechanism, especially for the iron production, is supposed to occur mainly in binary systems of intermediate-mass stars. There are two main progenitor models for \ac{sn1a}, depending on the number of \ac{co}-\ac{wd}s involved. They are called \emph{single-} or \emph{double-degenerate} channel. For the former, the primary star is already an electron-degenerate \ac{co}-\ac{wd} and the secondary starts donating material to the other, most likely because it enters the \ac{rgb} phase. The mass of the \ac{wd} will exceed the \emph{Chandrasekhar mass}\marginpar{\emph{Chandrasekhar mass} is the maximum stellar mass that can be supported by degenerate electron pressure \citep{Chandrasekhar1931}}, $\mathrm{m}\approx1.38\,\mathrm{M}_\odot$, igniting a thermonuclear runaway of carbon burning, which was first proposed by \citet{Hoyle1960} as a possible origin for \ac{sn1a}. In the latter case, the two stars are already \ac{co}-\ac{wd}s and spiral into each other, loosing angular momentum due to gravitational waves. When they coalesce, they exceed the Chandrasekhar limit and explode as a \ac{sn1a}. 
A successful model, reproducing the observed abundances, was calculated by \citet{Nomoto1984} using C-deflagration\footnote{The flame speed of the explosion propagating with subsonic speed.} and refined by \citet{Iwamoto1999}. It synthesises around $0.27\,\mathrm{M}_\odot$ of silicon to calcium elements and $0.21\,\mathrm{M}_\odot$ of carbon to aluminium (with $0.14\,\mathrm{M}_\odot$ of oxygen). Iron-peak elements make up $0.90\,\mathrm{M}_\odot$, with $0.75\,\mathrm{M}_\odot$ in the form of $^{56}$Ni, subsequently decaying to iron, giving the \ac{sn1a} light curves their characteristic shape. Chemical evolution models predict that around $30\,\%-50\,\%$ of the overall iron in the Galaxy is produced from \ac{sn1a}.\\
The two progenitor models of \ac{sn1a} also have different \ac{dtd} functions, depending on the stellar evolution and the angular momentum loss. A mix of both functions is usually assumed in chemical evolution models, with a peak at approximately $1$\,Gyr.
\subsection{High-mass stars}
Stars more massive than $8\,\mathrm{M}_\odot$ retain enough mass during hydrogen- and helium-burning in order to ignite subsequent fusion processes, though up to a mass of $11\,\mathrm{M}_\odot$, the core mass is not sufficient to fuse all elements up to iron. These stars will end their lives as oxygen-neon-magnesium \ac{wd}s, with typical masses between $1.2$\,M$_\odot$ and $1.4$\,M$_\odot$.\\
Cores of stars with an initial mass heavier than $11\,\mathrm{M}_\odot$ will exceed the Chandrasekhar mass limit and are able to fuse iron-peak elements, in nuclear statistical equilibrium at the end of their lives. With the energy source ceasing, a core-collapse supernova, \ac{sn2}, will ignite, as already outlined in section\,\ref{sec:first_stars}. This enriches the \ac{ism}, mainly with synthesised $\alpha$-elements, a bit of iron-peak elements and \ac{rprocess}-elements, leaving behind a \ac{ns} or a stellar-mass \ac{bh}.
\subsection{Chemical enrichment - a simple model}
\label{sec:simple_chemistry}
The elemental feedback from massive stars occurs on the order of $\sim10\,$Myr after their formation, enriching the \ac{ism} via \ac{sn2} on relatively short time scales. Intermediate mass stars, on the other hand, take on the order of $\sim100\,$Myr for \ac{agb} feedback or $\sim1,000\,$Myr for \ac{sn1a} explosions. Since the synthesised elements differ for each process, the signature of the \emph{delayed} enrichment can be detected in the chemical composition of subsequent stellar generations.\\
When talking about elemental abundances, there are two choices of how to communicate those, either by mass or by abundance. The composition of the Solar photosphere in mass fraction for example is $\mathrm{X}_\odot=73.9\,\%$ hydrogen, $\mathrm{Y}_\odot=24.69\,\%$ helium and $\mathrm{Z}_\odot=1.41\,\%$ metals using abundance determinations from \citet{Lodders2009}. The most abundant metals are oxygen Z$_{\mathrm{O,\,}\odot}=0.63\,\%$, carbon Z$_{\mathrm{C,\,}\odot}=0.22\,\%$, neon Z$_{\mathrm{Ne,\,}\odot}=0.17\,\%$ and iron Z$_{\mathrm{Fe,\,}\odot}=0.12\,\%$, already making up $80\,\%$ of all the metals in mass.\\
The observables in spectroscopic analysis are more related to number ratios of different elements. That is why observers usually express abundances via the elements logarithmic relative number ratio to hydrogen:
\begin{equation}
\mathrm{A}(\mathrm{Element})=\log_{10}\left(\frac{\mathrm{N}_\mathrm{Element}}{\mathrm{N}_\mathrm{H}}\right)+12.
\end{equation}
Oxygen for example has an abundance of $\mathrm{A}(\mathrm{O})_\odot=8.73$, meaning that for every oxygen atom there are $1,862$\,hydrogen atoms in the photosphere of the Sun. To relate the abundance to the mass fraction, we need to multiply each atom by its nucleon number, $\mathrm{m}_\mathrm{H}=1$ for hydrogen and $\mathrm{m}_\mathrm{O}=16$ for oxygen:
\begin{equation}
\frac{\mathrm{N}_{\mathrm{O,\,}\odot}\mathrm{m}_\mathrm{O}}{\mathrm{N}_{\mathrm{H,\,}\odot}\mathrm{m}_\mathrm{H}}=\mathrm{m}_\mathrm{O}10^{\mathrm{A}(\mathrm{O})_\odot-12}=\frac{16}{1,862}=\frac{\mathrm{Z}_{\mathrm{O,\,}\odot}}{\mathrm{X_\odot}}.
\end{equation}
Abundances of other stars are usually given in the \emph{bracket} notation, which is given in units of \emph{dex}\marginpar{dex means \emph{decimal exponential} and if a star has a [Fe/H] of $-1$ (or $-2$) then its iron atoms are 10 (or 100) times less abundant than in the Sun}, with the Sun as reference
\begin{equation}
[\mathrm{Fe}/\mathrm{H}]=\mathrm{A}(\mathrm{Fe})-\mathrm{A}(\mathrm{Fe})_\odot\quad\left(\simeq\log_{10}\left(\frac{\mathrm{Z}_\mathrm{Fe}}{\mathrm{Z}_{\mathrm{Fe,\,}\odot}}\right)\right).
\label{eq:feh}
\end{equation}
This is the iron abundance relative to Solar (with the equation in brackets only being exact for stars that have hydrogen mass fraction similar to Solar, X$_\odot=0.739$, which is a good approximation for most of the stars). Since iron lines are easy to observe, [Fe/H] is often used to indicate the metallicity of a star.\\

As an illustrative example for the signatures of chemical evolution, which can be traced by stellar elemental abundances, we will now consider the time-dependence of the oxygen to iron ratio
\begin{equation}
[\mathrm{O}/\mathrm{Fe}]=[\mathrm{O}/\mathrm{H}]-[\mathrm{Fe}/\mathrm{H}]
\label{eq:ofe}
\end{equation}  
in the Galactic \ac{ism}.\\
Neither of those elements is produced during the primordial nucleosynthesis. After a short eruptive period of first stars - which we do not consider -, the Galaxy settles and forms new stars at a - more or less - constant rate. The feedback from the \ac{sn2}, of short-lived massive stars, consists mainly of $\alpha$- and iron-peak-elements, producing a constant oxygen to iron ratio of about $[\mathrm{O}/\mathrm{Fe}]=0.5$. Substituting equation\,\ref{eq:feh} in equation\,\ref{eq:ofe}, we can calculate the corresponding mass fraction.
\begin{equation}
0.5=\log_{10}\left(\frac{\mathrm{Z}_\mathrm{O}}{\mathrm{Z}_{\mathrm{O,\,}\odot}}\right)-\log_{10}\left(\frac{\mathrm{Z}_\mathrm{Fe}}{\mathrm{Z}_{\mathrm{Fe,\,}\odot}}\right)\Leftrightarrow\frac{\mathrm{Z}_\mathrm{O}}{\mathrm{Z}_{\mathrm{Fe}}}=10^{0.5}\frac{\mathrm{Z}_{\mathrm{O,\,}\odot}}{\mathrm{Z}_{\mathrm{Fe,\,}\odot}}
\end{equation}
With the Solar ratio being $\tfrac{\mathrm{Z}_{\mathrm{O,\,}\odot}}{\mathrm{Z}_{\mathrm{Fe,\,}\odot}}\approx5$, it follows that $\tfrac{\mathrm{Z}_{\mathrm{O,\,SN\,II}}}{\mathrm{Z}_{\mathrm{Fe,\,SN\,II}}}\approx16$ in \ac{sn2} feedback, and equivalently $\tfrac{\mathrm{N}_{\mathrm{O,\,SN\,II}}}{\mathrm{N}_{\mathrm{Fe,\,SN\,II}}}\approx56$.\\
With some time delay, the \ac{sn1a} start exploding, with iron being their bulk contribution to the chemical enrichment. In fact, from section\,\ref{sec:sn1a} we know that $\tfrac{\mathrm{Z}_{\mathrm{O,\,SN\,Ia}}}{\mathrm{Z}_{\mathrm{Fe,\,SN\,Ia}}}\approx\tfrac{1}{5}$. This decreases the oxygen to iron abundance asymptotically towards the Solar value, $[\mathrm{O}/\mathrm{Fe}]\,$=$\,0$, and produces a \emph{knee} in the $[\mathrm{O}/\mathrm{Fe}]$ over $[\mathrm{Fe}/\mathrm{H}]$ distribution. The onset of the \emph{knee} occurs at higher $[\mathrm{Fe}/\mathrm{H}]$ values for larger systems, as they produce more \ac{sn2} and can retain most of their ejecta (contrary to smaller systems like dwarf galaxies). Since \ac{agb} feedback is not significantly altering the oxygen to iron ratio in the \ac{ism}, $\alpha$-enhancement\marginpar{$\alpha$-enhancement is measured from the abundance of an $\alpha$-element or an average of several $\alpha$-elements relative to iron, e.g. [O/Fe]} can be used as an age indicator for stars.

What follows is a list of issues that complicate this simple model and the interpretation of chemical abundance data in general:
 \begin{itemize}
 \item The relative abundance of intermediate- to high-mass stars is not fully constrained nor its dependence on metallicity and environmental effects.
 \item The exact feedback from various enrichment processes is not known and they vary with stellar mass, metallicity, binary evolution and rotation.
 \item The \ac{dtd} of \ac{sn1a} is not precisely known and might also vary with metallicity and environment.
 \item Neither the \ac{sfh} nor the available gas mass at a given time (infall, outflow, gas at start) is well constrained.
 \item The dynamics of stars and gas need to be considered.
 \item The Galactic gravitational potential is not fully known.
 \item Further exotic feedback processes probably contribute, like neutron star mergers producing lots of the early \ac{rprocess} elements.
 \item Synthesising abundance data to compare the model to observations has several caveats:
 \begin{itemize}
 \item The selection function of the stars needs to be reproduced.
 \item Usually, stars are assumed to preserve the abundance of the \ac{ism} from which they were formed in their photospheres, but it can change, due to dredge-up events.
 \item Stellar ages can only be roughly constrained from the commonly available data.
 \end{itemize}
  \end{itemize}
 
 As can be seen from that list, chemical, as well as dynamical, processes need to be taken into account, in order to constrain the evolution of our Galaxy. The field is guided by high-redshift observations of \ac{mw}-like spiral galaxies and also benefits from cosmological simulations, becoming able to reproduce observations. The interplay of all those fields, is constraining a consistent picture of the \ac{mw}-evolution and is called Galactic archaeology. In this introduction recent advances in the field will be summarised.
\section{Structure of the Milky Way}
\begin{figure}
\caption[Milky Way schematic]{Schematic Milky Way}
\centering
\def\svgwidth{\columnwidth}
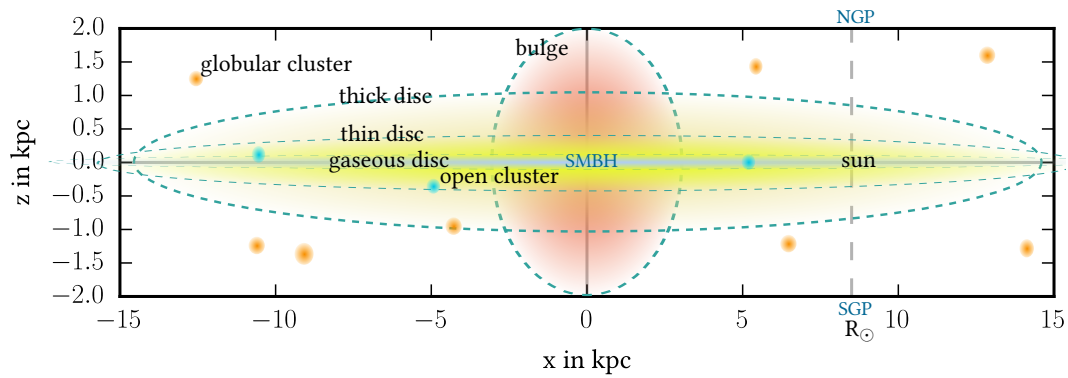
\label{fig:mw}
\end{figure}

The \ac{mw} disc sits in a \ac{cdm} halo, which extends out to about $200$\,kpc and dominates the mass budget, with $\sim10^{12}$\,M$_\odot$ \citep{McMillan2011}. Satellite galaxies orbit through its gravitational potential, with the most prominent being the two Magellanic clouds, at a distance of $55$\,kpc, and the Sagittarius dwarf, $15$\,kpc away from the Galactic centre,  which produces the prominent Sagittarius stream. Most of the baryonic mass is within $15$\,kpc radius, where the outer stellar halo is separated from the inner stellar halo. The inner structure, depicted in figure\,\ref{fig:mw} and containing most of the \ac{mw}'s stars and cold gas, is composed of a disc and a central bulge with a bar.\\
This thesis will mainly focus on the disc, but here I will briefly discuss the other major components as well. The interested reader is referred to \citet{Sparke2007} for an introduction to the \ac{mw} and to \citet{Scheffler1992} for a comprehensive overview on Galactic astronomy. Galactic dynamics are thoroughly covered in \citet{Binney2008} and a further reference for chemical evolution is \citet{Pagel2009}.
\subsection{Halo}
The Galactic stellar halo envelopes the disc and the bulge and reaches out to $150$\,kpc, with a stellar mass content of about $4\times10^8$\,M$_\odot$ \citep{Bell2008}. It is an important diagnostic for the \ac{mw} evolution, with its old field star population and the globular clusters. It can roughly be dissected into an inner oblate spheroidal halo, comprising mainly in-situ formed stars and an outer spherical halo, where accreted stars, from satellite galaxy debris, are prevalent \citep{Carollo2007}. The orbits around the Galactic centre are not ordered, with the outer halo showing a slight retrograde rotation, with respect to the disc rotation. Stars have a broad range of metallicities, with the bulk of stars being between $-3<[\mathrm{Fe}/\mathrm{H}]<-1$. The outer halo is more metal-poor and has a negative metallicity gradient with Galactocentric distance. There is no dust or ongoing star formation in the halo, though high velocity clouds are observed, which contribute to the gas infall of the disc \citep{Putman2012}.

\subsection{Barred bulge}
\begin{figure}
\caption[Milky Way centre in the optical]{Milky Way centre in the optical with Galactic coordinates. Bulge. Credit: NASA / A.\,Mellinger}
\centering
\def\svgwidth{\columnwidth}
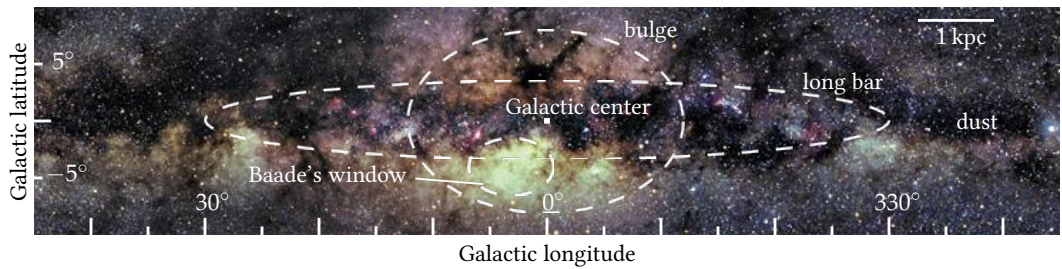
\label{fig:bulge}
\end{figure}
\begin{figure}
\caption[Milky Way face-on]{Milky Way face-on view from above the plane with Galactic latitude in degrees and distances in light years. The viewing angle from figure\,\ref{fig:bulge} is indicated in orange. Known spiral structure is annotated and was traced by H\,I, H\,II and CO surveys. The two bars have been constrained by \ac*{ir} and microlensing surveys. References are given in \citet{Churchwell2009} from which the picture is taken. Credit: NASA/ JPL-Caltech/ ESO/ R. Hurt}
\centering
\def\svgwidth{\columnwidth}
\input{gfx/chapter1/spiral.pdf_tex}
\label{fig:face-on}
\end{figure}
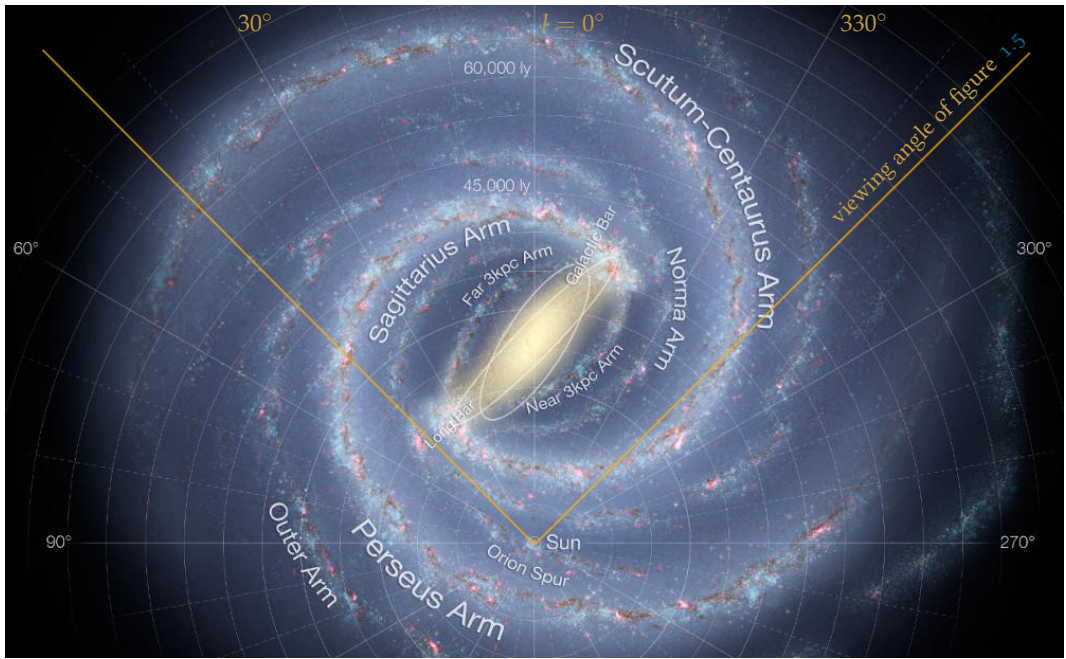

The bulge is the stellar nucleus of our Galaxy, with a boxy shape and a radius of about $3$\,kpc \citep{Wegg2015}. It is hard to observe the inner part in the optical, because of heavy extinction by dust in the Galactic plane, which absorbs selectively, with shorter wavelengths being stronger absorbed \citep{Trumpler1930}. This is clearly visible in the dark patches of figure\,\ref{fig:bulge}, which is an optical image into the direction of the Galactic centre. The first to observe the bulge was \citet{Baade1946}, using a dust-free window, classifying the stars as population\,II stars. Subsequent observations, in the \ac{nir} and radio, revealed that the very centre of the bulge hosts an \ac{smbh}, named Sagittarius A$^*$, with a mass of about $4\times10^6$\,M$_\odot$ \citep{Balick1974,Gillessen2009}. The overall stellar mass of the bulge region is approximately $10^{10}$\,M$_\odot$ \citep{Kafle2014}. It rotates ,but not as ordered as the disc and with a high velocity dispersion. The metallicity distribution of surveyed stars peaks at the Solar value and they are usually $\alpha$-enhanced, indicating a rapid chemical enrichment \citep{Zoccali2007}. An elongated bar, made up of a variety of stable stellar orbits, has been detected \citep{Blitz1991} and is believed to rotate rigidly at a pattern speed of $\sim50$\,km\,s$^{-1}$\,kpc$^{-1}$, corresponding to a corotation radius of $\sim4$\,kpc \citep{Gerhard2011}. The bar induces orbital resonances in the disc and its outer Lindblad resonance, which is just inside the Solar Galactocentric radius, R$_\odot$, is believed to be responsible for several moving groups in the Solar Neighbourhood \citep{Dehnen2000}. Together with the spiral arms, they induce radial migration of stars through the disc, facilitating mass inflow, chemical homogenisation and triggering star formation. Both structures seem to be dynamically decoupled, since the pattern speed of the spiral waves are much slower, with a corotation radius slightly outside of R$_\odot$ \citep{Gerhard2011}.\\  
The bulge seems to be a pseudo-bulge, meaning that it mainly consists of an in-situ formed bar \citep{Shen2010}, contrary to a classical bulge, which would have build up from mergers. \citet{Hammersley2000} even detected a second thinner and longer bar structure, which is indicated in figure\,\ref{fig:face-on}. The bar, together with the spiral density waves, probably have profound effects on the Galactic evolution and might also be responsible for the molecular ring, the biggest reservoir of molecular gas, which is necessary for star formation \citep{Ragan2009}, observed at $4$\,kpc from the Galactic centre.
\subsection{Disc(s)}
The disc is the main baryonic component of the \ac{mw}, with a mass of around $10^{11}$\,M$_\odot$ \citep{Kafle2014}. It is of almost circular shape, with a radius $\sim15$\,kpc and the Sun sitting at R$_\odot\sim8.5$\,kpc to the Galactic centre. That it is a system of its own and has an ordered rotation was discovered by \citet{Lindblad1927,Oort1927,Oort1928}. It has a flat rotation curve, rotating clockwise when viewed from the \ac{ngp}, with a circular rotation speed of $\mathrm{v}_\mathrm{c}\sim240\,\tfrac{\mathrm{km}}{\mathrm{s}}$\marginpar{a good approximation is $\tfrac{\mathrm{km}}{\mathrm{s}}\approx\tfrac{\mathrm{pc}}{\mathrm{Myr}}$, with a deviation $<3\,\%$} at the Solar radius \citep{Reid2014}, so that the Galactic year takes about $220$\,Myr. The orbits of the disc stars have only small deviations, with respect to the \emph{local standard of rest}, with the Sun having an offset of about U$_\odot=10\,\tfrac{\mathrm{km}}{\mathrm{s}}$, V$_\odot=3\,\tfrac{\mathrm{km}}{\mathrm{s}}$ and W$_\odot=7\,\tfrac{\mathrm{km}}{\mathrm{s}}$ \citep{Au09,Golubov2013}.\\
New stars are formed from the inhomogeneous gas and dust layer, which is closely confined to the Galactic plane. Therefore groups of young stars, usually have a small velocity dispersion and they are called dynamically \emph{cold}. Spiral arms are small overdensities in the form of transient waves, where star formation is preferentially taking place, which gives them their bluish glow (from massive and therefore hot, bright and short-lived stars), when seen in external face-on galaxies. In the outskirts of the disc a warp has been detected \citep{May1993}, bending the disc down from its plane at Galactic longitude $l=0^\circ$ and bending it up close to the anticentre. Another observed effect is called flaring \citep{Momany2006}, describing an increase in scale height with increasing Galactocentric distance.\\

When inspecting the density distribution of disc stars, \citet{Gilmore1983} found that the disc can be decomposed into two components, with different scale heights. The thin disc with a scale height of about $300$\,pc and the thick disc with $\sim1300$\,pc. Already in his PhD thesis \citet{Oort1926} saw kinematic traces of both discs, with the thick disc having a higher velocity dispersion, being dynamically hotter and therefore older than the thin disc stars. Similarly, in chemical abundance space the thick disc population shows a lower metallicity and higher $\alpha$-enhancement compared to the thin disc \citep{Gratton1996,Fuhrmann1998}.\\

The stellar density of the discs drop exponentially vertically, as well as radially, with controversial results for the scale-length. \citet{Juric2008} finds from \ac{sdss}, using \emph{M}-dwarfs\marginpar{M-dwarfs are stars of spectral type M, the spectral type sequence being O, B, A, F, G, K, M, L, T from hot to cold, with the last two already being \acl{bd}s} that the thick disc has a larger scale-length than the thin disc, which is usually also observed in edge-on views of external spiral galaxies (see for example the \ac{mw} analogue NGC\,$891$ \citep{Ibata2009}). When dissecting in abundance space, like \citet{Bensby2011}, the thick disc scale-length seems to be shorter compared to the thin disc. \citet{Bovy2012} even finds that, using \ac{segue} mono-abundance\footnote{Stellar populations binned in [Fe/H] and [$\alpha$/Fe].} populations, a smooth transition from old and spherical to young and elongated, takes place without the necessity of a distinct thick disc. The oldest population (with $[\mathrm{Fe}/\mathrm{H}]=-1$ and $[\alpha/\mathrm{Fe}]=0.5$), having a scale-length of $2$\,kpc and a scale height of $1$\,kpc and the youngest population (with about Solar abundances), having $\approx4.5$\,kpc and $\approx0.2$\,kpc respectively. Recent \ac{mw} modelling approaches, combining chemistry and dynamics, are also able to produce an apparently bimodal distribution from a simple one-disc model by kinematic selection effects \citep{Schonrich2009b,Minchev2013a}. The question arising is how both discs formed and whether or not they share a common formation mechanism.
\subsection{Formation scenarios}
As already mentioned, traces of the formation history of the \ac{mw} are imprinted in the chemical abundances and the kinematic information of its stars. When formed in the gaseous disc with low velocity dispersion, stars are believed to be randomly perturbed during their lifetime and to deviate increasingly from their initial nearly circular orbit. This makes their orbits more eccentric and also increases their deviations from the Galactic plane. The relation between the age of a stellar population and its kinematic state is called \emph{\ac{avr}}. Because of the difficulty to determine stellar ages, it is observationally not clear if the velocity dispersion increases infinitely with time \citep{Wielen1977} or already saturates after $\sim5$\,Gyr of heating \citep{Soubiran2008}.\\
The seminal \citet{Eggen1962} proposed a monolithic and very short collapse of a protogalaxy as a formation scenario of the \ac{mw}, which they derive from thorough analysis of stellar orbits and their correlations with \ac{ir}-excess (metal-abundance). \citet{Searle1978} investigated the globular cluster metal abundances in the outer halo and concluded that fragment accretion for a longer period is necessary to reproduce their distribution. Another paradigm, supported by the upcoming \ac{cdm} cosmology, is hierarchical clustering, with mergers of smaller structures accumulating to larger systems \citep{White1978}.\\
The true evolution will most likely be a mixture of all those scenarios. But how important are those mechanisms and at which epochs do they dominate? How is the secular evolution of the individual parts of the \ac{mw} and how do they interact with each other?\\
\section{Deciphering the Milky Way evolution}
\subsection{Extragalactic observations}
Because of our perspective from the inside of our Galaxy and the fact that we can only observe its present-day state, galaxy surveys can help us to gain insight into the formation history of the \ac{mw}. Not only that we can distinguish the different galaxy components easier from outside of a galaxy, but we can also look into the past of \ac{mw}-like galaxies when targeting high-redshift progenitors. \citet{VanDokkum2013} found that $90\%$ of the stellar mass of their \ac{mw}-like galaxy sample is build up after redshift of $2.5$ and that bulges and discs form in lockstep until redshift $\approx1$, when the bulge mass assembly ceases.\\
\subsection{Simulations}
The laboratory experiment in galaxy evolution can only be done via hydrodynamical N-body simulations. The \emph{Millenium runs} \citep{Springel2005,Boylan-Kolchin2009}, which are purely gravitational cosmological simulations, were able to show the success of \ac{lcdm} cosmology in reproducing the cosmic web, represented in the filamentary structure of galaxy distribution.
Meanwhile, with development of codes, which allow for solving fluid dynamic equations, for the purpose of dealing with gas, and also significant improvements of our understanding about baryonic physics, the \emph{Illustris} simulation \citep{Vogelsberger2014}, was able to implement baryons and their feedback processes in their cosmological calculation. The obtained galaxy morphologies and their frequency distribution quite successfully reproduces the observed ones \citep{Genel2014}.\\
Specialised simulations that reproduce \ac{mw}-like galaxies within realistic \ac{dm} halos are also able to synthesise structural parameters of our Galaxy \citep{Stinson2010,Guedes2011}. The bulge to disc ratio and the chemical evolution seem notoriously hard to reproduce. As these simulations are computationally demanding, they can not easily test a large parameter space. 
\subsection{Analytical models}
The exploration of a large parameter space can be better done with fast to compute (analytical) models that concentrate on a few major physical processes, which can be tested with discriminating observations, that are easy to synthesise. Usually these models are quite specialised, so that they only use chemical \citep{Ch97} or dynamical constraints \citep{JJ} or target a specialised topic, as for example radial migration \citep{Schonrich2009}. Recently also a hybrid approach by \citet{Minchev2013a}, combining an N-body simulation \citep{Martig2012} with a chemical evolution model, was very successful in reproducing multiple chemodynamical constraints. 
\subsection{Milky Way observation}
Future surveys are designed in order to pin down the assembly history of the \ac{mw} and understanding the processes governing its evolution. In the following I will outline the major \ac{mw} surveys, their specifications and prospective scientific yield from upcoming surveys.
\subsubsection{Astrometry}
One major contribution will be from the astrometric satellite mission, \ac{gaia}. \ac{gaia} will deliver precise parallaxes, proper motions and photometry of all stars brighter than G\,$=20$\,mag. For a sub sample, of these $10^9$ stars, it will also measure the radial velocity from low-resolution spectroscopy, mapping the complete phasespace distribution for an unprecedented number of stars, homogeneously. The precision will be about $15\,\mu$as for stars at $15$\,\ac{avmag}.\\
For comparison, the successful \ac{hip} mission, which was magnitude limited at $\ac{avmag}\approx7.5$, yielded $\approx10^6$ stars with a median precision of $\sim1\,$mas and had major impact on astronomy.
\subsubsection{Spectroscopy}
Spectroscopic follow-up mission are under way with \ac{ges} and \ac{galah}. These will supplement the position and velocity data from \ac{gaia} with stellar parameters and chemical abundances from high-resolution spectroscopy for about $10^5-10^6$ stars. Significant statistical number of stellar spectra have been provided by the \ac{gcs}, \ac{rave}, \ac{segue} and \ac{apogee}. The \ac{rc} sample from the latter is very valuable when probing a large Galactic volume, because it provides almost $20,000$\,stars with precise distances, stellar parameters, and abundances for up to $15$ elements.
\subsubsection{Asteroseismology}
In order to increase the accuracy of age determination asteroseismology can be facilitated, the major missions being \ac{kepler} and \ac{corot}. So far only $\sim10^{3}$ stars with spectroscopic data benefit from added seismological data.





\acresetall

%% file: gfx/chapter1/hrd.pdf_tex
\begingroup%
  \makeatletter%
  \providecommand\color[2][]{%
    \errmessage{(Inkscape) Color is used for the text in Inkscape, but the package 'color.sty' is not loaded}%
    \renewcommand\color[2][]{}%
  }%
  \providecommand\transparent[1]{%
    \errmessage{(Inkscape) Transparency is used (non-zero) for the text in Inkscape, but the package 'transparent.sty' is not loaded}%
    \renewcommand\transparent[1]{}%
  }%
  \providecommand\rotatebox[2]{#2}%
  \ifx\svgwidth\undefined%
    \setlength{\unitlength}{353bp}%
    \ifx\svgscale\undefined%
      \relax%
    \else%
      \setlength{\unitlength}{\unitlength * \real{\svgscale}}%
    \fi%
  \else%
    \setlength{\unitlength}{\svgwidth}%
  \fi%
  \global\let\svgwidth\undefined%
  \global\let\svgscale\undefined%
  \makeatother%
  \begin{picture}(1,0.52691218)%
    \put(0,0){\includegraphics[width=\unitlength]{hrd.pdf}}%
    \put(0.5827567,0.17800404){\color[rgb]{0,0,0}\rotatebox{-27.33457091}{\makebox(0,0)[lb]{\smash{\acs{zams}}}}}%
    \put(0.12411348,0.36131101){\color[rgb]{0,0,0}\makebox(0,0)[lb]{\smash{$8\,\text{M}_\odot$}}}%
    \put(0.21094934,0.31459136){\color[rgb]{0,0,0}\makebox(0,0)[lb]{\smash{$5\,\text{M}_\odot$}}}%
    \put(0.32383784,0.24642456){\color[rgb]{0,0,0}\makebox(0,0)[lb]{\smash{$3\,\text{M}_\odot$}}}%
    \put(0.4414458,0.19185548){\color[rgb]{0,0,0}\makebox(0,0)[lb]{\smash{$2\,\text{M}_\odot$}}}%
    \put(0.63980539,0.10162111){\color[rgb]{0,0,0}\makebox(0,0)[lb]{\smash{$1\,\text{M}_\odot$}}}%
    \put(0.79417709,0.51340072){\color[rgb]{0.2,0.2,0.2}\makebox(0,0)[lb]{\smash{{\tiny Hayashi
 limit}}}}%
    \put(0.83866969,0.36158134){\color[rgb]{0,0.50196078,0.50196078}\rotatebox{45.29172846}{\makebox(0,0)[lb]{\smash{{\tiny RGB}}}}}%
    \put(0.86956218,0.41973452){\color[rgb]{0,0.50196078,0.50196078}\rotatebox{44.57238044}{\makebox(0,0)[lb]{\smash{{\tiny AGB}}}}}%
    \put(0.37154948,0.34829997){\color[rgb]{0,0.50196078,0.50196078}\makebox(0,0)[lb]{\smash{$0.1$\,Gyr}}}%
    \put(0.5178163,0.27759275){\color[rgb]{1,0,0}\makebox(0,0)[lb]{\smash{$0.3$\,Gyr}}}%
    \put(0.64173422,0.21986867){\color[rgb]{0,0.50196078,0}\makebox(0,0)[lb]{\smash{$1$\,Gyr}}}%
    \put(0.25540494,0.40348361){\color[rgb]{0.50196078,0,0.50196078}\makebox(0,0)[lb]{\smash{$0.04$\,Gyr}}}%
    \put(0.52489239,0.35798902){\color[rgb]{0,0.50196078,0.50196078}\rotatebox{-1.82620708}{\makebox(0,0)[lb]{\smash{{\tiny subgiant branch}}}}}%
    \put(0.28756658,0.33818424){\color[rgb]{0,0.50196078,0.50196078}\rotatebox{18.59949471}{\makebox(0,0)[lb]{\smash{{\tiny MS}}}}}%
    \put(0.11524388,0.30221802){\color[rgb]{0.2,0.2,0.2}\makebox(0,0)[lb]{\smash{{\tiny $10^0$ R$_\odot$}}}}%
    \put(0.11692531,0.42510709){\color[rgb]{0.2,0.2,0.2}\makebox(0,0)[lb]{\smash{{\tiny $10^1$ R$_\odot$}}}}%
    \put(0.32005688,0.48835023){\color[rgb]{0.2,0.2,0.2}\makebox(0,0)[lb]{\smash{{\tiny $10^2$ R$_\odot$}}}}%
    \put(0.73563322,0.31164825){\color[rgb]{0,0,0}\makebox(0,0)[lb]{\smash{\acs{rc}}}}%
    \put(0.67839845,0.39256852){\color[rgb]{0,0.50196078,0.50196078}\rotatebox{0.90272491}{\makebox(0,0)[lb]{\smash{{\tiny core He exhaustion}}}}}%
    \put(0.27659027,0.37469608){\color[rgb]{0,0.50196078,0.50196078}\rotatebox{12.64941882}{\makebox(0,0)[lb]{\smash{{\tiny H shell ignition}}}}}%
    \put(0.70283818,0.133822){\color[rgb]{0.19607843,0.09803922,0.85490196}\rotatebox{0.85802606}{\makebox(0,0)[lb]{\smash{{\tiny H core burning}}}}}%
    \put(0.78674694,0.18397652){\color[rgb]{0.19607843,0.09803922,0.85490196}\rotatebox{0.11273582}{\makebox(0,0)[lb]{\smash{{\tiny H shell 
burning $1$\,Gyr}}}}}%
    \put(0.8213667,0.23896042){\color[rgb]{0.19607843,0.09803922,0.85490196}\rotatebox{0.11273582}{\makebox(0,0)[lb]{\smash{{\tiny He core 
burning
 $0.1$\,Gyr}}}}}%
    \put(0.85511691,0.27951213){\color[rgb]{0.19607843,0.09803922,0.85490196}\rotatebox{43.42616158}{\makebox(0,0)[lb]{\smash{{\tiny He shell 
burning $<1$\,
Myr}}}}}%
    \put(0.805951,0.13140121){\color[rgb]{0,0,1}\makebox(0,0)[lb]{\smash{$9$\,Gyr}}}%
    \put(0.10856255,0.25528072){\color[rgb]{0.30196078,0.30196078,0.30196078}\rotatebox{-63.58752564}{\makebox(0,0)[lb]{\smash{$\leftarrow$}}}}%
    \put(0.11128932,0.20842894){\color[rgb]{0.30196078,0.30196078,0.30196078}\makebox(0,0)[lb]{\smash{ \acs{pn} {\tiny ($\approx10^5$ K)}}}}%
    \put(0.29047645,0.09885773){\color[rgb]{0.30196078,0.30196078,0.30196078}\rotatebox{120.60016943}{\makebox(0,0)[lb]{\smash{$\leftarrow$}}}}%
    \put(0.29373958,0.10044645){\color[rgb]{0.30196078,0.30196078,0.30196078}\makebox(0,0)[lb]{\smash{\acs{wd} {\tiny ($\approx10^{-2}$ L$_\odot$)}}}}%
    \put(0.83161087,0.48838825){\color[rgb]{0,0.50196078,0.50196078}\rotatebox{0.90272491}{\makebox(0,0)[lb]{\smash{{\tiny post
-AGB}}}}}%
  \end{picture}%
\endgroup%

%% file: gfx/chapter1/mw_coordinate.pdf_tex
\begingroup%
  \makeatletter%
  \providecommand\color[2][]{%
    \errmessage{(Inkscape) Color is used for the text in Inkscape, but the package 'color.sty' is not loaded}%
    \renewcommand\color[2][]{}%
  }%
  \providecommand\transparent[1]{%
    \errmessage{(Inkscape) Transparency is used (non-zero) for the text in Inkscape, but the package 'transparent.sty' is not loaded}%
    \renewcommand\transparent[1]{}%
  }%
  \providecommand\rotatebox[2]{#2}%
  \ifx\svgwidth\undefined%
    \setlength{\unitlength}{349.75bp}%
    \ifx\svgscale\undefined%
      \relax%
    \else%
      \setlength{\unitlength}{\unitlength * \real{\svgscale}}%
    \fi%
  \else%
    \setlength{\unitlength}{\svgwidth}%
  \fi%
  \global\let\svgwidth\undefined%
  \global\let\svgscale\undefined%
  \makeatother%
  \begin{picture}(1,0.36070407)%
    \put(0,0){\includegraphics[width=\unitlength]{mw_coordinate.pdf}}%
    \put(0.79173836,0.0377278){\color[rgb]{0,0,0}\makebox(0,0)[lb]{\smash{{\footnotesize R$_\odot$}}}}%
    \put(0.47993501,0.3043445){\color[rgb]{0,0,0}\makebox(0,0)[lb]{\smash{{\footnotesize bulge}}}}%
    \put(0.31292934,0.25850537){\color[rgb]{0,0,0}\makebox(0,0)[lb]{\smash{{\footnotesize thick disc}}}}%
    \put(0.31451868,0.22216171){\color[rgb]{0,0,0}\makebox(0,0)[lb]{\smash{{\footnotesize thin disc}}}}%
    \put(0.30329591,0.19845193){\color[rgb]{0,0,0}\makebox(0,0)[lb]{\smash{{\footnotesize gaseous disc}}}}%
    \put(0.52733354,0.19845193){\color[rgb]{0,0,0}\makebox(0,0)[lb]{\smash{{\footnotesize \ac{smbh}}}}}%
    \put(0.78786536,0.19845193){\color[rgb]{0,0,0}\makebox(0,0)[lb]{\smash{{\footnotesize sun}}}}%
    \put(0.40824078,0.18294906){\color[rgb]{0,0,0}\makebox(0,0)[lb]{\smash{{\footnotesize open cluster}}}}%
    \put(0.18207254,0.28907416){\color[rgb]{0,0,0}\makebox(0,0)[lb]{\smash{{\footnotesize globular cluster}}}}%
    \put(0.78384044,0.33408252){\color[rgb]{0,0,0}\makebox(0,0)[lb]{\smash{{\footnotesize \acs{ngp}}}}}%
    \put(0.78526558,0.05815494){\color[rgb]{0,0,0}\makebox(0,0)[lb]{\smash{{\footnotesize \acs{sgp}}}}}%
  \end{picture}%
\endgroup%

%% file: gfx/chapter1/bulge_new.pdf_tex
\begingroup%
  \makeatletter%
  \providecommand\color[2][]{%
    \errmessage{(Inkscape) Color is used for the text in Inkscape, but the package 'color.sty' is not loaded}%
    \renewcommand\color[2][]{}%
  }%
  \providecommand\transparent[1]{%
    \errmessage{(Inkscape) Transparency is used (non-zero) for the text in Inkscape, but the package 'transparent.sty' is not loaded}%
    \renewcommand\transparent[1]{}%
  }%
  \providecommand\rotatebox[2]{#2}%
  \ifx\svgwidth\undefined%
    \setlength{\unitlength}{594.35bp}%
    \ifx\svgscale\undefined%
      \relax%
    \else%
      \setlength{\unitlength}{\unitlength * \real{\svgscale}}%
    \fi%
  \else%
    \setlength{\unitlength}{\svgwidth}%
  \fi%
  \global\let\svgwidth\undefined%
  \global\let\svgscale\undefined%
  \makeatother%
  \begin{picture}(1,0.25153529)%
    \put(0,0){\includegraphics[width=\unitlength]{bulge_new.pdf}}%
    \put(0.22761701,0.08197793){\color[rgb]{1,1,1}\makebox(0,0)[lb]{\smash{{\footnotesize Baade's window}}}}%
    \put(0.58282827,0.21856478){\color[rgb]{1,1,1}\makebox(0,0)[lb]{\smash{{\footnotesize bulge}}}}%
    \put(0.47045781,0.14644728){\color[rgb]{1,1,1}\makebox(0,0)[lb]{\smash{{\footnotesize Galactic center}}}}%
    \put(0.8968521,0.13206888){\color[rgb]{1,1,1}\makebox(0,0)[lb]{\smash{{\footnotesize dust}}}}%
    \put(0.87739314,0.2121832){\color[rgb]{1,1,1}\makebox(0,0)[lb]{\smash{{\footnotesize $1$\,kpc}}}}%
    \put(0.7517829,0.17001109){\color[rgb]{1,1,1}\makebox(0,0)[lb]{\smash{{\footnotesize long bar}}}}%
    \put(0.04200592,0.18901919){\color[rgb]{1,1,1}\makebox(0,0)[lb]{\smash{{\footnotesize $5^\circ$}}}}%
    \put(0.03576275,0.0822575){\color[rgb]{1,1,1}\makebox(0,0)[lb]{\smash{{\footnotesize $-5^\circ$}}}}%
    \put(0.1777564,0.05523399){\color[rgb]{1,1,1}\makebox(0,0)[lb]{\smash{{\footnotesize $30^\circ$}}}}%
    \put(0.81989212,0.05523399){\color[rgb]{1,1,1}\makebox(0,0)[lb]{\smash{{\footnotesize $330^\circ$}}}}%
    \put(0.42624723,0.00688326){\color[rgb]{0,0,0}\makebox(0,0)[lb]{\smash{{\footnotesize Galactic longitude}}}}%
    \put(0.01731183,0.06358509){\color[rgb]{0,0,0}\rotatebox{90}{\makebox(0,0)[lb]{\smash{{\footnotesize Galactic latitude}}}}}%
    \put(0.50534386,0.05523399){\color[rgb]{1,1,1}\makebox(0,0)[lb]{\smash{{\footnotesize $0^\circ$}}}}%
  \end{picture}%
\endgroup%

%% file: gfx/chapter1/spiral.pdf_tex
\begingroup%
  \makeatletter%
  \providecommand\color[2][]{%
    \errmessage{(Inkscape) Color is used for the text in Inkscape, but the package 'color.sty' is not loaded}%
    \renewcommand\color[2][]{}%
  }%
  \providecommand\transparent[1]{%
    \errmessage{(Inkscape) Transparency is used (non-zero) for the text in Inkscape, but the package 'transparent.sty' is not loaded}%
    \renewcommand\transparent[1]{}%
  }%
  \providecommand\rotatebox[2]{#2}%
  \ifx\svgwidth\undefined%
    \setlength{\unitlength}{1920bp}%
    \ifx\svgscale\undefined%
      \relax%
    \else%
      \setlength{\unitlength}{\unitlength * \real{\svgscale}}%
    \fi%
  \else%
    \setlength{\unitlength}{\svgwidth}%
  \fi%
  \global\let\svgwidth\undefined%
  \global\let\svgscale\undefined%
  \makeatother%
  \begin{picture}(1,0.61708333)%
    \put(0,0){\includegraphics[width=\unitlength]{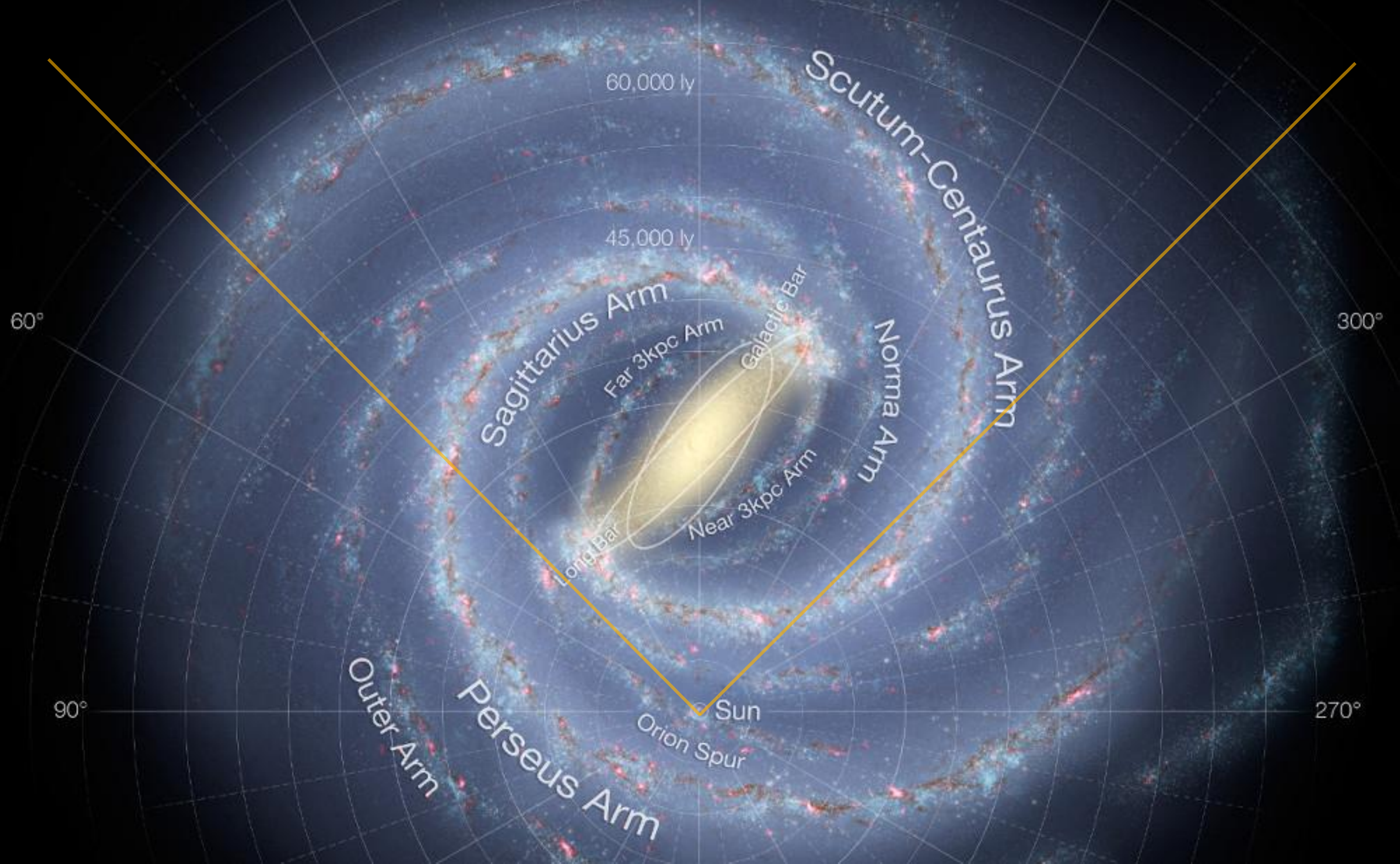}}%
    \put(0.78835653,0.40927018){\color[rgb]{0.76470588,0.61176471,0.23529412}\rotatebox{45}{\makebox(0,0)[lb]{\smash{{\footnotesize viewing angle of
 figure\,
\ref{fig:bulge}}}}}}%
    \put(0.21963982,0.59249005){\color[rgb]{0.76470588,0.61176471,0.23529412}\makebox(0,0)[lb]{\smash{{\footnotesize $30^\circ$}}}}%
    \put(0.78955953,0.59249005){\color[rgb]{0.76470588,0.61176471,0.23529412}\makebox(0,0)[lb]{\smash{{\footnotesize $330^\circ$}}}}%
    \put(0.50633331,0.59249005){\color[rgb]{0.76470588,0.61176471,0.23529412}\makebox(0,0)[lb]{\smash{{\footnotesize $l=0^\circ$}}}}%
  \end{picture}%
\endgroup%

%% file: Chapters/Chapter02.tex
\chapter{Statistical inference with astronomical data}\label{ch:methods}

This chapter is about the interplay of models and data. In section\,\ref{sec:bayesian_inference} a Bayesian framework, to infer model parameters from observations, will be introduced and illustrated with the common problem of inferring distances from parallaxes. The second part will be on the construction of volume-complete stellar samples, in order to infer stellar number densities. It builds up on the results derived in the first section and discusses further complications, as the effect of interstellar extinction or binaries.
\section{Bayesian inference}
\label{sec:bayesian_inference}
In this section the statistical framework will be introduced, which connects models with data in order to handle inference problems. \\
First, conditional probability will be illustrated with the Monty Hall problem. Then Bayes' theorem will be used for parameter inference and the involved steps will be elaborated. The section concludes with a short discussion of parameter space sampling and model comparison.
\subsection{Conditional probability}
\begin{figure}

\caption[Conditional probability]{The conditional probability of the host opening door three when the candidate picked door one. It can be seen how the occurrence of event $H_3$ changes the probability space.}
\centering
\def\svgwidth{\columnwidth}
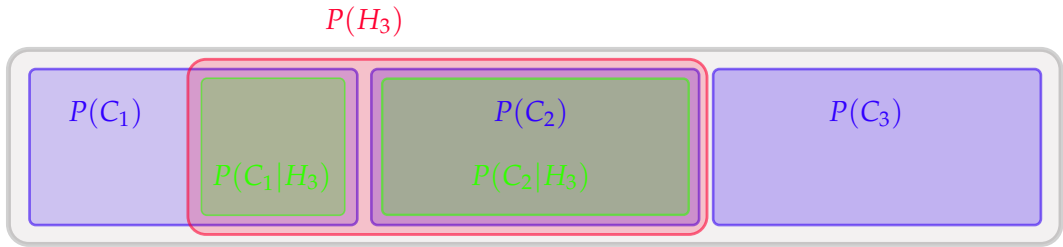
\label{fig:conditional_probability}
\end{figure}

The concept of conditional probability, which will motivate Bayes' theorem, will be introduced by solving the famous \emph{Monty Hall problem}:
\begin{quote}
\emph{Suppose\marginpar{this question to the "Ask Marilyn" column of Parade magazine in 1990 was seeking advice for the common situation a candidate would find him or herself in the show "Let's make a deal", hosted by Monty Hall} you're on a game show, and you're given the choice of three doors: Behind one door is a car; behind the others, goats. You pick a door, say No. 1, and the host, who knows what's behind the doors, opens another door, say No. 3, behind which there is a goat. He then says to you, "Do you want to pick door No. 2?" Is it to your advantage to switch?}
\end{quote}
The questioner is looking for the conditional probability of the car being behind a door, given the host has opened another door with a goat, after the initial choice of the candidate.\\
Before the host opens a door, the probability of the events (C$_i$) of the car being behind door $i\in1,2,3$, are equally likely and sum up to unity,
\begin{equation}
P(C_1\cup C_2\cup C_3) = P(C_1) + P(C_2) + P(C_3) = 1.
\end{equation}
Without loss of generality we follow the path of the question and want to know the probability of winning the car when switching from the chosen door one to door number two, in the event ($H_3$) that the host opens door three with a goat behind it. This is referred to as the conditional probability,
\begin{equation}
P(C_2|H_3) = \frac{P(C_2\cap H_3)}{P(H_3)},
\label{eq:conditional_probability}
\end{equation}
 which is defined as the \emph{intersection} of both events normed by the probability of $H_3$, illustrated as the red rectangle in figure\,\ref{fig:conditional_probability}. Because these two events are not statistically independent, as the host would not open door three when the car is behind it, the conditional probability does not commute.
 \begin{equation}
 P(C_2|H_3)\ne P(H_3|C_2) = \frac{P(H_3\cap C_2)}{P(C_2)}.
 \label{eq:commute}
 \end{equation}
 They can however be related to each other, when replacing the intersection from equation\,\ref{eq:conditional_probability} with the conditional probability from equation\,\ref{eq:commute}.
 \begin{equation}
 P(C_2|H_3) = \frac{P(H_3|C_2)P(C_2)}{P(H_3)}
 \label{eq:small_bayes}
 \end{equation}

 This is a simple form of Bayes' theorem.
  The probability of the host opening door three, $P(H_3)$ can be partitioned with help of the events $C_i$ (because they are mutually exclusive and span the whole probability space)
 \begin{equation}
P(H_3)=P(H_3\cap C_1) + P(H_3\cap C_2) + P(H_3\cap C_3),
 \end{equation}
 which can be transformed, using conditional probability (equation\,\ref{eq:conditional_probability}),
 \begin{equation}
 P(H_3) = P(H_3|C_1)P(C_1)+P(H_3|C_2)P(C_2)+P(H_3|C_3)P(C_3).
 \end{equation}
Plugging this into equation\,\ref{eq:small_bayes} leads to:
\begin{equation}
P(C_2|H_3) = \frac{P(H_3|C_2)P(C_2)}{P(H_3|C_1)P(C_1)+P(H_3|C_2)P(C_2)+P(H_3|C_3)P(C_3)}.
\label{eq:bayes}
\end{equation}
For the right hand side all probabilities can be deduced from the game rules. In the event of $C_1$ the host has free choice, whether to open door two or door three (equation\,\ref{eq:h1}), if $C_2$, he has to chose door three (equation\,\ref{eq:h2}), and if $C_3$ he can not open door three (equation\,\ref{eq:h3}).
\begin{subequations}
\begin{align}
P(H_3|C_1)& = \tfrac{1}{2} \label{eq:h1}\\
P(H_3|C_2)& = 1 \label{eq:h2} \\
P(H_3|C_3)& = 0 \label{eq:h3}
\end{align}
\label{eq:host}
\end{subequations}
Therefore $P(C_2|H_3)=\tfrac{2}{3}$ and $P(C_1|H_3)=\tfrac{1}{3}$, as indicated in the green rectangles of figure\,\ref{fig:conditional_probability}, which doubles the chances to win the car for the changing candidate.\\
\subsection{Parameter estimation using Bayes' theorem}
\emph{Bayes' theorem} states in its general form:
\label{th:bayes}
\begin{quote}
Fix a probability space and an event A. If the set $B_1,...B_k$ are a partition, i.e. mutually exclusive events, spanning the whole probability space, then
\end{quote}
\begin{equation}
P(B_i|A) = \frac{P(A|B_i)P(B_i)}{\sum_{j=1}^k P(A|B_j)P(B_j)}=\frac{P(A|B_i)P(B_i)}{P(A)}.
\end{equation}
It can be facilitated to determine the best parameters ($\theta_i$) for specific models ($M_i$) given some data ($D$),
\begin{equation}
P(\theta_i|D,M_i)=\frac{P(D|\theta_i,M_i)P(\theta_i|M_i)}{P(D|M_i)}.
\label{eq:parameter_bayes}
\end{equation}
The left-hand side is called \emph{posterior} giving the probability distribution over the parameter space for the data assuming the given model $M_i$. Bayes' theorem is used to express the posterior as a product of \emph{prior} and \emph{likelihood}, divided by (what can be seen to be) a normalisation constant. The \emph{prior}, $P(\theta_i|M_i)$, incorporates the a priori knowledge of the possible model parameter values, independent of the data. The \emph{likelihood}, $P(D|\theta_i,M_i)$, representing the probability of the data given the model parameters, can also be referred to as \emph{measurement model}. The denominator, $P(D|M_i)$, is called the \emph{evidence} and gives the probability of observing the data assuming the model to be true. It is given by marginalizing (integrating) the nominator over the parameter space
\begin{equation}
P(D|M_i)=\int_{\theta_i}P(D|\theta_i,M_i)P(\theta_i|M_i)d\theta_i.
\end{equation}
Generally, $D$ and $\theta_i$ are vectors. Since the evidence is not depending on the parameter it can be seen as a normalization constant, $Z$, which is only a scaling factor that does not change the shape of the posterior but will get important when comparing different models in section\,\ref{sec:bayes_factor}.\\
\subsection{An example - Inferring distances from parallaxes}
On the following pages, the properties of the different terms in Bayes' theorem are illustrated using a common astrophysical inference problem presented in \citet{Coryn2015}. The paper is on estimating the true distance, $d$ in pc, from a measured parallax, $\varpi$ in arcsec, and its associated error, $\sigma_\varpi$, for a single star, using a measurement model (likelihood) and prior information about distances. 
\begin{equation}
P(d|\varpi,\sigma_\varpi,M_\varpi)=\frac{1}{Z}P(\varpi|d,\sigma_\varpi,M_\varpi)P(d|M_\varpi)
\label{eq:parallax_model}
\end{equation}
with $Z$ being the evidence for our model 
\begin{equation}
\label{eq:normalisation}
Z=P(\varpi|\sigma_\varpi,M_\varpi)=\int\limits_{d=0}^{d=\infty}P(\varpi|d,\sigma_\varpi,M_\varpi)P(d|M_\varpi)\mathrm{d}d.
\end{equation}
Since the physical model, $M_\varpi$, of the parallax is well understood and we are only interested in the distribution of the true distance, $d$, we will drop $M_\varpi$ in the notation. The $\sigma_\varpi$ is treated as independent data, which is always \emph{given}, be it in the posterior, or in the likelihood, or in the evidence. This is possible because the parallax is independent of its error measurement and the variance of the error is negligible, compared to the parallax error.\\
Here the focus will be on the distance estimation for a \emph{single} star. In section\,\ref{sec:volume_completeness} the inference of distances for a sample of stars will be discussed. The following paragraphs until subsection\,\ref{sec:posterior} are using and summarising content from \citet{Coryn2015}.
\subsubsection{Data}
Data are fundamentally some empirical evidence. In astronomy those are usually observations. It can be used to test or deduce hypotheses.\\
The observation ($D$), used to infer distances from, are parallaxes ($\varpi$), which are angle measurements that will ultimately be acquired from an interplay of optics and photodetectors. In automated surveys like \ac{hip} and \ac{gaia} a model of the optics and electrical signals will interpret the detections and yield two measurements, the parallax, $\varpi$, and the parallax error, $\sigma_\varpi$. These are independent as the error primarily depends on the number of photons arriving at the detector, which in turn depends on the star's apparent magnitude and on the position in the sky through the satellite's scanning law \citep{DeBruijne2012,Arenou1995}. For an ensemble of stars the parallax and its associated error might be correlated since the more distant stars are also more likely to be fainter \citep[fig.\,9]{Luri2014}.\\
Because of the noise, $\sigma_\varpi$, in the parallax measurement, $\varpi$, we can never measure the exact distance of a star but can only derive a posterior \ac{pdf}, $P(d|\varpi, \sigma_\varpi)$, on the possible distances, $d$, given a specific measured parallax and its associated error.\\
Since the error is Gaussian, negative parallaxes occur naturally \citep{Lee1943} and with the inference model it will be possible to deduce valuable information from them (see figure\,\ref{fig:posterior_error} for an illustration).

\subsubsection{Model}
\begin{figure}
\caption[Parallax model]{Geometry of parallax measurement. For $\varpi=1\text{arcsec}\rightarrow d=1\text{pc}\approxeq2\times 10^6 AU$. In the course of $6$\,month the baseline is $2$\,AU with the maximum angular shift being two times the parallax, $\varpi$.}
\centering
\def\svgwidth{\columnwidth}
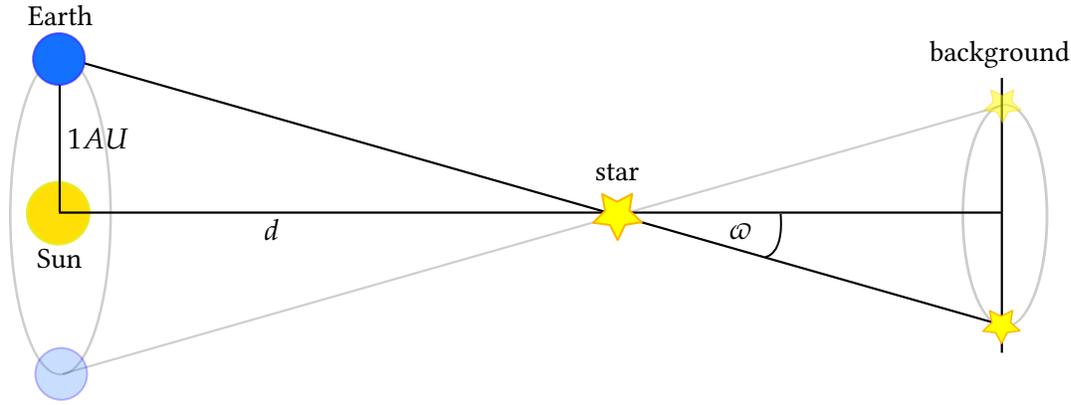
\label{fig:parallax_model}
\end{figure}

The physical model, $M_\varpi$, that we have in mind when deriving true distances, $d$, from measured parallaxes, $\varpi$, is simple geometry and based on the knowledge of Solar system mechanics.\\
A star which is observed will change its position on the sky with respect to a fixed background (e.g. quasars) with Earth's motion around the Sun as indicated in figure\,\ref{fig:parallax_model}. The tangent of this parallax, $\tan(\varpi)$, equals the distance from the Earth to the Sun, $1AU$, divided by the distance from the Sun to the distant star, $d$. Because of the huge distances between stars in the Galaxy the small angle approximation applies and we can write 
\begin{equation}
d^{-1} = \tan(\varpi)\approxeq \varpi
\end{equation}
In reality there are possible complications, for example binaries, but to first-order it is sufficient to account for the star's proper motion and the parallax to incorporate its movement with respect to the background.
\subsubsection{Likelihood}
\begin{figure}
\caption[Measurement distribution]{Probability distribution for true distance of 100\,pc (equivalently a true parallax of 0.01\,arcsec) and a relative error, $\tfrac{\sigma_\varpi}{\varpi}$, of 0.2}
\centering
\def\svgwidth{\columnwidth}
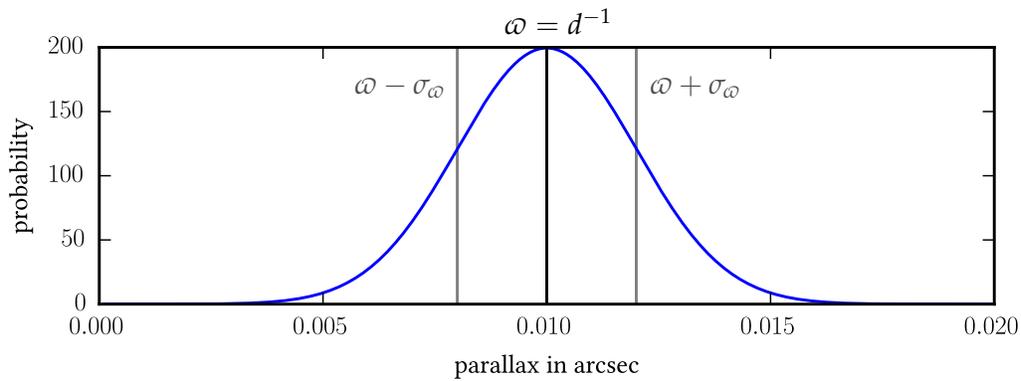
\label{fig:parallax_probability}
\end{figure}
\begin{figure}
\caption[Likelihood over distance]{Likelihood (equation\,\ref{eq:likelihood}) in dashed magenta as a function of distance for a measured parallax $\varpi=0.01$ and a parallax error $\sigma_\varpi=0.002$. Since the likelihood as a function of distance is no longer a probability measure its mode is arbitrarily normalised to the mode of the transformed likelihood \ac{pdf} (equation\,\ref{eq:l_in_d}). For the transformed \ac{pdf} the mode and the mean are given, with both no longer coinciding with the median as in figure\,\ref{fig:parallax_probability}. The inverted 1$\sigma$ error intervals from figure\,\ref{fig:parallax_probability} are also indicated to illustrate the loss of a symmetric error distribution when transforming the likelihood \ac{pdf} from parallax to distance.}
\centering
\def\svgwidth{\columnwidth}
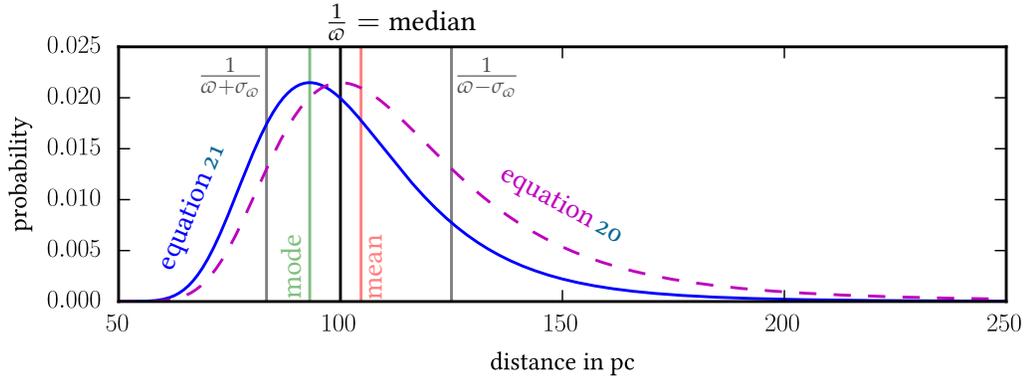
\label{fig:distance_probability}
\end{figure}

Because of the data acquired and the physics involved, a measurement distribution can be deduced for a specific true distance, $d$. The unknown true parallax would be $\tfrac{1}{d}$ and we assume the measurement of $\varpi$ to be normally distributed according to the known positive parallax error\footnote{The error of the noise measurement is neglected (error in apparent magnitude), since its impact on the posterior is small compared to the parallax error.}, $\sigma_\varpi\geq 0$, around the true parallax, $\tfrac{1}{d}$,
\begin{equation}
P(\varpi|d,\sigma_\varpi)=\frac{1}{\sqrt{2{\pi} \sigma_\varpi^2}}\exp\left(-\frac{\left(\varpi-\tfrac{1}{d}\right)^2}{2\sigma_\varpi^2}\right).
\label{eq:likelihood}
\end{equation} 
This is the probability of observing a parallax for a specific true distance with our measurement model. With coordinate transformation we can transform the \ac{pdf} per unit parallax, into a \ac{pdf} per unit distance. Because of the inverse relation, $\varpi=\tfrac{1}{d}$, and the Jacobian, $\tfrac{1}{d^2}$, the resulting probability will no longer be normal but skewed as shown in figure\,\ref{fig:distance_probability}. This probability function in units of distance, $f(d|\varpi,\sigma_\varpi)$, has the median in the same position as in equation\,\ref{eq:likelihood} because the probability transforms monotonically but the mode (peak) and the mean (expected value) are shifted. It can also be seen that the symmetric error in parallax transforms into an asymmetric error in distance
\begin{equation}
f(d|\varpi,\sigma_\varpi)=\frac{1}{d^2\sqrt{2{\pi} \sigma_\varpi^2}}\exp\left(-\frac{\left(\tfrac{1}{d}-\varpi\right)^2}{2\sigma_\varpi^2}\right).
\label{eq:l_in_d}
\end{equation} 
However, this probability function of the distance, $f(d|\varpi,\sigma_\varpi)$, is not helpful in determining the posterior, because it is just the \ac{pdf} of $\varpi^{-1}$.\\

It might be insightful to mention that \citet[sec.2.3]{Francis2013} is drawing samples from $f(d|\varpi,\sigma_\varpi)$ in order to correct for a bias referred to as \emph{parallax bias}, which is attributed to the skewed probability distribution of equation\,\ref{eq:l_in_d} (see figure\,\ref{fig:distance_probability}). A true parallax $\varpi_\mathrm{true}=0.01$ together with a fractional error are assumed. Since the resulting mean is higher than $\varpi_\mathrm{true}^{-1}$ \citet[eq.2.11]{Francis2013} deduces that the distance $d=\varpi^{-1}$ needs to be decreased in order to have a better approximation of the true distance. Beside $\varpi_\mathrm{true}\ne\varpi$, I would argue that starting from a measured parallax $\varpi=0.01$, the inverted distance $d=\varpi^{-1}$ should be increased, since the mean of the \ac{pdf} (over units of distance) is larger than 100\,pc, as depicted in figure\,\ref{fig:distance_probability}.\\
Probably this is more of a philosophical question, but definitely an interesting one, as the corrections are of opposite sign and it also illustrates the different approaches of Bayesian and frequentist statistics. In Bayesian statistics such a bias correction would not be applied, because it is considered in the prior. The prior contains all the information about the scale of $d$, for the purpose of integration, and translates the likelihood into the posterior, which is (after normalisation) a \ac{pdf} over $d$. 
\subsubsection{Prior}
\begin{figure}
\caption[Distance prior]{Normed distance priors}
\centering
\def\svgwidth{\columnwidth}
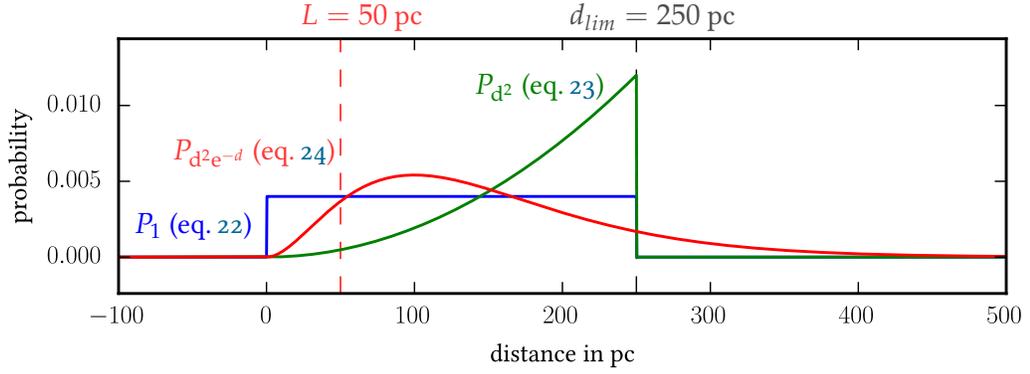
\label{fig:priors}
\end{figure}

The prior, $P(d)$, or $P(\theta_i,M_i)$, incorporates the assumption about the distribution of true distances, before a measurement is obtained (and it can also be updated with new measurements or independent observations). All distances are positive, by definition. In order to normalise the prior we need to use a limiting distance, $d_\mathrm{lim}$ which could be the maximum distance for which we expect to observe a star,
\begin{equation}
P_1(d)=
\left\{ \begin{array}{l}
 \frac{1}{d_\mathrm{lim}}\\
 0
 \end{array} \mbox{\ if \ } 
 \begin{array}{l}
 0<d<d_\mathrm{lim}\\
 d\leq 0\vee d\geq d_\mathrm{lim}. \end{array} \right.
\label{eq:1}
\end{equation}
This flat prior incorporates implicitly the assumption that stellar density varies as $d^{-2}$ because short distances have the same probability as large distances though more stars are expected at larger distances. An improvement is the constant volume density prior,
\begin{equation}
P_{\mathrm{d}^2}(d)=
\left\{ \begin{array}{l}
 \frac{3}{d_\mathrm{lim}^3}d^2\\
 0
 \end{array} \mbox{\ if \ } 
 \begin{array}{l}
 0<d<d_\mathrm{lim}\\
 d\leq 0\vee d\geq d_\mathrm{lim}.  \end{array} \right.
\label{eq:d2}
\end{equation}
This has the highest expectation at $d_\mathrm{lim}$ with no probability of stars at larger distances, which is a discontinuity that results in a bias for measurements with large errors. An improved prior allows for arbitrarily large distances, has no (unphysical) sharp cut-off at $d_\mathrm{lim}$ and should still be normalisable.\\
An easy way to implement that mathematically is an exponentially decreasing volume density prior, which is also physically motivated (our instrument not being able to see faint/distant stars and the components of the Galaxy having exponentially decreasing stellar densities\footnote{In principle the prior could also be dependent on Galactic latitude and Galactic longitude with a more sophisticated Galaxy model using independent data in the background.}), 
\begin{equation}
P_{\mathrm{d}^2\mathrm{e}^{-d}}(d)=
\left\{ \begin{array}{l}
 \frac{1}{2L^3}d^2 e^{-\tfrac{d}{L}}\\
 0
 \end{array} \mbox{\ if \ } 
 \begin{array}{l}
 d>0\\
 d\leq 0.\end{array} \right.
\label{eq:d2eL}
\end{equation}
with $L>0$ being a length scale though the mode of the prior\,\ac{pdf} occurs at somewhat larger true distance values, as can be seen in figure\,\ref{fig:priors} where all priors are depicted. This is just a short summary of possible priors for this inference problem, which is elaborated in greater detail in \citet{Coryn2015}.\\

The exponentially decreasing volume density prior, $P_{\mathrm{d}^2\mathrm{e}^{-d}}$, has the favourable properties of being simple, physically motivated, continuous and normalisable. This does not mean that it is necessarily the best prior, but without better knowledge about the distribution of stars or the survey selection function, it will yield robust estimates for the distance of a star, even when its associated parallax error is large.\\
\subsubsection{Evidence}
\label{sec:Evidence}
In the distance inference problem the evidence, $P(\varpi|\sigma_\varpi)$, is hidden as the normalisation constant, $Z$, because a preferred model, M$_\varpi$, already exists and the unnormalised posterior already yields the probability distribution of the true distance. But when interested in calculating the evidence for model comparison, the likelihood and the prior would have to be marginalised over all possible distances (equation\,\ref{eq:normalisation}) and compared to the evidence of other models.\\
\subsubsection{Posterior}
\label{sec:posterior}
\begin{figure}
\caption[Posterior]{The prior, in green (with a length scale of $L=100$\,pc), multiplied by the likelihood, in blue, yields the unnormalised posterior over the distances, in red, for $\varpi=0.01$ and $\sigma_\varpi=0.002$. For the posterior the median and $90\,\%$ confidence intervals are indicated with red dashed lines. Only the posterior and the prior are probability measures and integrate to unity contrary to the likelihood over distance. Therefore the mode of the likelihood, marked by the dashed blue line, is normalised to the mode of the posterior, indicated in dashed magenta.}
\centering
\def\svgwidth{\columnwidth}
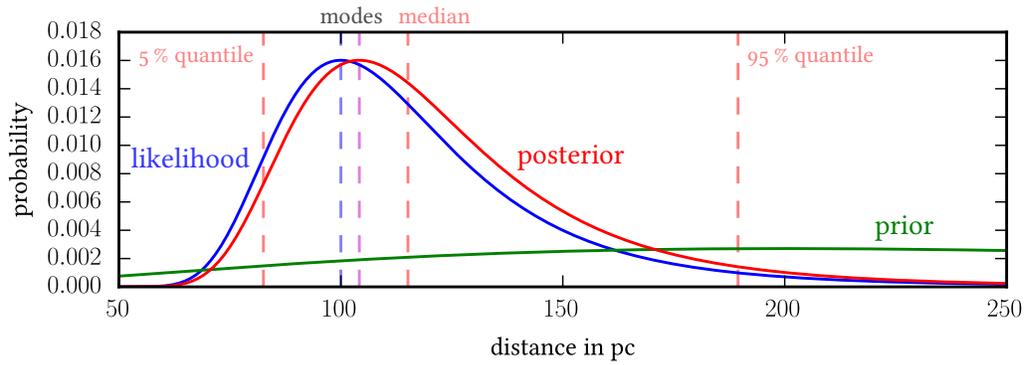
\label{fig:posterior}
\end{figure}

\begin{figure}
\caption[Posterior for negative parallaxes]{Likelihood in blue and posterior in red for negative parallax, $\varpi=-0.01$, and different relative parallax errors, $\tfrac{\sigma_\varpi}{\varpi}=10\,\%$ in dashed lines and $40\,\%$ in solid lines. The prior in green stays unaffected. What is visible of the likelihoods as functions of distance is arbitrarily normalised to unity.}
\centering
\def\svgwidth{\columnwidth}
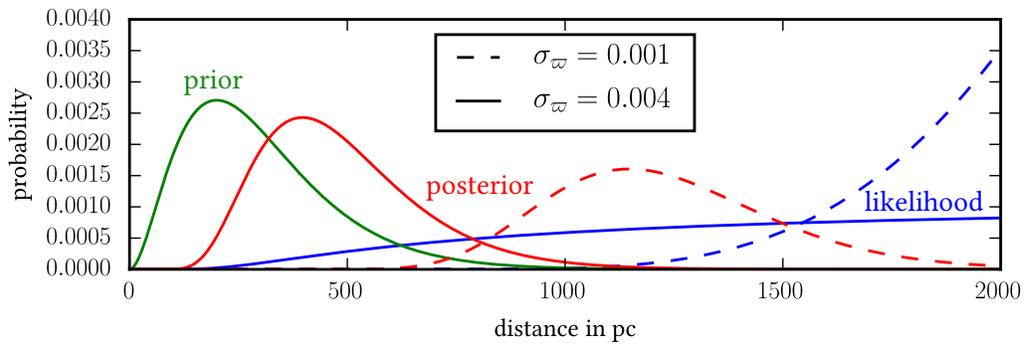
\label{fig:posterior_error}
\end{figure}

After obtaining the likelihood, $P(\varpi|d,\sigma_\varpi)$, and choosing a prior, $P_{\mathrm{d}^2\mathrm{e}^{-d}}(d)$, the posterior, $P(d|\varpi,\sigma_\varpi)$, can be calculated. By multiplying the likelihood, which is a probability measure of $\varpi$, with the prior, which contains all the a priori information about the scale of $d$, the posterior is yielded, which is an unnormalised \ac{pdf} over distance. After normalisation with the evidence (which is not necessary, if only interested in the relative probability distribution of the true distance) the posterior is the (relative) probability of different true distances given a specific parallax measurement and its associated error.\\
In figure\,\ref{fig:posterior}, the posterior is depicted for a parallax measurement $\varpi=0.01$, associated error $\sigma_\varpi=0.002$, and a length scale $L=100$\,pc for the prior $P_{\mathrm{d}^2\mathrm{e}^{-d}}(d)$. Beware that naively inverting the parallax would yield 100\,pc. Compared to the median of the posterior $d\approx$115\,pc, this is significantly less, which is mainly due to the error distribution not being taken into account.\\
As a summary of the posterior \ac{pdf}, the mode or the median can be reported together with associated errors. Because in our case the posterior is asymmetric, it is preferable to give confidence intervals (as shown in figure\,\ref{fig:posterior}) instead of a standard deviation.\\
In figure\,\ref{fig:posterior_error} the likelihood and the posterior are displayed for a negative parallax of $\varpi=-0.01$ and two different relative parallax errors of $\tfrac{\sigma_\varpi}{\varpi}=0.1$ and $0.4$. First, it should be noticed that, as a function of distance, the likelihood is no longer a probability measure. However, even though it is increasing steadily towards infinity (for negative parallaxes), the product with the prior yields a reasonable posterior distance distribution. Secondly, for larger parallax errors the likelihood in solid blue is broader and its corresponding posterior resembles more the prior, which makes sense as the measurement is less definite, so the prior knowledge is weighted stronger in the posterior \ac{pdf}. And lastly, for increasing precision the negative parallax measurement indicates increasingly low true parallaxes, which shifts the posterior, in dashed red, to much higher distance values in figure\,\ref{fig:posterior_error}.
\subsection{Sampling the parameter space}
For models involving many parameters the evaluation of the posterior over the whole parameter space will become computationally expensive. In order to sample the posterior distribution efficiently \ac{mcmc} algorithms can be facilitated. For an increasing number of steps, their sample density will converge towards the posterior probability density. The commonly used Metropolis-Hastings algorithm achieves this by doing a probability-guided random walk on the parameter space. It rejects proposed steps with a certain probability when the outcome is lower compared to the previous step, and accepts those steps that increase the outcome. Depending on the initial position in the parameter space, it can take a while until the algorithm finds the volume of highest density. This phase is referred to as burn-in and steps during that process should be thrown away, in order not to bias the inferred \ac{pdf}. Once it has reached the volume of highest density, it will sample this space and the sampling density will converge towards the \ac{pdf}. In order to test if this is the global maximum and not only a local maximum, the starting points should be varied and it should be checked whether each such Markov chain is equilibrating in the same volume of highest density.\\
With multiple chains exchanging step information the convergence towards the \ac{pdf} can be accelerated and the probability evaluation of single steps can be distributed on different processors for each chain, which makes the calculation much faster on multi-core systems, especially when single probability evaluations take a lot of time. We will use an efficient algorithm with Python implementation in chapter\,\ref{ch:imf}.\\

Sometimes the likelihood is impossible to construct or computationally not feasible to evaluate. Still in astrophysics it is usually possible to simulate mock data sets through forward modelling. In that case an \ac{abc} \ac{mcmc} method can be used, in order to approximate the posterior distribution. This is achieved by quantifying the differences of an observed data set, to simulated ones via distance metrics, which are usually applied on summary statistics of the data sets (see \citet{Robin2014} for an example). Summary statistics can be bins or quantiles of a distribution, which then can be measured using Euclidean distance or more sophisticated metrics. To compare one-dimensional cumulative distributions the Anderson-Darling test \citep{Anderson1952} can also be used as a minimum-distance-estimate between two data sets, and should be preferred to the (less discriminative but more often used) Kolmogorov–Smirnov test.
\subsection{Model comparison}
\label{sec:bayes_factor}
When interested in the probability of a specific model given the data, $P(M_i|D)$, the evidence ($P(D|M_i)$) from equation\,\ref{eq:parameter_bayes} needs to be evaluated. With Bayes' theorem it follows that
\begin{equation}
P(M_i|D)=\frac{P(D|M_i)P(M_i)}{P(D)}.
\label{eq:HypoT}
\end{equation}
Using exhaustive and mutually exclusive models, $M_j,j\in 1...k$, the evidence for the data can be rewritten as
\begin{equation}
P(D)=\sum_{j=1}^k P(D|M_j)P(M_j).
\label{eq:span}
\end{equation}
Unlike in the simple Monty Hall problem, see equation\,\ref{eq:bayes}, it usually is unfeasible to find an appropriate set of models covering all possibilities. Since $P(D)$ is a model-independent value (i.e. the same for all models) two models, $M_1,M_2$, can be compared to each other, by the ratio of their probability given the data, $P(M_i,D)$ (equation\,\ref{eq:HypoT}). Under the assumption that no model is a priori preferred, $P(M_1)=P(M_2)$ (i.e. they cancel out in equation\,\ref{eq:bayes_factor}) this ratio is called \emph{Bayes factor}, $R$, and involves the integration of the unnormalised posterior (i.e. likelihood, $P(D|\theta_i,M_i)$, times the prior, $P(\theta_i|M_i)$,) over the whole parameter space, $\theta_i$
\begin{equation}
\label{eq:bayes_factor}
R = \frac{P(M_1|D)}{P(M_2|D)}= \frac{P(D|M_1)\cancel{P(M_1)}}{P(D|M_2)\cancel{P(M_2)}}=\frac{\int_{\theta_1} P(D|\theta_1,M_1)P(\theta_1|M_1)d\theta_1}{\int_{\theta_2} P(D|\theta_2,M_2)P(\theta_2|M_2)d\theta_2}.
\end{equation}
As can be seen, the Bayes factor is comparing the ability of different models to explain a common dataset. Because it involves marginalising over the whole parameter space it inherently punishes complex models with many parameters, which will not be favoured for having a high maximum likelihood at a specific set of parameters (very small volume of the parameter space) but for their ability to explain the data over the whole parameter space\footnote{More complex models should be motivated by more complex data.}. Since the priors for the parameters, $P(\theta_i|M_i)$, are involved in the calculation of the evidence, $P(D|M_i)$, the sensitivity of the Bayes factor to different priors should be tested. Due to this and also the assumption of equal model priors, the preference of one model over the other should only be stated for sufficiently\footnote{What \emph{sufficient} means is subjective and different authors state different values, $R>10$ is an indication for the preference of one model over the other.} large $R$. Given a large Bayes factor it can be concluded that one model is better than the other, but in order to state that \emph{the} best model is found, the evidence has to be calculated for all possible models, or at least all plausible models.\\
\section{Stellar number densities from star counts}
\label{sec:volume_completeness}
In Galaxy modelling, an important parameter is the stellar mass density. To derive this quantity, a complete census of stars in a given volume is needed. In order to attribute stars to a specific volume accurate distances are required. As was argued in the previous section, parallaxes are well suited to infer distances from. Parallax measurements are most reliable for close-by stars. The best set of parallaxes to date was measured during the \ac{hip} survey. Keep in mind that the methods described in section\,\ref{sec:bayesian_inference} do not allow the derivation of exact distances, but only a posterior probability distribution, whose width scales with the parallax errors. Another complication arises because stars of the same \ac{vmag} appear fainter the further they are away. At a certain distance they are no longer visible to the detectors depending on the apparent magnitude limit of the survey. 
\begin{figure}
\caption[Apparent magnitude limit]{The \ac{avmag} limit is plotted in the luminosity distance plane for the \ac{hip} 100\,\% completeness limit of $7\,\ac{avmag}$. The distance is given in pc and the luminosity in Solar luminosity, L$_\odot$. The relation to absolute V magnitude, $\ac{vmag}$, is indicated on the left in magenta. It should be noted that the stellar density indicated in the figure represents a line of sight view for a homogeneous stellar distribution. If considering the whole volume, within a specific distance, the number of stars should increase with the distance squared.}
\centering
\def\svgwidth{\columnwidth}
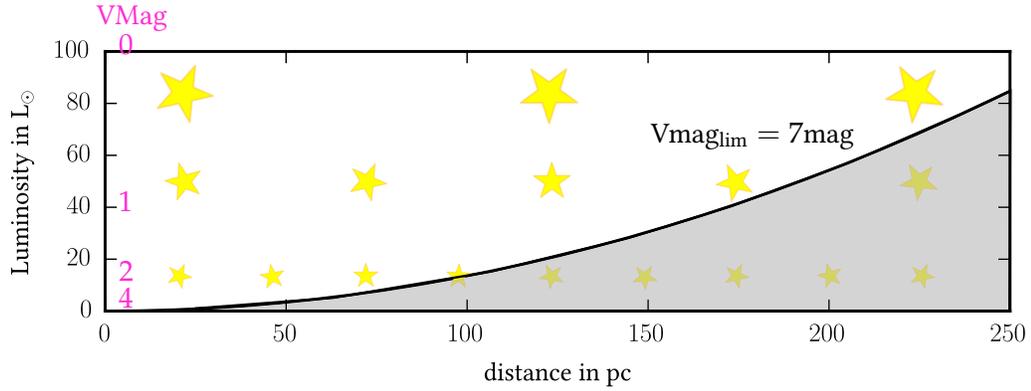
\label{fig:inverse_square_law}
\end{figure}
In figure\,\ref{fig:inverse_square_law} this situation is depicted for the \ac{hip} survey, which is only complete for stars brighter than \ac{avmag}$_\text{lim}=7$mag\footnote{Though \ac{hip} has no sharp cut-off, but rather a slow decline in completeness.}. This implies that a Sun-like star with \ac{vmag}$=4.83$\,mag will be seen to a distance modulus, $\mu$, of 
\begin{equation}
\mu\equiv Vmag-VMag=7-4.83=2.17,
\label{eq:distance_modulus}
\end{equation}
where $\mu$ is a way of expressing distances in magnitudes. Equation\,\ref{eq:distance_modulus} can be translated to a distance,
\begin{equation}
d = 10^{1+\tfrac{\mu}{5}}\approx 27pc,
\end{equation}
corresponding to a limiting distance until which the Sun would be visible for a survey with a magnitude limit of $Vmag=7$. Brighter stars, e.g. Vega with $VMag=0$, will be visible up to a distance of approximately 250\,pc. This means that for stars with different intrinsic brightnesses the limiting distances, up to which they will be observable, are also different, as shown by the inverse square law (black curve) in figure\,\ref{fig:inverse_square_law}. Because of this the inferred mean luminosity of the visible stars in figure\,\ref{fig:inverse_square_law} will be overestimated, which is referred to as Malmquist bias.\\

In the following sections it is explained how to obtain volume-complete samples of stars, beginning with the approach used in \ac{paper1}. The methods in sections \ref{sec:binary_splitting} and \ref{sec:prob_distances} have been implemented by the author. The sections following thereafter can be seen as ideas or advises that could further increase the accuracy of stellar number density determinations. This is why the last sections are less elaborate and of a more conceptual nature.
\subsection{Magnitude and distance cuts}
\begin{figure}
\caption[Distance cuts]{Same as figure\,\ref{fig:inverse_square_law} but now the \acl{avmag} limit is plotted in the \acl{vmag} over distance plane. $7\,\ac{avmag}$ is the $100\,\%$ completeness limit of the \ac{hip} catalogue. The distance cuts for the different absolute magnitude bins are indicated as boxes, it can be seen that the first magnitude bin has an relatively conservative distance cut. For reference the luminosity scale is given in magenta on the left.}
\centering
\def\svgwidth{\columnwidth}
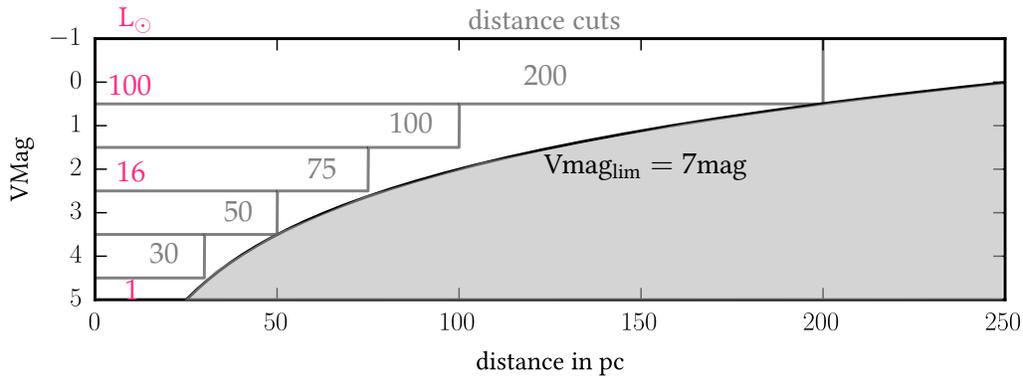
\label{fig:distance_cuts}
\end{figure}
The method of distance and absolute magnitude cuts was applied in \ac{paper1}. They used the vertical velocity distributions of volume-complete \ac{ms} star samples, as observational constraint for their dynamical model. The local mass density, however, was taken from \citep{Ja97}.\\
In their approach distances are derived directly from inverted parallaxes, $d=\tfrac{1}{\varpi}$, and stars with relative parallax error larger than 15\,\% are excluded, $\tfrac{\sigma_\varpi}{\varpi}>0.15$, because their distance error becomes too large to assign them reliably to a volume\footnote{The number of stars, that do not fulfil these quality cuts, is usually below $2$\,\% for stars brighter than the limiting \acl{avmag} for completeness in the \ac{hip} catalogue. Beware that stars with higher relative parallax errors are more likely to have faint apparent magnitudes and large distances. So a bias is introduced when throwing out these stars.}. In a next step \acl{vmag}s are calculated from the measured distances and apparent magnitudes. Stars are binned in \ac{vmag} and each bin has a specific distance cut, such that \ac{avmag} of the stars never fall below the \ac{avmag}-limit for which the \ac{hip} survey is complete, Vmag$=7$mag, as illustrated in figure\,\ref{fig:distance_cuts}.
\subsection{Binary splitting and analytic extinction correction}
\label{sec:binary_splitting}
In \citet{Rybizki2015} we used volume-complete samples as observational constraint to infer \ac{imf} parameters from the \ac{jj}. The method is basically the same as in the previous subsection but with the additional correction for binary systems and extinction.\\

\begin{figure}
\caption[Extinction and binary correction]{A zoom-in of figure\,\ref{fig:distance_cuts} is shown. The stars' \acl{vmag} is shown on the left y-axis and the extinction in V-Band on the right which is referring to the red curves illustrating our adopted extinction model. The upper red curve is the extinction for stars in the Galactic plane and the lower red curve has no extinction referring to stars with Galactic latitude larger $52^\circ$. The extinction correction is illustrated for a star in the Galactic plane with a distance of approximately $175$\,pc. On the left hand-side at about $55$\,pc the effect of splitting up an equally bright binary is shown.}
\centering
\def\svgwidth{\columnwidth}
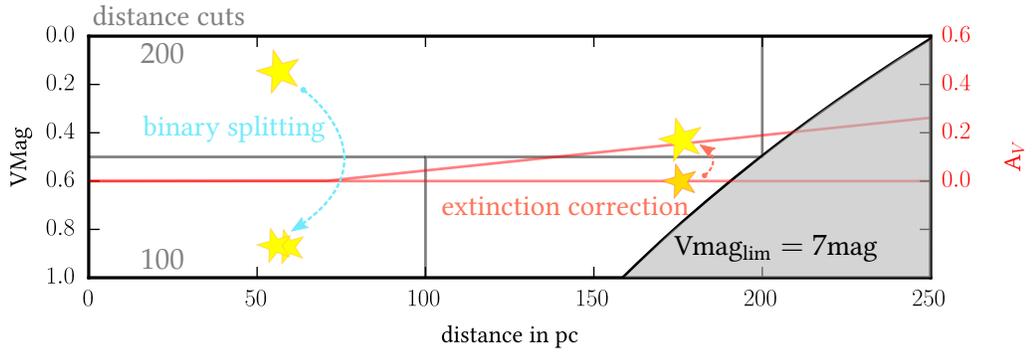
\label{fig:extinction}
\end{figure}

The known binary systems, which have a magnitude difference entry in the \ac{hip} input catalogue, were split up so that the formerly unresolved stellar systems changed their magnitudes. This can be seen in figure\,\ref{fig:extinction} where it is illustrated for an equal mass binary system with a \ac{vmag}\,$\approx0.15$ at about $55$\,pc. When splitting the system we now have two stars each being half as bright as the combined binary system with approximately $\ac{vmag}\approx0.75$ larger magnitude. As a consequence the unresolved binary system is lost from the $200$\,pc volume and two stars are added to the $100$\,pc volume. This is because the binary system in the depicted example of figure\,\ref{fig:extinction} is shown for a distance smaller than $100$\,pc. It should be mentioned that if the binary distance was between $100$ and $200$\,pc then it would be lost from the $200$ and the $100$\,pc volume. This happens more often than the case of figure\,\ref{fig:extinction}, because the volume between $0$ to $100$\,pc distance is seven times smaller than the volume between $100$ and $200$\,pc distance.\\
The changes in star counts for each magnitude bin after correcting for binaries are on the two percent level (see table\,\ref{tab:corrections}). This does not reflect the true magnitude of the effect, because binaries with a massive primary are hard to resolve, due to blurred lines from fast rotation, and are most likely not listed in the \ac{hip} input catalogue and hence excluded from our correction.\\

For the extinction correction (dereddening) we used an analytical homogeneous dust model based on \citet{Ve98} with no dust in the local bubble up to $70$\,pc distance and a constant dust density within $55$\,pc distance from the Galactic plane. The extinction is depicted as red lines in figure\,\ref{fig:extinction}. For stars above $52^\circ$ it is zero (the lower red line) and for stars in the Galactic plane it linearly increases for stars with distances above $70$\,pc and reaches a maximum value of $0.26$\,A$_V$ at $250$\,pc (the upper red curve). In figure\,\ref{fig:extinction} the extinction correction is illustrated for an hypothetical star with $\ac{vmag}=0.6$ (which was calculated from the \ac{avmag} and the distance without dereddening) and a distance of $175$\,pc in the Galactic plane. Because the true \ac{vmag} would be around $0.15$\,mag brighter we increase its \ac{vmag} by that amount, so that the star will be associated to the right \ac{vmag} bin, when applying the magnitude and distance cuts. In our example, as depicted in figure\,\ref{fig:extinction}, the star would not be included into our volume-complete sample without extinction correction. Since its true \ac{vmag} is below $0.5$\,mag including it into the $200$\,pc sphere is more accurate.\\
Overall, the star counts only change for the $200$\,pc sphere, however it should be noted that the $10$\,\% increase is much more significant than the binary correction. The assumption of a homogeneous dust distribution is inaccurate, at least when considering the Solar Neighbourhood. Therefore extinction data should be preferred over simple analytical models when correcting the light absorbing effect of interstellar dust.\\

In order to get a local number density from the star counts in different \ac{vmag} bins we divided each magnitude bin with the corresponding volume, $\tfrac{4}{3}\pi \mathrm{d}_{_{\mathrm{lim}}}^3$ (cf. equation\,\ref{eq:local_model} and figure\,\ref{fig:SFH}), and also correct for the scale height dilution (young stars being confined to the Galactic plane because they are formed from the gaseous disc which has a very low scale height). The correction for scale height dilution introduces a model-dependency, but otherwise the large $200$\,pc volume can not be correctly interpreted in terms of local stellar number densities. On larger scales, structures like the spiral arms or the radial density distribution of stars across the Galactic disc, need to be taken into account as well.
\subsection{Probabilistic distances}
\label{sec:prob_distances}
As argued in section\,\ref{sec:bayesian_inference}, parallaxes are very useful measurements, when inferring distances, because they make no assumptions about the intrinsic properties of the observed object. Still, inferring distances can only be done probabilistically and the error is also propagated into the \acl{vmag}\footnote{The error of the measured \acl{avmag} should also be propagated into the distance error, however it is so small that its effect on the distance error can be neglected compared to the parallax error.}. Since distance and \ac{vmag} are used to limit the star counts in the different volume-complete samples, it is important to account for the parallax error. The following method was used in \citet{Just2015} to assess the systematic error that was made compared to the previous two methods.\\

\begin{figure}
\caption[Distance sampling]{$100$ realisations of two hypothetical stars with measured parallax of $0.01$, associated error and \acl{avmag}, are shown in the \acl{vmag} and distance space. The blue star has a \ac{avmag} of $5$\,mag and a relative parallax error of $10\%$. The yellow stars is one magnitude fainter and its error is twice as large. The posterior \ac{pdf} from which the samples are drawn were calculated using the exponentially decreasing volume density prior, $P_{\text{d}^2\text{e}^{-d}}$ from equation\,\ref{eq:d2eL}, with a length scale of $L=100$.}
\centering
\includegraphics[width=\textwidth]{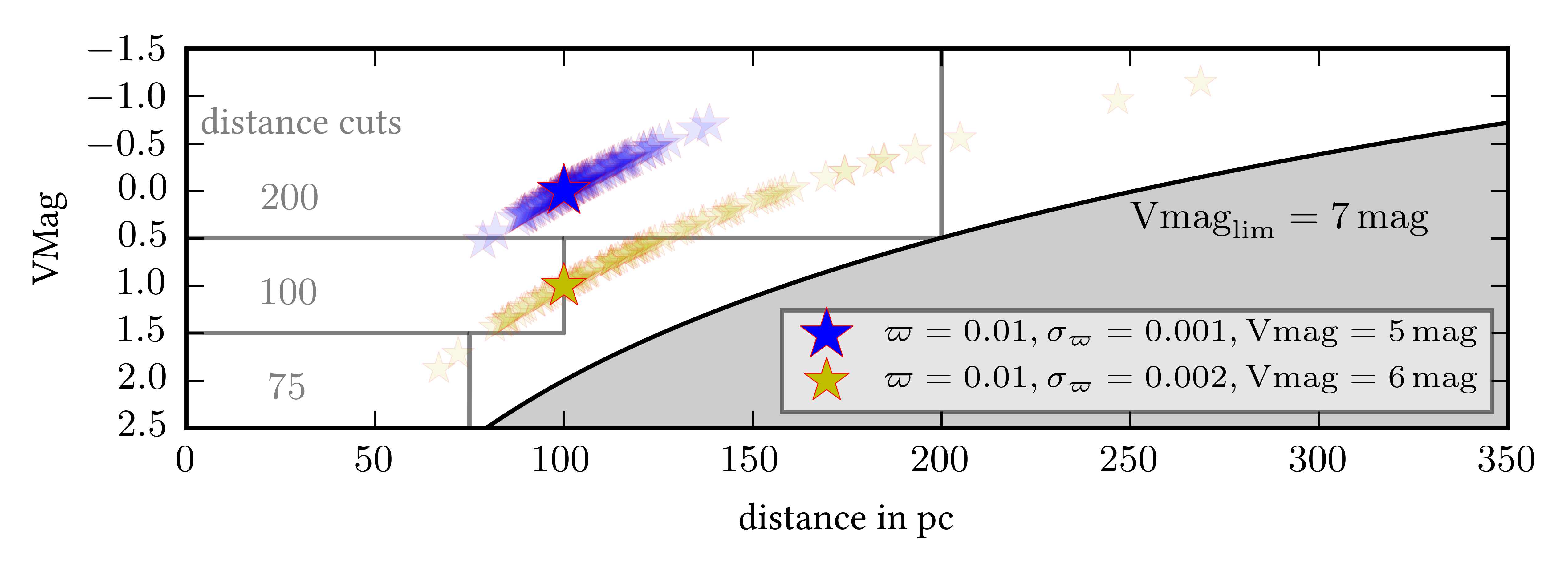}
\label{fig:sampling}
\end{figure}

First, the posterior distance distribution is calculated for each star in the \ac{hip} catalogue from the measured parallax and its associated error as described in section\,\ref{sec:bayesian_inference}.\\
Then, in order to propagate the error to \acl{vmag}, an appropriate number (we used $100$) of sample distances is drawn from the posterior.\\
From each star's \acl{avmag} and its $100$ sampled distances $100$ associated \ac{vmag}s are calculated (as depicted in figure\,\ref{fig:sampling} for two hypothetical stars with a measured parallax of $0.01$, associated relative parallax errors of $10$ and $20$\,\% and \acl{avmag}s of $5$ and $6$\,mag).\\
In the next step the distance and \acl{vmag} cuts are applied to the $100$ times oversampled \ac{hip} catalogue. Figure\,\ref{fig:sampling} illustrates that for stars with high relative error (like the yellow star) the $100$ realisations are distributed across several magnitude bins with significant fractions falling into the second, first, zeroth \acl{vmag} bin and considerable parts are even outside of the sample, either because they are too faint and (or) too far. In this context it is important to note that even for stars with relatively small relative error (blue star) the borders to other magnitude bins are easily crossed.\\
In a last step the star counts of each magnitude bin are divided by $100$ in order to normalise the star counts to the correct level.\\

The advantage of this method is that the sampled probability distribution of all stars is considered and weighted into the final result. There is no need to exclude stars with high fractional parallax error. Also the \citet{Trumpler1953} bias\footnote{The bias is due to the error volume outside of the limiting distance being larger than inside. So more stars are scattered into the sample, if we assume constant stellar density, than out of it.} is accounted for, because the error contribution from all stars of the \ac{hip} catalogue (outside and inside the limiting distances) is sampled, and the catalogue does not have a sharp apparent magnitude limit\footnote{Otherwise, not many stars outside of the distance limit will be included in the catalogue data, which is why we encourage conservative distance cuts.}.\\
This method can of course be combined with the binary and extinction correction from the previous subsection, but we will now show improved methods to account for these two effects.  
\subsection{Using inhomogeneous extinction data}
\label{sec:inhomogeneous_extinction}
\begin{figure}
\caption[Inhomogeneous extinction]{This is a zoom-in of figure\,\ref{fig:distance_cuts} with an additional extinction scale on the right indicating the effect of two dust clouds at $130$\,pc and $180$\,pc in the line of sight between the observer and a hypothetical star of $0.3\,\ac{vmag}$ at $190$\,pc. The first, smaller, cloud has an extinction of $A_V=0.2\,\text{mag}$ and the second an extinction of $A_V=0.3\,\text{mag}$. Their cumulative effect on the line of sight extinction is shown in dashed red and on the decreased volume covered by the magnitude limit in dashed black. The star would usually be bright enough to be seen by the survey and included into the $200$\,pc sample. Because of the strong extinction its apparent magnitude is above $\ac{avmag}_\text{lim}=7\,\text{mag}$ and therefore no longer in the $100\,\%$ completeness limit of the survey.}
\centering
\def\svgwidth{\columnwidth}
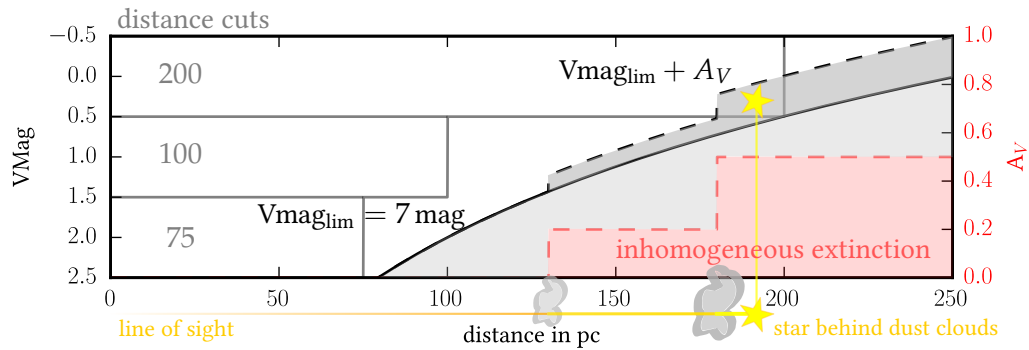
\label{fig:inhomogeneous_extinction}
\end{figure}

Extinction maps have long been available \citep{Schlegel1998}, but in 3D only recently high resolution maps were published. An early attempt was a 3D map from \citet{Marshall2006} but they were interested mainly in the bulge so the sky coverage is confined to a small stripe along the Galactic plane and the local distance resolution is too coarse to be useful for our assessment. In \citet{La14} excellent 3D maps with high local resolution were presented but data were not published yet. A very good extinction data base was released recently by \citet{Green2015} and put up online. It uses the \ac{ps1} photometry of $800$\,million stars and matches \ac{2mass} photometry of $200$\,million stars. The \ac{ps1} survey covers three quarters of the sky with major parts of the Galactic plane (with most of the extinction) being included.\\
When combining this method with the probabilistic distances, the \acl{vmag} of every realisation needs to be dereddened separately. Similarly, \citet{Green2015} is not giving single estimates of the extinction in a specific distance bin but samples of $100$ or $500$ from their posterior \ac{pdf}.\\
In figure\,\ref{fig:inhomogeneous_extinction}, the effect of light-absorbing dust clouds is illustrated on the visibility of a star. Usually, the hypothetical star of $0.3\,\ac{vmag}$ would be visible out to $200\,$pc distance for a magnitude limited survey with $\ac{avmag}_\text{lim}=7\,\text{mag}$. But because the light of the star is weakened, additionally to the inverse square law, by the extinction within dust clouds, the visibility volume for stars with $0.3\,\ac{vmag}$ shrinks below $180$\,pc, right before the second dust cloud. This loss of volume, which depends on the line of sight extinction (can be different in all directions), should be accounted for when determining stellar densities.\\
Another effect is indicated (by the reddened colour of the light ray) in the sketch at the bottom of figure\,\ref{fig:inhomogeneous_extinction}, namely the differential extinction of dust \citep[pp.187]{Scheffler1992}. Shorter wavelength are preferentially scattered which is why light retains a larger fraction of red light when passing through a cloud of dust. This is why sometimes extinction and reddening are used synonymously in astronomy.
\subsection{Probabilistic binary correction}
If we want to infer the stellar number density, with binaries resolved into individual stars, as opposed to the system number density, we need to correct for binaries. Once they are known and the magnitude difference is determined, we can split them up as mentioned in subsection\,\ref{sec:binary_splitting}. The problem, however, is that for O/B-stars the detection rate of binaries is very low (because of their fast rotation their spectra are broadened) though the binary rate itself is almost $100\,\%$.\\
In this case again, we can use a probabilistic approach for stars where binaries are hard to detect. We can assign probabilities to a star of being an undetected binary (or even higher multiple stellar system) and sample the probability space of their mass ratios. When assuming that they have the same age and metallicity, isochrones can be used to assign new \ac{vmag}s and $B-V$ colours.\\
Of course these additional samples increase the data volume we are working with. For \ac{hip} this might still be feasible, for \ac{gaia} it will probably not. Maybe a median effect of binaries on the stellar densities can be determined probabilistically and then corrected for with simpler methods. Forward modelling could also be more feasible (section\,\ref{sec:forward_modelling}), where binaries are modelled and the data is matched with synthesised observations.
\subsection{Relaxing completeness}
\begin{figure}
\caption[Completeness]{Zoom-in of figure\,\ref{fig:distance_cuts} now with different \acl{avmag} limits and the corresponding completeness in percent as calculated for \ac{hip} compared to \ac{tycho2}.}
\centering
\def\svgwidth{\columnwidth}
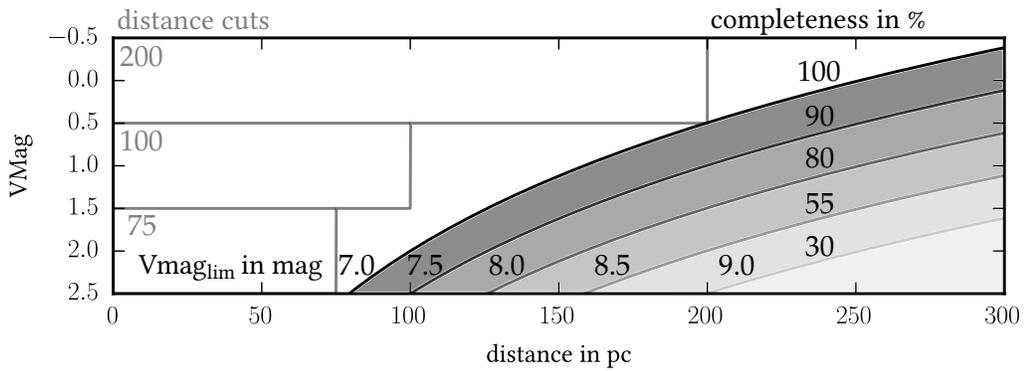
\label{fig:completeness}
\end{figure}
When comparing \ac{hip} and \ac{tycho2}\footnote{Which is complete up to 11\,\ac{avmag}.} star counts in bins of \ac{avmag}, the completeness according to \ac{hip} for different apparent magnitudes can be determined. As displayed in figure\,\ref{fig:completeness} the \ac{hip} catalogue is complete until $\ac{avmag}=7\,\mathrm{mag}$ (and not as the quoted $\ac{avmag}=7.3-9\,\mathrm{mag}$ from \citet{Pe97}), slowly declining to $80\,\%$ completeness at $8\,\ac{avmag}$, and going down to $30\,\%$ at $9\,\ac{avmag}$\footnote{With a slight dependence on colour and Galactic latitude, but not as strong as claimed by \citet{Pe97}.}. When extrapolating star counts up from $55\,\%$, corresponding to $\ac{avmag}=8.5\,\mathrm{mag}$, the limiting distance is doubled compared to the $100\,\%$ completeness, which results in an eightfold volume coverage. Of course this gain will only be possible where there is no extinction. In reality the distance-completeness depends on the extinction, as was explained in section\,\ref{sec:inhomogeneous_extinction} and can be seen in figure\,\ref{fig:inhomogeneous_extinction}. A hybrid method could be used to account for the distorted visibility volume, due to interstellar extinction, and also to increase the magnitude limit, by extrapolating star counts.\\
Loss of completeness is stronger for stars that are redder and for stars closer to the Galactic plane, which indicates a preferential loss of stars with higher reddening. A possibility to account for this bias is to make the completeness comparison with \ac{tycho2} not only in bins of \ac{avmag} but also in colour and Galactic latitude.
\subsection{Modelling a magnitude limited sample}
\label{sec:forward_modelling}
With a Galaxy or disc model in the background, a magnitude limited sample can be modelled theoretically and compared directly to the observations. The latest \ac{besancon} is also able to synthesise binaries. With a forward modelling approach, the local stellar density could be used as a free parameter and changed until a best fit between the mock observations and the data is found. The downside of this method is that there are many systematics and degeneracies due to model dependencies. We have the \ac{imf}, \ac{sfh}, \ac{avr}, and stellar evolutionary tracks, which can all bias our determination. Moreover, our mock observations, as well as the real data, will be a random realisation of an underlying true \ac{pdf}, which in the case of the model can be approximated via averaging, and in the case of the real data will always include Poisson noise.\\

This concludes the section on deriving stellar number densities and the chapter on statistical inference. In the next chapter volume-complete stellar samples will be used to infer properties of the stellar \ac{imf}, using a vertical local Galaxy model.
\acresetall

%% file: gfx/chapter2/montyhall2.pdf_tex
\begingroup%
  \makeatletter%
  \providecommand\color[2][]{%
    \errmessage{(Inkscape) Color is used for the text in Inkscape, but the package 'color.sty' is not loaded}%
    \renewcommand\color[2][]{}%
  }%
  \providecommand\transparent[1]{%
    \errmessage{(Inkscape) Transparency is used (non-zero) for the text in Inkscape, but the package 'transparent.sty' is not loaded}%
    \renewcommand\transparent[1]{}%
  }%
  \providecommand\rotatebox[2]{#2}%
  \ifx\svgwidth\undefined%
    \setlength{\unitlength}{299.525bp}%
    \ifx\svgscale\undefined%
      \relax%
    \else%
      \setlength{\unitlength}{\unitlength * \real{\svgscale}}%
    \fi%
  \else%
    \setlength{\unitlength}{\svgwidth}%
  \fi%
  \global\let\svgwidth\undefined%
  \global\let\svgscale\undefined%
  \makeatother%
  \begin{picture}(1,0.22880955)%
    \put(0,0){\includegraphics[width=\unitlength]{montyhall2.pdf}}%
    \put(0.05878143,0.11889545){\color[rgb]{0.19607843,0,1}\makebox(0,0)[lb]{\smash{$P(C_1)$}}}%
    \put(0.46113271,0.11889545){\color[rgb]{0.19607843,0,1}\makebox(0,0)[lb]{\smash{$P(C_2)$}}}%
    \put(0.77801547,0.11889545){\color[rgb]{0.19607843,0,1}\makebox(0,0)[lb]{\smash{$P(C_3)$}}}%
    \put(0.30144783,0.20648775){\color[rgb]{1,0,0.19607843}\makebox(0,0)[lb]{\smash{$P(H_3)$}}}%
    \put(0.19495944,0.05882507){\color[rgb]{0.19607843,1,0}\makebox(0,0)[lb]{\smash{$P(C_1|H_3)$}}}%
    \put(0.43970597,0.05882507){\color[rgb]{0.19607843,1,0}\makebox(0,0)[lb]{\smash{$P(C_2|H_3)$}}}%
  \end{picture}%
\endgroup%

%% file: gfx/chapter2/parallax.pdf_tex
\begingroup%
  \makeatletter%
  \providecommand\color[2][]{%
    \errmessage{(Inkscape) Color is used for the text in Inkscape, but the package 'color.sty' is not loaded}%
    \renewcommand\color[2][]{}%
  }%
  \providecommand\transparent[1]{%
    \errmessage{(Inkscape) Transparency is used (non-zero) for the text in Inkscape, but the package 'transparent.sty' is not loaded}%
    \renewcommand\transparent[1]{}%
  }%
  \providecommand\rotatebox[2]{#2}%
  \ifx\svgwidth\undefined%
    \setlength{\unitlength}{396.0645852bp}%
    \ifx\svgscale\undefined%
      \relax%
    \else%
      \setlength{\unitlength}{\unitlength * \real{\svgscale}}%
    \fi%
  \else%
    \setlength{\unitlength}{\svgwidth}%
  \fi%
  \global\let\svgwidth\undefined%
  \global\let\svgscale\undefined%
  \makeatother%
  \begin{picture}(1,0.37261181)%
    \put(0,0){\includegraphics[width=\unitlength]{parallax.pdf}}%
    \put(0.01795512,0.35573088){\color[rgb]{0,0,0}\makebox(0,0)[lb]{\smash{Earth}}}%
    \put(0.02646746,0.12575392){\color[rgb]{0,0,0}\makebox(0,0)[lb]{\smash{Sun}}}%
    \put(0.05546703,0.23915539){\color[rgb]{0,0,0}\makebox(0,0)[lb]{\smash{$1AU$}}}%
    \put(0.68133909,0.15309438){\color[rgb]{0,0,0}\makebox(0,0)[lb]{\smash{$\varpi$}}}%
    \put(0.87048553,0.32182586){\color[rgb]{0,0,0}\makebox(0,0)[lb]{\smash{background}}}%
    \put(0.55408705,0.20900154){\color[rgb]{0,0,0}\makebox(0,0)[lb]{\smash{star}}}%
    \put(0.24129534,0.15241633){\color[rgb]{0,0,0}\makebox(0,0)[lb]{\smash{$d$}}}%
  \end{picture}%
\endgroup%

%% file: gfx/chapter2/parallax_error.pdf_tex
\begingroup%
  \makeatletter%
  \providecommand\color[2][]{%
    \errmessage{(Inkscape) Color is used for the text in Inkscape, but the package 'color.sty' is not loaded}%
    \renewcommand\color[2][]{}%
  }%
  \providecommand\transparent[1]{%
    \errmessage{(Inkscape) Transparency is used (non-zero) for the text in Inkscape, but the package 'transparent.sty' is not loaded}%
    \renewcommand\transparent[1]{}%
  }%
  \providecommand\rotatebox[2]{#2}%
  \ifx\svgwidth\undefined%
    \setlength{\unitlength}{365bp}%
    \ifx\svgscale\undefined%
      \relax%
    \else%
      \setlength{\unitlength}{\unitlength * \real{\svgscale}}%
    \fi%
  \else%
    \setlength{\unitlength}{\svgwidth}%
  \fi%
  \global\let\svgwidth\undefined%
  \global\let\svgscale\undefined%
  \makeatother%
  \begin{picture}(1,0.36438356)%
    \put(0,0){\includegraphics[width=\unitlength]{parallax_error.pdf}}%
    \put(0.34682795,0.28936352){\color[rgb]{0.30196078,0.30196078,0.30196078}\makebox(0,0)[lb]{\smash{$\varpi-\sigma_\varpi$}}}%
    \put(0.62632664,0.28936352){\color[rgb]{0.30196078,0.30196078,0.30196078}\makebox(0,0)[lb]{\smash{$\varpi+\sigma_\varpi$}}}%
    \put(0.48838158,0.34601453){\color[rgb]{0,0,0}\makebox(0,0)[lb]{\smash{$\varpi=d^{-1}$}}}%
  \end{picture}%
\endgroup%

%% file: gfx/chapter2/parallax_distance_rescaled_test.pdf_tex
\begingroup%
  \makeatletter%
  \providecommand\color[2][]{%
    \errmessage{(Inkscape) Color is used for the text in Inkscape, but the package 'color.sty' is not loaded}%
    \renewcommand\color[2][]{}%
  }%
  \providecommand\transparent[1]{%
    \errmessage{(Inkscape) Transparency is used (non-zero) for the text in Inkscape, but the package 'transparent.sty' is not loaded}%
    \renewcommand\transparent[1]{}%
  }%
  \providecommand\rotatebox[2]{#2}%
  \ifx\svgwidth\undefined%
    \setlength{\unitlength}{368bp}%
    \ifx\svgscale\undefined%
      \relax%
    \else%
      \setlength{\unitlength}{\unitlength * \real{\svgscale}}%
    \fi%
  \else%
    \setlength{\unitlength}{\svgwidth}%
  \fi%
  \global\let\svgwidth\undefined%
  \global\let\svgscale\undefined%
  \makeatother%
  \begin{picture}(1,0.36141304)%
    \put(0,0){\includegraphics[width=\unitlength]{parallax_distance_rescaled_test.pdf}}%
    \put(0.44118632,0.29602214){\color[rgb]{0.30196078,0.30196078,0.30196078}\makebox(0,0)[lb]{\smash{$\tfrac{1}{\varpi-\sigma_\varpi}$}}}%
    \put(0.19872446,0.29602214){\color[rgb]{0.30196078,0.30196078,0.30196078}\makebox(0,0)[lb]{\smash{$\tfrac{1}{\varpi+\sigma_\varpi}$}}}%
    \put(0.31939983,0.34630121){\color[rgb]{0,0,0}\makebox(0,0)[lb]{\smash{$\tfrac{1}{\varpi}=\text{median}$}}}%
    \put(0.29626177,0.09178672){\color[rgb]{0.49803922,0.74901961,0.49803922}\rotatebox{90}{\makebox(0,0)[lb]{\smash{mode}}}}%
    \put(0.37291394,0.09143838){\color[rgb]{1,0.49803922,0.49803922}\rotatebox{90}{\makebox(0,0)[lb]{\smash{mean}}}}%
    \put(0.17425014,0.11795317){\color[rgb]{0,0,1}\rotatebox{67.63069294}{\makebox(0,0)[lb]{\smash{equation\,\ref{eq:l_in_d}}}}}%
    \put(0.4836465,0.20634234){\color[rgb]{0.74901961,0,0.74901961}\rotatebox{-29.57494913}{\makebox(0,0)[lb]{\smash{equation\,\ref{eq:likelihood}}}}}%
  \end{picture}%
\endgroup%

%% file: gfx/chapter2/parallax_prior.pdf_tex
\begingroup%
  \makeatletter%
  \providecommand\color[2][]{%
    \errmessage{(Inkscape) Color is used for the text in Inkscape, but the package 'color.sty' is not loaded}%
    \renewcommand\color[2][]{}%
  }%
  \providecommand\transparent[1]{%
    \errmessage{(Inkscape) Transparency is used (non-zero) for the text in Inkscape, but the package 'transparent.sty' is not loaded}%
    \renewcommand\transparent[1]{}%
  }%
  \providecommand\rotatebox[2]{#2}%
  \ifx\svgwidth\undefined%
    \setlength{\unitlength}{368bp}%
    \ifx\svgscale\undefined%
      \relax%
    \else%
      \setlength{\unitlength}{\unitlength * \real{\svgscale}}%
    \fi%
  \else%
    \setlength{\unitlength}{\svgwidth}%
  \fi%
  \global\let\svgwidth\undefined%
  \global\let\svgscale\undefined%
  \makeatother%
  \begin{picture}(1,0.34782609)%
    \put(0,0){\includegraphics[width=\unitlength]{parallax_prior.pdf}}%
    \put(0.29609164,0.3415971){\color[rgb]{1,0.21568627,0.21568627}\makebox(0,0)[lb]{\smash{$L=50\text{ pc}$}}}%
    \put(0.55050373,0.34192987){\color[rgb]{0.30196078,0.30196078,0.30196078}\makebox(0,0)[lb]{\smash{$d_{lim}=250\text{ pc}$}}}%
    \put(0.17533398,0.21430271){\color[rgb]{1,0.21568627,0.21568627}\makebox(0,0)[lb]{\smash{$P_{\mathrm{d}^2\mathrm{e}^{-d}}$ (eq.\,\ref{eq:d2eL})}}}%
    \put(0.4607801,0.27574472){\color[rgb]{0,0.50196078,0}\makebox(0,0)[lb]{\smash{$P_{\mathrm{d}^2}$ (eq.\,\ref{eq:d2})}}}%
    \put(0.13870542,0.1446737){\color[rgb]{0,0,1}\makebox(0,0)[lb]{\smash{$P_1$ (eq.\,\ref{eq:1})}}}%
  \end{picture}%
\endgroup%

%% file: gfx/chapter2/posterior.pdf_tex
\begingroup%
  \makeatletter%
  \providecommand\color[2][]{%
    \errmessage{(Inkscape) Color is used for the text in Inkscape, but the package 'color.sty' is not loaded}%
    \renewcommand\color[2][]{}%
  }%
  \providecommand\transparent[1]{%
    \errmessage{(Inkscape) Transparency is used (non-zero) for the text in Inkscape, but the package 'transparent.sty' is not loaded}%
    \renewcommand\transparent[1]{}%
  }%
  \providecommand\rotatebox[2]{#2}%
  \ifx\svgwidth\undefined%
    \setlength{\unitlength}{368bp}%
    \ifx\svgscale\undefined%
      \relax%
    \else%
      \setlength{\unitlength}{\unitlength * \real{\svgscale}}%
    \fi%
  \else%
    \setlength{\unitlength}{\svgwidth}%
  \fi%
  \global\let\svgwidth\undefined%
  \global\let\svgscale\undefined%
  \makeatother%
  \begin{picture}(1,0.36141304)%
    \put(0,0){\includegraphics[width=\unitlength]{posterior.pdf}}%
    \put(0.8387899,0.13909603){\color[rgb]{0,0.50196078,0}\makebox(0,0)[lb]{\smash{prior}}}%
    \put(0.13523676,0.20213413){\color[rgb]{0.20784314,0.20784314,1}\makebox(0,0)[lb]{\smash{likelihood}}}%
    \put(0.50065779,0.20631599){\color[rgb]{1,0,0}\makebox(0,0)[lb]{\smash{posterior}}}%
    \put(0.14218109,0.30259633){\color[rgb]{1,0.49803922,0.49803922}\makebox(0,0)[lb]{\smash{{\footnotesize 5\,\% quantile}}}}%
    \put(0.71817896,0.30259633){\color[rgb]{1,0.49803922,0.49803922}\makebox(0,0)[lb]{\smash{{\footnotesize 95\,\% quantile}}}}%
    \put(0.38811692,0.33846687){\color[rgb]{1,0.49803922,0.49803922}\makebox(0,0)[lb]{\smash{{\footnotesize median}}}}%
    \put(0.31408588,0.33846687){\color[rgb]{0.30196078,0.30196078,0.30196078}\makebox(0,0)[lb]{\smash{{\footnotesize modes}}}}%
  \end{picture}%
\endgroup%

%% file: gfx/chapter2/posterior_error.pdf_tex
\begingroup%
  \makeatletter%
  \providecommand\color[2][]{%
    \errmessage{(Inkscape) Color is used for the text in Inkscape, but the package 'color.sty' is not loaded}%
    \renewcommand\color[2][]{}%
  }%
  \providecommand\transparent[1]{%
    \errmessage{(Inkscape) Transparency is used (non-zero) for the text in Inkscape, but the package 'transparent.sty' is not loaded}%
    \renewcommand\transparent[1]{}%
  }%
  \providecommand\rotatebox[2]{#2}%
  \ifx\svgwidth\undefined%
    \setlength{\unitlength}{375bp}%
    \ifx\svgscale\undefined%
      \relax%
    \else%
      \setlength{\unitlength}{\unitlength * \real{\svgscale}}%
    \fi%
  \else%
    \setlength{\unitlength}{\svgwidth}%
  \fi%
  \global\let\svgwidth\undefined%
  \global\let\svgscale\undefined%
  \makeatother%
  \begin{picture}(1,0.35466667)%
    \put(0,0){\includegraphics[width=\unitlength]{posterior_error.pdf}}%
    \put(0.18538405,0.25926068){\color[rgb]{0.08627451,0.54509804,0.08627451}\makebox(0,0)[lb]{\smash{prior}}}%
    \put(0.41459383,0.15954142){\color[rgb]{1,0,0}\makebox(0,0)[lb]{\smash{posterior}}}%
    \put(0.82901668,0.14331219){\color[rgb]{0,0,1}\makebox(0,0)[lb]{\smash{likelihood}}}%
  \end{picture}%
\endgroup%

%% file: gfx/chapter2/magnitude_limit_in_luminosity.pdf_tex
\begingroup%
  \makeatletter%
  \providecommand\color[2][]{%
    \errmessage{(Inkscape) Color is used for the text in Inkscape, but the package 'color.sty' is not loaded}%
    \renewcommand\color[2][]{}%
  }%
  \providecommand\transparent[1]{%
    \errmessage{(Inkscape) Transparency is used (non-zero) for the text in Inkscape, but the package 'transparent.sty' is not loaded}%
    \renewcommand\transparent[1]{}%
  }%
  \providecommand\rotatebox[2]{#2}%
  \ifx\svgwidth\undefined%
    \setlength{\unitlength}{361bp}%
    \ifx\svgscale\undefined%
      \relax%
    \else%
      \setlength{\unitlength}{\unitlength * \real{\svgscale}}%
    \fi%
  \else%
    \setlength{\unitlength}{\svgwidth}%
  \fi%
  \global\let\svgwidth\undefined%
  \global\let\svgscale\undefined%
  \makeatother%
  \begin{picture}(1,0.36842105)%
    \put(0,0){\includegraphics[width=\unitlength]{magnitude_limit_in_luminosity.pdf}}%
    \put(0.62258903,0.25100011){\color[rgb]{0,0,0}\makebox(0,0)[lb]{\smash{Vmag$_\text{lim}=7$mag}}}%
    \put(0.09819625,0.36266886){\color[rgb]{1,0.16470588,0.83137255}\makebox(0,0)[lb]{\smash{VMag}}}%
    \put(0.11930642,0.33407643){\color[rgb]{1,0.16470588,0.83137255}\makebox(0,0)[lb]{\smash{$0$}}}%
    \put(0.11908623,0.18533135){\color[rgb]{1,0.16470588,0.83137255}\makebox(0,0)[lb]{\smash{$1$}}}%
    \put(0.1196278,0.1198841){\color[rgb]{1,0.16470588,0.83137255}\makebox(0,0)[lb]{\smash{$2$}}}%
    \put(0.11938974,0.09442544){\color[rgb]{1,0.16470588,0.83137255}\makebox(0,0)[lb]{\smash{$4$}}}%
  \end{picture}%
\endgroup%

%% file: gfx/chapter2/magnitude_limit.pdf_tex
\begingroup%
  \makeatletter%
  \providecommand\color[2][]{%
    \errmessage{(Inkscape) Color is used for the text in Inkscape, but the package 'color.sty' is not loaded}%
    \renewcommand\color[2][]{}%
  }%
  \providecommand\transparent[1]{%
    \errmessage{(Inkscape) Transparency is used (non-zero) for the text in Inkscape, but the package 'transparent.sty' is not loaded}%
    \renewcommand\transparent[1]{}%
  }%
  \providecommand\rotatebox[2]{#2}%
  \ifx\svgwidth\undefined%
    \setlength{\unitlength}{359bp}%
    \ifx\svgscale\undefined%
      \relax%
    \else%
      \setlength{\unitlength}{\unitlength * \real{\svgscale}}%
    \fi%
  \else%
    \setlength{\unitlength}{\svgwidth}%
  \fi%
  \global\let\svgwidth\undefined%
  \global\let\svgscale\undefined%
  \makeatother%
  \begin{picture}(1,0.37047354)%
    \put(0,0){\includegraphics[width=\unitlength]{magnitude_limit.pdf}}%
    \put(0.12387331,0.35021843){\color[rgb]{1,0.16470588,0.49803922}\makebox(0,0)[lb]{\smash{L$_\odot$}}}%
    \put(0.11379257,0.2858121){\color[rgb]{1,0.16470588,0.49803922}\makebox(0,0)[lb]{\smash{$100$}}}%
    \put(0.12940014,0.09167017){\color[rgb]{1,0.16470588,0.49803922}\makebox(0,0)[lb]{\smash{$1$}}}%
    \put(0.12159635,0.2041311){\color[rgb]{1,0.16470588,0.49803922}\makebox(0,0)[lb]{\smash{$16$}}}%
    \put(0.5259821,0.21090083){\color[rgb]{0,0,0}\makebox(0,0)[lb]{\smash{Vmag$_\text{lim}=7$mag}}}%
    \put(0.45405205,0.34750146){\color[rgb]{0.50196078,0.50196078,0.50196078}\makebox(0,0)[lb]{\smash{distance cuts}}}%
    \put(0.50662516,0.29536724){\color[rgb]{0.50196078,0.50196078,0.50196078}\makebox(0,0)[lb]{\smash{$200$}}}%
    \put(0.37950819,0.24949088){\color[rgb]{0.50196078,0.50196078,0.50196078}\makebox(0,0)[lb]{\smash{$100$}}}%
    \put(0.30235652,0.20698611){\color[rgb]{0.50196078,0.50196078,0.50196078}\makebox(0,0)[lb]{\smash{$75$}}}%
    \put(0.22311221,0.16652567){\color[rgb]{0.50196078,0.50196078,0.50196078}\makebox(0,0)[lb]{\smash{$50$}}}%
    \put(0.15295572,0.12648979){\color[rgb]{0.50196078,0.50196078,0.50196078}\makebox(0,0)[lb]{\smash{$30$}}}%
  \end{picture}%
\endgroup%

%% file: gfx/chapter2/extinction.pdf_tex
\begingroup%
  \makeatletter%
  \providecommand\color[2][]{%
    \errmessage{(Inkscape) Color is used for the text in Inkscape, but the package 'color.sty' is not loaded}%
    \renewcommand\color[2][]{}%
  }%
  \providecommand\transparent[1]{%
    \errmessage{(Inkscape) Transparency is used (non-zero) for the text in Inkscape, but the package 'transparent.sty' is not loaded}%
    \renewcommand\transparent[1]{}%
  }%
  \providecommand\rotatebox[2]{#2}%
  \ifx\svgwidth\undefined%
    \setlength{\unitlength}{388bp}%
    \ifx\svgscale\undefined%
      \relax%
    \else%
      \setlength{\unitlength}{\unitlength * \real{\svgscale}}%
    \fi%
  \else%
    \setlength{\unitlength}{\svgwidth}%
  \fi%
  \global\let\svgwidth\undefined%
  \global\let\svgscale\undefined%
  \makeatother%
  \begin{picture}(1,0.34278351)%
    \put(0,0){\includegraphics[width=\unitlength]{extinction.pdf}}%
    \put(0.09743242,0.324245){\color[rgb]{0.50196078,0.50196078,0.50196078}\makebox(0,0)[lb]{\smash{distance cuts}}}%
    \put(0.14293823,0.28837855){\color[rgb]{0.50196078,0.50196078,0.50196078}\makebox(0,0)[lb]{\smash{$200$}}}%
    \put(0.14293823,0.0933539){\color[rgb]{0.50196078,0.50196078,0.50196078}\makebox(0,0)[lb]{\smash{$100$}}}%
    \put(0.64770956,0.1063679){\color[rgb]{0,0,0}\makebox(0,0)[lb]{\smash{Vmag$_\text{lim}=7$mag}}}%
    \put(0.14501304,0.21934759){\color[rgb]{0.44313725,0.91764706,0.98823529}\makebox(0,0)[lb]{\smash{binary splitting}}}%
    \put(0.42726985,0.14452492){\color[rgb]{1,0.44705882,0.34901961}\makebox(0,0)[lb]{\smash{extinction correction}}}%
  \end{picture}%
\endgroup%

%% file: gfx/chapter2/inhomogeneous_extinction.pdf_tex
\begingroup%
  \makeatletter%
  \providecommand\color[2][]{%
    \errmessage{(Inkscape) Color is used for the text in Inkscape, but the package 'color.sty' is not loaded}%
    \renewcommand\color[2][]{}%
  }%
  \providecommand\transparent[1]{%
    \errmessage{(Inkscape) Transparency is used (non-zero) for the text in Inkscape, but the package 'transparent.sty' is not loaded}%
    \renewcommand\transparent[1]{}%
  }%
  \providecommand\rotatebox[2]{#2}%
  \ifx\svgwidth\undefined%
    \setlength{\unitlength}{388bp}%
    \ifx\svgscale\undefined%
      \relax%
    \else%
      \setlength{\unitlength}{\unitlength * \real{\svgscale}}%
    \fi%
  \else%
    \setlength{\unitlength}{\svgwidth}%
  \fi%
  \global\let\svgwidth\undefined%
  \global\let\svgscale\undefined%
  \makeatother%
  \begin{picture}(1,0.34278351)%
    \put(0,0){\includegraphics[width=\unitlength]{inhomogeneous_extinction.pdf}}%
    \put(0.11828159,0.32126231){\color[rgb]{0.50196078,0.50196078,0.50196078}\makebox(0,0)[lb]{\smash{distance cuts}}}%
    \put(0.15775636,0.268901){\color[rgb]{0.50196078,0.50196078,0.50196078}\makebox(0,0)[lb]{\smash{$200$}}}%
    \put(0.15775636,0.19320808){\color[rgb]{0.50196078,0.50196078,0.50196078}\makebox(0,0)[lb]{\smash{$100$}}}%
    \put(0.16497689,0.11419579){\color[rgb]{0.50196078,0.50196078,0.50196078}\makebox(0,0)[lb]{\smash{$75$}}}%
    \put(0.53639107,0.27329988){\color[rgb]{0,0,0}\makebox(0,0)[lb]{\smash{Vmag$_\text{lim}+A_V$}}}%
    \put(0.58986514,0.10369753){\color[rgb]{1,0.33333333,0.33333333}\makebox(0,0)[lb]{\smash{inhomogeneous extinction}}}%
    \put(0.2512572,0.1365665){\color[rgb]{0,0,0}\makebox(0,0)[lb]{\smash{Vmag$_\text{lim}=7\,\text{mag}$}}}%
    \put(0.11994069,0.03008964){\color[rgb]{1,0.8,0}\makebox(0,0)[lb]{\smash{\footnotesize{line of sight}}}}%
    \put(0.7390342,0.0310363){\color[rgb]{1,0.8,0}\makebox(0,0)[lb]{\smash{\footnotesize{star 
behind dust clouds}}}}%
  \end{picture}%
\endgroup%

%% file: gfx/chapter2/completeness.pdf_tex
\begingroup%
  \makeatletter%
  \providecommand\color[2][]{%
    \errmessage{(Inkscape) Color is used for the text in Inkscape, but the package 'color.sty' is not loaded}%
    \renewcommand\color[2][]{}%
  }%
  \providecommand\transparent[1]{%
    \errmessage{(Inkscape) Transparency is used (non-zero) for the text in Inkscape, but the package 'transparent.sty' is not loaded}%
    \renewcommand\transparent[1]{}%
  }%
  \providecommand\rotatebox[2]{#2}%
  \ifx\svgwidth\undefined%
    \setlength{\unitlength}{366.551bp}%
    \ifx\svgscale\undefined%
      \relax%
    \else%
      \setlength{\unitlength}{\unitlength * \real{\svgscale}}%
    \fi%
  \else%
    \setlength{\unitlength}{\svgwidth}%
  \fi%
  \global\let\svgwidth\undefined%
  \global\let\svgscale\undefined%
  \makeatother%
  \begin{picture}(1,0.36304361)%
    \put(0,0){\includegraphics[width=\unitlength]{completeness.pdf}}%
    \put(0.12252023,0.34069069){\color[rgb]{0.50196078,0.50196078,0.50196078}\makebox(0,0)[lb]{\smash{distance cuts}}}%
    \put(0.12501981,0.30516328){\color[rgb]{0.50196078,0.50196078,0.50196078}\makebox(0,0)[lb]{\smash{$200$}}}%
    \put(0.12501981,0.22543271){\color[rgb]{0.50196078,0.50196078,0.50196078}\makebox(0,0)[lb]{\smash{$100$}}}%
    \put(0.13266283,0.14528718){\color[rgb]{0.50196078,0.50196078,0.50196078}\makebox(0,0)[lb]{\smash{$75$}}}%
    \put(0.14293668,0.10868798){\color[rgb]{0,0,0}\makebox(0,0)[lb]{\smash{Vmag$_\text{lim}\,\text{in mag}$}}}%
    \put(0.6835712,0.34153987){\color[rgb]{0,0,0}\makebox(0,0)[lb]{\smash{completeness in \%}}}%
    \put(0.33211772,0.10762124){\color[rgb]{0,0,0}\makebox(0,0)[lb]{\smash{$7.0$}}}%
    \put(0.76568249,0.29178434){\color[rgb]{0,0,0}\makebox(0,0)[lb]{\smash{$100$}}}%
    \put(0.77332556,0.248866){\color[rgb]{0,0,0}\makebox(0,0)[lb]{\smash{$90$}}}%
    \put(0.77332556,0.20944459){\color[rgb]{0,0,0}\makebox(0,0)[lb]{\smash{$80$}}}%
    \put(0.77332556,0.16624386){\color[rgb]{0,0,0}\makebox(0,0)[lb]{\smash{$55$}}}%
    \put(0.77332556,0.12618462){\color[rgb]{0,0,0}\makebox(0,0)[lb]{\smash{$30$}}}%
    \put(0.39693999,0.10762124){\color[rgb]{0,0,0}\makebox(0,0)[lb]{\smash{$7.5$}}}%
    \put(0.47400005,0.10762124){\color[rgb]{0,0,0}\makebox(0,0)[lb]{\smash{$8.0$}}}%
    \put(0.57382436,0.10762124){\color[rgb]{0,0,0}\makebox(0,0)[lb]{\smash{$8.5$}}}%
    \put(0.69143252,0.10762124){\color[rgb]{0,0,0}\makebox(0,0)[lb]{\smash{$9.0$}}}%
  \end{picture}%
\endgroup%

%% file: Chapters/Chapter03.tex
\ctparttext{Using star counts and elemental abundances together with the \acs{jj} to redetermine the stellar \acs{imf} and chose the best set of stellar yields.}

\part{Milky Way disc model inference}

\chapter{The local initial mass function, as derived from star counts}
\label{ch:imf} 
In this chapter, building up on the methods described in the previous chapter, I will infer the stellar \acl{imf} parameters using the vertical \acl{mw} disc model of \citet{JJ}, together with local Solar Neighbourhood stars with accurate parallaxes. Most of the content has been published in \citet{Rybizki2015}, but a few post-publication findings are added (mainly into the discussion, section\,\ref{sec:disc}, with two new subsections), as well as further references.
\section{Introduction}
The stellar \ac{imf} has seen numerous re-determinations since the seminal work of \citet{Sa55}. The most popular \ac{imf} forms are broken power-laws \citep{Scalo1986,Kr93}, lognormal with a Salpeter-like high-mass slope \ac{chabrier}, or, more recently, tapered power-laws \citep{Parravano2011}, which are similar to \ac{chabrier} but result in more low-mass stars. Originally, the \ac{imf} was derived by the \ac{lf} of the Solar Neighbourhood via luminosity-mass relations and stellar lifetimes. More sophisticated derivations account for the \ac{sfh} \citep{Sc59} and the scale height dilution \citep{Mi79} of the observational sample.\\

Another, more direct approach is to look at young stellar clusters and derive an \ac{imf} from these spatially and temporally confined starburst populations. This yields a range of different \ac{imf}s \citep{Va60,Di14}, which are not necessarily in conflict \citep{We05}, with a global \ac{imf} being able to describe the field stars in the \ac{mw} disc. Here we get hold on the theoretical concept of a time-independent local \ac{mw} \ac{imf} by exploring its effect within our disc model in the realm of observables.\\

This chapter is based on a series of papers, which started with \ac{paper1}, where a local, vertical \ac{mw} disc model was constructed, using dynamical constraints of \ac{ms} stars. The main advantage of this method is its weak dependence on the assumed \ac{imf}, as only the integrated mass-loss enters the dynamical model. Other determinations, as for example \citet{Ha97a}, have a degeneracy of their derived \ac{sfh} and the high-mass slope of their \ac{imf}. As a result, the old Besan\c{c}on model \citep{Robin2003}, using the constant \ac{sfh} from \citet{Ha97a}, had to be revised recently \citep{Cz14} to a decreasing one, in line with different determinations from chemical evolution models \citep{Ch97}, extragalactic trends \citep{Ly2007,VanDokkum2013} and dynamical constraints \citep{Au09,JJ}.\\

In \ac{paper2}, the disc model was further constrained by comparing \ac{sdss} data with predicted star counts at the \ac{ngp}. This yielded the \ac{jj} with a fixed \ac{sfh} and a well-defined local thick disc model. The agreement between the model and the \ac{ngp} data is very good, which was also confirmed in \citet{Cz14}.\\

After fixing the \ac{sfh}, \ac{avr} and \ac{amr}, we now want to further build up a consistent disc model by constraining the fiducial \ac{imf}, using local star counts, and taking into account turn-off stars and giants as well. In \ac{paper1}, the \ac{pdmf} was already converted into an \ac{imf}, taking into account scale height correction and finite stellar lifetimes. Binaries and reddening were not accounted for, producing a very steep high-mass slope with $\alpha$\,=\,4.16, which is also related to a relatively high break in the two-slope power-law, at 1.72\,$\mathrm{M}_\odot$.\\

In this chapter, we select volume-complete stellar samples from the revised \ac{hip} \citep{le07}, and an updated version of the \ac{cns}4 \citep{Ja97} (which will soon be published as \ac{cns}5), also giving us a sound statistical sample to investigate the local stellar mass density (see discussion in sec\,\ref{sec:LF}). We use \ac{galaxia} to create mock observations from the \ac{jj} with arbitrary \ac{imf}s. We find a new fiducial \ac{imf}, describing the data by sampling the two-slope broken power-law parameter space, using a \ac{mcmc} technique. Given the data, the model likelihood is approximated, assuming discrete Poisson probabilities for different magnitude bins, differentiating between dwarfs and giants.\\

A similar determination of the \ac{imf} was done in the Besan\c{c}on update of \citet{Cz14}, with matching results. Instead of local \ac{lf}s, they used the colour projections of \ac{tycho2} data in different directions to determine their best fit. Their modelling machinery is very modular, being able to incorporate different stellar evolutionary and atmosphere models, as well as accounting for binarity or reddening from the model side. On the other hand, they do not calculate a likelihood, but use a $\chi^2$ summary statistic. Since they have so many degenerate free parameters, they do not explore the full \ac{imf} parameter space, but test pre-determined ones from the literature together with other parameters of their Galaxy model.\\

Overall, the sophisticated approach with a concordance Galaxy model in the background able to produce mock observations to test theoretical concepts, like the \ac{imf}, is a very promising technique. Especially with the underlying physical models getting more and more refined with increasing observational evidence. One unsolved problem, which is also pointed out by \citet{Cz14} is missing detailed 3D extinction data of the local \ac{ism}.\\

In section\,\ref{sec:model}, we explain how to create mock observations from our vertical disc model using \ac{galaxia}. Section\,\ref{sec:obs} describes the reduction of \ac{hip} and \ac{cns} data to obtain our observational sample, followed by section\,\ref{sec:stats} on the likelihood determination in the \ac{imf} parameter space. The results are then presented in section\,\ref{sec:results} together with a comparison to widely used \ac{imf}s. In section\,\ref{sec:disc}, the results are put into context, followed by the conclusions in section\,\ref{sec:conclusion}.

\section{Synthesising a local disc model}
\label{sec:model}
In \ac{paper1}, a self-consistent dynamical, vertical \ac{mw} disc model was developed. Combined with constraints from star counts of the \ac{ngp} in \ac{paper2}, a fiducial disc model (\acs{jj}) was chosen. The \ac{jj} fixes the \ac{sfh}, the \ac{avr}, a simple \ac{cem}, and from that predicts the stellar vertical disc structure in terms of kinematics, star counts, ages and metallicities as functions of distance to the Galactic plane.\\
In order to obtain mock observations of the Solar Neighbourhood, we turn our vertical disc model into a local representation, used as an input for \ac{galaxia} to synthesise stellar samples.    
\subsection{The disc model locally}
\label{sec:localmodel}

\begin{figure}
\caption[Local 25\,pc disc model representation]{The upper panel shows in dashed blue the global \acs{sfh} and in solid red the effective scale height of the \acs{jj} (at Solar Galactocentric distance).

In the lower panel the input N-body data, which is passed to \acs{galaxia} to create the 25\,pc sphere, is displayed. The age, mass and metallicity of each \acs{ssp} are visualised.}
\label{fig:SFH}
\includegraphics[width=\textwidth]{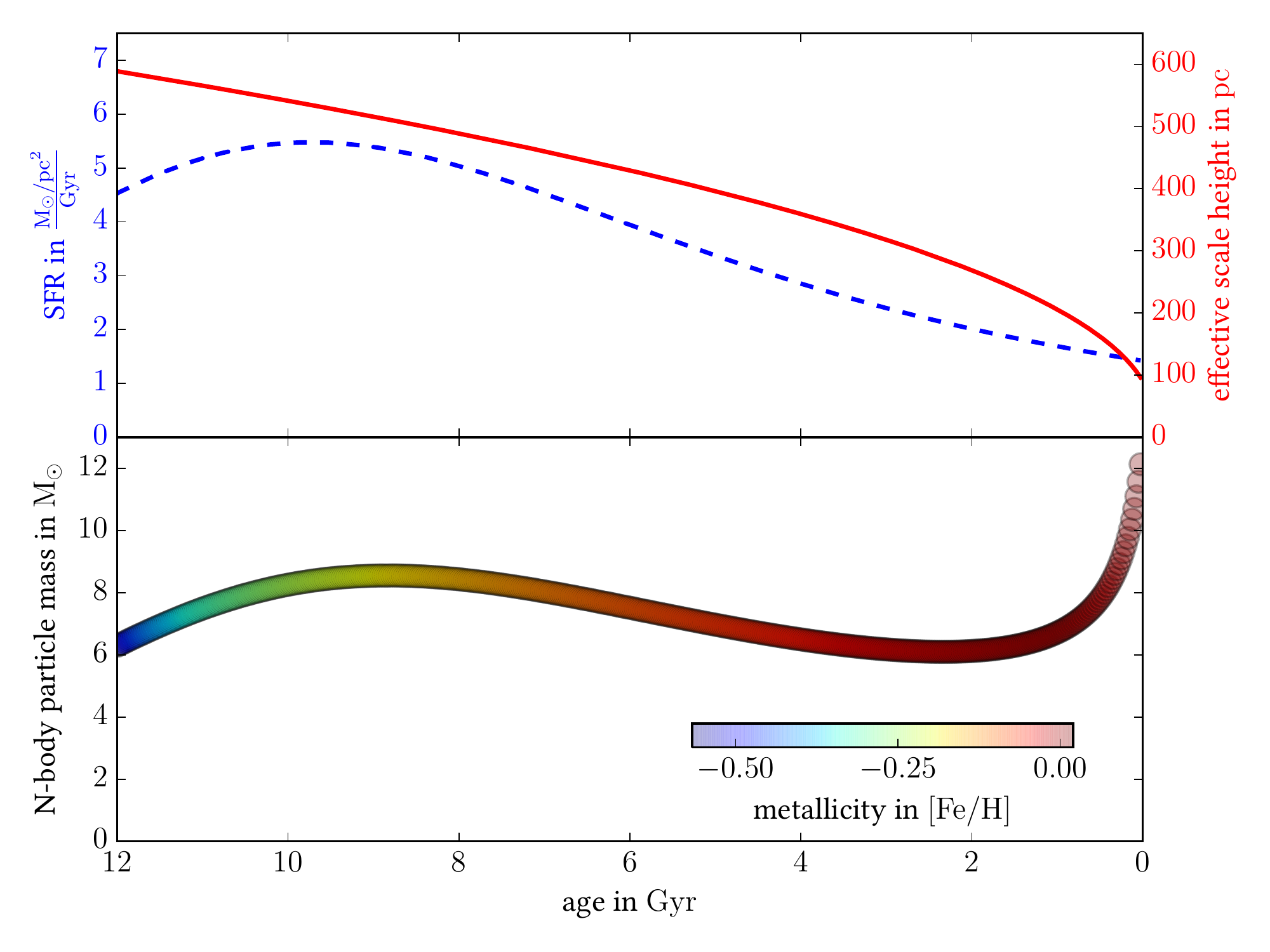}
\end{figure}

As we want to compare our \ac{jj} to volume-complete stellar samples in the Solar Neighbourhood, we construct seven\,spheres that are disjunct in \ac{vmag}, and have heliocentric distances from 20 to 200\,pc (see table\,\ref{tab:hipparcos_dwarf}\,\&\,\ref{tab:hipparcos_giant} for detailed limits). In the following, we explain how we prepare N-body particle representations of our model for each sphere such that \ac{galaxia} can turn them into mock observations.\\

As our model is evolved over 12\,Gyr in 25\,Myr steps, we construct 480\,N-body particles (plus one for the thick disc) with specific ages, masses and metallicities for each sphere separately. These are passed to \ac{galaxia}, where each particle is turned into a \ac{ssp} according to a prescribed \ac{imf}. The two youngest \ac{ssp}s are split up age-wise, so that the youngest stars coming from \ac{galaxia} are 6.25\,Myr old. We assign arbitrary phase-space information to the particles, since the final observable is a local \ac{cmd} based on volume-complete subsamples.\\

We will now illustrate the construction of a local N-body representation of our model for the 25\,pc sphere following figure\,\ref{fig:SFH}.\\
The upper panel depicts the global \ac{sfh} in units of surface density per time and the effective scale height, $\mathrm{h}_\mathrm{eff}$, over age for the \ac{jj}, respectively. From that a local mass density, $\rho_0$, for each \ac{ssp} can be calculated, which then is multiplied with the volume of the 25\,pc sphere, $\mathrm{V_{25}}$, to obtain
\begin{equation}
\label{eq:local_model}
\mathrm{M}_{25\,\mathrm{pc}}(t_i) = \mathrm{V}_{25\,\mathrm{pc}}\,\rho_0(t_i) = \frac{4}{3}\pi \mathrm{r}_{_{25\,\mathrm{pc}}}^3\frac{\mathrm{SFH}(t_i)\cdot\,0.025\,\mathrm{Gyr}}{2\,\mathrm{h}_\mathrm{eff}(t_i)}
\end{equation}
which is shown in the lower panel of figure\,\ref{fig:SFH} as resulting N-body particle masses. Colour-coded, the \ac{amr} is depicted, which is also coming from the \ac{jj} as an analytic function, but with an added Gaussian scatter of 0.13\,dex standard deviation (tested with \ac{gcs} data, cf. fig. 15 of \ac{paper1}).

For the thin disc, 3552\,M$_\odot$ ($\mathcal{M}_\mathrm{IMF,thin,25\,pc}$) in particle mass is passed to \ac{galaxia} for the 25\,pc sphere
\begin{equation}
\mathcal{M}_\mathrm{IMF,thin,25\,pc}=\int\limits_{0\,\mathrm{Gyr}}^{12\,\mathrm{Gyr}}\mathrm{M}_{25\,\mathrm{pc}}(t)\mathrm{d}t\simeq\sum_{i=1}^{480}\mathrm{M}_{25\,\mathrm{pc}}(t_i).
\end{equation}
This is by construction of the \ac{jj} the gas mass that was used to create the thin disc stars (and in the meanwhile also remnants) still residing in the 25\,pc sphere. Nowadays, only a fraction of that is left in stars, due to stellar evolution ($\mathcal{M}_\mathrm{PDMF,25\,pc}$, cf. section\,\ref{sec:LF}). In the mass-age distribution of the N-body data, the peak from the global \ac{sfh} (dashed blue line) around 10\,Gyr can still be recognised. The increase for younger stellar populations stems from the decreasing effective scale height confining these stars closer to the Galactic plane (i.e. a bigger fraction of them is found in the local sphere, cf. fig. 14 of paper\,I).

The thick disc is implemented by inserting a single starburst (i.e. one \ac{ssp} and equivalently one N-body particle) with 6.5\,\% of the thin disc mass 
\begin{equation}
\mathcal{M}_\mathrm{IMF,discs}=\mathcal{M}_\mathrm{IMF,thin}+\mathcal{M}_\mathrm{IMF,thick}=1.065\cdot\mathcal{M}_\mathrm{IMF,thin}
\end{equation}
12\,Gyr ago and with a metallicity of [Fe/H]\,=\,-0.7, resulting in a present-day thick disc mass fraction of around 5\,\%. Since the local density of the stellar halo is negligible compared to the disc, we do not consider it as a separate component in this work. 

A change in the \ac{imf} will affect the mass fraction remaining in the stellar component (cf. sec. 2.5 \ac{paper1}), therefore we introduce a mass factor ($\mathrm{mf}$), scaling our model's total mass ($\mathcal{M}_\mathrm{IMF}$), which is distributed between $\mathrm{m}_\mathrm{low}=0.08$\,M$_\odot$ and $\mathrm{m}_\mathrm{up}=100$\,M$_\odot$, to fit the observed stellar mass density, $\mathcal{M}_\mathrm{PDMF}$, such that
\begin{equation}
\mathcal{M}_\mathrm{IMF}=\mathrm{mf}\cdot\mathcal{M}_\mathrm{IMF,discs}.
\label{eq:totalmass}
\end{equation}
This is necessary because in \ac{paper1} a Scalo-like \ac{imf} \citep{Scalo1986} was used, with a different stellar evolutionary model \citep{Fioc1997} and also an independent local stellar mass density normalisation \citep{Ja97}.

When going to larger heights above the Galactic plane, we also need to correct for decreasing vertical density profiles. For the 25\,pc sphere, the deviation from a homogeneous density distribution is still negligible but this changes with larger spheres and especially with young stellar populations. For example when rescaling the star count in the highest magnitude bin in table\,\ref{tab:hipparcos_dwarf} with a mean age of 0.1\,Gyr, the number must be increased by 54\,\% to account for the low scale height of these stars (cf. figure\,2 in \ac{paper2}).

\subsection{Mock observations with Galaxia}
\label{galaxia}
\ac{galaxia} is a tool to generate mock catalogues from either analytic models or N-body data. It already has a default model, which is similar to the old Besan\c{c}on model \citep{Robin2003}, but the updates \citep{Ro12,Cz14,Robin2014} are not implemented yet.

In the previous subsection, we constructed particles representing our model locally, which we now pass for each sphere separately to \ac{galaxia}, together with the disjunct \ac{vmag} limits building up a \ac{cmd} successively. 

So far \ac{galaxia} uses Padova isochrones \citep{Ma08}, which have problems in reproducing the lower end of the \ac{ms} and the \ac{rc}. We include their revised templates \citep[PARSEC version 1.2\,S\footnote{\url{http://stev.oapd.inaf.it/cgi-bin/cmd}}]{Bressan2012}, where only minor differences at low-mass stars persist (cf. lower \ac{ms} in figure\,\ref{fig:division}). This remaining discrepancy is also pointed out in \citet[fig.\,A3]{Ch14}, but should have only a negligible effect on the star counts in our used magnitude range.\\

Binaries, \ac{wd}s or other remnants are not implemented in \ac{galaxia} yet but an update is being planned (private communication, Sharma 2015). When inspecting the \ac{cmd} in figure\,\ref{fig:division} a second blue-shifted \ac{ms} of subdwarfs in the synthesised catalogue is visible which comes from our distinct thick disc metallicity.

Beside being able to change the \ac{imf} from which \ac{galaxia} is distributing the particle masses into stars, we are using it as a black box. Specifying a photometric system will already yield a detailed stellar catalogue\footnote{See \url{http://galaxia.sourceforge.net/Galaxia3pub.html} for detailed instructions} in terms of a random realisation of our local model representation as for example depicted in figure\,\ref{fig:division} for our newly determined \ac{imf}.

\begin{table}
\begin{adjustwidth*}{-0cm}{-2.0cm}
\begin{threeparttable}

\begin{footnotesize}
\begin{tabular}{c|c c c c c c c c c c c}
\hline
Catalogue&d & M$_V$-limits & N$_{fin}$ & $\frac{\sigma_\pi}{\pi} >$\,15\,\% & \ac{cns}5 & N$_{25}$ & JJ$_{25}$ & log-likelihood & Mean mass & Mean age \\
&(pc) & (Mag) & \# & \# lost & 25\,pc &\multicolumn{2}{c}{rescaled$^\dagger$ to 25\,pc}
& $\ln\left(\mathcal{L}/\mathrm{P}_\mathrm{max}\right)$ & M$_\odot$ & Gyr \\
\hline
&200 & ],-1.5]    & 98  & 0  & 0    & 0.28      & 0.13    	& -21.9	& 6.4   & 0.1\\
&200 & [-1.5,-0.5]  & 233   & 3  & 1    & 0.62      & 0.63    & \phantom{-}0.00	& 4.0   & 0.2 \\
&200 & [-0.5,0.5]   & 901   & 12 & 4    & 2.26      & 2.60    & -9.20	& 2.9   & 0.3 \\
\acs{hip}&100 & [0.5,1.5]& 520& 2  & 15   & 8.61      & 9.09  & -0.76	& 2.2   & 0.5 \\
&75 & [1.5,2.5]     & 677   & 1  & 27   & 25.6      & 24.4    & -0.81	& 1.7   & 1.0 \\
&50 & [2.5,3.5]     & 518   & 1  & 62   & 65.1      & 59.9    & -1.79	& 1.4   & 2.3 \\
&30 & [3.5,4.5]     & 200   & 0  & 110  & 115.7     & 146.5   & -5.90	& 1.1   & 5.3 \\
\hline
&25 & [4.5,5.5]     & 191   & 4  & 191  & 191       & 190.3   & \phantom{-}0.00	& 0.9   & 6.1 \\
&25 & [5.5,6.5]     & 198   & 11 & 198  & 198     & 207.9   	& -0.22	& 0.8   & 6.4 \\

\ac{cns}5&25 & [6.5,7.5] & 193   & 16 & 193  & 193       & 196.0 & -0.02		& 0.7   & 6.5 \\
&25 & [7.5,8.5]     & 207   & 13 & 207  & 207     & 205.3   	& -0.01		& 0.6   & 6.6 \\
&20 & [8.5,9.5]     & 139   & 15 & \phantom{$^\star$}245$^\star$ & 271.5  & 258.0 & -0.20 & 0.5   & 6.7 \\
\hline
total &  & ],9.5]  & 4075  & 78 & 1253 & 1278.8  & 1300.8  	& -40.8	& 0.8 & 6.0 \\
\hline
\end{tabular}
\end{footnotesize}

\footnotesize
\begin{tablenotes}
\item \textit{Catalogue} gives the source catalogue, \textit{d} gives the heliocentric distance of stars included, \textit{M$_V$-limits} gives the magnitude range of each bin, \textit{Nfin} 
is the final star count in each bin, \textit{$\frac{\sigma_\pi}{\pi} >$\,15\,\%} is the number of stars thrown out due to high distance errors, \textit{\ac{cns}5} gives the number of stars within the volume-complete 25\,pc sphere, the next two columns give the star counts of the observations and our \ac{jj} rescaled to 25\,pc. The \ac{jj} with newly determined \ac{imf} is averaged over 400 random realisations, \textit{log-likelihood} shows the probability of each bin after equation\,\ref{eq:chi} normed with the maximal possible probability (cf. section\,\ref{sec:likelihood}) in natural logarithm which indicates each bin's impact on the likelihood function, the \textit{mean mass} and \textit{mean age} show the values for the corresponding JJ magnitude bins where the sum at the bottom is an average of all stars within 25\,pc.
\item $^\star$for this magnitude bin volume-completeness is not given so the 139\,stars from the 20\,pc sphere have been rescaled to 25\,pc yielding 271.5\,stars which is 10\,\% more than the 245\,stars observed in the 25\,pc sphere
\item $^\dagger$rescaling the volume and accounting for the density profile of the magnitude bin's mean age population
\end{tablenotes}
\end{threeparttable}
\end{adjustwidth*}
\caption[Summary table - dwarfs]{Observational sample and mock catalogues - dwarf stars}
\label{tab:hipparcos_dwarf}
\end{table}

\begin{table}
\begin{adjustwidth*}{-0cm}{-2.0cm}
\begin{footnotesize} 
\begin{tabular}{c|ccccccccccc}
\hline
Catalogue&d & M$_V$-limits & N$_{fin}$ & $\frac{\sigma_\pi}{\pi} >$\,15\,\% &\ac{cns}5 & N$_{25}$ & JJ$_{25}$ & log-likelihood & Mean mass & Mean age \\
&(pc) & (Mag) & \# & \# lost & 25\,pc &\multicolumn{2}{c}{rescaled$^\dagger$ to 25\,pc}
& $\ln\left(\mathcal{L}/\mathrm{P}_\mathrm{max}\right)$ & M$_\odot$ & Gyr \\
\hline
&200 & ],-1.5]    & 74  & 1  & 0  & 0.16  & 0.14      	& -0.87	& 3.4 & 1.6\\
&200  & [-1.5,-0.5] & 375   & 2  & 1  & 0.77  & 0.84      & -1.47	& 1.8 & 4.4 \\
&200 & [-0.5,0.5]   & 1341  & 20 & 3  & 2.78  & 2.29      & -23.4	& 1.7 & 3.9 \\
\acs{hip}&100 & [0.5,1.5] &526& 0 & 6  & 8.33  & 9.27    & -3.03	& 1.5 & 4.8 \\
&75 & [1.5,2.5]     & 126   & 0  & 5  & 4.70  & 3.76      & -2.99	& 1.2 & 6.3 \\
&50 & [2.5,3.5]     & 62  & 0  & 7  & 7.77  & 11.6      	& -5.55	& 1.2 & 6.6 \\
&30 & [3.5,4.5]     & 9   & 0  & 6  & 5.21  & 7.11      	& -0.34	& 1.0 & 8.8\\
\hline
total &  & ],4.5]   & 2513  & 23 & 28 & 29.7    & 35.0    & -37.7	& 1.3 & 6.3\\
\hline
\end{tabular}
\end{footnotesize}
\end{adjustwidth*}
\caption[Summary table - giants]{Observational sample and mock catalogues - giant stars}
\label{tab:hipparcos_giant}
\end{table}

\section{Observations}
\label{sec:obs}
The anchoring point for every Galaxy model in terms of observational constraints is the stellar distribution in the Solar Neighbourhood since detailed and volume-complete samples can only be obtained here. After using the vertical component of the velocity distribution of \ac{ms} stars in \ac{paper1} (dynamical constraint) and the \ac{sdss} \ac{ngp} star counts in \ac{paper2} (vertical density distribution constraint) we now use absolute local stellar densities for dwarf and giant stars.\\ 
This is achieved by constructing different samples combining absolute magnitude cuts and heliocentric distances such that the selected stars represent a volume-complete sphere. At the bright end we go up to 200\,pc in order to obtain enough massive stars and giants to have reliable statistics. Our observational sample consists of stars from the extended \ac{hip} catalogue \citep{an12} and the \ac{cns}. A statistically more robust technique to construct volume-complete samples (taking into account the parallax error and without the necessity to exclude stars with poor parallaxes) was tested for \citet{Just2015} and is presented in section\,\ref{sec:volume_completeness}.\\
Fundamentally, we would like to implement all observational biases on the models side and compare the synthesised mock observations to unaltered observables. In this respect the updated Besan\c{c}on model \citep{Cz14} has pushed the link between model and data in the right direction by implementing extinction models and a scheme for binary systems into their model. As \ac{galaxia} is not able to account for binaries yet and we are not providing positional information with the N-body particles we have to treat binaries and dereddening from the observational side. 
\subsection{Hipparcos}
We use the extended \ac{hip} compilation (117955 entries) which cross-matches the original stars from the revised \ac{hip} catalogue \citep{le07} with other data sources.\\
By using heliocentric distance and \ac{vmag} cuts, similar to table\,$1$ of \ac{paper1} we obtain volume-complete observational spheres for different stellar magnitudes going down to $4.5$\,\ac{vmag}. The only further selection criteria to eliminate misidentifications is a distance error below 15\,\% which reduces the sample insignificantly as visible in tables\,\ref{tab:hipparcos_dwarf}\,\&\,\ref{tab:hipparcos_giant}.\\
Before the \ac{vmag} and distance cuts are applied, all stars (or better \emph{stellar systems} represented by only a single entry in \ac{hip}), which have both a binary flag and a \ac{deltam} entry in the original \ac{hip} catalogue \citep{Hi97}, are split up into two components. The binary correction changes the number of stars in each distance bin slightly (see table\,\ref{tab:corrections}).\\
\begin{table}
\begin{tabular}{ l| c c c c c }
\hline
  Radius of sphere [pc] & 30 & 50 & 75 & 100 & 200 \\
  Magnitude bin [VMag] & 4th	& 3rd	& 2nd	& 1st & <0.5	\\
\hline  
  No correction & 204 & 591 & 793 & 1052 & 2798 \\
  Binary correction & 209 & 580 & 803 & 1046 & 2756 \\
  Extinction correction & 204 & 591 & 793 & 1053 & 3060 \\
  Both corrections (N$_{fin}$) & 209 & 580 & 803 & 1046 & 3022 \\
\hline
\end{tabular}
\caption{Effect of binarity and dereddening on the star counts}
\label{tab:corrections}
\end{table}

With the binary correction we have 5394 stars in the \ac{hip} sample of which 552 entries come from split up binary systems.
On average the bins lose stars when correcting for binaries because the single components get fainter than the system bringing these stars below the magnitude limit. As the fainter magnitude bins have smaller limiting radii only a fraction of 'lost' stars fall into the fainter magnitude bins.\\
On scales of the investigated volume the \ac{ism} is in-homogeneously distributed with Ophiuchus and Taurus molecular cloud being the biggest absorbers in the 200\,pc sphere \citep{Sc14}. 3D extinction maps with appropriate resolution are getting published \citep{La14,Green2015} but the data are not available yet or do not cover the whole sky.
In order to deredden our stars we adopt an analytic model from \citet{Ve98} describing a homogeneous extinction depending on the distance and Galactic latitude. Due to the local bubble we set the extinction to zero below 70\,pc distance and above 52$^\circ$ Galactic latitude (cf. \citet[fig. 4, 11]{Ve98}).

\citet{Au09} also adopted this model but only within 40\,pc of the Galactic mid-plane resulting in 4\,\% less stars in the 200\,pc sphere compared to our adaptation of the extinction model.
 We use the \citet[p.548]{Ve98} cosecant law for the color excess,
\begin{equation}
E_{B-V}\left(d,b\right)=
\left\{ \begin{array}{l}
 0, \\
 \mathrm{E}_0(d-\mathrm{d}_0), \\
 \mathrm{E}_0\left(\frac{\mathrm{h}_0}{|\sin(b)|}-\mathrm{d}_0\right),
 \end{array} \mbox{\ if \ } 
 \begin{array}{l}
 d < \mathrm{d}_0\,\vee\, b > \text{52}^\circ\\
 d < \frac{\mathrm{h}_0}{|\sin(b)|}\\
 d > \frac{\mathrm{h}_0}{|\sin(b)|}\end{array} \right.
\end{equation}

with $\mathrm{h}_0$\,=\,55\,pc, $\mathrm{d}_0$\,=\,70\,pc, $\mathrm{E}_0$\,=\,0.47\,$\tfrac{\text{mag}}{\text{kpc}}$ and $d,b$ being the heliocentric distance and the Galactic latitude. To transform from colour excess to extinction ($A_V$) we adopt the canonical $\tfrac{A_V}{E_{B-V}}=R_V=3.1$ \citep[p.190]{Scheffler1992}, which results in a maximum extinction $A_V\approx$\,0.19\,mag in a distance of 200\,pc, close the Galactic plane.\\
In essence, the dereddening leaves the closer samples unaltered only increasing the 200\,pc sample star count by $\sim$10\,\%.
This is due to the photometry of stars, with a heliocentric distance between 100 and 200\,pc, becoming brighter after extinction correction and thereby satisfying the magnitude limits.\\
As the \ac{ism} is highly inhomogeneous, adopting an analytic model is only a crude approximation but the best we can do at the moment. In the future we want to redo the analysis with upcoming $3$D extinction maps, like for example \citet{Green2015}.
\subsection{Catalogue of Nearby Stars\,$5$}
The \ac{hip} sample is supplemented with stars from \ac{cns}$5$ ($7251$\,entries) for fainter magnitudes ($4.5 - 9.5$\,\ac{vmag}), which is a volume-complete catalogue for stars brighter than $8.5$ ($9.5$) \ac{vmag} up to a distance of $25$ ($20$)\,pc.\\
The distances of the stars are calculated from parallaxes (when advisable the photometric parallaxes were preferred to the trigonometric) with a correction for the parallax bias \citep{Francis2013} according to equation\,1 in \citet{an12}, which was also used for stars in the \ac{hip} catalogue.\\

The $20$\,pc sample includes $139$\,stars from $8.5$ to $9.5$\,\ac{vmag}, after excluding $2$\,\ac{wd}s and $15$\,stars with relative distance errors above 15\,\%. For the $25$\,pc sphere the \ac{vmag} ranges from $4.5$ to $8.5$. Here $44$ stars have a too high error so that $789$ stars remain in the final sample.\\
All together $928$ stars originate from the \ac{cns}$5$, of which $372$ are in resolved multiple stellar systems ($24$ of the remaining $556$ 'single systems' are detected spectroscopic binaries). Of the $372$ stars in multiple stellar systems $117$ mostly primary and $87$ mostly secondary components share a joint B-V value but have individual \ac{vmag} entries so that they can be corrected by putting them on the \ac{ms}. The \ac{ms} was empirically assigned from \ac{ms} stars with low parallax error.

In the end three stars reside in the \ac{cns}$5$ sample as well as in the \ac{hip} sample because of slightly different \ac{vmag} values directly at the \ac{vmag} borders of $4.5$\,\ac{vmag}. Those stars are excluded from the \ac{hip} sample so that in our joint sample every star has a unique entry.\\
As a comparison to the \ac{hip} sample we also list the $247$ volume-complete (within $25$\,pc) \ac{cns}$5$ stars which are brighter than $4.5$\,\ac{vmag} in table\,\ref{tab:hipparcos_dwarf} and\,\ref{tab:hipparcos_giant}. The only peculiarity is the $2\sigma$ outlier with $15$ against $8.65$ expected stars in the $100$\,pc bin of the dwarf sample. Even the \ac{cns}$5$ sample can be seen as a valid random realisation of the enlarged \ac{hip} sample.

\section{Statistical analysis}
\label{sec:stats}
\begin{figure}
\includegraphics[width=\textwidth]{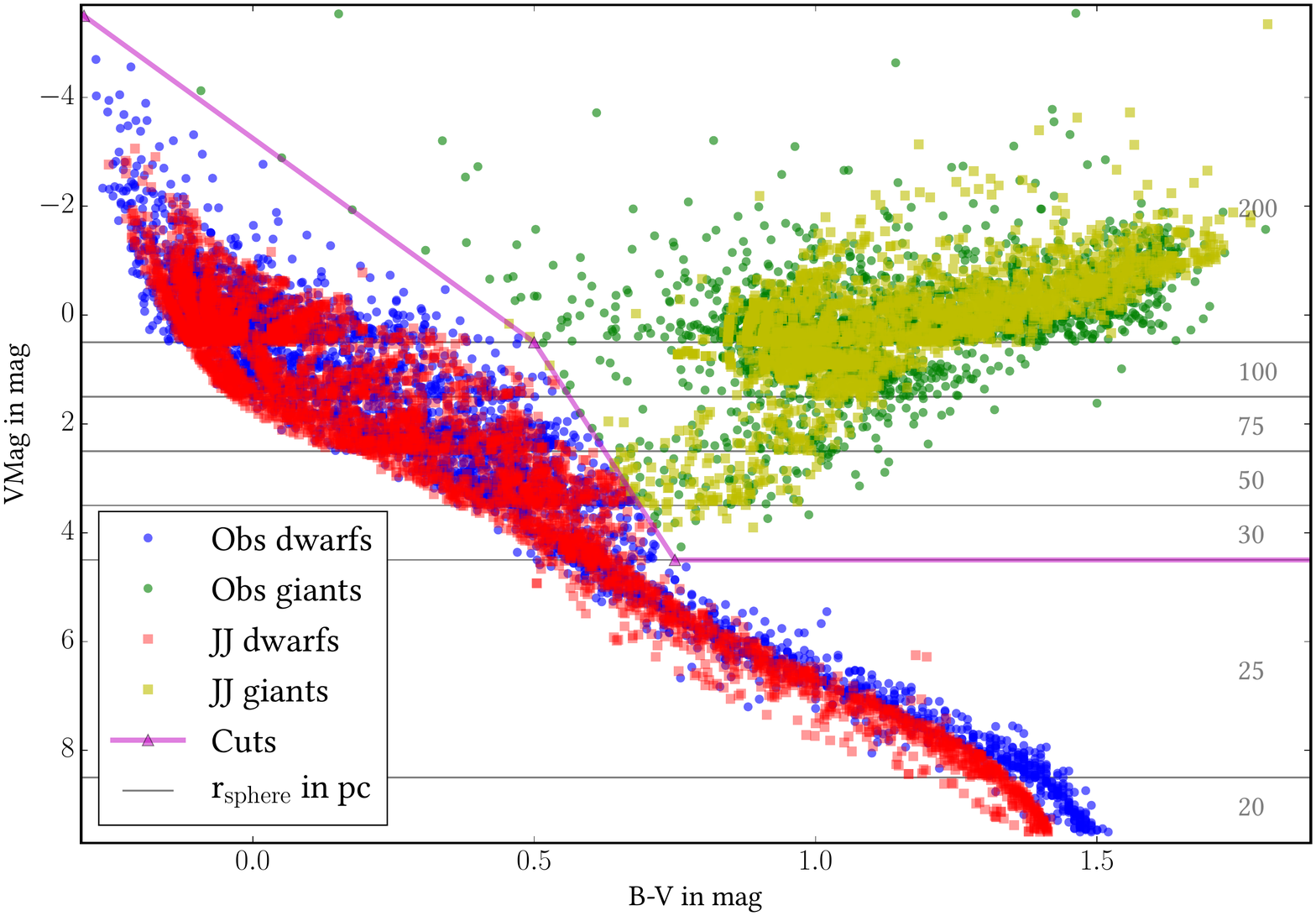}
\caption[Cuts in the colour-magnitude diagram]{Colour-magnitude diagram of the observational sample (\ac{hip} stellar systems are not split up) and one random realisation from the \acs{jj} with the newly determined \acs{imf}. The cuts with connected triangles at (B-V,M$_\text{V}$) = (-0.3,-5.5), (0.5,0.5), (0.75,4.5) and (3,4.5) for the division into dwarf and giant sample are indicated in magenta. On the right, the different sphere radii are written in units of pc.}
\label{fig:division}
\end{figure}
For our analysis we divide the derived \ac{cmd}s into dwarf and giant stars with the cuts specified in figure\,\ref{fig:division}. Since \ac{ms} stars have a tight correlation between luminosity and stellar mass the dwarf sample contains information on the \ac{pdmf}. The giant sample adds constraints for the integrated \ac{sfh} and the \ac{imf} of higher mass stars (above 0.9 M$_\odot$).\\
To compare our observable (the \ac{lf}) to a theoretical \ac{imf} we feed local representations of our model together with different \ac{imf}s to \ac{galaxia} leaving us with synthesised star counts from which we construct a likelihood assuming discrete Poisson processes (section\,\ref{sec:likelihood}). This is implemented into a \ac{mcmc} scheme to obtain a representation of the \ac{pdf} in the two-slope \ac{imf} parameter space (section\,\ref{sec:bestfit}).
\subsection{Likelihood calculation}
\label{sec:likelihood}
We approximate the likelihood of our model given the data by dividing each \ac{cmd} into 12\,magnitude bins for the dwarf sample and 7 for the giants (see table\,\ref{tab:hipparcos_dwarf}\,\&\,\ref{tab:hipparcos_giant}) and calculate the discrete Poisson probability distribution. The expected value is coming from our model ($m_i$) and the number of occurrences is the star count observed in each bin ($d_i$) leading to the likelihood
\begin{equation}
\mathcal{L}_\mathrm{total}=\prod\limits_{i=1}^{12+7}\mathcal{L}_i,
\mbox{\ where \ }\mathcal{L}_i=\frac{m_i^{d_i}e^{-m_i}}{d_i!}.
\label{eq:chi}
\end{equation}
The log-likelihood then follows as:
\begin{equation}
\log\mathcal{L}_\mathrm{total}=\sum\limits_{i=1}^{12+7}\big(d_i\log(m_i)-\log(d_i!)-m_i\big).
\end{equation}
For the calculation of the log factorial a very accurate approximation for $n>0$ from \citet[p. 339]{Ai88} is used:
\begin{equation}
\log n! \approx n\log n - n + \frac{\log\Big(n\big(1+4n(1+2n)\big)\Big)}{6}+\frac{\log(\pi)}{2}.
\end{equation}
We will normalise the log-likelihood with its maximal possible value, P$_\mathrm{max} = -68.5$, occurring when the observed sample is tested with itself.\\
It should be kept in mind that a linear increase in star counts results in exponentially increasing penalties for our likelihood function, when the relative deviation remains constant. For example, if the expected value is $10$ and the number of occurrences is $9$, then we have a $\ln\left(\mathcal{L}/\mathrm{P}_\mathrm{max}\right)$ of $-0.04$. For $100$ expected stars and $90$ occurrences $\ln\left(\mathcal{L}/\mathrm{P}_\mathrm{max}\right)$ equals $-0.46$, for $1000$ and $900$ it is $-5.1$ and so forth. This on the one hand takes into account that bins with a lot of stars get a higher statistical weight but could also be dangerous when small systematic errors (which could come from a bad \ac{avr} or connectedly isochrones indicating wrong ages) result in large penalties for the likelihood potentially pointing our \ac{mcmc} simulation to a biased equilibrium \ac{imf} parameter configuration.

\subsection{Sampling the likelihood distribution}
\label{sec:bestfit}
The variability of the outcome of \ac{galaxia} is twofold. First the \ac{imf} parameters can be varied changing the laws according to which the mock observations are produced. Second for fixed parameters the random seed of \ac{galaxia} can be changed yielding different random realisations. The latter can be minimised by averaging over many realisations. We over-sample each point in parameter space 400\,times so that this noise is reduced by a factor of 20 (cf. tab\,\ref{tab:bootstrap}) and should be a second order effect compared to the Poisson noise in the data (if we assume the observed stars have been randomly realised from an underlying probability distribution).\\

To sample the \ac{pdf} of our parameter space we use a Python implementation \citep{Foreman2013} of an affine invariant ensemble sampler for \ac{mcmc} \citep{Go10} where step proposals using the information of multiple walkers reduce the autocorrelation time significantly.\\
Since the overall mass turned into stars $(\mathcal{M}_\mathrm{IMF,discs})$ could be slightly different to the \ac{jj} we add the mass factor $(\mathrm{mf})$ as a fourth free parameter besides the three two-slope \ac{imf} parameters, low-mass index $(\alpha_1)$, high-mass index $(\alpha_2)$, and the power-law break $(\mathrm{m}_1)$. The functional form of the \ac{imf} is\\
\begin{equation}
\frac{\mathrm{d}n}{\mathrm{d}m}=k_{\alpha} m^{-\alpha}
\left\{ \begin{array}{l}
 \alpha = \alpha_1, \\
 \alpha = \alpha_2,
 \end{array} \mbox{\ if \ } \begin{array}{l}\mathrm{m}_\mathrm{low}<m<\mathrm{m}_1\\ \mathrm{m}_1<m<\mathrm{m}_\mathrm{up}\end{array} \right.
\label{eq:func}
\end{equation}
with the lower and upper mass limit of the \ac{imf} $\mathrm{m}_\mathrm{low}$ = 0.08\,M$_\odot$ and $\mathrm{m}_\mathrm{up}$ = 100\,M$_\odot$ being fixed. The $k_{\alpha_\mathrm{i}}$ ensure that the \ac{imf} is a continuous function at the power-law break $(\mathrm{m}_1)$ and normalises to unity in mass
\begin{equation}
\int_{\mathrm{m}_\mathrm{low}}^{\mathrm{m}_\mathrm{up}} m\frac{\mathrm{d}n}{\mathrm{d}m}\mathrm{d}m=1,
\label{eq:mass_normalisation1}
\end{equation}
such that it can be multiplied with the mass factor, $\mathrm{mf}$, and the \ac{jj} mass, $\mathcal{M}_\mathrm{IMF,discs}$, to represent the gas mass turned into stars
\begin{equation}
\mathcal{M}_\mathrm{IMF} = \mathrm{mf}\cdot\mathcal{M}_\mathrm{IMF,discs}\cdot\int_{\mathrm{m}_\mathrm{low}}^{\mathrm{m}_\mathrm{up}} k_\alpha m^{-\alpha+1}\mathrm{d}m.
\label{eq:mass_normalisation2}
\end{equation}

For each set of parameters we use the product of the likelihoods from the dwarf bins and the giant bins
\begin{equation}
\log\mathcal{L}_\mathrm{total} = \log\mathcal{L}_\mathrm{dwarf} + \log\mathcal{L}_\mathrm{giant}
\end{equation}
to sample the parameter space. In this way the probability for each bin (12 from the dwarf sample and 7 from the giant sample), being an independent discrete Poisson process, is weighted equally into the final likelihood ($\mathcal{L}_\mathrm{total}$).\\

\section{Results}
\label{sec:results}
Here we present the newly determined fiducial \ac{imf} for the \ac{jj}. For comparison we also show the log-likelihood of the observational data with synthesised data generated from our model but using common \ac{imf}s from the literature.
\subsection{New IMF parameters}
\label{sec:our_IMF}
 \begin{figure}
 \includegraphics[width=1\textwidth]{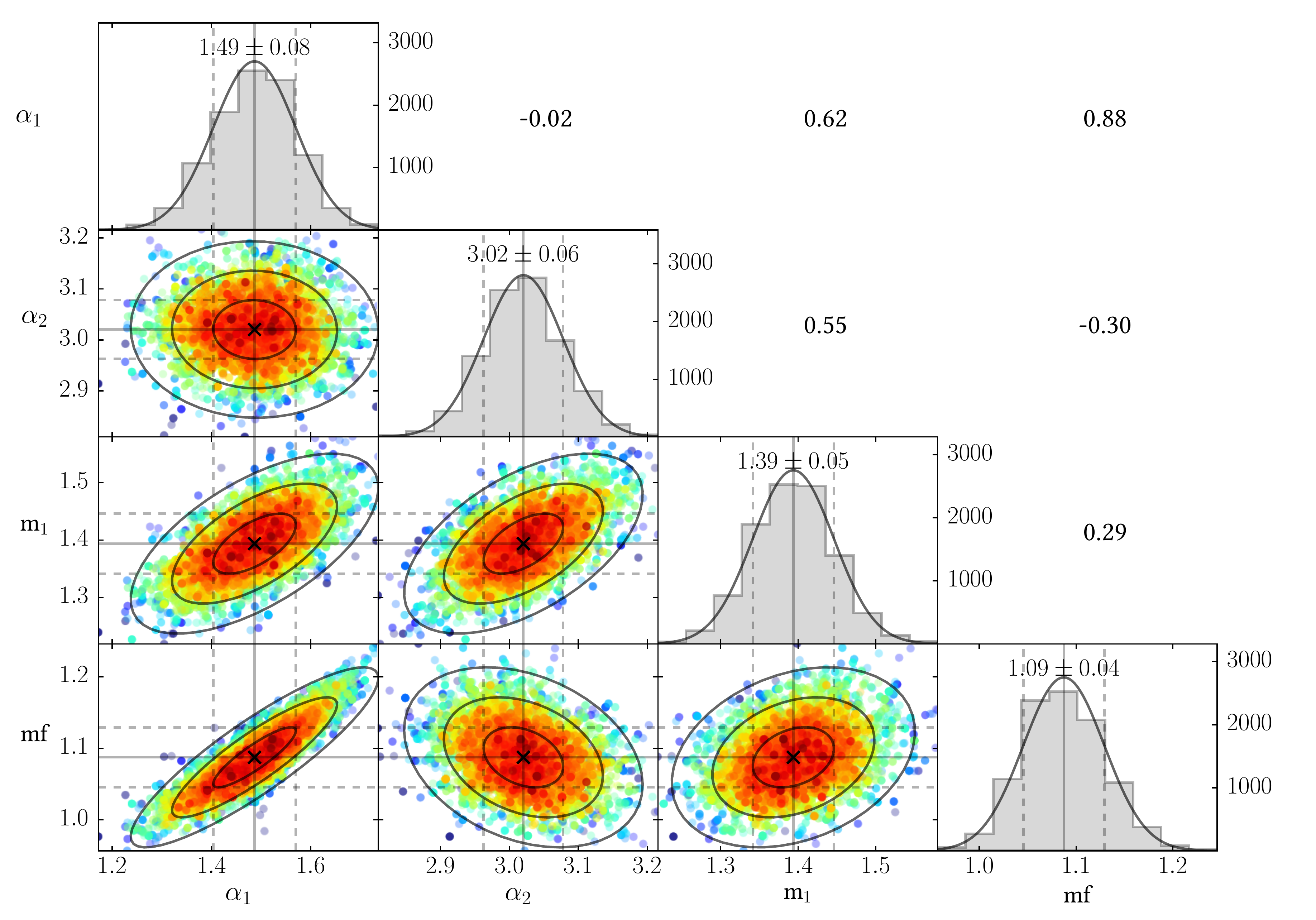}
 
 \caption[Probability density function of the parameter space]{Marginalised parameter distribution of the \ac{mcmc} run exploring the equilibrium distribution of the parameter space with respect to our log-likelihood using 10$^5$ evaluations. Each scatter plot shows the projected 2D parameter distribution with points coloured by likelihood increasing from blue to red. Crosses indicate the mean values and ellipses encompass the 1$\sigma$-3$\sigma$ regions. The respective correlation coefficients are given at the position mirrored along the diagonal. Gaussian fits and histograms of the marginalised parameter distribution are given on the diagonal. The mean and standard deviation of each parameter is written and also indicated by solid and dashed grey lines.}
 \label{fig:parameter}
 \end{figure}
  \begin{figure}
  \caption[Luminosity function]{Luminosity function of the observations and the new \ac{imf} of the \ac{jj} (cf. table\,\ref{tab:hipparcos_dwarf} and\,\ref{tab:hipparcos_giant}). Error bars indicate Poisson noise in the observational sample and the standard deviation for the 400\,times over-sampled synthesised catalogue. Star counts are not normalised for the different distance limits. The limiting radii of the corresponding magnitude bins are written in the top.}
 \centering
 \def\svgwidth{\columnwidth}
 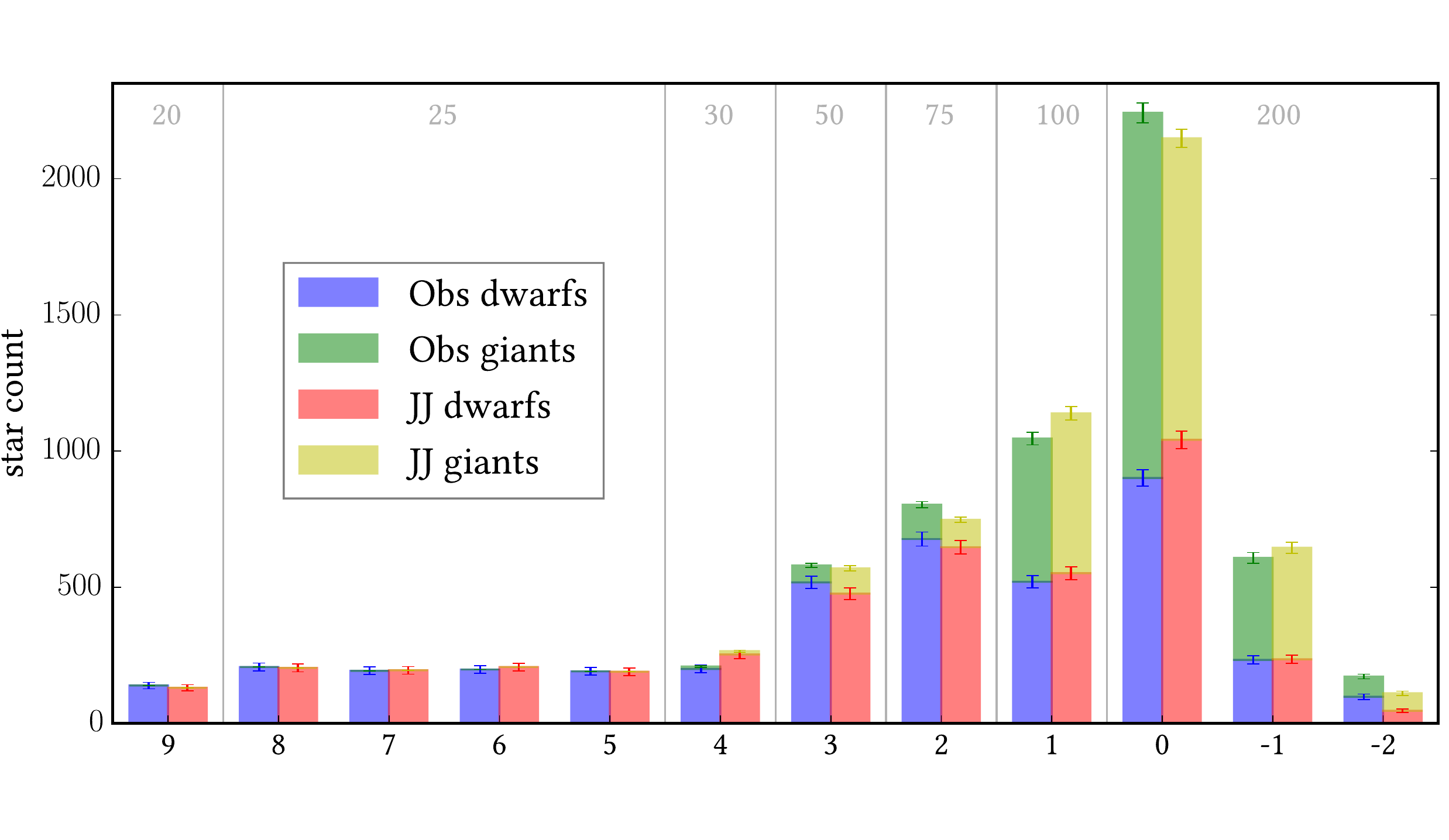
 \label{fig:lf}
 \end{figure}
Multiple burn-ins from different starting points all settling in the same equilibrium configuration suggest a well-behaved parameter space with respect to our likelihood calculation. We use 20\,walkers each sampling 500\,steps to generate the point cloud representing the \ac{pdf} of the parameter space. For each step we average over 400\,realisations. Figure\,\ref{fig:parameter} shows the marginalised likelihood distribution of each parameter as histograms on the diagonal and as point clouds for each parameter pair in the lower left. Just for illustrative reasons each dot is coloured to indicate high ($\mathrm{red}_\mathrm{max}$\,=\,-77.2) and low ($\mathrm{blue}_\mathrm{min}$\,=\,-86.7) log-likelihood values with grey dots being below this 3$\sigma$ range. In the histograms the mean and the standard deviation of the number density of each marginalised parameter is given which represents our central result and defines our new fiducial \ac{imf}: 
\begin{equation}
\begin{array}{rcl}
\alpha_1 &=& \text{1.49}\pm \text{0.08},\\
\alpha_2 &=& \text{3.02}\pm \text{0.06},\\
\mathrm{m}_1 &=& \text{1.39}\pm \text{0.05},\\
\mathrm{mf} &=& \text{1.09}\pm \text{0.04}.\\
\end{array}
\end{equation} 
Furthermore Pearson's correlation coefficient for each parameter pair is given in the upper right of figure\,\ref{fig:parameter} which is also represented in the 1$\sigma$-3$\sigma$\,ellipses of the projected point clouds. The only two parameter which are uncorrelated are the power-law indices. Almost positive linear is the correlation of the mass factor with the low-mass slope. This is due to more mass being put into stars which are not represented in our data ($m$\,<\,0.5\,M$_\odot$) for high $\alpha_1$ which can be counterbalanced with a high-mass factor. Similarly but less strong the anti-correlation of the mass factor with the high-mass slope is due to mass being shifted out of our represented data domain when $\alpha_2$ is getting lower. The reason for both power-law indices being positively correlated with the power-law break is due to the shape of the \ac{imf} which is sharply decreasing at the position of the break and the number densities of the faintest and brightest magnitude bin which need to be matched when shifting $\mathrm{m}_1$. This reasoning can be best visualised when looking at the green dotted line (our \ac{imf}) and the blue error bars (our observational sample) in figure\,\ref{fig:imf}. Additionally when increasing both $\alpha_1$ and $\alpha_2$ the mass that is gained by the steeper low-mass slope will be counterbalanced by the mass-loss of the steeper high-mass slope.\\

Figure\,\ref{fig:lf} shows the binned \ac{lf} of our fiducial \ac{imf} (averaged over 400 realisations) compared to the observations in absolute numbers. The \ac{cmd} representation of the \ac{lf}s can be seen in figure\,\ref{fig:division} where the same colours have been used for the different samples. The deviations in each bin look small and systematics are not apparent neither in the dwarf nor the giant sample which shows that the whole machinery, consisting of the disc model and \ac{galaxia} producing the mock observations, works well and the \ac{mcmc} simulation has likely converged towards the equilibrium configuration.\\
When inspecting the detailed likelihood contribution of each bin in table\,\ref{tab:hipparcos_dwarf}\,\&\,\ref{tab:hipparcos_giant} we see that the largest penalty comes from the zeroth\,\ac{vmag} bin. Especially the giants with $\ln\left(\mathcal{L}/\mathrm{P}_\mathrm{max}\right)$ = -23.4 have a huge impact. The reason for that is likely a \ac{rc} that is too faint in our mock catalogue. In the dwarf sample too many stars are in the synthesised zeroth\,\ac{vmag} bin with JJ$_{25}\cdot$(N$_{fin}$/N$_{25})$\,=\,1036 compared to 901\,stars in the \ac{hip} catalogue.\\
A weakness of our modelling machinery is apparent in the high $\ln\left(\mathcal{L}/\mathrm{P}_\mathrm{max}\right)$ value of the brightest dwarf bin indicating that too few bright stars are produced. The reason is most likely that we are not accounting for binaries and that we use a two-slope power-law. Minor effects could be missing high-metallicity stars and that our synthesised stars are not younger than 6.25\,Myr. It could also indicate a change of the high-mass power-law index for stars more massive than contained in our data.
\subsection{Tested IMFs}
\label{sec:otherIMF}
Because for other shapes of the \ac{imf} part of the mass could also be hidden in the mass range not represented by our observational sample (0.5 to 8\,M$_\odot$) we determine the factor with which the total mass needs to be rescaled in order to maximise the likelihood for a particular \ac{imf} when it is used with our \ac{jj} against the observational sample. We again average over 400\,realisations and get the standard deviation as the enclosing 68\,\% of the likelihood.\\
In table\,\ref{tab:IMFcomp} the log-likelihoods of the different \ac{imf}s (using our disc model but adjusting for each \ac{imf}'s best fit mass factor) are listed with JJ$_{3\sigma}$ being the log-likelihood value which is lower than 99.7\,\% of the points representing the \ac{pdf} in figure\,\ref{fig:parameter}.\\
Table\,\ref{tab:bootstrap} illustrates a few properties of our log-likelihood. JJ$_{400}$ and JJ$_{1}$ show the mean and the standard deviation of 100 log-likelihood determinations with different seeds which are averaged over 400 in the former and 1 realisation in the latter case. This shows that the averaging is important for the \ac{mcmc} simulation in order to smoothen the likelihood distribution. The deteriorated mean for single realisations is due to a skewed distribution since P$_\mathrm{max}$ is a lower limit and the penalty increases for extreme values which are not smoothed out as in JJ$_{400}$.\\

The last column of table\,\ref{tab:bootstrap} (JJ$_\mathrm{ideal}$) gives an ideal log-likelihood which is obtained when the data is indeed represented by the model. For that we draw 100 single random samples (JJ$_1$) from JJ$_{400}$ with replacement and evaluate their log-likelihood with the parent distribution (JJ$_{400}$). Each sample fulfilling the observational constraint of having dwarf and giant star counts fixed to N$_\mathrm{Obs,dwarf}$\,=\,4075 and N$_\mathrm{Obs,giant}$\,=\,2513. This means that a $\ln\left(\mathcal{L}/\mathrm{P}_\mathrm{max}\right)$ of around -13.8 would indicate a perfect model. An even lower log-likelihood value close to $\mathrm{P}_\mathrm{max}$ ($\ln\left(\mathcal{L}/\mathrm{P}_\mathrm{max}\right)$\,=\,0) would be unrealistic since there is a natural scatter to Poisson processes.\\
Related to that we also inspected the distribution of star counts in individual magnitude bins for random realisations (with the same parameters but different seeds) which indeed is Poissonian.
\begin{table}
\caption{Likelihoods of the different IMFs}
\begin{center}
\begin{tabular}{ c|c c c c c}
\hline
  \ac{imf} &  JJ$_\mathrm{}$ & JJ$_{3\sigma}$  & \acs{besb} & \acs{ktg} & \acs{chabrier}  \\
\hline  
  $\ln\left(\frac{\mathcal{L}}{\mathrm{P}_\mathrm{max}}\right)$ & -79.3 & -86.7 & -96.7 & -195.6 & -216.1  \\
\hline
\end{tabular}
\end{center}
\label{tab:IMFcomp}
\end{table}

\begin{table}
\caption{Variability of the log-likelihood}
\begin{threeparttable}

\begin{tabular}{ c|c c c}
\hline
    Sample &  JJ$_{400}$ & JJ$_{1}$  & JJ$_\mathrm{ideal^\star}$  \\
\hline  
  $\ln\left(\frac{\mathcal{L}}{\mathrm{P}_\mathrm{max}}\right)$ & -79.3\,$\pm$\,0.6 & -89.5\,$\pm$\,14  & -13.8\,$\pm$\,2.8 \\
\hline
\end{tabular}
\begin{tablenotes}
\item$^\star$if data was coming from our model (see text)
\end{tablenotes}
\end{threeparttable}

\label{tab:bootstrap}
\end{table}
\subsubsection*{\acs{ktg}}
The widely used \ac{ktg} \ac{imf} is a three-slope broken power-law with $\alpha_1$\,=\,1.3, $\alpha_2$\,=\,2.2, $\alpha_3$\,=\,2.7, $\mathrm{m}_1$\,=\,0.5 and $\mathrm{m}_2$\,=\,1. \begin{equation}
\frac{\mathrm{d}n}{\mathrm{d}m}=k_{\alpha} m^{-\alpha}
\left\{ \begin{array}{l}
 \alpha = \alpha_1, \\
 \alpha = \alpha_2, \\
 \alpha = \alpha_3,
 \end{array} \mbox{\ if \ } 
 \begin{array}{l}
 \mathrm{m}_\mathrm{low}<m<\mathrm{m}_1\\
 \mathrm{m}_1<m<\mathrm{m}_2\\
 \mathrm{m}_2<m<\mathrm{m}_\mathrm{up}\end{array} \right.
\label{KTG}
\end{equation}
Again the $k_{\alpha_\mathrm{i}}$ ensure continuity and normalisation of the \ac{imf} to unity in mass between $\mathrm{m}_\mathrm{low}$ = 0.08\,$\mathrm{M}_\odot$\,and\,$\mathrm{m}_\mathrm{up}$ = 100\,M$_\odot$.\\
The likelihood peak is obtained with a mass factor of 1.392 $\pm$ 0.019 which is quite high. The reason for that is too much mass being put into low-mass stars which are not represented in our observational sample (see table\,\ref{tab:sn2}).\\
With $\ln\left(\mathcal{L}/\mathrm{P}_\mathrm{max}\right)$ = -195.6 it scores poorly compared to our or the \ac{besb} \ac{imf} showing that its shape is not able to reproduce local star counts. This is visible in figure\,\ref{fig:imf} where it produces too few stars within 0.9 - 2\,M$_\odot$ and too many outside of this range compared to our \ac{imf}.
\subsubsection*{\acs{besb}}
This is one of the new fiducial Besan\c{c}on model \ac{imf}s from \citet{Cz14} tested with \ac{tycho2} all-sky colour distribution. It is also a three-slope broken power-law with $\alpha_1$ = 1.3, $\alpha_2$ = 1.8, $\alpha_3$ = 3.2, $\mathrm{m}_1$ = 0.5 and $\mathrm{m}_2$ = 1.53. The mass factor for this \ac{imf} is 1.107 $\pm$ 0.015 which is compatible with our own $\mathrm{mf}$. Also the shape of the \ac{imf} (cf. figure\,\ref{fig:imf}), the likelihood (see table\,\ref{tab:IMFcomp}) and the mass fractions (cf. table\,\ref{tab:sn2}) are similar. This is remarkable since they have different scale heights, \ac{sfh} and also an additional data set (\ac{tycho2} colour vs. \ac{hip}/\ac{cns} \ac{vmag}).\\
Compared to our \ac{imf} the \ac{besb} \ac{imf} is producing slightly less high-mass stars and more low-mass stars which could be partly due to their rigorous treatment of binaries (see figure\,\ref{fig:imf} and cf. section\,\ref{sec:binary}).
\subsubsection*{\ac{chabrier}}
Another widely used \ac{imf} comes from \citet{Ch03}. It is a mixture of a lognormal form in the low-mass and a Salpeter power-law in the high-mass regime
 \begin{equation}
 \frac{{\mathrm d}n}{{\mathrm d}m}=\left\{ \begin{array}{l}
 \frac{0.852464}{m}\,\exp\big(\frac{-\log^2\left(\frac{m}{0.079}\right)}{2\cdot 0.69^2}\big), \\
 0.237912\cdot m^{-2.3},
 \end{array} \mbox{\ if \ } \begin{array}{l}m<\mathrm{M}_\odot\\ m>\mathrm{M}_\odot .\end{array} \right.
\label{eq:Cha} 
 \end{equation} 
The mass factor for the best likelihood is: 1.317 $\pm$ 0.017. It is so high because a huge mass fraction is going into stars more massive than 8\,M$_\odot$ (cf. table\,\ref{tab:sn2}, highest supernova rate compared to any other standard \ac{imf}). The shape of the \ac{chabrier} \ac{imf} fits our model worst with respect to the data scoring a log-likelihood of $\ln\left(\mathcal{L}/\mathrm{P}_\mathrm{max}\right)$ = -216.1.\\

\subsection{From \acl{lf} to local stellar mass density}
\label{sec:LF}
  \begin{figure}

 \caption[Luminosity function translated into mass space]{25\,pc \ac{lf} of dwarf stars translated into mass space. Our fiducial \ac{imf} is plotted in thick blue and the \ac{besb}, \ac{ktg}, and \ac{chabrier} \ac{imf}s are plotted in cyan, magenta, and black, respectively. The green error bars represent the 12\,magnitude bins from the observational dwarf sample with their limiting radius given in parsec. The x-value and the x-error are associated with the median and the range of stellar masses in the corresponding mock magnitude bin synthesised with our fiducial \ac{imf}. The y-value represents the number of stars normalised to the mass range and the y-error is the Poisson noise. As the masses from different magnitude bins overlap the values are added up in the dashed green line representing a kind of 'observational' \ac{pdmf}. Below 0.7\,$\text{M}_\odot$ the yellow shaded area indicates the mass that is missing due to incompleteness by the magnitude cut. In dotted red the same effect is visible, as this line represents the synthesised dwarf mass function of our \ac{imf} averaged over 100\,realisations. From 0.9\,M$_\odot$ upwards the \ac{imf} and the local \ac{imf} in dashed blue deviate since locally (i.e. close to the Galactic plane) young and therefore massive stars are over-represented, which is indicated by the blue arrow. The change from the \ac{imf} to the \ac{pdmf} due to stellar evolution is indicated by the red arrow.}
  \centering
  \def\svgwidth{\columnwidth}
  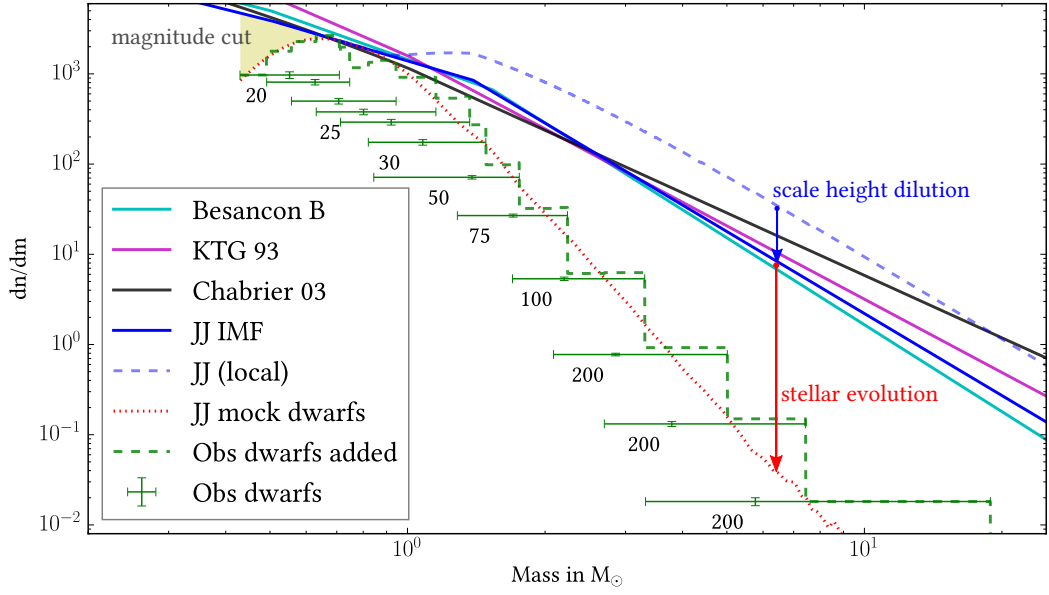
 \label{fig:imf}
 \end{figure}

In figure\,\ref{fig:imf} the \ac{imf}s are normalised such that their integrated mass is representing the mass of gas that was turned into stars (of which a few already turned into remnants) still residing in the 25\,pc sphere (i.e. $\mathcal{M}_\mathrm{IMF,25\,pc}$ including the thick disc and each \ac{imf}'s mass factor cf. equation\,\ref{eq:totalmass}). Beware that in figure\,\ref{fig:imf} actually the number of stars per mass interval is displayed though normalisation happens in mass space (see discussion in section\,\ref{sec:massfactor}).\\

The yellow shaded area shows the incompleteness of our observations in low-mass stars due to the magnitude cut. This comes from high-metallicity stars being redder and fainter than their equal-mass low-metallicity counterparts so they are excluded and do not contribute to the smallest mass bin leading to a decrease of the observed mass compared to the theoretical \ac{imf}. The same is true for the \ac{jj} mock dwarfs in dotted red (as we apply the same magnitude cuts) which were spawned using the blue \ac{imf}. We could drop the magnitude cut for faint stars in \ac{galaxia} and the red dotted line would be perfectly aligned with the \ac{jj} \ac{imf} in the low-mass regime, which corresponds to the local \ac{pdmf}. In the 'observational' \ac{pdmf} represented by the dashed green line the little bump compared to the \ac{jj} mock dwarfs at around 0.8\,M$_\odot$ coincides with the \emph{Wielen dip} \citep{Wi74}, observed in the local luminosity function at \ac{vmag}\,$\sim7$.\\
In the high-mass regime, the blue arrow indicates the over-representation of massive stars in the local \ac{imf} of the 25\,pc sphere, as their vertical distribution is more confined to the Galactic plane (if one would not account for the scale height dilution the deduced \ac{imf} would look like this). The red arrow indicates the conversion from the \ac{imf} to the \ac{pdmf} due to stellar evolution.\\
We can quantify the difference of \ac{imf} and \ac{pdmf} by looking at the integrated mass (between $\mathrm{m}_\mathrm{low}$ and $\mathrm{m}_\mathrm{up}$) of the two functions represented by 'JJ\,IMF' and 'JJ\,mock\,dwarfs' from figure\,\ref{fig:imf} (the latter being made equal to 'JJ\,IMF' for the low-mass part not affected by stellar evolution). The outcome is that 55.8\,\% of the mass originally turned into stars is still present in dwarf stars today ($\mathcal{M}_{\mathrm{PDMF,25\,pc}}$ / $\mathcal{M}_{\mathrm{IMF,25\,pc}}$). Including giants this value only changes slightly to 56.9\,\% which is consistent with the integrated stellar mass of our disc model (figure\,1 of \ac{paper1} adapted for 25\,pc age distribution yields 56.5\,\% of stars increasing to 70.8\,\% when remnants are included which is their g$_\mathrm{eff}$) confirming that the new \ac{imf} still fits within our model's framework as \ac{sfh} and \ac{avr} are only dependent on the integrated mass-loss.\\ 

In order to clarify our remnants which \ac{galaxia} is not synthesising we derive from our \ac{pdmf} and \ac{imf} that we should have 707\,stellar remnants for stars between 1 and 8\,M$_\odot$ (potentially \ac{wd}s) and 17 heavier ones (potentially black holes or neutron stars) in the 25\,pc sphere. Compared to \citet{Si14} (or similarly \citet{Just2015}), we have over a factor of 2 more since they expect 344\,\ac{wd}s within the same limits. But their assumed volume-completeness for the 13\,pc sample is probably more like a lower limit, at least when speaking of the cool end of the \ac{wd} cooling sequence. \citet{Ho08} propose a \ac{wd} mean mass of 0.665\,M$_\odot$ which for us results in a \ac{wd} mass density of 0.0072\,$\mathrm{M}_\odot/\mathrm{pc}^3$ quite close to the Besan\c{c}on value based on \citet{Wi74} which was already corrected downwards in \citet{Ja97} because one out of 5\,\ac{wd}s left the 5\,pc sphere. Our high number of \ac{wd}s is partly due to the functional form of the two-slope \ac{imf} because it has an over-abundance where the power-law break lies (when expecting a concave function describing the underlying distribution in log-space, see our \ac{imf} in figure\,\ref{fig:imf}) which slightly exaggerates the mass fraction of the \ac{imf} going into \ac{pn} (cf. table\,\ref{tab:sn2}).\\

With these values (summarised in table\,\ref{tab:Holmberg}) the overall present-day mass fraction of stars and stellar remnants ($\mathcal{M}_\mathrm{PDMF,total,\,25\,pc}$) is
\begin{equation}
g_\mathrm{eff} = \frac{\mathcal{M}_\mathrm{PDMF,\,25\,\mathrm{pc}}+\mathcal{M}_\mathrm{BD\,\& WD\,25\,\mathrm{pc}}}{\mathcal{M}_\mathrm{IMF,25\,pc}} = 71.7\,\%
\end{equation}
which is close to 70.8\,\% for the 25\,pc sample of \ac{paper1}. This also implies a combined \ac{bd} and \ac{wd} mass fraction of about 20\,\% of the local mass budget ($\mathcal{M}_\mathrm{PDMF,total,\,25\,pc}$) consistent with the \citet[tab.\,3]{Ja97} value and the stellar evolution of \ac{paper1}. Then again we also could have chosen the proposed \ac{wd} local mass density of \citet{Ho08} $\mathrm{M}_\odot/\mathrm{pc}^3$ = 0.0032 which would decrease the local disc mass density and change g$_\mathrm{eff}$ and the remnant fraction. Here we chose to stay self-consistent with our \ac{imf}'s remnant fraction though we could have used those independent observations as a prior yielding tighter constraints for our \ac{imf}.\\

\begin{table}
\caption[Local stellar mass density of thin disc components]{Local stellar mass density of different thin disc components in $10^{-4}\text{M}_\odot/\text{pc}^3$}
\begin{threeparttable}
\begin{tabular}{ c|c c c}
\hline
Mass component  & \acl{fh06} & \ac{jj} & \phantom{$^*$}\ac{jj}$^*$ \\
\hline
giants  & 6   & 7  & 7\\
\phantom{2...5<}$\mathrm{M}_\mathrm{V}$ < 2.5  & 31  & 11 & 11\\
2.5 < $\mathrm{M}_\mathrm{V}$ < 3\phantom{.5} & 15  & 5 & 5\\
3 < $\mathrm{M}_\mathrm{V}$ < 4   & 20  & 19 & 19\\
4 < $\mathrm{M}_\mathrm{V}$ < 5   & 22  & 26 & 25\\
5 < $\mathrm{M}_\mathrm{V}$ < 8  & 70  & 66 & 65\\
8 < $\mathrm{M}_\mathrm{V}$\phantom{5..<}  & 135   & 205 & \phantom{$^*$}176$^*$\\
\hline
$\rho_\mathrm{thin\,disc}$(z = 0\,pc, t = 12\,Gyr) & 299 & 338 & 309\\
$\rho_\mathrm{thick\,disc}$(z = 0\,pc, t = 12\,Gyr) & 35 & 17 & 15\\
brown dwarfs    & 20  & \phantom{$^*$}20$^\star$  & \phantom{$^*$}20$^\star$\\
white dwarfs    & 60  & \phantom{$^*$}72$^\dagger$ & \phantom{$^*$}30$^*$\\
\hline
$\rho_{\text{\ac{pdmf},\,total}}$ & 414 & 447 & 374 \\
\hline
\end{tabular}
\begin{tablenotes}
\item $^*$using \citet{Just2015} (empirical extension from section\,\ref{sec:empirical_fix})
\item $^\star$taking the same value as \ac{fh06} 
\item $^\dagger$derived implicitly from our \ac{pdmf} (see text)
\end{tablenotes}
\end{threeparttable}
\label{tab:Holmberg}
\end{table}
For the following local mass density test we use our disc model and the fiducial \ac{imf} to produce all stars within 25\,pc down to 0.08\,M$_\odot$ without any magnitude cuts and analyse the sample's properties. In this sample we find a local stellar mass density of 0.034\,$\mathrm{M}_\odot/\mathrm{pc}^3$ compared to 0.030\,$\mathrm{M}_\odot/\mathrm{pc}^3$ for the same selection of stars in \citet[tab.\,2]{Fl06} (including giants, excluding \ac{bd}s, \ac{wd}s, other remnants and the thick disc component) who use a similar method. In table\,\ref{tab:Holmberg} the detailed comparison reveals that especially the two brightest bins of the \ac{fh06} sample are 3\,times denser than the corresponding bins in the \ac{jj} sample. For \ac{vmag}$>3$ the densities match quite well except for the faintest mag bin. The over-abundance of bright stars could be caused by their large value of n in \citet[eq.\,3]{Holmberg1997}, which is resulting in nearly exponential vertical density distribution with high local mass densities though this then should apply to the other mag bins as well. Another indication for an over-estimation of their model's local star counts becomes evident when comparing the number densities from \citet{Holmberg2000} (upon which \ac{fh06} is based) with those from \ac{cns}5. For $\mathrm{M}_\mathrm{V} < 2.5$ they have $0.0013$\,star\,pc$^{-3}$ whereas our volume-complete sample has half of this with $0.0007$\,star\,pc$^{-3}$ and this despite the fact that the $25$\,pc from the \ac{cns}5 seem to have an over-representation of the upper \ac{ms} compared to the larger sample (cf. table\,\ref{tab:hipparcos_dwarf}). The next mag bin $2.5 < \mathrm{M}_\mathrm{V} < 3$ is three times denser according to \citet{Holmberg2000} with $0.0010$\,star\,pc$^{-3}$ compared to the $0.0003$\,star\,pc$^{-3}$ we measure for our $25$\,pc sample, which strongly indicates a necessary revision. Fainter mag bins are much better fit in star counts as well as in stellar mass density.\\
Comparing our thin disc stellar mass density (excluding thick disc stars and \ac{wd}s) of $0.0338$ $\mathrm{M}_\odot/\mathrm{pc}^{3}$ to the default \ac{besb} model that has $0.0330$\,$\mathrm{M}_\odot/\mathrm{pc}^{3}$, reveals a similar discrepancy. Our disc model has $12$\,Gyr of evolution compared to $10$\,Gyr, and in table\,\ref{tab:mass_density} we make a detailed comparison, keeping in mind the different \ac{sfh} and vertical profiles of each disc model \citep[tab.\,7 and fig. 4]{Cz14}. The \ac{besb} model uses the same local mass density as we do from \citet{Ja97} based on $25$\,pc, though they add the thick disc on top of this value, which is an inconsistency, as \citet[table\,3]{Ja97} accounts for all local stars, not only those of the thin disc.\\
\begin{table}
\caption[Local mass density over stellar age]{Local mass density over stellar age in $10^{-4}\text{M}_\odot/\text{pc}^3$}
\begin{threeparttable}
\begin{tabular}{ c|c c c c}
\hline
Age [Gyr] & Besan\c{c}on\,A & \ac{besb} & \ac{jj} & \phantom{$^*$}\ac{jj}$^*$ \\
\hline  
\phantom{0.15}0 - 0.15\phantom{0} & 20 & 19 & 9 & 8\\ 
\phantom{1}0.15 - 1\phantom{0.15} & 55 & 50 & 36 & 34\\ 
1 - 2 & 46 & 41 & 30 & 29\\
2 - 3 & 33 & 28 & 27 & 25\\
3 - 5 & 58 & 49 & 52 & 48\\
5 - 7 & 61 & 50 & 54 & 49\\ 
\phantom{10}7 - 10\phantom{7} & 117 & 93 & 84 & 76\\
\phantom{12}10 - 12\phantom{10} & - & - & 46 & 41\\
\hline
Thin disc & 390 & 330 & 338 & 309\\
Thick disc & 29 & 29 & 17 & 15\\
WD & 71 & 71 & \phantom{$^\dagger$}92$^\dagger$ & \phantom{$^\dagger$$^*$}50$^\dagger$$^*$\\
\hline
$\sum$ & 490 & 430 & 447 & 416\\
\hline
\end{tabular}
\begin{tablenotes}
\item $^*$using \citet{Just2015} (empirical extension from section\,\ref{sec:empirical_fix})
\item $^\dagger$ including \ac{bd}s
\end{tablenotes}
\end{threeparttable}
\label{tab:mass_density}
\end{table}

Overall, for the other disc mass models more mass seems to be sitting in stellar mass bins of bright stars, which might be partly due to the local dwarf sample being almost 2\,times denser than the 200 and 100\,pc sample at the upper \ac{ms} (cf. table\,\ref{tab:hipparcos_dwarf} 'Observations' \& '\ac{cns}5'). Since we use volume-complete samples to fit the luminosity function and take into account the scale height dilution according to the stellar ages, we trust our mass distribution in the mass range that our observational evidence samples (0.5 - 8\,M$_\odot$).\\
From the local \ac{lf} the high-mass end of the \ac{imf} for stars more massive than 8\,M$_\odot$ can not be constraint, due to too few massive stars. The impact on the stellar mass density is negligible. In chapter\,\ref{ch:chemistry} we will constrain the high-mass slope, using abundance data and a chemical extension of the \ac{jj}.\\
Ideally we should have included a constraint for the low-mass stellar mass density from other observations (which we do empirically in section\,\ref{sec:empirical_fix}) because now the mass factor and the low-mass power-law index are strongly correlated and certainly a bit too high. Better values for a more realistic two-slope \ac{imf} would probably be in the lower left of the 1$\sigma$ ellipse in figure\,\ref{fig:parameter} (e.g. $\mathrm{mf}\simeq$ 1.05 and $\alpha_1\simeq$ 1.4).\\

\section{Discussion}
\label{sec:disc}
We discuss our findings with respect to their model dependencies as well as our analysis method. Then we provide a comparison of the mass distribution from various \ac{imf}s and end this section with an empirically driven adaptation of our \ac{imf} to account for missing low-mass star representation in our data.

\subsection{Isochrones}
\label{sec:isochrones}
A crucial ingredient for our investigation is the used set of isochrones since it translates our analytical disc model into the realm of observables. When we were using the default option \citep{Ma08} provided by \ac{galaxia}, the highest likelihood we could score was $\ln\left(\mathcal{L}/\mathrm{P}_\mathrm{max}\right)$ = -109 with slightly different \ac{imf} parameters. With the latest PARSEC \citep{Bressan2012} isochrones the observations are much better fit by the model increasing the normed log-likelihood by 30. Still a few discrepancies are visible when inspecting figure\,\ref{fig:division}.\\
The M dwarf V-Band problem was already mentioned in section\,\ref{galaxia} and is discussed in \citet{Ch14}.\\
Section\,\ref{sec:our_IMF} discusses the huge likelihood penalty from the brightest magnitude bin in table\,\ref{tab:hipparcos_dwarf} which could be due to unaccounted binaries in our sample or due to missing turn-off stars (which could be related to our discrete time steps) or due to missing super Solar metallicity stars or a high-mass power-law index which is too large which is due to the functional form of our two-slope \ac{imf}.\\
Table\,\ref{tab:hipparcos_giant} shows missing stars in the zeroth mag bin and an over-abundance in the first for the giants. This is an indication for the \ac{rc} being too faint compared to the observations.\\
Another striking feature is the complete absence of synthesised stars bluewards from the \ac{rc} whereas \ac{hip} shows several dozens. One reason is probably that population\,II stars are not synthesised, such that the horizontal branch is not reproduced by our model.
Also the extrapolated high-mass slope produces too few massive stars, which would populate this region of the \ac{cmd}.\\
Of course not all differences are linked to the isochrones as the \ac{jj} \ac{sfh}, \ac{amr} or \ac{imf} also affect the distribution of stars in the \ac{cmd}. Another reason for mismatch is our used reddening law not accounting for inhomogeneous \ac{ism} which explains the unmatched faint giants on the red end of the giant sequence in figure\,\ref{fig:division}.

\subsection{Mass factor}
\label{sec:massfactor}
The overall mass with our newly determined \ac{imf} compared to the \ac{jj} \ac{sfh} increased by the thick disc fraction ($6.5$\,\%) and the mass factor ($\mathrm{mf} = 1.09$). The normalisation in \ac{paper1} was done using $\rho_\mathrm{PDMF,total}$ = $0.039$\,M$_\odot$pc$^{\text{-}3}$ from \citet{Ja97}. Since we utilise new isochrones together with number densities derived from volume-complete star counts, a change of about $10\,\%$ is not unexpected but our value of $0.045$\,M$_\odot$pc$^{\text{-}3}$ is probably exaggerated. The present-day mass fraction (g$_\mathrm{eff}$) stays similar and also the remnant fraction is compatible with \ac{paper1} values as shown in section\,\ref{sec:LF}, though \ac{wd} and faint stellar local mass density are arguably too high. The problem is that we do not have observational constraints for the whole mass range resulting in a degeneracy of the mass factor and the low-mass slope (see figure\,\ref{fig:parameter}, also valid for the high-mass slope). We propose a solution to this in section\,\ref{sec:empirical_fix}.

\subsection{Binarity}
\label{sec:binary}
In our observational sample we tried to account for all binaries that have \ac{deltam}\,entries, so that we could split them up. Apart from these another 365\,stellar entries in our observational sample are listed in the \textit{Washington Double Star Catalogue} \citep{Ma01} or the \textit{Catalogue of Components of Double and Multiple Stars} \citep{Do02}, as multiple systems but are not resolved, i.e. they are not split up for this analysis but just kept as one 'star'.\\
As mentioned in section\,\ref{sec:our_IMF}, we have a problem with binaries in massive stars, which are hard to detect, as lines are blurred, and increased luminosity could also be due to ageing of a single star. Since binary fraction in massive stars is expected to be up to $80$\,\% and our likelihood gets a large penalty from the brightest dwarf mag bin not being matched well (other important reasons for too few synthesised stars in that bin are our two-slope functional form and our rough analytic extinction model), we believe that we miss quite a few high-mass binaries.
Other than that, listings of binary stars in the \ac{hip} catalogue are said to be 'fairly complete' \citep{Li97} for $\Delta Hp$ < 3.5\,mag and an angular separation bigger than 0.12 to 0.3\,arcsec (increasing with $\Delta Hp$). 
In \ac{cns}5 all stellar systems from the literature with resolved magnitude differences are split into their respective components.\\

The Besan\c{c}on group is doing the favourable approach of accounting for binaries from the model's side \citep{Cz14,Ro12,Ar11}. They use an angular resolution limit of 0.8\,arcsec for resolved binaries in their \ac{tycho2} data. In \citet[fig. A.3]{Cz14} the relative difference in stellar mass frequency produced with binarity treatment in single, primary and secondary stars compared to the same sample excluding secondary stars is shown. With B-components included the \ac{imf} produces around 6\,\% more stars below 1.1\,M$_\odot$ and around 6\,\% less above with a short transition in between. Overall the effect of binarity seems to play a secondary but not negligible role especially in massive dwarf stars.
\subsection{Different functional forms of the IMF}
\label{sec:functional_forms}
A crude investigation of the three-slope \ac{imf} after the publication of \citet{Rybizki2015} yielded an improved $\ln\left(\tfrac{\mathcal{L}}{\mathrm{P}_\mathrm{max}}\right)$ = -46.2 compared to -79.3 for the best two-slope \ac{imf}, but with two more free parameters: $\alpha_1$ = 1.51, $\alpha_2$ =  3.47, $\alpha_3$ = 1.47, $\mathrm{m}_1$ = 1.55, $\mathrm{m}_2$ = 3.60 and $\mathrm{mf}$ = 2.1.\\
This \ac{imf} would need further observational constraints from stars lighter than 0.5\,M$_\odot$ (and heavier than 8\,M$_\odot$) which are not (well) represented in our data. For the two-slope \ac{imf} this is done in section\,\ref{sec:empirical_fix} by introducing another low-mass break in the power-law. Obtaining a constraint for the high-mass stars is challenging (see section\,\ref{sec:high_mass_stars_constraint}) but would certainly yield a steeper high-mass slope compared to $\alpha_3$ = 1.47.\\
The assessment whether we are over-fitting or adding significant parameters would need the comparison of \ac{bf}s (see section\,\ref{sec:bayes_factor}) for each functional form (and replaces the \emph{reduced} $\chi^2$ methodology). In that case we should not only test the parameter space of multicomponent power-law functions, but also of other functional forms (e.g. \citet{Ch03}, \citet{Parravano2011}). In our analysis we only used the fixed literature values and adjusted for the mass factor which just shows that these \ac{imf}s yield lower likelihoods (within our model and likelihood determination) than our optimised two-slope power-law.
\subsection{Splitting the CMD}
Before exploiting the full 2D information of the \ac{cmd} and adapting the statistical machinery as well as dealing with colour issues of the isochrones (not speaking of enhanced sensitivity to reddening), the easiest way to increase the data constraints is to split up the \ac{cmd} into dwarf and giant samples as we do here. Weighting both into the final likelihood is a valuable gain since they represent stellar populations with different ages and masses (check the last columns of table\,\ref{tab:hipparcos_dwarf}\,\&\,\ref{tab:hipparcos_giant}), and are still build from the same \ac{imf} (as well as \ac{sfh}, \ac{avr} and \ac{amr}). Interestingly the likelihood penalties from both samples are similar though the dwarfs contribute 12 and the giants only 7\,bins. In fact, the faint mag bins from the \ac{cns} are well matched with a one-slope \ac{imf} in the low-mass range (0.4 < M$_\odot$ < 1.0). This mass range is only represented in the dwarf stars, so no trade-off with the giants is necessary leading to a small cumulative penalty of $\ln\left(\mathcal{L}/\mathrm{P}_\mathrm{max}\right)$ = -0.45 from these 5\,bins.\\

Exemplary for the insight gained from splitting up the \ac{cmd}, the zeroth mag bin can be inspected where a common sample would have balanced our model predicting too few giants and too many dwarfs. As these two bins have well distinct age and mass, constraints are put on different parts of the \ac{imf}. On the other hand the over-abundance of stars in the third mag bin in giants and the 4th mag bin in both giants and dwarfs is indicating too many stars for the \ac{imf} around 1.1\,M$_\odot$. This is in balance with the depletion of stars from around 1.3\,M$_\odot$ (second mag in giants and third in dwarfs) in order to fit the two-slope power-law with respect to our constructed likelihood.
\subsection{Binning}
\label{sec:binning}
As pointed out in section\,\ref{sec:likelihood} (and also in a blog entry\footnote{\url{https://astrostatistics.wordpress.com/2015/01/15/some-unwitting-pdbil-mcmc/}} written on our statistical method) the importance of likelihood is unevenly assigned between different bins. On the one hand, this is wanted, because higher star counts in a bin give a stronger constraint for the assumed Poisson process, on the other hand this is problematic when small systematic errors, translate into larger penalties for densely populated mag bins. Similarly for the \ac{mcmc} when it is trading-off log-likelihood penalties to \emph{find} its equilibrium position. With this in mind the relatively high $\ln\left(\mathcal{L}/\mathrm{P}_\mathrm{max}\right)$ values of the third\,mag giant bin, the 4th\,mag and the brightest dwarf bin, (though having comparatively low star counts) indicate an even stronger deviation from the data than represented in the log-likelihood penalty. The deviations are not compatible with expected Poisson noise, as the third\,mag giant bin and the 4th\,mag dwarf bin are close to $4\,\sigma$ off their expected value and the brightest dwarf bin is off by more than $5\,\sigma$.

This means that our \ac{imf} produces too many stars around 1.1\,M$_\odot$ and too few high-mass stars, which shows that our adopted functional form of a two-slope \ac{imf} is probably too inflexible to match the underlying \ac{imf} represented by the data. A test for the \ac{bf} of different functional forms together with additional low- and high-mass constraints (priors) could clarify this hypothesis (cf. discussion in sec\,\ref{sec:functional_forms}).\\

The best solution would be, to avoid binning in the first place what we did not do because we wanted a high quality observational sample with little model-dependency leading to the choice of volume-complete samples and their conservative magnitude cuts (a more robust way of constructing volume-complete samples is discussed in section\,\ref{sec:volume_completeness}). On the other hand modelling the complete magnitude-limited \ac{hip} catalogue would have made more use of available data. The complication would have been, to simulate the smooth density distribution of our \ac{jj} in \ac{galaxia} and account for increasing contribution of thick disc stars with increasing height above Galactic plane. The integration of a semi-analytic form of our \ac{jj} in a future version of \ac{galaxia} could make this endeavour feasible and also simplify the integration of a 3D extinction model.
\subsection{High-mass slope}
\label{sec:highmass}
Our high-mass power-law index of $\alpha_2$ = 3.02 is at the higher end of the literature values which can mostly be attributed to the functional form and should also not be extrapolated to O and early B type stars, because these are not very well sampled in our 200\,pc volume. For studies using star counts, it will always be challenging to account for high-mass stars, because of them being rare and confined to the Galactic plane, where extinction complicates things. This (and a similar functional form) is the reason why we and also the new default Besan\c{c}on model \citep{Cz14} propose steep high-mass slopes for A and late B type stars. Only with \ac{gaia} the surface density of O,B stars will be available for a large area, yielding viable constraints on the high-mass slope of the \ac{imf} from the \ac{lf}.\\

An important measure for an \ac{imf} in terms of Galaxy simulation is the fractional mass going into stars heavier than 8\,M$_\odot$ as they supposedly explode as \ac{sn2} and have the highest and fastest stellar feedback to the Galactic evolution in terms of gross elemental synthesis and heating of the gas phase. In table\,\ref{tab:sn2} we list the fraction of total mass and total star count going into the \ac{sn2} mass bin for all investigated \ac{imf}s adding Salpeter (with $\alpha$ = 2.3) for comparison.\\
\begin{table}
\begin{threeparttable}
\caption[Mass distribution of different IMFs]{Mass distribution, \acs{pn}- and \acs{sn2}-occurrences for different \acs{imf}s}
\begin{tabular}{ c|c c c c c c}
\hline
\acs{imf}:    &\acs{jj}& \acs{jj}$^*$&\acs{besb}&\acs{ktg}&\acs{chabrier}& Salpeter\,55\\
Mass range in M$_\odot$  & \multicolumn{6}{c}{mass fraction in\,\%} \\   
\hline  
\phantom{100}8 - 100\phantom{8}   & 6  & 7 & 4   & 8   & 22  & 15  \\
\phantom{8}1.4 - 8\phantom{1.4}  & 32 & 34 & 30  & 24  & 29  & 20  \\
\phantom{1.4}1 - 1.4\phantom{1}  & 13 & 13 & 12  & 9  & 7   & 5   \\
\phantom{1}0.5 - 1\phantom{0.5}   & 20 & 21 & 23  & 22  & 16  & 12  \\
\phantom{0.5}0.08 - 0.5\phantom{0.08}  & 29 & 25 & 31  & 36  & 26  & 48  \\
\hline
\acs{pn}$^\star$ (1 - 8)   & 994 & 1044 & 951   & 707   & 687   & 467   \\
\acs{sn2}$^\star$ (> 8)  & 17 & 18 & 12  & 21  & 47  & 32  \\
\hline
\end{tabular}
\begin{tablenotes}
\item $^\star$number of occurrences for an \acs{ssp} with mass M$_\text{\ac{imf},25\,pc}\approx$ 4100\,M$_\odot$ at t = $\infty$
\item $^*$using \citet{Just2015} (empirical extension from section\,\ref{sec:empirical_fix})
\end{tablenotes}
\label{tab:sn2}
\end{threeparttable}
\end{table}
The fraction of mass going into stars ending their lives as \ac{sn2} ranges from 4\,\% for \ac{besb} to 22\,\% for \ac{chabrier}. In order to obtain sensible results, modellers usually tune their feedback physics according to their used \ac{imf}. In general \ac{cem}s and their predicted element ratios are especially sensitive to the number of \ac{sn2} events. So using a \ac{chabrier} \ac{imf} will need substantially different stellar yields to obtain similar results (i.e. reproduce observations) compared to a \ac{cem} using the new \ac{besb} \ac{imf} (see chapter\,\ref{ch:chemistry}).\\
Same is true for cosmological hydrodynamical simulations reproducing the number of dwarf galaxies around spirals and also galaxy morphologies. Usually the necessary feedback is high, so that top-heavy mass functions are used like the \ac{chabrier} \ac{imf}. To get even higher feedback the limiting mass for \ac{sn2}-explosion can be reduced below the fiducial value of 8\,M$_\odot$ as for example in \citet[with a limiting mass of 6\,M$_\odot$]{Vogelsberger2013}. Depending on the sub-grid physics of wind mass loading and wind metal loading which they treat separately the observed stellar masses of galaxies and especially low-mass systems are reproduced (this is achieved with large wind mass loading and generally a huge amount of \ac{sn2}). But at the same time this over-predicts the metallicity of massive galaxies and underestimates the metallicity of dwarf galaxies \citep[fig. 11 \& sec. 4.2.7]{Vogelsberger2013}, which can not be counterbalanced by the wind metal loading factor. This tension not only suggests that too much stellar mass is going supernova but might as well point to a metallicity - or more general: an environment-dependent \ac{imf} \citep{Conroy2012}.
\subsection{Empirically motivated three-slope IMF extension}
\label{sec:empirical_fix}
Our observational sample not constraining the high (volume too small) and low-mass stars (sparse V\,Band data and no reliable isochrones) is deteriorating our inferred \ac{imf} parameters. A good solution would be to include observationally based priors into the likelihood determination directing the \ac{mcmc} simulation to a more physically motivated solution. Possible constraints directly connected to the shape of the \ac{imf} and the mass normalisation could be the \ac{sn2} rate and H\,II\,regions the \ac{pn} rate the \ac{wd} number density and low-mass stellar density. But it is not trivial to account for the uncertainty of those observations in the prior function.\\
As an empirical fix we introduce a second power-law break at 0.5\,M$_\odot$ leaving the shape of the \ac{imf} above the same, but changing it for lower masses. With a look at table\,\ref{tab:Holmberg} all bins should stay the same except for the 8 < $\mathrm{M}_\mathrm{V}$ bin which should decrease to 0.017\,M$_\odot\mathrm{pc}^{\text{-}3}$ being a new value derived from volume-complete near-\ac{ir} data of the \ac{cns}5 \citep{Just2015}. This shrinks our mass factor and still uses our high quality data for the higher mass bins. The proposed \ac{imf} parameters fulfilling this additional constraint are: $\alpha_0$ = 1.26, $\mathrm{m}_0$ = 0.5, and $\mathrm{mf}$ = 1.03. Compared to our low-mass slope of $\alpha_1$ = 1.49 the extension to lower masses is a bit shallower and quite similar to \ac{ktg} also having its low-mass power-law break at 0.5\,M$_\odot$.
\section{Conclusion}
\label{sec:conclusion}
We use Solar Neighbourhood stars to determine a new fiducial \ac{imf}, in the mass range $0.5-8$\,M$_\odot$, within the framework of our local vertical \ac{mw} disc model (\ac{jj}).
 For that we carefully select volume-complete samples based on dereddened and binary corrected \ac{hip} and \ac{cns}5 data. Then we use \ac{galaxia} to create the corresponding mock observations from our \ac{jj}. We construct a likelihood by assuming a discrete Poisson process for the star count in magnitude bins differentiating between dwarfs and giants. With \ac{mcmc} simulations we sample the \ac{pdf} of the two-slope \ac{imf} parameters. The derived \ac{imf} has a low-mass power-law index of $\alpha_1$ = 1.49 $\pm$ 0.08, a power-law break at $\mathrm{m}_1$ = 1.39 $\pm$ 0.05\,M$_\odot$, a high-mass index of $\alpha_2$ = 3.02 $\pm$ 0.06, a mass factor of $\mathrm{mf}$ = 1.09 $\pm$ 0.04 with respect to our \ac{paper1} mass normalisation. Except for physics not accurately represented in our model (binaries, inhomogeneous \ac{ism}, variable stars, low-mass stellar atmospheres, metal enrichment law and thick disc) these findings are robust in our observationally backed mass range from 0.5 to 8\,M$_\odot$. An empirically driven low-mass extension adds $\alpha_0$ = 1.26 and $\mathrm{m}_0$ = 0.5 and decreases the mass factor to 1.03.\\
Independently from us, the Besan\c{c}on model using \ac{tycho2} colour projections as observational constraints favours similar \ac{imf}s. The steep high-mass slopes decrease the number of \ac{sn2} ejected by an \ac{ssp} compared to classical \ac{imf}s like Salpeter or \ac{chabrier} by a factor of about 3.\\
The future of analytic Galaxy modelling will see increasing modularity to incorporate up-to-date theoretical progress in stellar atmospheres and evolutionary tracks. Observational biases like binarity, selection effects or reddening will be accounted for from the models side and the observational samples will get diversified to overcome degeneracies in the various model ingredients like \ac{sfh}, \ac{avr} and \ac{cem}. With ever more realistic \ac{mw} models traces of theoretical concepts like the \ac{cdm} halo or chemical yields can be mapped into the space of observables putting tighter constraints on the model ingredients.\\
To achieve that, tools to measure the probability of model predictions given the data and schemes to optimise for the various data sets in an automated hierarchical fashion need to be implemented. On the other hand high-quality data, able to discriminate between different model parameters, is needed. With the second data-release of \ac{gaia}, precise parallaxes of up to a billion stars will be published so that an exciting era for Galaxy modelling is lying ahead.
\acresetall

%% file: gfx/chapter3/plot1.pdf_tex
\begingroup%
  \makeatletter%
  \providecommand\color[2][]{%
    \errmessage{(Inkscape) Color is used for the text in Inkscape, but the package 'color.sty' is not loaded}%
    \renewcommand\color[2][]{}%
  }%
  \providecommand\transparent[1]{%
    \errmessage{(Inkscape) Transparency is used (non-zero) for the text in Inkscape, but the package 'transparent.sty' is not loaded}%
    \renewcommand\transparent[1]{}%
  }%
  \providecommand\rotatebox[2]{#2}%
  \ifx\svgwidth\undefined%
    \setlength{\unitlength}{713.07460938bp}%
    \ifx\svgscale\undefined%
      \relax%
    \else%
      \setlength{\unitlength}{\unitlength * \real{\svgscale}}%
    \fi%
  \else%
    \setlength{\unitlength}{\svgwidth}%
  \fi%
  \global\let\svgwidth\undefined%
  \global\let\svgscale\undefined%
  \makeatother%
  \begin{picture}(1,0.57966095)%
    \put(0,0){\includegraphics[width=\unitlength]{plot1.pdf}}%
    \put(0.24410763,0.52970162){\color[rgb]{0.50196078,0.50196078,0.50196078}\makebox(0,0)[lb]{\smash{{\footnotesize distance cuts}}}}%
    \put(0.09576271,0.03624359){\color[rgb]{1,0.50196078,0.50196078}\makebox(0,0)[lb]{\smash{0.5}}}%
    \put(0.17265022,0.03624359){\color[rgb]{1,0.50196078,0.50196078}\makebox(0,0)[lb]{\smash{0.6}}}%
    \put(0.24953774,0.03624359){\color[rgb]{1,0.50196078,0.50196078}\makebox(0,0)[lb]{\smash{0.7}}}%
    \put(0.32642526,0.03624359){\color[rgb]{1,0.50196078,0.50196078}\makebox(0,0)[lb]{\smash{0.8}}}%
    \put(0.40331278,0.03624359){\color[rgb]{1,0.50196078,0.50196078}\makebox(0,0)[lb]{\smash{0.9}}}%
    \put(0.4802003,0.03623264){\color[rgb]{1,0.50196078,0.50196078}\makebox(0,0)[lb]{\smash{1.1}}}%
    \put(0.55708777,0.03623264){\color[rgb]{1,0.50196078,0.50196078}\makebox(0,0)[lb]{\smash{1.4}}}%
    \put(0.63397529,0.03623264){\color[rgb]{1,0.50196078,0.50196078}\makebox(0,0)[lb]{\smash{1.7}}}%
    \put(0.71086281,0.03608473){\color[rgb]{1,0.50196078,0.50196078}\makebox(0,0)[lb]{\smash{2.2}}}%
    \put(0.78775038,0.03624359){\color[rgb]{1,0.50196078,0.50196078}\makebox(0,0)[lb]{\smash{2.9}}}%
    \put(0.86463781,0.03624359){\color[rgb]{1,0.50196078,0.50196078}\makebox(0,0)[lb]{\smash{4.0}}}%
    \put(0.94152533,0.03624359){\color[rgb]{1,0.50196078,0.50196078}\makebox(0,0)[lb]{\smash{6.4}}}%
    \put(0.63636265,0.00571057){\color[rgb]{1,0.50196078,0.50196078}\makebox(0,0)[lb]{\smash{dwarf mean mass in M$_\odot$}}}%
    \put(0.46771859,0.00539832){\color[rgb]{0,0,0}\makebox(0,0)[lb]{\smash{VMag in mag}}}%
    \put(0.47911209,0.53144637){\color[rgb]{0.8745098,0.8745098,0.50196078}\makebox(0,0)[lb]{\smash{1.0}}}%
    \put(0.55495956,0.5312875){\color[rgb]{0.8745098,0.8745098,0.50196078}\makebox(0,0)[lb]{\smash{1.2}}}%
    \put(0.63080702,0.5312875){\color[rgb]{0.8745098,0.8745098,0.50196078}\makebox(0,0)[lb]{\smash{1.2}}}%
    \put(0.70665444,0.53159427){\color[rgb]{0.8745098,0.8745098,0.50196078}\makebox(0,0)[lb]{\smash{1.5}}}%
    \put(0.7825019,0.53143542){\color[rgb]{0.8745098,0.8745098,0.50196078}\makebox(0,0)[lb]{\smash{1.7}}}%
    \put(0.85834936,0.53144637){\color[rgb]{0.8745098,0.8745098,0.50196078}\makebox(0,0)[lb]{\smash{1.8}}}%
    \put(0.93419683,0.53144637){\color[rgb]{0.8745098,0.8745098,0.50196078}\makebox(0,0)[lb]{\smash{3.4}}}%
    \put(0.60036774,0.55688211){\color[rgb]{0.8745098,0.8745098,0.50196078}\makebox(0,0)[lb]{\smash{giant mean mass in M$_\odot$}}}%
  \end{picture}%
\endgroup%

%% file: gfx/chapter3/IMF10a.pdf_tex
\begingroup%
  \makeatletter%
  \providecommand\color[2][]{%
    \errmessage{(Inkscape) Color is used for the text in Inkscape, but the package 'color.sty' is not loaded}%
    \renewcommand\color[2][]{}%
  }%
  \providecommand\transparent[1]{%
    \errmessage{(Inkscape) Transparency is used (non-zero) for the text in Inkscape, but the package 'transparent.sty' is not loaded}%
    \renewcommand\transparent[1]{}%
  }%
  \providecommand\rotatebox[2]{#2}%
  \ifx\svgwidth\undefined%
    \setlength{\unitlength}{720bp}%
    \ifx\svgscale\undefined%
      \relax%
    \else%
      \setlength{\unitlength}{\unitlength * \real{\svgscale}}%
    \fi%
  \else%
    \setlength{\unitlength}{\svgwidth}%
  \fi%
  \global\let\svgwidth\undefined%
  \global\let\svgscale\undefined%
  \makeatother%
  \begin{picture}(1,0.575)%
    \put(0,0){\includegraphics[width=\unitlength]{IMF10a.pdf}}%
    \put(0.10677616,0.52709581){\color[rgb]{0.30196078,0.30196078,0.30196078}\makebox(0,0)[lb]{\smash{{\footnotesize magnitude cut}}}}%
    \put(0.73197361,0.38165459){\color[rgb]{0,0,1}\makebox(0,0)[lb]{\smash{{\footnotesize scale height dilution}}}}%
    \put(0.7395891,0.18925743){\color[rgb]{1,0,0}\makebox(0,0)[lb]{\smash{{\footnotesize stellar evolution}}}}%
  \end{picture}%
\endgroup%

%% file: Chapters/Chapter04.tex
\chapter{Chemical enrichment model parameter estimation with APOGEE data}
\label{ch:chemistry} 
In section\,\ref{sec:chempy} Chempy, a new numerical chemical evolution model, will be presented. In a proof-of-concept, important parameters of the chemical model will be determined, as inferred from Solar Neighbourhood abundances of \ac{rc} stars. The statistical techniques to map the model into the space of observables and to construct a suitable statistical measure, for comparing the mock data to the real data, are discussed in sections\,\ref{sec:mapping}\,\&\,\ref{sec:lik}. With an \ac{mcmc} the posterior of the parameter space will be explored and the resulting new fiducial values are presented in section\,\ref{sec:result}. Different yield sets are tested in section\,\ref{sec:yieldsets} and the results indicate that the chemical model can put constraints on the high-mass index of the \ac{imf}.\\
The chapter is beginning by a short introduction to the concept of analytical chemical enrichment models following the derivation of \cite[chapter 3]{Matteucci2012}. 
\section{The simple one zone closed box model}
For the simple model, the following assumptions are made:
\begin{itemize}
\item the system consists of one zone and is closed
\item the initial gas is primordial
\item the \ac{imf} $\phi(m)$ is constant in time
\item the gas is well mixed at any time
\end{itemize}
First, the fractional gas mass is defined,
\begin{equation}
\mu=\frac{M_{gas}}{M_{tot}},
\end{equation}
which relates the gas mass to the total mass, containing stellar and gas mass,
\begin{equation}
M_{tot}=M_\star+M_{gas}.
\end{equation}
The system begins with $\mu=1$ and a metallicity $Z=0$, which is defined as the metal mass fraction of the gas,
\begin{equation}
Z=\frac{M_Z}{M_{gas}}.
\end{equation}
The equation governing the evolution of the gas in the system is
\begin{equation}
\frac{dM_{gas}}{dt}=-\psi(t)+E(t),
\label{eq:evo}
\end{equation}
where $\psi(t)$ is the \ac{sfr} and $E(t)$ is the rate at which dying stars restore both enriched and unenriched material into the ISM at time $t$, 
\begin{equation}
E(t)=\int\limits_{m(t)}^{\infty}(m-M_R)\psi(t-\tau_m)\phi(m)\mathrm{d}m.
\label{eq:et}
\end{equation}
Here $M_R$ is the remnant mass and $\tau_m$ is the lifetime for a star of mass $m$. Equation\,\ref{eq:evo} can only be solved analytically, when simplifying equation\,\ref{eq:et}, by assuming that all stars above a specific mass ($\mathrm{m}_\mathrm{IRA}$) die instantaneously and all stars below that limit live forever. This approximation is referred to as \ac{ira} and yields
\begin{equation}
\label{eq:r}
R(\mathrm{m}_\mathrm{IRA})=\int\limits_{\mathrm{m}_\mathrm{IRA}}^{\infty}(m-M_R)\phi(m)\mathrm{d}m,
\end{equation}
the total mass fraction restored into \ac{ism} by a stellar generation. $R$ depends on the \ac{imf} and on metallicity and varies between 0.2 and 0.5 for the usually adopted $\mathrm{m}_\mathrm{IRA}=$\,1\,M$_\odot$.\\
Another important quantity is the \emph{net yield per stellar generation}:
\begin{equation}
y_Z=\frac{1}{1-R}\int\limits_{1}^{\infty}mp_{Zm}\phi(m)dm
\end{equation}
where $p_{Zm}$ is the fraction of newly produced and ejected metals by a star of mass $m$, which is referred to as \emph{net stellar yields}.\\
With \ac{ira} equation\,\ref{eq:et} can be rewritten,
\begin{equation}
E(t)=\psi(t)R,
\end{equation}
which also simplifies equation\,\ref{eq:evo}
\begin{equation}
\frac{dM_{gas}}{dt}=-\psi(t)(1-R).
\label{eq:3.12}
\end{equation}
Focusing on metals only the equation governing the evolution writes 
\begin{equation}
\frac{d(ZM_{gas})}{dt}=-Z\psi(t)+E_Z(t),
\label{eq:3.13}
\end{equation}
and again, under the assumption of \ac{ira}, the convolution in the metal feedback,   
\begin{equation}
E_Z(t)=\int\limits_{m(t)}^{\infty}[(m-M_R)Z(t-\tau_m)+mp_{Zm}]\psi(t-\tau_m)\phi(m)dm,
\end{equation}
can be avoided by replacing with $R$,
\begin{equation}
E_Z(t)=\psi(t)RZ(t)+y_Z(1-R)\psi(t).
\end{equation}
This can be substituted in equation\,\ref{eq:3.13} and, after some algebraical manipulation, yields
\begin{equation}
\frac{dZ}{dt}=y_Z(1-R)\psi(t).
\end{equation}
Dividing by equation\,\ref{eq:3.12} it follows
\begin{equation}
\frac{dZ}{dM_{gas}}M_{gas}=-y_Z,
\end{equation}
which,  after integrating over $M_{gas}(0)=M_{tot}$ and $M_{gas}(t)$ and $Z(0)=0$ and $Z(t)$., has the following analytic solution
\begin{equation}
Z=y_Z\ln\left(\frac{1}{\mu}\right).
\end{equation}
Remarkably, the resulting metal enrichment is independent from the \ac{sfr}.\\

Due to the crude approximation, which is dominated by the \ac{ira}, the simple model is not a good representation of the metal evolution of our \ac{mw}. One major shortcoming is the \emph{G-dwarf problem} which was discovered by \citet{VandenBergh1962} and \citet{Schmidt1963} and refers to the overprediction of metal-poor stars by the model, which is not observed in Solar Neighbourhood stars. One obvious solution to this problem is to start from pre-enriched gas. The simple model can be extended to account for \emph{secondary elements}\marginpar{the production of \emph{secondary} elements is depending on the initial abundance of a star, contrary to \emph{primary} elements}, gas infall, or outflow. But it is not able to reproduce the $\alpha$-enhancement of old stars, see discussion in section\,\ref{sec:simple_chemistry}, unless the \ac{ira} is relaxed.\\
\begin{figure}
\caption[SSP mass fractions]{The time evolution of the mass fractions of an \ac{ssp} with Z = Z$_\odot$ and \acsu{jj} \ac{imf} from 0.025 to 12\,Gyr on a logarithmic time scale. The lifetime of a Solar mass star with metallicity Z$_\odot$ and 0.01\,Z$_\odot$ are indicated with the dashed magenta line and the dash-dotted green line respectively.}
\centering
\def\svgwidth{\columnwidth}

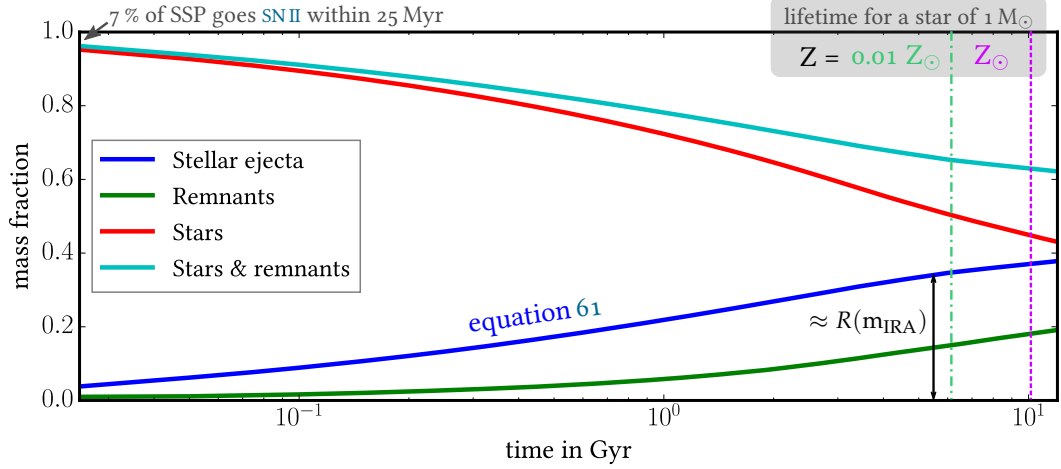
\label{fig:lifetime}
\end{figure}

To illustrate the impact of the \ac{ira}, the mass budget over time for a \ac{ssp} with Z = Z$_\odot$ and \acsu{jj} \ac{imf} is displayed in figure\,\ref{fig:lifetime}. The stellar lifetimes ($\tau(m,Z)$) are taken from \citet{Argast2000} and for simplicity it is assumed that they eject all their feedback at the end of their life, meaning that stellar wind feedback is delayed. For the remnant mass ($M_R$ from equation\,\ref{eq:et}) left from \acs{agb} stars and \ac{sn2} the values of \citet{Karakas2010} and \citet{Nomoto2013} are adopted respectively.\\
If the metal-dependency of $\tau(m)$ is neglected, so that it can be inverted $\tau^{-1}(t)$, the cumulative stellar feedback, which is depicted as the blue curve in figure\,\ref{fig:lifetime}, can be stated in terms of $R$ (c.f. equations\,\ref{eq:et}\,\&\,\ref{eq:r}),
\begin{equation}
\int\limits_{0}^{t}E(t)\mathrm{d}t=R(\tau^{-1}(t)).
\label{eq:ejecta}
\end{equation}
\indent For Z = Z$_\odot$ and $\mathrm{m}_\mathrm{IRA}=1$\,M$_\odot$ this means that the instantaneously returned material is about 38\,\% of the \ac{ssp} mass. Without \ac{ira} the feedback would be distributed over a time span of 10,3\,Gyr and feedback from low-mass and high-mass stars would be well separated.\\
For \ac{sn2} feedback \ac{ira} is a well justified approximation, because the lifetime of massive stars is too short to be resolved with 25\,Myr, which is the time resolution of the \ac{jj}. But \ac{sn2} only constitutes about 7\,\% of the \ac{ssp} and their overall feedback sums to 4\,\%, with the missing mass being confined in compact remnants.\\
It should be mentioned that $R$ is highly dependant on the assumed \ac{imf}. For \ac{ktg} \ac{imf} the \ac{sn2} feedback is 5\,\% and $R(\mathrm{m}_\mathrm{IRA})=$\,0.3. The respective quantities for \ac{chabrier} \ac{imf} are 0.14 and 0.43.\\
Another issue that can hardly be implemented into an analytical model is metal-dependency. As shown in the grey box of figure\,\ref{fig:lifetime} stellar lifetimes change considerably with metallicity. Therefore the yield of stars depends on the initial abundance of the star. With numerical integration the metal-dependence can be taken into account and the assumption of \ac{ira} can be relaxed.
\section{Chempy - a numerical chemical enrichment code}
\label{sec:chempy}
Chempy is a Python implementation of a simple chemical enrichment model, written by the author and intended for open-source publication. It was designed to be modular and fast with the aim to explore a huge parameter space with \ac{mcmc} simulations. After minor cleaning of the source code, Chempy will be staged publicly\footnote{The code will be available at github: \url{https://github.com/SIKz17/Chempy}.} for everybody to use and contribute. This is intended to help comparing and reproducing results of chemical evolution modelling.\\
So far a one zone model with gas infall is implemented so that it follows the chemical evolution of the thin disc. It can be easily extended for other purposes, e.g. mixing of different zones or Galactic outflows.\\
In the following the various routines of the code will be explained and put into their astrophysical context. The possible choices will be indicated together with the default model parameters.
\subsection{Solar abundances}
Solar abundances are very important because most observations are normalised with respect to the Sun. As the model calculates mass fractions, Solar abundances are necessary in order to compare the outcome with observations. So far the user can chose between two different sets of Solar abundances, namely \citet{Asplund2009} and \citet[tab.6]{Lodders2009}, with minor differences in the metallicity ($Z_\odot=0.0134$ and $Z_\odot=0.0141$, respectively) and also in element ratios for example in [O/Mg].\\ 
\subsection{Star formation history}
Chemical evolution models usually facilitate a Schmidt law, which connects the gas surface density with the rate of star formation \citep{Ch97}. In Chempy, the \ac{sfh} and the infall rate can be fixed separately, because the \ac{jj} already provides constraints for the thin disc star formation. The default is the global \ac{sfh} as depicted in the upper panel of figure\,\ref{fig:SFH}, which begins 12\,Gyr ago, peaks around 10\,Gyr look-back time and slowly declines until the present day. The time resolution is 25\,Myr, resulting in an array of 481 mass values, which are provided in units of $\tfrac{\mathrm{M}_\odot/\mathrm{pc}^2}{\mathrm{Gyr}}$. The total mass formed into stars is
\begin{equation}
\mathrm{M}_\mathrm{SFR}=\int\limits_{0\,\mathrm{Gyr}}^{12\,\mathrm{Gyr}}\mathrm{SFR}(t)\mathrm{d}t\simeq\sum_{i=0}^{480}\mathrm{SFR}(t_i).
\end{equation}
The functional form of the \ac{jj} \ac{sfh} can be directly manipulated and other forms can be easily included.
\subsection{Gas infall}
For the gas infall, the user has the choice between different functional forms. The default is an exponential function,
\begin{equation}
\dot{\mathrm{G}}(t)\propto e^{\lambda_\mathrm{infall} t}
\end{equation}
with $\lambda_\mathrm{infall}$ being the \emph{infall decay rate}. The overall infalling mass is normalised with respect to $\mathrm{M}_\mathrm{SFR}$ multiplied by the \emph{infall scaling factor} $\gamma_\mathrm{infall}$,
\begin{equation}
\mathrm{M}_\mathrm{infall}=\gamma_\mathrm{infall}\mathrm{M}_\mathrm{SFR}.
\end{equation}
Also the elemental composition of the gas infall can be chosen from primordial over $\alpha$-enhanced to Solar. The default being [Fe/H]=-3\,dex and an [$\alpha$/Fe]=0.4\,dex.
\subsection{Yield tables}
In order to calculate the feedback, yield tables are needed in machine readable form. The user can chose between different authors for the various feedback processes. The yield should be given as the feedback of both enriched and unenriched material, as in equation\,\ref{eq:et}, and will be normalised to the stellar mass.\\
\subsubsection{Supernova of type II}
The default is from the review yield set of \citet{Nomoto2013}. It provides feedback for metallicity $Z\in[0,0.001,0.004,0.008,0.02,0.05]$, the masses $\mathrm{M}/\mathrm{M}_\odot$ $\in$\\ $[13,15,18,20 , 25,30,40]$, and all the elements up to germanium-32. They also provide feedback for \ac{hn}, assumed to occur at 50\,\% rate for stars more massive than 20\,M$_\odot$, which slightly changes the chemical composition of the overall ejecta.\\
Other yield sets for massive stars are from \citet{Chieffi2004}, \citet{Francois2004} and \citet{Pignatari2013}.
\subsubsection{AGB stars}
For \ac{agb} stars the default yields were taken from \citet{Pignatari2013}. They provide feedback for metallicity $Z\in[0.01,0.02]$ ,the masses $\mathrm{M}/\mathrm{M}_\odot\in[1.65,2,3,5]$, and all the elements up to bismuth-83. The alternative yield set is coming from \citet{Karakas2010}
\subsubsection{Supernova of type Ia}
For \ac{sn1a} only one yield set is available coming from \citet{Iwamoto1999}. They provide the feedback for metallicities $Z\in[0.0001,0.02]$ including all elements from carbon-6 to zinc-30.
\subsection{IMF}
The \ac{imf} ($\phi(m)$) is one of the central parameters as it determines the mass fractions of stars going into the individual feedback processes. The choice includes the \acp{imf} from chapter\,\ref{ch:imf} and all parameters can be freely varied. The default is the new \ac{jj} \ac{imf} determined in \citet{Rybizki2015}, but with an additional break at $\mathrm{m}_2$ = 6\,M$_\odot$ and a \emph{high-mass index}, $\alpha_3$.\\
Not only the mass but also the number fraction of stars can be queried. The \ac{imf} is calculated in linear steps between $\mathrm{m}_\mathrm{low}$ = 0.08\,M$_\odot$ and $\mathrm{m}_\mathrm{up}$ = 100\,M$_\odot$ and is normalised to unity,
\begin{equation}
\int_{\mathrm{m}_\mathrm{low}}^{\mathrm{m}_\mathrm{up}} m\phi(m)\mathrm{d}m=1.
\end{equation}
The mass limits and the number of steps can be set at will. A good compromise between accuracy and speed is reached at $10^5$ mass steps. 
\subsection{Yield table of an SSP}
\begin{figure}
\caption[Delay time distribution of SNIa]{The occurrences of \ac{sn1a} for an \ac{ssp} of one M$_\odot$ over time in with logarithmic axes. The default Chempy model prescription with the fiducial parameters from equation\,\ref{eq:result_chemistry} are used. For comparison the data of \citet{Maoz2010} and \citet{Maoz2012} are plotted.}
\centering
\def\svgwidth{\columnwidth}
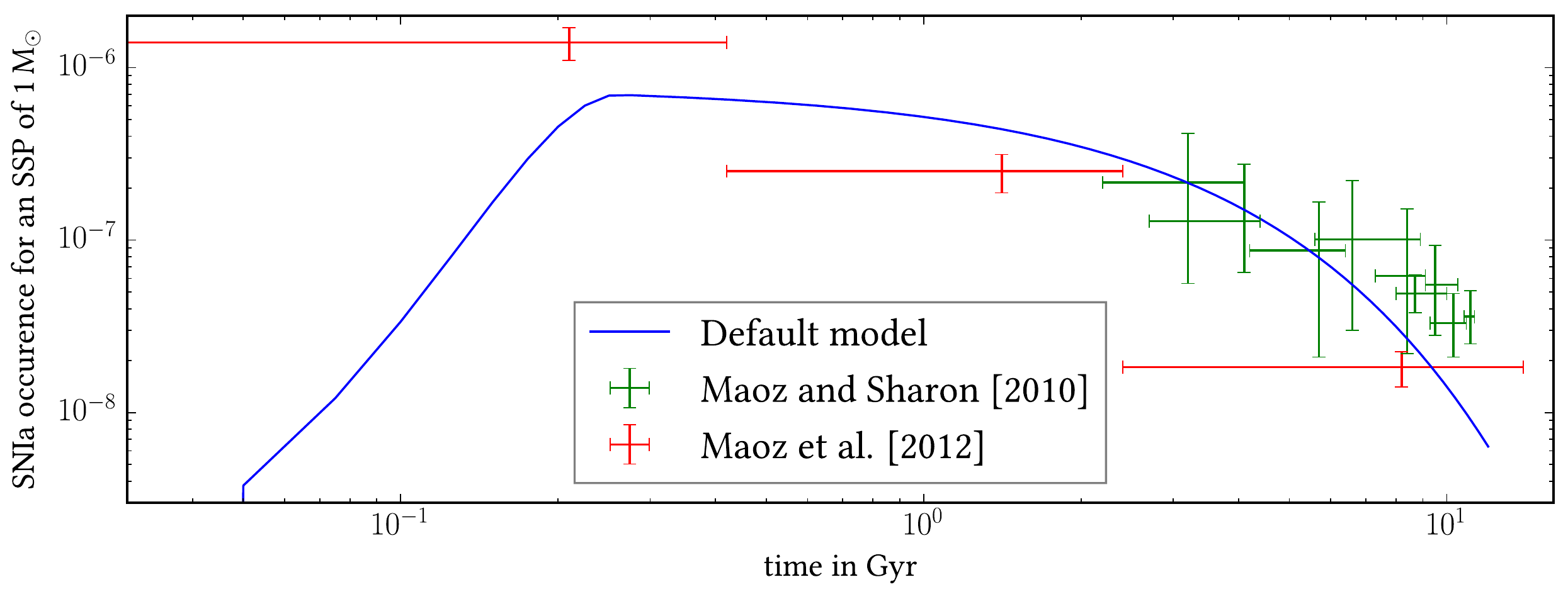
\label{fig:sn1a}
\end{figure}
In order to calculate the yield of an \ac{ssp} it is necessary to declare the mass range in which a specific feedback process acts. In the default model of Chempy, mass ranges similar to the \citet{Nomoto2013} chemical enrichment model are used, but can also be varied freely.
From $\mathrm{m}_\mathrm{low}$ = 0.08\,M$_\odot$ to $\mathrm{m}_\mathrm{AGB,low}$ = 0.8\,M$_\odot$, no feedback is necessary because stars live longer than 12\,Gyr. \ac{agb} feedback will be calculated for stars between $\mathrm{m}_\mathrm{AGB,low}$ and $\mathrm{m}_\mathrm{AGB,up}$ = 8\,M$_\odot$. The feedback of \ac{sn2} is calculated for masses of $\mathrm{m}_\mathrm{SNII,low}$ = 10\,M$_\odot$ and $\mathrm{m}_\mathrm{SNII,up}$ = 50\,M$_\odot$ and \ac{sn1a} are assumed to come from stars with $\mathrm{m}_\mathrm{SNIa,low}=\mathrm{m}_\mathrm{AGB,low}$ and $\mathrm{m}_\mathrm{SNIa,up}=\mathrm{m}_\mathrm{SNII,low}$. The time delay function of \ac{sn1a},
\begin{equation}
DTD(t)=
\left\{ \begin{array}{l}
 \exp\left(\tfrac{1}{2} \left(\tfrac{t-\mathrm{t}_\mathrm{peak}}{\sigma_\mathrm{SNIa}}\right)^2 \right), \\
 \exp\left(-\tfrac{(t-\mathrm{t}_\mathrm{peak})}{\tau_\mathrm{SNIa}} \right),
 \end{array} \mbox{\ if \ } \begin{array}{l}t<\mathrm{t}_\mathrm{peak}\\ t\ge\mathrm{t}_\mathrm{peak}\end{array} \right.
\label{eq:DTD}
\end{equation}
is parametrised using a Gaussian before the \emph{time of peaking \ac{sn1a} occurrence} ($\mathrm{t}_\mathrm{peak}$) and an exponential with decay time $\tau_\mathrm{SNIa}=$\,2.5\,Gyr thereafter, which can be inspected in figure\,\ref{fig:sn1a} for the parameters derived in equation\,\ref{eq:result_chemistry}, together with data from \citet{Maoz2010} and \citet{Maoz2012}. Noteworthy, the data has not been used in the \ac{mcmc} simulation to obtain the \ac{sn1a} parameters, though they guided the prescription of the \ac{dtd} functional form.\\
In the default model $\sigma_\mathrm{SNIa}$ is related to the peak time as $\sigma_\mathrm{SNIa}=\tfrac{1}{4}\mathrm{t}_\mathrm{peak}$. The number of \ac{sn1a} events ($\mathrm{n}_\mathrm{SNIa}$), distributed with the \ac{dtd} over a time span of 12\,Gyr, is determined by the \emph{\ac{sn1a} number fraction} $\beta_\mathrm{SNIa}$, of stars in the \ac{sn1a} mass range that explode,\\
\begin{equation}
\mathrm{n}_\mathrm{SNIa} = \beta_\mathrm{SNIa} \int\limits_{\mathrm{m}_\mathrm{SNIa,low}}^{\mathrm{m}_\mathrm{SNIa,up}}\phi(m)\mathrm{d}m.
\end{equation}
For mass conservation, the mean remnant mass ($\overline{M_R}$) per \ac{sn1a} is subtracted from the remnants of the \ac{ssp}. Then the missing mass, for reaching the Chandrasekhar limit, is subtracted from the hydrogen in the \ac{ism} and turned, together with the remnant, into the \ac{sn1a} feedback, which goes back into the \ac{ism}. In this way the single-degenerate progenitor model is reproduced, though the tail of the \ac{dtd} also accounts for the longer time needed for the in-spiral of the double-degenerate model.\\
The feedback from stars of the mass range between $\mathrm{m}_\mathrm{AGB,up}$ and $\mathrm{m}_\mathrm{SNII,low}$, referred to as super-\ac{agb}, is taken into account by feeding back the initial stellar composition into the \ac{ism} and leaving behind a compact remnant, made up from 13\,\% of the initial stellar mass, as in \citet[sec.2.1]{Kobayashi2011}. Similarly for stars more massive than $\mathrm{m}_\mathrm{SNII,up}$ the initial composition is given back via stellar winds and a \ac{bh} with 25\,\% of the initial mass of the star is created.\\

\begin{figure}
\caption[Net yields]{Net yields of an \ac{ssp} with Z=Z$_\odot$ normed to Solar for the different feedback processes. The default model of Chempy, with the parameter values of equation\,\ref{eq:result_chemistry}, is used and the feedback is plotted over time. Masses of stars dying are indicated in magenta dashed lines.}
\centering
\def\svgwidth{\columnwidth}
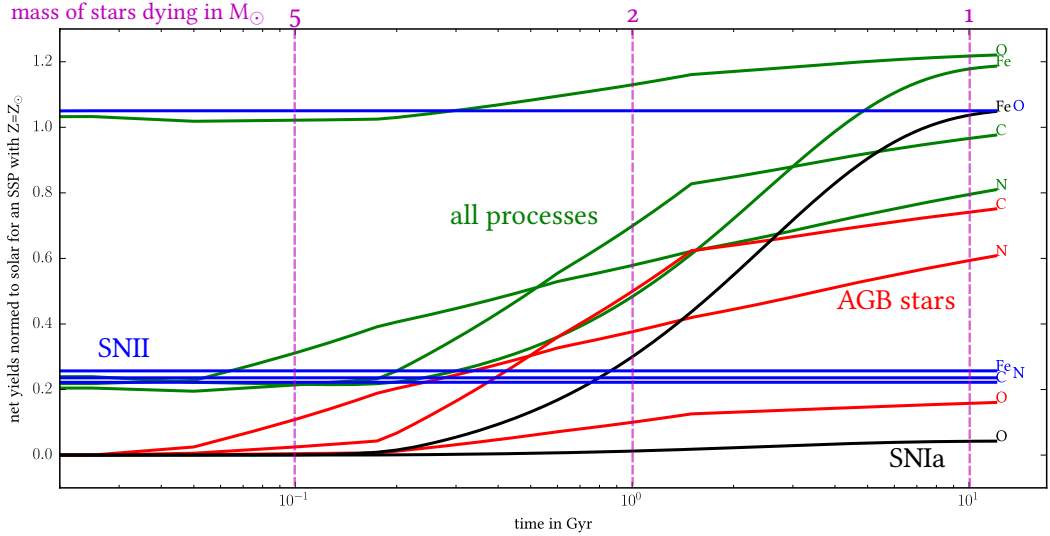
\label{fig:net_yields}
\end{figure}

After fixing the feedback processes, the mass return per time needs to be determined from \ac{imf}, using stellar lifetimes. In Chempy the metal and mass-dependent lifetimes of \citet[eq.3]{Raiteri1996} and \citet{Argast2000} are included, with the latter as default and only small differences between the two. Because the metallicity of an \ac{ssp} is fixed, the mass of dying stars ($\Delta\mathrm{m}$) can be calculated for each time step ($\mathrm{t}_\mathrm{i}$),
\begin{equation}
\Delta\mathrm{m}(\mathrm{t}_\mathrm{i})=\int\limits_{\tau^{-1}(\mathrm{t}_\mathrm{i})}^{\tau^{-1}(\mathrm{t}_\mathrm{i-1})}m\phi(m)\mathrm{d}m.
\end{equation}  
In order to distribute $\Delta\mathrm{m}(\mathrm{t}_\mathrm{i})$ across the elemental and remnant mass fractions of the total \ac{ssp} feedback, the yield tables are interpolated linearly in mass and metallicity for the corresponding processes at work for stars with initial mass between $\tau^{-1}(\mathrm{t}_\mathrm{i-1})$ and $\tau^{-1}(\mathrm{t}_\mathrm{i})$. The resulting table is also keeping track of the supernova events taking place and the \acp{wd} being produced over time.\\
In order to investigate the feedback from specific processes only, they can be switched on and off at will and plotted in different representations, like total mass fractions or net yields normed to Solar as depicted in figure\,\ref{fig:net_yields}. The yields of oxygen, carbon, nitrogen and iron are shown for a Solar metallicity \ac{ssp} over time, using the default Chempy model with the fiducial values from equation\,\ref{eq:result_chemistry}. The \ac{sn2} yield in blue contribute the bulk of oxygen and does not change with time, as all \ac{sn2} explode after the first time step. \ac{sn1a} contribute most of the iron and the long decay time ($\tau_\mathrm{SNIa}$) is dominating the time-dependence of the iron feedback.\\
An interesting feature of the \ac{agb} yields is that carbon is mostly produced in low-mass \ac{agb} stars (it takes more time) and nitrogen in high-mass \ac{agb} stars. This means that the carbon over nitrogen ratio is potentially sensitive to the \ac{imf}, which can be investigated in a future study.\\
In the chemical evolution of the thin disc all of these feedback processes work at the same time, which is indicated in green in figure\,\ref{fig:net_yields} for a single \ac{ssp}. Remarkably, the major contribution of the $\alpha$ elements, represented by oxygen here, can be attributed to \ac{sn2}, for iron it is \ac{sn1a}, and \ac{agb} stars produce the bulk of carbon and nitrogen. This means that these elements are sensitive to the feedback of the respective processes, which is why they are used in section\,\ref{sec:result} to guide the \ac{mcmc} simulation.
\subsection{Initial conditions}
As a last step, before the chemical evolution model can be started, the initial gas mass and its elemental abundances need to be set. The default is an amount of $\mathrm{G}_0=3\tfrac{\,\mathrm{M}_\odot}{\mathrm{pc}^2}$ with [Fe/H]=-0.7\,dex and an $\alpha$-enhancement of 0.4\,dex. This should simulate conditions of an \ac{ism} that has experienced fast enrichment by a pronounced initial burst of star formation.\\

The arising question is, what part and evolution of the \ac{mw} is the model actually reproducing. The \ac{jj} \ac{sfr} is only valid for the thin disc at Solar Galactocentric distance. When speaking of chemical enrichment, gas flows and radial migration of stars, which are not well constrained, should be taken into account. Otherwise the elemental abundances of Solar Neighbourhood stars are impossible to reconcile with the present-day cosmic abundance standard as inferred from young O and B-stars \citep{Nieva2012}. Excellent approaches to model the chemical enrichment of the whole disc can be found in \citet{Schonrich2009,Minchev2013a,Kubryk2015}.\\

So far the Chempy model is too simplistic incorporate radial migration but it should be extended in the future. The approach here is to look how far this simple model can approximate the reality as inferred from observations.
\subsection{Time integration}
The main program evolves the chemical model in the following way, which is also represented in the triangular  matrix of equation\,\ref{eq:cube} where the diagonal represents the chemical abundance state of the model and the upper right holds the yield tables of the different \acp{ssp}.\\
The simulation starts at $t=0$\,Gyr, column $\mathrm{s}_0$ and row $\mathrm{t}_0$, with the initial gas mass and adds the gas from the infall. Then the mass $\mathrm{SFR}(\mathrm{t}_0)$ of the first \ac{ssp} is subtracted and turned into stars, inheriting the elemental abundance of the gas. Lastly, the feedback table for the first \ac{ssp}, $\Delta\mathrm{m}_0$, is calculated for all times ($s_i, i\in0,1,...,480$), which fills up the first row.\\
Now the next time step ($\mathrm{t}_1,\mathrm{s}_1$) can be iterated. First the feedback $\Delta\mathrm{m}_0(\mathrm{s}_1-\mathrm{t}_0)$ is calculated, with remnants being kept separately. Then the gas infall of that time step is added, after which the second \ac{ssp} is formed. Lastly, the feedback for that \ac{ssp} until $\mathrm{s}_{480}-\mathrm{t}_1$ is calculated and the next iteration takes place.\\
For each of those the feedback just needs to be added up from the values in the same column above. This is done until $\mathrm{t}_{480}=\mathrm{s}_{480}=12$\,Gyr, except the gas turns negative, due to wrongly prescribed infall, which causes the simulation to prompt an error.
\begin{footnotesize}
\begin{equation}
\label{eq:cube}
\bordermatrix{
& \mathrm{s}_\mathrm{0}	& \mathrm{s}_\mathrm{1}   & \mathrm{s}_\mathrm{2}& ...  & \mathrm{s}_\mathrm{480} \cr
\mathrm{t}_\mathrm{0} & \mathrm{G}_0+\dot{\mathrm{G}}(\mathrm{t}_0)-\mathrm{SFR}(\mathrm{t}_\mathrm{0}) & +\Delta\mathrm{m}_0(\mathrm{s}_1-\mathrm{t}_0) & +\Delta\mathrm{m}_0(\mathrm{s}_2-\mathrm{t}_0) & ...& +\Delta\mathrm{m}_0(\mathrm{s}_{480}-\mathrm{t}_0) \cr
\mathrm{t}_\mathrm{1} & 0 & \dot{\mathrm{G}}(\mathrm{t}_1)-\mathrm{SFR}(\mathrm{t}_\mathrm{1}) & +\Delta\mathrm{m}_1(\mathrm{s}_2-\mathrm{t}_1) & ...& +\Delta\mathrm{m}_1(\mathrm{s}_{480}-\mathrm{t}_1)\cr
\mathrm{t}_\mathrm{2} & \vdots & \ddots & \dot{\mathrm{G}}(\mathrm{t}_2)-\mathrm{SFR}(\mathrm{t}_\mathrm{2}) & ...&+\Delta\mathrm{m}_2(\mathrm{s}_{480}-\mathrm{t}_2) \cr
\vdots & \vdots & 0 & \ddots & \ddots  & \vdots\cr
\mathrm{t}_\mathrm{480}	& 0 	&\dots	& \dots	& 0	& \dot{\mathrm{G}}(\mathrm{t}_{480})-\mathrm{SFR}(\mathrm{t}_\mathrm{480}) \cr
}
\end{equation} 

\end{footnotesize}
Thanks to the Python library numpy \citep{VanderWalt2011}, these array manipulations are reasonably fast. On a modern processor a single simulation, including all elements, takes about 20\,s.
\subsection{Output}
The outcome of a simulation can be analysed with different plotting routines that are implemented. Also the different yield sets can be investigated. The program is written in such a way that it can be called by an \ac{mcmc} in order to explore the parameter space of input functions. For that also a probability measure is needed, the construction of which is explained in section\,\ref{sec:lik}.\\
Since a large simulation should not be interrupted, the program, by default, gives a prompt upon failure, which usually happens because of negative gas, and returns a probability of $-\infty$.
\section{Mapping model outcome into the space of observables}
\label{sec:mapping}
Before a likelihood can be constructed a data set with discriminative power for the problem under investigation needs to be found and the model needs to be prepared in such a way that it represents the observations as good as possible. First the data is presented and then the way, in which the model is mapped into the space of observables.\\
The method has also been described in a submitted conference proceedings paper \citep{Just}.
\subsection{Data}
A homogeneous set of high quality abundances of around $10^5$ stars have been published in the \ac{apogee} survey. A catalogue for a sub-sample of 20,000 \ac{rc} stars with added distances by \citet{Bovy2014} is used here, because spatial cuts can be applied to the stars. All stars with 7.5\,kpc < $R_{\mathrm{Gal}}$ < 8.5\,kpc and -150\,pc < $Z_{\mathrm{Gal}}$ < 150\,pc are selected in order to stay within the locus of the \ac{jj}. This yields a sample of 412\,\ac{rc} stars, for which up to 15 elemental abundances are available.\\
\subsection{Model}
\begin{figure}
\caption[Colour-magnitude diagram selection]{Selection of specific stars in the \ac{cmd}. The stars have been synthesised with a flat \ac{sfr} and the \ac{jj} parameters using \ac{galaxia} and their binned density is shown in logarithmic colour code. The \ac{rc} selection is shown in magenta and the old \ac{ms} in cyan. The corresponding age distribution of these selected stellar samples can be inspected in figure\,\ref{fig:selection_age}}
\centering
\def\svgwidth{\columnwidth}

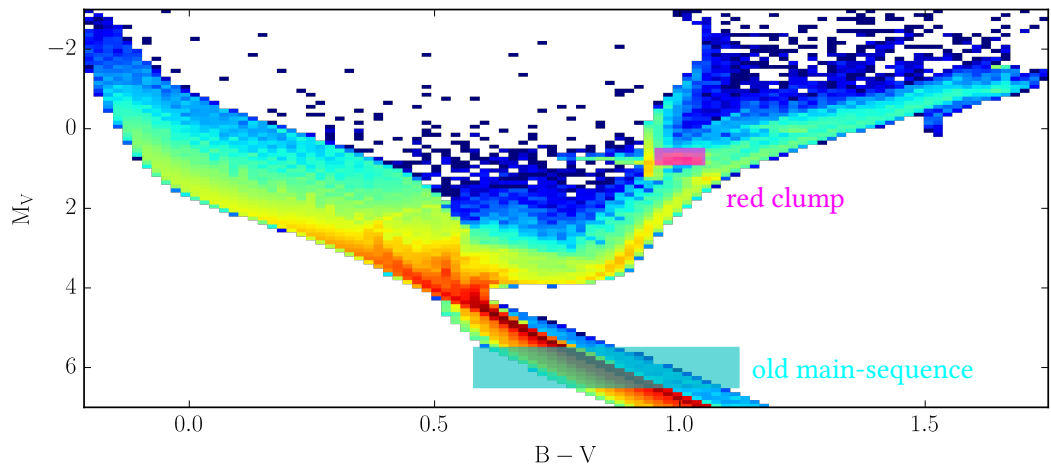
\label{fig:selection}
\end{figure}
\begin{figure}
\caption[Age distribution of specific stellar types]{Distribution functions of \ac{rc} stars in magenta, old \ac{ms} stars in cyan as selected in figure\,\ref{fig:selection}, the global \ac{jj} in dashed green, and a local \ac{jj} representation in dashed blue. The upper panel shows the age distribution of the different selections and the lower panel the resulting iron distribution functions with an added Gaussian scatter of 0.1\,dex. For comparison the local \ac{apogee} \ac{rc} sample distribution convoluted with its error is plotted. The chemical model parameter are the same as in figure\,\ref{fig:selection_height}, which are not the fiducial values.}
\centering
\def\svgwidth{\columnwidth}

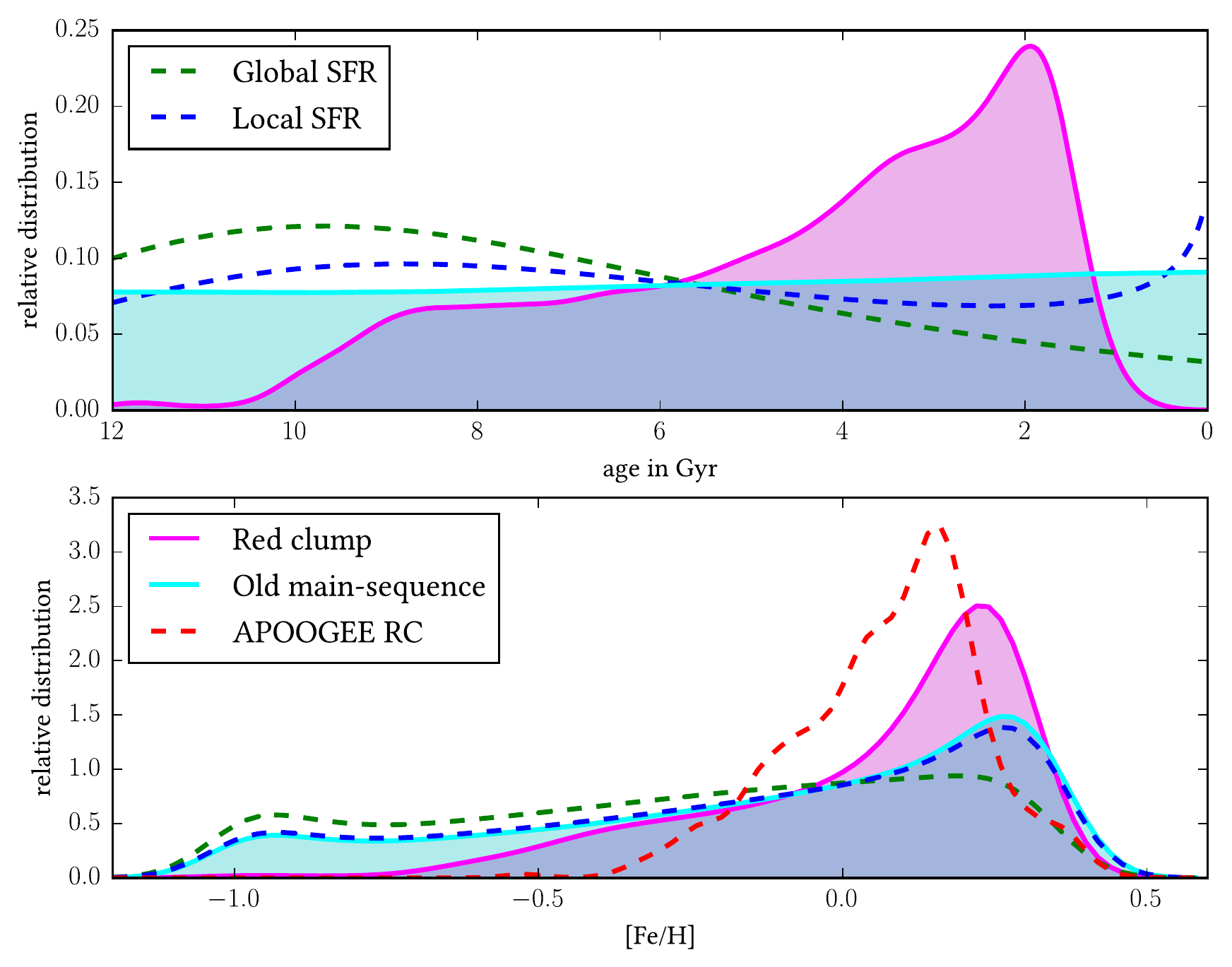
\label{fig:selection_age}
\end{figure}

With Chempy it is possible to reproduce the selections, as applied to the data in the previous section. This is realised by taking the relative chemical abundances of the gas at each time-step and weight them with the age-distribution of the stellar sample. In order to reproduce Solar Neighbourhood stars with a lifetime > 12\,Gyr the global \ac{sfr} of the \ac{jj} needs to be corrected for the scale height dilution (see discussion in section\,\ref{sec:localmodel}). These relative weighting factors are represented in the N-body masses of the lower panel of figure\,\ref{fig:SFH} and also by the blue dashed curve in the upper panel of figure\,\ref{fig:selection_age}.\\
Additionally, the age distribution of a specific stellar type, e.g. \ac{rc} stars, can be reproduced with \ac{galaxia}, using PARSEC \citep{Bressan2012} isochrones. It is realised by feeding the \ac{jj} parameters and a flat \ac{sfr} into \ac{galaxia} and apply the (e.g. photometric) cuts, as visualised in figure\,\ref{fig:selection}, to the resulting stellar sample. For \ac{rc} and old \ac{ms} stars the relative age distributions are shown in the upper panel of figure\,\ref{fig:selection_age}.\\
Multiplying these stellar type age distributions (e.g. the magenta line from figure\,\ref{fig:selection_age}, for \ac{rc} stars) with the age distribution of the \ac{jj}, which represents a spatial selection (e.g. the blue dashed curve of figure\,\ref{fig:selection_age}, for the Solar Neighbourhood), and using the resulting relative age distributions to weight the abundances from the chemical evolution model, gives a very good approximation of the abundance distribution derived from stars with similar (spatial and stellar type) cuts.\\
\begin{figure}
\caption[Abundances with height]{The changing $\alpha$-enhancement with Galactic height is shown with the model in magenta and the 50\,star average of the \ac{apogee} \ac{rc} sample. From upper to lower panel the height above the Galactic plane increases from 0 to 600\,pc. The chemical model stays the same but its error-convoluted and age-weighted density distribution changes. The age distributions for \ac{rc} stars and different heights above the Galactic plane from the \ac{jj} are used. In cyan the time evolution of the model is indicated. The model parameters are not the fiducial values from equation\,\ref{eq:result_chemistry}, but were randomly chosen, to illustrate the effect of the different age-distributions representing the spatial and stellar-type selection functions.}
\centering
\def\svgwidth{\columnwidth}
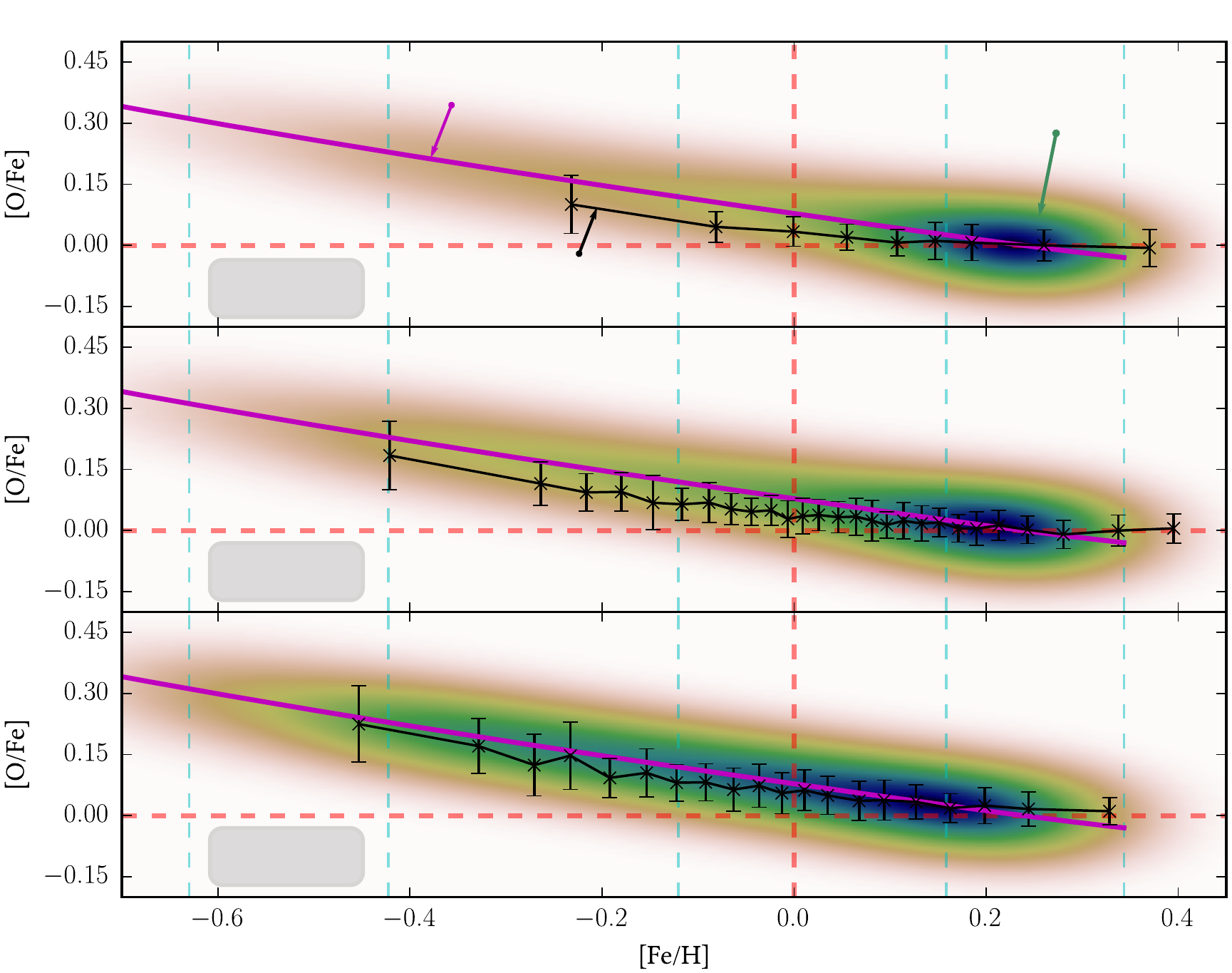
\label{fig:selection_height}
\end{figure}
To illustrate this effect, the model is compared to \ac{rc} stars for different heights above the Galactic plane in figure\,\ref{fig:selection_height}. It is worth noting that, even though the model stays the same for all three heights and the \ac{apogee} \ac{rc} stars become metal-poorer with height, the age-weighted model tracks the difference quite well. One caveat is that the thick disc is not modelled, which is contributing with an increasing stellar fraction with increasing height above the Galactic plane. On the other hand, since the \ac{rc} stars are quite young (see figure\,\ref{fig:selection_age}), only few of those should come from the thick disc.\\
The difference for different stellar populations in iron abundance can be inspected for the Solar Neighbourhood in the lower panel of figure\,\ref{fig:selection_age}. For comparison the local \ac{apogee} \ac{rc} sample is also plotted. Without the stellar type selection, the local model abundance distribution would be incompatible with the data. The \ac{rc} model distribution is still off but resembles the data much better. Beware that the model used to plot figures\,\ref{fig:selection_age}\,\&\,\ref{fig:selection_height}, is not containing the fiducial parameter values yet and just used here to illustrate the selection effects. The peak at -0.9\,dex for the old \ac{ms} is due to the global \ac{sfr} peak and the pre-enriched gas mass $\mathrm{G}_0$ from which the simulation starts.\\

These two selection effects should be the ones with strongest effect on the \ac{apogee} \ac{rc} sample, as the survey is designed to minimise observational biases in age and metallicity \citep{Zasowski2013} and represent the Galactic giant star distribution.
\section{Posterior prescription}
\label{sec:lik}
As Chempy is intended to extend the \ac{jj} and to further constrain its input parameters (especially the infall function ($\gamma_\mathrm{infall},\lambda_\mathrm{infall}$), the high-mass index ($\alpha_3$), and the \ac{sn1a} parameters ($\mathrm{t}_\mathrm{peak},\beta_\mathrm{SNIa}$), the used data sets need to be chosen accordingly.\\
\subsection{Likelihood from data}
Very few data are used in this first approach to determine the chemical evolution model parameters, because of the many subtleties involved in order to set the \ac{mcmc}.
In order to have a constraint on the infall and also on the supernova rates, the local present day gas mass fraction $\mathrm{G}_\mathrm{p}=10.3\pm3$ \citep{Kubryk2015} is used and the \ac{sn2} over \ac{sn1a} ratio for \ac{mw} like galaxies $\mathrm{SNII}/\mathrm{SNIa}=5\pm3$ based on \citet{Mannucci2005}.\\
These quantities are assumed to be Gaussian distributed, which admittedly is an oversimplification, especially in the supernova rate case. In order to balance the likelihood penalty from different data sets as well as the penalty contribution from the priors easier, the Gauss approximation was chosen. The normalised log-likelihood (as well as the prior probability) for a model value ($m_i$) to represent the data value ($d_i$) with a standard deviation ($\sigma_i$), will be constructed as
\begin{equation}
\mathcal{P}_i=\ln\left(\tfrac{\mathcal{L}_i}{\mathrm{p}_{i\mathrm{,max}}}\right)=\ln\left(\frac{\mathcal{N}(m_i,d_i,\sigma_i)}{\mathcal{N}(d_i,d_i,\sigma_i)}\right) = \frac{(m_i-d_i)^2}{2\sigma^2},
\end{equation}  
with $\mathcal{N}(x,\mu,\sigma)$ being the Gaussian distribution. The overall likelihood becomes a summation
\begin{equation}
\mathcal{P}=\sum\limits_{i=1}^{n}\mathcal{P}_i
\end{equation}
over all the data points. For reference the penalties for different standard deviations offset of observed values are given in table\,\ref{tab:sigmas}.\\
\begin{table}
\begin{tabular}{ c| c c c c c c c}
\hline
$\tfrac{m_i-d_i}{\sigma_i}$ &$ 0$ &$ 1$ & $2$ & $3$ & $4$ &$ 5$ &$ 6$ \\
\hline 
 
 $\ln\left(\tfrac{\mathcal{L}_i}{\mathrm{p}_{i\mathrm{,max}}}\right)$ &$ 0$ &$ -0.5$ &$ -2 $& $-4.5 $& $-8$ & $-12.5$ & $-18$ \\
 $\tfrac{\mathcal{L}_i}{\mathrm{p}_{i\mathrm{,max}}}$ & $1$	&$ 0.61$	& $0.14$	&$ 0.01$ &$ 3.4\times 10^{-4}$ & $3.7\times 10^{-6}$ &	$1.5\times10^{-8}$ \\

\hline
\end{tabular}
\caption{The log-likelihood penalty for the model differing from the data, normed to standard deviations.}
\label{tab:sigmas}
\end{table}

For the local sample of the \ac{apogee} \ac{rc} elemental abundances only four dimensions of the chemical space are used. The metal content ([M/H]) is the only abundance that is externally calibrated \citep{Holtzman2015}. It is fitted, together with [$\alpha$/M], [C/M] and [N/M], to the whole spectrum and treated as a stellar parameter, contrary to the other elements. Carbon and nitrogen are added because dredge up alters the C/N ratios, but the sum of both should remain relatively constant over the stellar lifetime of the star \citep{Hawkins2015}.\\
$\text{[C+N/M]}$ constrains the \ac{agb} contribution to the chemical enrichment. [$\alpha$/M] is mainly related to \ac{sn2}, therefore [Fe/H] is added in order to have an element that is selectively sensitive to \ac{sn1a} feedback.\\

The data points (for the Gaussian likelihood evaluation) are constructed from the element distribution function quantiles [5,25,50,75,95]. This yield five data points per elemental dimension, which is compared to the corresponding model value. For $\sigma$ the median error given for the \ac{rc} sample is taken.\\
Overall four times five correlated data points from elemental abundances are used to establish a statistic constraining the parameters. Together with the two data points $G_0$ and the supernova ratio they are intended to guide the \ac{mcmc} parameter exploration into a position motivated by the observational data.
\subsection{Assigning priors}
Contrary to chapter\,\ref{ch:imf}, where flat priors are used, here the free parameters are constrained with relatively\footnote{The weight of the priors are five broad Gaussians, which compared to the likelihood with its 22\,Gaussians from the data, contribute comparatively little to the final posterior probability.} relaxed Gaussian priors, taken into account realistic assumptions. For parameter like $\beta_\mathrm{SNIa}$ a prior distribution is not easy to anticipate a priori, so they were determined empirically through testing and also large $\sigma$ values were chosen, in order avoid too tight parameter constraints. The \acp{pdf} representing the parameter priors are,
\begin{equation}
\begin{array}{rclr}
\alpha_3 &=& \mathcal{N}\left(\mu = \text{2.5},\, \sigma = \text{0.5}\right),&\text{(high-mass index)}\\
\beta_\mathrm{SNIa} &=& \mathcal{N}\left(\text{0.004},\, \text{0.002}\right),&\text{{\footnotesize (number fraction of stars in the \ac{sn1a} mass range that explode)}}\\
\mathrm{t}_\mathrm{peak} &=& \mathcal{N}\left(\text{0.8\,Gyr},\,\text{0.3\,Gyr}\right),&\text{(time of peaking \ac{sn1a} occurrence)}\\
\gamma_\mathrm{infall} &=& \mathcal{N}\left(\text{0.8},\, \text{0.2}\right),&\text{(infall mass scaling factor)}\\
\lambda_\mathrm{infall} &=& \mathcal{N}\left(\text{-0.3},\, \text{0.2}\right).&\text{(infall decay rate)}\\
\end{array}
\end{equation} 

The \ac{mcmc} is free to chose from the parameters, except when unphysical limits are violated, which results in Chempy returning a posterior of $-\infty$. Therefore the range of parameters is effectively confined to
\begin{equation}
\begin{array}{rcl}
\alpha_3 &\ge& 0\\
\beta_\mathrm{SNIa} &>& 0\\
\mathrm{t}_\mathrm{peak} &\ge& \text{7.5\,Myr}\\
\mathrm{t}_\mathrm{peak} &\le& \text{6\,Gyr}\\

\gamma_\mathrm{infall} &\ge& 0.\\
\end{array}
\end{equation}
As mentioned before $-\infty$ will also be returned when the gas mass becomes negative during the calculation.\\

\subsection{Remarks on the posterior sampling}
The log-posterior is obtained by adding the log-prior and the log-likelihood. It is returned to the \ac{mcmc}, which requires the probability of a parameter position in logarithmic scale.\\ As in chapter\,\ref{ch:imf}, \emph{emcee}\marginpar{emcee is a Python implementation \citep{Foreman2013} of an affine invariant ensemble sampler for \ac{mcmc} \citep{Go10}} is used to sample the posterior \ac{pdf} of the parameter space. In this specific case multiprocessing with 60 threads and walkers was used to speed up the calculation. On a small cluster with 16 modern processing units it takes about five hours to sample $10^5$ values from the posterior, sufficient to approximate its distribution if the burn-in phase is not too long.\\
The initial parameters are assigned to each of the 60\,walkers by drawing randomly from the Gaussian, $\mathcal{N}(\mu=\mu_\mathrm{prior},\sigma = \tfrac{1}{4}\mu_\mathrm{prior})$, for each of the five parameters. Few walkers start with negative gas values, but the whole sample converges quickly to the equilibrium position of highest posterior probability.
\section{Results}
\label{sec:result}
\begin{figure}
\caption[Posterior of the parameter space]{Marginalised parameter distribution of the \ac{mcmc} sampling from the posterior. Each scatter plot shows the projected 2D parameter distribution with points colour-coded according to their posterior probability. The black cross indicates the position with highest posterior and ellipses encompass the 1$\sigma$ and 2$\sigma$ regions. The respective correlation coefficients are given at the position mirrored along the diagonal. Gaussian fits and histograms of the marginalised parameter distribution are given on the diagonal. The mean and standard deviation of each parameter is written above the histogram and are also indicated by solid and dashed grey lines.}
\centering
\def\svgwidth{\columnwidth}
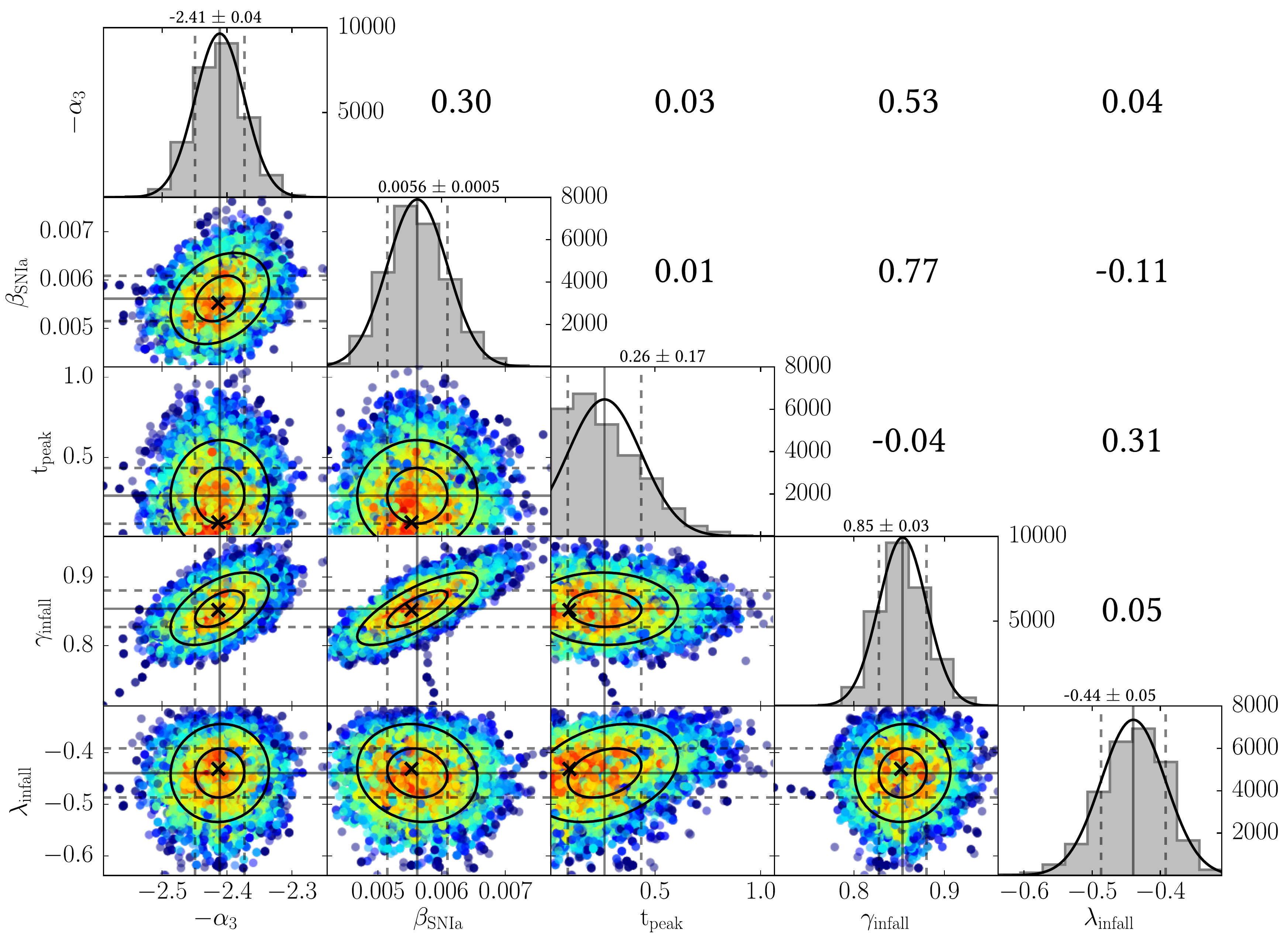
\label{fig:default_posterior}
\end{figure}

The default model posterior \ac{pdf} is approximated by the \ac{mcmc} that is depicted in figure\,\ref{fig:default_posterior}. Excluding the burn-in phase, which typically takes 50 steps with 60 walkers, a total of 26,000 samples were drawn from the posterior. The highest value is marked with a black cross in the 2D projections. The colour code spans from the best log-posterior value, -43.7 in red, to -51.7 in blue, which is the equivalent of a 4$\sigma$ offset.\\

\begin{figure}
\caption[Metal distribution]{Metal distribution function of the fiducial model and the local \ac{apogee} \ac{rc} stars. The construction of the likelihood, from the difference of the quantiles, is indicated at the top.}
\centering
\def\svgwidth{\columnwidth}
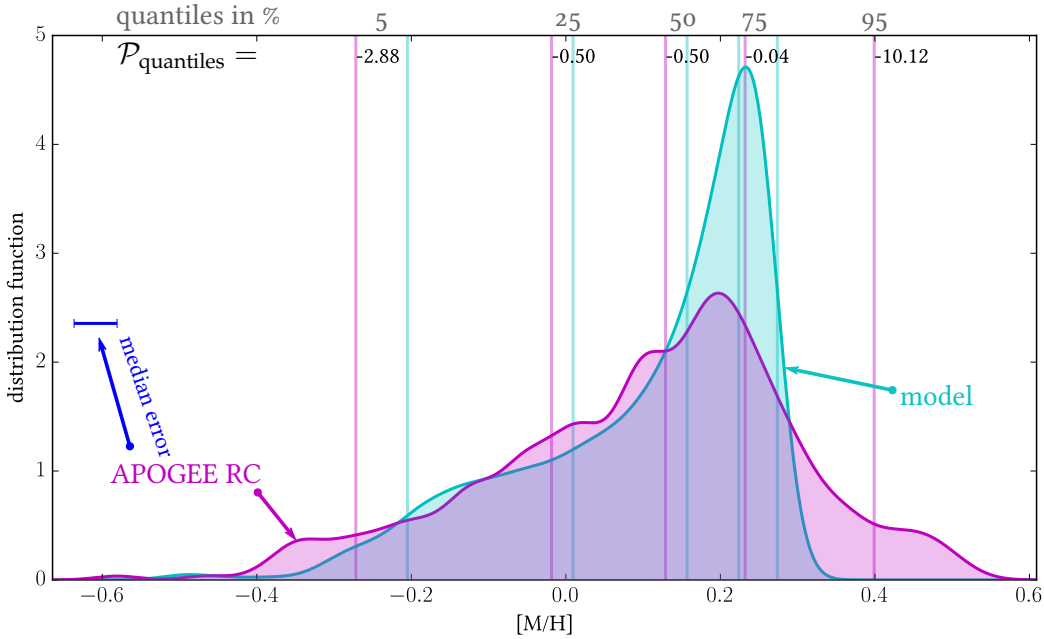
\label{fig:apogee_metals}
\end{figure}

Overall the posterior distribution is well-behaved. Except for $\mathrm{t}_\mathrm{peak}$ all marginalised parameter distributions are well approximated by a Gaussian with the highest value falling onto the mean of the distribution.\\
The strongest correlation is between $\beta_\mathrm{SNIa}$ and $\gamma_\mathrm{infall}$ because with increasing primordial gas mass also an increasing number of \ac{sn1a} is needed to match the metal and iron distribution. Similarly, but not as pronounced, $\alpha_3$ is also correlated with the infalling gas mass. Minor correlations are between the number of \ac{sn1a} and the high-mass index, which is essentially the number of \ac{sn2}. These correlate because the [$\alpha$/M] distribution depends on their overall feedback matching. Also the infall time scale and $\mathrm{t}_\mathrm{peak}$ are positively correlated, probably because if more gas is falling in earlier (e.g. $\lambda_\mathrm{infall}$ is smaller) then the metal enrichment need to act earlier as well.\\
The \ac{sn1a} peak time is, contrary to the other values, not tightly constraint by the used data. Maybe because the \ac{rc} stars only sample a relatively young stellar population. Probably also because the \ac{dtd} is a wide spread distribution. Shorter values are preferred, but the best posterior is obtained at 0.1\,Gyr, though also shorter values could have been possible. In order to be easily comparable with other \ac{mcmc} runs the Gaussian mean values of each parameter are adopted for the new fiducial model.
\begin{equation}
\label{eq:result_chemistry}
\begin{array}{rcl}
\alpha_3 &=& \text{2.41}\pm\text{0.04},\\
\beta_\mathrm{SNIa} &=& \text{0.0056}\pm\text{0.0005},\\
\mathrm{t}_\mathrm{peak} &=& \text{0.26\,Gyr}\pm\text{0.17\,Gyr},\\
\gamma_\mathrm{infall} &=& \text{0.85}\pm\text{0.03},\\
\lambda_\mathrm{infall} &=& \text{-0.44}\pm\text{0.05}.\\
\end{array}
\end{equation}
\begin{figure}
\caption[Model mass fractions]{The fiducial model mass evolution over time. Gas mass, living stellar mass and remnant mass are shown in blue, green and red, respectively. The cumulative values of the stellar feedback, the infall and the \ac{sfr} are indicated in cyan, magenta and yellow. The \ac{sfr} is fixed by the \ac{jj} but the infall function parameters could vary freely (with respect to the priors) in the \ac{mcmc}. The data constraint is indicated and the simulation settles at the upper end of the \citet{Kubryk2015} value.}
\centering
\def\svgwidth{\columnwidth}
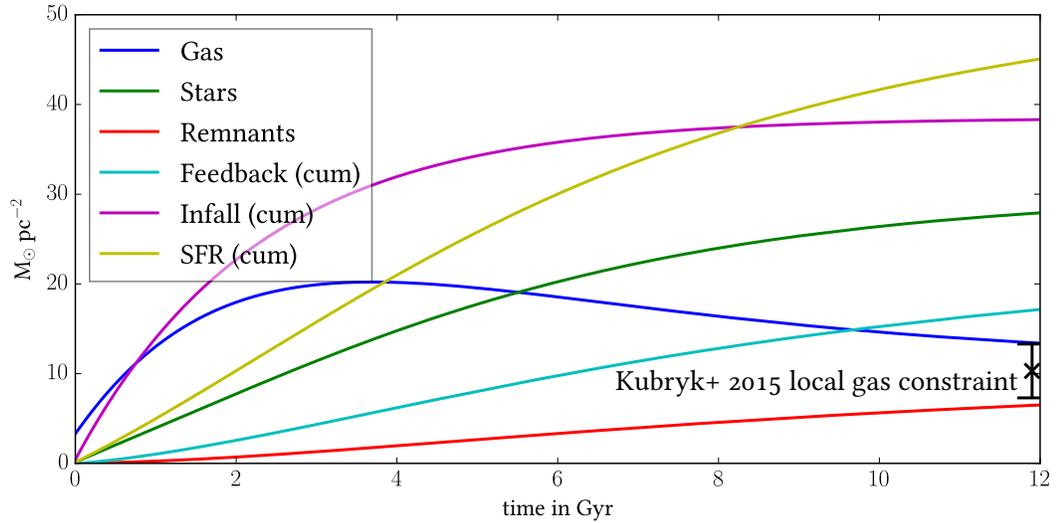
\label{fig:model}
\end{figure}
When using these parameters the posterior obtains the following contributions. The log-priors add up to -2.2. The log-likelihood in total is -42.6. The gas mass at end, as indicated in figure\,\ref{fig:model} in blue together with other global model parameters, is 13.4\,M$_\odot$\,pc$^\text{-2}$ contributing $\mathcal{P}_{\mathrm{gas}}$\,=\,-0.53. The SNII/SNIa ratio at the present day is also roughly one sigma off, with a value of 1.6 resulting in $\mathcal{P}_{\mathrm{SN}}$\,=\,-0.64. The abundance distributions from \ac{apogee} \ac{rc} stars yield log-likelihoods of
\begin{equation}
\begin{array}{rclr}
\mathcal{P}_\mathrm{[Fe/H]}& =& -14.82,\\
\mathcal{P}_\mathrm{[M/H]} &=& -14.05,\\
\mathcal{P}_{\mathrm{[}\alpha\mathrm{/M]}} &=& -5.78,\\
\mathcal{P}_\mathrm{[C+N/M]} &=& -6.73.
\end{array}
\end{equation}
From these values it is clear that the main penalty, guiding the \ac{mcmc}, is originating from the abundance data. This illustrates that the priors are not tightly constraining the parameter space, instead the abundance data has by far the largest impact on the posterior.\\
The compared metal distributions, together with the broken down likelihood contributions from each quantile, are illustrated in figure\,\ref{fig:apogee_metals}. The first four quantiles match quite well, with only minor likelihood penalties, as indicated at the top of the figure. The major penalty is contributed by the last 95\,\% quantile, ultimately originating from a metal-rich population, which could coincide with the $\alpha$-rich young thin disc population found in recent spectroscopic surveys \citep{Chiappini2015,Bergemann2014}. That indicates that the chemical enrichment of the \ac{mw} disc is more complex than can be reproduced by the one zone model. Interestingly, the [Fe/H] matches quite well in the metal-rich part and also the [$\alpha$/M] distribution fits very good.\\

\begin{figure}
\caption[Sulphur over iron]{The [S/Fe] - [Fe/H] for the model and the local \ac{apogee} \ac{rc} stars is plotted. The age-weighted model, convolved with the observational error, is colour-coded in the background.}
\centering
\def\svgwidth{\columnwidth}
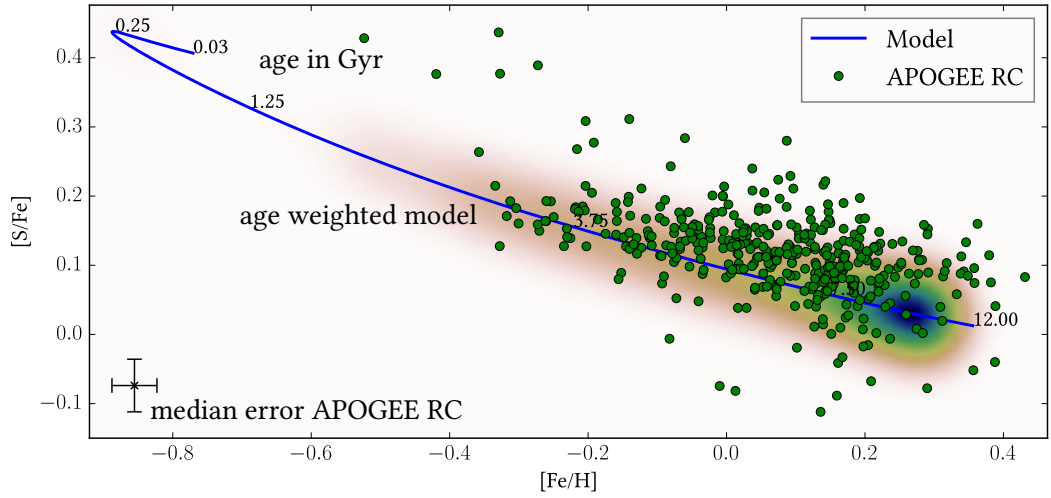
\label{fig:sulphur}
\end{figure}

\begin{figure}
\caption[Manganese over magnesium]{Same as figure\,\ref{fig:sulphur} but for the [Mn/Mg] - [Mg/H] abundance plane.}
\centering
\def\svgwidth{\columnwidth}
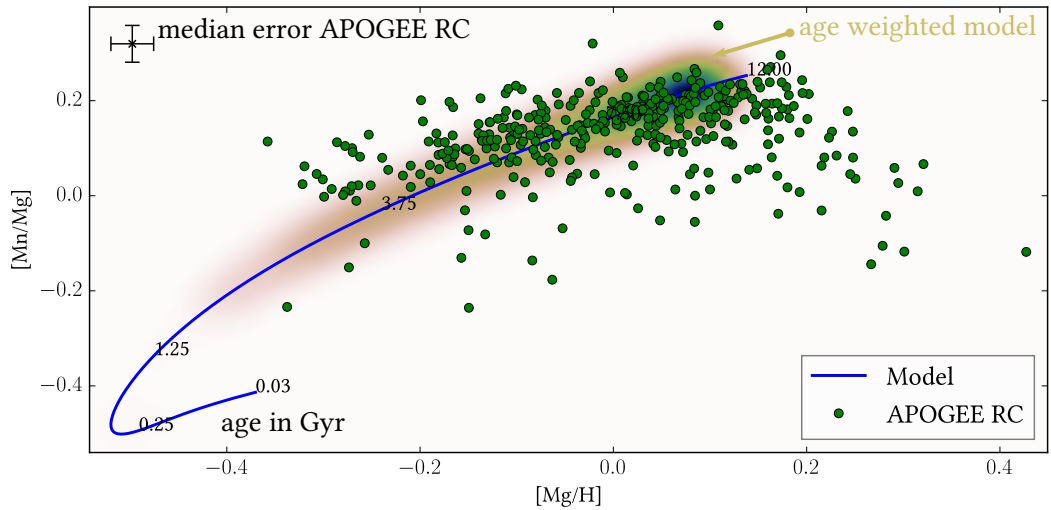
\label{fig:manganese}
\end{figure}

Even though, other elemental abundances are not used to equilibrate the \ac{mcmc}, the used metal, iron, $alpha$ and C+N distributions should constrain the three feedback processes well enough, to have reasonable abundance patterns for other elements, provided that the yield set is realistic. Two example elemental distributions, where it works well, are shown in figure\,\ref{fig:sulphur}\,\&\,\ref{fig:manganese}. In figure\,\ref{fig:sulphur} the sulphur-enhancement is shown for the age-weighted model and the local \ac{apogee} \ac{rc} sample. The correlation in the [S/Fe] - [Fe/H] plane is reproduced well, though the iron enrichment seems to be a little too strong in the model. A spurious effect can be seen in the beginning of the model, when the initial abundances for the gas at the beginning are lowered and then increase quickly. This happens due to the primordial gas infall and the metal-dependent stellar yields. After about one\,Gyr the dependence on the initial conditions seems to play a minor role and the model evolves smoothly.\\
Manganese is an element which, is also mainly produced in \ac{sn1a}. When plotting it with respect to the $\alpha$-element magnesium, in the [Mn/Mg] - [Mg/H] plane, it should represent the increasing contribution of \ac{sn1a}. In figure\,\ref{fig:manganese} the model and the data are fitting quite well. Since the model was fitted using the iron and $\alpha$ over metal distribution, it is comforting that other elements representing the same processes are also fit quite well. Of course this is highly dependent on the yield sets used. Also not all elements are matched as good, since yields for some elements are not well determined and also the derived abundances from the \ac{apogee} \ac{rc} stars do have systematic biases. The hope is that, using modelling techniques like the one presented here, these discrepancies can be resolved. 
\subsection{Different yields}
\label{sec:yieldsets}
A major uncertainty, and also a source of potential bias, are the used yield sets for the different feedback processes. Therefore the \ac{mcmc} was done repeatedly for various combinations of yields, leaving all the other parameters, observational constraints, initial conditions, and priors the same. Each run samples at least 10,000 posterior values, excluding the burn-in. The results are listed in table\,\ref{tab:yields}.
\begin{table}
\begin{threeparttable}
\begin{footnotesize}
\begin{tabular}{c c| c c c c c c c c c c c}
\hline
\multicolumn{2}{c|}{yield set} & log-posterior & \multicolumn{2}{c}{$\alpha_3$} & \multicolumn{2}{c}{$\beta_\mathrm{SNIa}$} & \multicolumn{2}{c}{$\mathrm{t}_\mathrm{peak}$} & \multicolumn{2}{c}{$\gamma_\mathrm{infall}$} & \multicolumn{2}{c}{$\lambda_\mathrm{infall}$} \\
\ac{agb}&\ac{sn2} & maximum & $\mu$ & $\sigma$& $\mu$ & $\sigma$& $\mu$ & $\sigma$& $\mu$ & $\sigma$& $\mu$ & $\sigma$\\
\hline
Nugrid & Nomoto & -43.7  & 2.41  & 0.04  & 0.0056    & 0.0005      & 0.26    	& 0.17	& 0.85   & 0.03 & -0.44 & 0.05\\

Nugrid & Chieffi & -51.0  & 2.25  & 0.04  & 0.0056    & 0.0005      & 0.35    	& 0.21	& 0.95   & 0.03 & -0.51 & 0.06\\

Karakas & Chieffi & -63.2  & 2.15  & 0.04  & 0.0060    & 0.0007      & 0.35    	& 0.21	& 0.96   & 0.04 & -0.48 & 0.06\\

Karakas & Nomoto & -66.4  & 2.31  & 0.03  & 0.0062    & 0.0006      & 0.26    	& 0.17	& 0.87   & 0.03 & -0.44 & 0.05\\

Nugrid & Francois & -83.1  & 2.82  & 0.03  & 0.0041    & 0.0004      & 0.39    	& 0.23	& 0.68   & 0.02 & -0.35 & 0.03\\

Nugrid & Nugrid & -106.1  & 1.14  & 0.08  & 0.0085    & 0.0012      & 0.39    	& 0.23	& 0.87   & 0.02 & -0.58 & 0.07\\
\hline
\multicolumn{3}{c}{median and sigma of all sets:} & 2.28 & 0.5 & 0.0058 & 0.0013 & 0.35 & 0.05 & 0.87 & 0.09 & -0.46 & 0.07\\
\hline
\multicolumn{13}{c}{Varying statistics for the model with Nugrid and Nomoto yields}\\
\hline
\multicolumn{2}{c}{priors = $\sigma$/2} & -47.9  & 2.42  & 0.04  & 0.0054    & 0.0004      & 0.57    	& 0.15	& 0.84   & 0.02 & -0.40 & 0.04\\
\multicolumn{2}{c}{quantiles at [5,15,50,85,95]} & -46.0  & 2.41  & 0.04  & 0.0056    & 0.0005      & 0.21    	& 0.15	& 0.84   & 0.03 & -0.43 & 0.04\\
\multicolumn{2}{c}{Only [M/H] and [$\alpha$/M]} & -11.7  & 2.60  & 0.07  & 0.0044    & 0.0015      & 0.91    	& 0.29	& 0.75   & 0.04 & -0.60 & 0.08\\
\end{tabular}
\end{footnotesize}
\end{threeparttable}
\caption[Summary table - different yield sets]{Equilibrium parameter positions and posterior values for \ac{mcmc} runs using different yield sets}
\label{tab:yields}
\end{table}
The best representation of the data is obtained when using the Nugrid \citep{Pignatari2013} \ac{agb} yields together with \citet{Nomoto2013} yields for the high-mass stars.\\
When inspecting the whole table a good overview is gained about the preferred values of the parameters. The peak of the \ac{dtd} at about 0.35\,Gyr seems to be at lower values than a priori assumed. Maybe this is also due to the fixed delay time of 2.5\,Gyr and the functional form of the \ac{dtd}, which should be investigated in the future.\\
The high-mass index, which is most interesting with respect to the results from chapter\,\ref{ch:imf} seems to favour the \citet{Sa55} value at about 2.3, where the four best yield set combinations are pointing. The last two yield sets, which are a bit off in terms of maximum posterior probability, favour a value at 2.8 and 1.1, respectively.
Also for those the different correlations are at at work, as for example a lower \ac{sn1a} rate for less \ac{sn2} as with the \citet{Francois2004} yields and the opposite with the Nugrid high-mass yields.\\

Overall, this shows that the model can find reasonable parameter values with different yield sets, without the necessity to tune the initial conditions of the \ac{mcmc}. Other model ingredients should be tested as well and also the statistics should be further investigated. As a first small test, the \ac{mcmc} runs were made, using the default model, but with changed quantiles, a reduced data set with only metallicity and $\alpha$-enhancement, and priors with halved standard deviations. The results of these are shown in the lower part of table\,\ref{tab:yields}.\\
The \emph{tightening} of the priors seems to have a minor effect, except for the loosely constraint $\mathrm{t}_\mathrm{peak}$. Changing the spacing of the quantiles has a very small effect on the best parameter values but decreases the maximum posterior obtainable. Probably because more weights are given on the tails of the distribution and getting more independent information from it, i. e. the summary statistic is more responsive, compared to two values that are directly next to each other. Only taking half of the elemental abundance data set significantly changes the parameter position of highest posterior. Of course the posterior also increases drastically as 10 values less are multiplied into the probability. In the $\alpha$ over iron plane a larger high-mass index and also later peak time for \ac{sn1a} is preferred, which traces the expected time dependence of \ac{sn2} to \ac{sn1a} feedback. This indicates that one of the other abundance distributions probably pushes the \ac{mcmc} to a faster chemical enrichment. It also shows that with less data at hand the prior gets more important as its fractional contribution to the posterior increases.\\
The statistical tests show how sensitive the \ac{mcmc} reacts on changing the used data sets. On the other hand, the priors and quantiles seemed to be well chosen as changing them has minor effects.
\section{Conclusion}
Chempy, a new, fast and versatile chemical enrichment code, intended for open-source publication, has been presented. A technique to use the output of Chempy and synthesise mock elemental abundances with the help of the \ac{jj} and \ac{galaxia} is introduced. An observational constraint is deduced, using the quantiles of abundance distributions as a summary statistic. This is then converted into a likelihood, assuming a Gaussian probability distribution for each quantile and applying it to the difference between mock and real data distributions. The four most important dimensions of the chemical abundance space data of \ac{apogee}, [Fe/H], [M/H], [$\alpha$/M] and [C+N/M], are turned into distribution functions, as inferred from Solar Neighbourhood \ac{rc} stars. Together with observational constraints about the present-day local gas mass and the \ac{sn2} over \ac{sn1a} ratio, they measure the model likelihood.\\
Five important parameters of the model, the \ac{imf} high-mass index ($\alpha_3$), the number of \ac{sn1a} exploding ($\beta_\mathrm{SNIa}$), the \ac{sn1a} peak time ($\mathrm{t}_\mathrm{peak}$), the infall mass fraction ($\gamma_\mathrm{infall}$) and the infall time scale ($\lambda_\mathrm{infall}$), are given a reasonable prior distribution and an \ac{mcmc} is used to sample the posterior \ac{pdf} of the parameter space. This is done for six different combinations of yield sets with Nugrid \citep{Pignatari2013} and \citet{Nomoto2013} giving the best results. Within that model the high-mass index of the \ac{imf} has a tightly constrained value of 2.41 with a standard deviation of only 0.04.
When taking into account the systematics, with respect to the other four yield sets, best reproducing the observations, a value of 2.28$\pm$0.09 is obtained.\\
In the future, other model parameters should also be tested for their impact on the obtained results. The model should be extended to track the thick disc evolution and to allow for multiple zones, which could be implemented using the isochrone formalism presented in \citet[eq.\,C3]{Kubryk2015}.\\

To the authors knowledge, this is the first time that the parameter space of a chemical enrichment model (in this case spanned by five main parameters) is explored, using Bayesian inference and an \ac{mcmc} simulation to sample the posterior \ac{pdf}, with respect to the likelihood of the chosen observational constraints. Similar approaches usually rely on a $\chi^2$ minimal-distance-estimate and only sample a set of discrete parameter values (cf. \citet{Molla2015}). The presented method here seems to be robust, as no change of initial parameters or further human supervision is necessary for the \ac{mcmc} runs to converge, even for different yield sets. The chemical enrichment model is a good first-order approximation to the chemical evolution of the thin disc, as inferred parameters seem reasonable and other model outcome compares well with independent observations.\\
Overall, the forward modelling approach, using simple physical models and comparing the outcome to real data, seems a promising way to infer crucial parameters of the \ac{mw} evolution. With upcoming spectroscopic surveys the model parameters can be further constrained and it is intended to use dwarf abundances from \ac{ges}, in order to be more sensitive to the early evolution of the \ac{mw} disc.

\acresetall

%% file: gfx/chapter4/lifetimes.pdf_tex
\begingroup%
  \makeatletter%
  \providecommand\color[2][]{%
    \errmessage{(Inkscape) Color is used for the text in Inkscape, but the package 'color.sty' is not loaded}%
    \renewcommand\color[2][]{}%
  }%
  \providecommand\transparent[1]{%
    \errmessage{(Inkscape) Transparency is used (non-zero) for the text in Inkscape, but the package 'transparent.sty' is not loaded}%
    \renewcommand\transparent[1]{}%
  }%
  \providecommand\rotatebox[2]{#2}%
  \ifx\svgwidth\undefined%
    \setlength{\unitlength}{705.40625bp}%
    \ifx\svgscale\undefined%
      \relax%
    \else%
      \setlength{\unitlength}{\unitlength * \real{\svgscale}}%
    \fi%
  \else%
    \setlength{\unitlength}{\svgwidth}%
  \fi%
  \global\let\svgwidth\undefined%
  \global\let\svgscale\undefined%
  \makeatother%
  \begin{picture}(1,0.44832322)%
    \put(0,0){\includegraphics[width=\unitlength]{lifetimes.pdf}}%
    \put(0.0983347,0.42113477){\color[rgb]{0.30196078,0.30196078,0.30196078}\makebox(0,0)[lb]{\smash{{\footnotesize 7\,\% of SSP goes \acs{sn2} within 25\,Myr}}}}%
    \put(0.73669326,0.4193904){\color[rgb]{0.30196078,0.30196078,0.30196078}\makebox(0,0)[lb]{\smash{{\footnotesize lifetime for a star of 1\,M$_\odot$}}}}%
    \put(0.91630029,0.38166831){\color[rgb]{0.8,0,1}\makebox(0,0)[lb]{\smash{Z$_\odot$}}}%
    \put(0.8006429,0.38166831){\color[rgb]{0.21568627,0.78431373,0.44313725}\makebox(0,0)[lb]{\smash{0.01 Z$_\odot$}}}%
    \put(0.75129389,0.37934251){\color[rgb]{0,0,0}\makebox(0,0)[lb]{\smash{Z =}}}%
    \put(0.43933463,0.12716153){\color[rgb]{0,0,1}\rotatebox{8.43845954}{\makebox(0,0)[lb]{\smash{equation\,\ref{eq:ejecta}}}}}%
    \put(0.76021512,0.13141411){\color[rgb]{0,0,0}\makebox(0,0)[lb]{\smash{{\footnotesize $\approx R(\mathrm{m}_\mathrm{IRA})$}}}}%
  \end{picture}%
\endgroup%

%% file: gfx/chapter4/sn1a.pdf_tex
\begingroup%
  \makeatletter%
  \providecommand\color[2][]{%
    \errmessage{(Inkscape) Color is used for the text in Inkscape, but the package 'color.sty' is not loaded}%
    \renewcommand\color[2][]{}%
  }%
  \providecommand\transparent[1]{%
    \errmessage{(Inkscape) Transparency is used (non-zero) for the text in Inkscape, but the package 'transparent.sty' is not loaded}%
    \renewcommand\transparent[1]{}%
  }%
  \providecommand\rotatebox[2]{#2}%
  \ifx\svgwidth\undefined%
    \setlength{\unitlength}{717bp}%
    \ifx\svgscale\undefined%
      \relax%
    \else%
      \setlength{\unitlength}{\unitlength * \real{\svgscale}}%
    \fi%
  \else%
    \setlength{\unitlength}{\svgwidth}%
  \fi%
  \global\let\svgwidth\undefined%
  \global\let\svgscale\undefined%
  \makeatother%
  \begin{picture}(1,0.38214784)%
    \put(0,0){\includegraphics[width=\unitlength]{sn1a.pdf}}%
  \end{picture}%
\endgroup%

%% file: gfx/chapter4/net_yield.pdf_tex
\begingroup%
  \makeatletter%
  \providecommand\color[2][]{%
    \errmessage{(Inkscape) Color is used for the text in Inkscape, but the package 'color.sty' is not loaded}%
    \renewcommand\color[2][]{}%
  }%
  \providecommand\transparent[1]{%
    \errmessage{(Inkscape) Transparency is used (non-zero) for the text in Inkscape, but the package 'transparent.sty' is not loaded}%
    \renewcommand\transparent[1]{}%
  }%
  \providecommand\rotatebox[2]{#2}%
  \ifx\svgwidth\undefined%
    \setlength{\unitlength}{699bp}%
    \ifx\svgscale\undefined%
      \relax%
    \else%
      \setlength{\unitlength}{\unitlength * \real{\svgscale}}%
    \fi%
  \else%
    \setlength{\unitlength}{\svgwidth}%
  \fi%
  \global\let\svgwidth\undefined%
  \global\let\svgscale\undefined%
  \makeatother%
  \begin{picture}(1,0.49642346)%
    \put(0,0){\includegraphics[width=\unitlength]{net_yield.pdf}}%
    \put(0.09412272,0.17559029){\color[rgb]{0,0,1}\makebox(0,0)[lb]{\smash{SNII}}}%
    \put(0.42711826,0.30117542){\color[rgb]{0,0.50196078,0}\makebox(0,0)[lb]{\smash{all processes}}}%
    \put(0.7925291,0.22146521){\color[rgb]{1,0,0}\makebox(0,0)[lb]{\smash{AGB stars}}}%
    \put(0.84235877,0.07118524){\color[rgb]{0,0,0}\makebox(0,0)[lb]{\smash{SNIa}}}%
    \put(0.27420712,0.49315978){\color[rgb]{0.74901961,0,0.74901961}\makebox(0,0)[lb]{\smash{5}}}%
    \put(0.59329424,0.49284684){\color[rgb]{0.74901961,0,0.74901961}\makebox(0,0)[lb]{\smash{2}}}%
    \put(0.91108723,0.49299772){\color[rgb]{0.74901961,0,0.74901961}\makebox(0,0)[lb]{\smash{1}}}%
    \put(0.01186469,0.49534482){\color[rgb]{0.74901961,0,0.74901961}\makebox(0,0)[lb]{\smash{{\footnotesize mass of stars dying in M$_\odot$}}}}%
  \end{picture}%
\endgroup%

%% file: gfx/chapter4/selection.pdf_tex
\begingroup%
  \makeatletter%
  \providecommand\color[2][]{%
    \errmessage{(Inkscape) Color is used for the text in Inkscape, but the package 'color.sty' is not loaded}%
    \renewcommand\color[2][]{}%
  }%
  \providecommand\transparent[1]{%
    \errmessage{(Inkscape) Transparency is used (non-zero) for the text in Inkscape, but the package 'transparent.sty' is not loaded}%
    \renewcommand\transparent[1]{}%
  }%
  \providecommand\rotatebox[2]{#2}%
  \ifx\svgwidth\undefined%
    \setlength{\unitlength}{612bp}%
    \ifx\svgscale\undefined%
      \relax%
    \else%
      \setlength{\unitlength}{\unitlength * \real{\svgscale}}%
    \fi%
  \else%
    \setlength{\unitlength}{\svgwidth}%
  \fi%
  \global\let\svgwidth\undefined%
  \global\let\svgscale\undefined%
  \makeatother%
  \begin{picture}(1,0.45588235)%
    \put(0,0){\includegraphics[width=\unitlength]{selection.pdf}}%
    \put(0.70820184,0.09562436){\color[rgb]{0,1,1}\makebox(0,0)[lb]{\smash{old main-sequence}}}%
    \put(0.6837597,0.25756292){\color[rgb]{1,0,1}\makebox(0,0)[lb]{\smash{red clump}}}%
  \end{picture}%
\endgroup%

%% file: gfx/chapter4/selection_age.pdf_tex
\begingroup%
  \makeatletter%
  \providecommand\color[2][]{%
    \errmessage{(Inkscape) Color is used for the text in Inkscape, but the package 'color.sty' is not loaded}%
    \renewcommand\color[2][]{}%
  }%
  \providecommand\transparent[1]{%
    \errmessage{(Inkscape) Transparency is used (non-zero) for the text in Inkscape, but the package 'transparent.sty' is not loaded}%
    \renewcommand\transparent[1]{}%
  }%
  \providecommand\rotatebox[2]{#2}%
  \ifx\svgwidth\undefined%
    \setlength{\unitlength}{502bp}%
    \ifx\svgscale\undefined%
      \relax%
    \else%
      \setlength{\unitlength}{\unitlength * \real{\svgscale}}%
    \fi%
  \else%
    \setlength{\unitlength}{\svgwidth}%
  \fi%
  \global\let\svgwidth\undefined%
  \global\let\svgscale\undefined%
  \makeatother%
  \begin{picture}(1,0.78486056)%
    \put(0,0){\includegraphics[width=\unitlength]{selection_age.pdf}}%
  \end{picture}%
\endgroup%

%% file: gfx/chapter4/height_selection.pdf_tex
\begingroup%
  \makeatletter%
  \providecommand\color[2][]{%
    \errmessage{(Inkscape) Color is used for the text in Inkscape, but the package 'color.sty' is not loaded}%
    \renewcommand\color[2][]{}%
  }%
  \providecommand\transparent[1]{%
    \errmessage{(Inkscape) Transparency is used (non-zero) for the text in Inkscape, but the package 'transparent.sty' is not loaded}%
    \renewcommand\transparent[1]{}%
  }%
  \providecommand\rotatebox[2]{#2}%
  \ifx\svgwidth\undefined%
    \setlength{\unitlength}{497.71875bp}%
    \ifx\svgscale\undefined%
      \relax%
    \else%
      \setlength{\unitlength}{\unitlength * \real{\svgscale}}%
    \fi%
  \else%
    \setlength{\unitlength}{\svgwidth}%
  \fi%
  \global\let\svgwidth\undefined%
  \global\let\svgscale\undefined%
  \makeatother%
  \begin{picture}(1,0.78928863)%
    \put(0,0){\includegraphics[width=\unitlength]{height_selection.pdf}}%
    \put(0.00074368,0.76838422){\color[rgb]{0,1,1}\makebox(0,0)[lb]{\smash{time in Gyr}}}%
    \put(0.14821792,0.76532337){\color[rgb]{0,1,1}\makebox(0,0)[lb]{\smash{2}}}%
    \put(0.30986401,0.76555097){\color[rgb]{0,1,1}\makebox(0,0)[lb]{\smash{3}}}%
    \put(0.54519399,0.76576288){\color[rgb]{0,1,1}\makebox(0,0)[lb]{\smash{5}}}%
    \put(0.76241559,0.76555097){\color[rgb]{0,1,1}\makebox(0,0)[lb]{\smash{8}}}%
    \put(0.89928395,0.76532337){\color[rgb]{0,1,1}\makebox(0,0)[lb]{\smash{12}}}%
    \put(0.36456626,0.70801168){\color[rgb]{0.74901961,0,0.74901961}\makebox(0,0)[lb]{\smash{model}}}%
    \put(0.34828801,0.55344482){\color[rgb]{0,0,0}\makebox(0,0)[lb]{\smash{APOGEE RC}}}%
    \put(0.71628857,0.68840637){\color[rgb]{0,0.50196078,0}\makebox(0,0)[lb]{\smash{model age-weighted}}}%
    \put(0.21484702,0.54603382){\color[rgb]{0,0,0}\makebox(0,0)[lb]{\smash{0 pc}}}%
    \put(0.19421381,0.31909896){\color[rgb]{0,0,0}\makebox(0,0)[lb]{\smash{300 pc}}}%
    \put(0.19431584,0.08602475){\color[rgb]{0,0,0}\makebox(0,0)[lb]{\smash{600 pc}}}%
  \end{picture}%
\endgroup%

%% file: gfx/chapter4/parameter_space_default.pdf_tex
\begingroup%
  \makeatletter%
  \providecommand\color[2][]{%
    \errmessage{(Inkscape) Color is used for the text in Inkscape, but the package 'color.sty' is not loaded}%
    \renewcommand\color[2][]{}%
  }%
  \providecommand\transparent[1]{%
    \errmessage{(Inkscape) Transparency is used (non-zero) for the text in Inkscape, but the package 'transparent.sty' is not loaded}%
    \renewcommand\transparent[1]{}%
  }%
  \providecommand\rotatebox[2]{#2}%
  \ifx\svgwidth\undefined%
    \setlength{\unitlength}{1005bp}%
    \ifx\svgscale\undefined%
      \relax%
    \else%
      \setlength{\unitlength}{\unitlength * \real{\svgscale}}%
    \fi%
  \else%
    \setlength{\unitlength}{\svgwidth}%
  \fi%
  \global\let\svgwidth\undefined%
  \global\let\svgscale\undefined%
  \makeatother%
  \begin{picture}(1,0.73233831)%
    \put(0,0){\includegraphics[width=\unitlength]{parameter_space_default.pdf}}%
  \end{picture}%
\endgroup%

%% file: gfx/chapter4/model.pdf_tex
\begingroup%
  \makeatletter%
  \providecommand\color[2][]{%
    \errmessage{(Inkscape) Color is used for the text in Inkscape, but the package 'color.sty' is not loaded}%
    \renewcommand\color[2][]{}%
  }%
  \providecommand\transparent[1]{%
    \errmessage{(Inkscape) Transparency is used (non-zero) for the text in Inkscape, but the package 'transparent.sty' is not loaded}%
    \renewcommand\transparent[1]{}%
  }%
  \providecommand\rotatebox[2]{#2}%
  \ifx\svgwidth\undefined%
    \setlength{\unitlength}{715bp}%
    \ifx\svgscale\undefined%
      \relax%
    \else%
      \setlength{\unitlength}{\unitlength * \real{\svgscale}}%
    \fi%
  \else%
    \setlength{\unitlength}{\svgwidth}%
  \fi%
  \global\let\svgwidth\undefined%
  \global\let\svgscale\undefined%
  \makeatother%
  \begin{picture}(1,0.50909091)%
    \put(0,0){\includegraphics[width=\unitlength]{model.pdf}}%
    \put(0.57898291,0.13346208){\color[rgb]{0,0,0}\makebox(0,0)[lb]{\smash{Kubryk+ 2015 local gas constraint}}}%
  \end{picture}%
\endgroup%

%% file: gfx/chapter4/apogee_m.pdf_tex
\begingroup%
  \makeatletter%
  \providecommand\color[2][]{%
    \errmessage{(Inkscape) Color is used for the text in Inkscape, but the package 'color.sty' is not loaded}%
    \renewcommand\color[2][]{}%
  }%
  \providecommand\transparent[1]{%
    \errmessage{(Inkscape) Transparency is used (non-zero) for the text in Inkscape, but the package 'transparent.sty' is not loaded}%
    \renewcommand\transparent[1]{}%
  }%
  \providecommand\rotatebox[2]{#2}%
  \ifx\svgwidth\undefined%
    \setlength{\unitlength}{701bp}%
    \ifx\svgscale\undefined%
      \relax%
    \else%
      \setlength{\unitlength}{\unitlength * \real{\svgscale}}%
    \fi%
  \else%
    \setlength{\unitlength}{\svgwidth}%
  \fi%
  \global\let\svgwidth\undefined%
  \global\let\svgscale\undefined%
  \makeatother%
  \begin{picture}(1,0.59771755)%
    \put(0,0){\includegraphics[width=\unitlength]{apogee_m.pdf}}%
    \put(0.10834887,0.15401656){\color[rgb]{0.74901961,0,0.74901961}\makebox(0,0)[lb]{\smash{APOGEE RC}}}%
    \put(0.85515017,0.22744311){\color[rgb]{0.04705882,0.76078431,0.76078431}\makebox(0,0)[lb]{\smash{model}}}%
    \put(0.11461698,0.55159424){\color[rgb]{0,0,0}\makebox(0,0)[lb]{\smash{$\mathcal{P}_\mathrm{quantiles}=$}}}%
    \put(0.11372155,0.58809727){\color[rgb]{0.4,0.4,0.4}\makebox(0,0)[lb]{\smash{quantiles in \%}}}%
    \put(0.35817809,0.58623609){\color[rgb]{0.4,0.4,0.4}\makebox(0,0)[lb]{\smash{5}}}%
    \put(0.52831265,0.58608563){\color[rgb]{0.4,0.4,0.4}\makebox(0,0)[lb]{\smash{25}}}%
    \put(0.63666155,0.58608563){\color[rgb]{0.4,0.4,0.4}\makebox(0,0)[lb]{\smash{50}}}%
    \put(0.70471535,0.58623609){\color[rgb]{0.4,0.4,0.4}\makebox(0,0)[lb]{\smash{75}}}%
    \put(0.81843687,0.58608563){\color[rgb]{0.4,0.4,0.4}\makebox(0,0)[lb]{\smash{95}}}%
    \put(0.11686439,0.29005265){\color[rgb]{0,0,1}\rotatebox{-71.974729}{\makebox(0,0)[lb]{\smash{{\footnotesize median error}}}}}%
  \end{picture}%
\endgroup%

%% file: gfx/chapter4/model_S.pdf_tex
\begingroup%
  \makeatletter%
  \providecommand\color[2][]{%
    \errmessage{(Inkscape) Color is used for the text in Inkscape, but the package 'color.sty' is not loaded}%
    \renewcommand\color[2][]{}%
  }%
  \providecommand\transparent[1]{%
    \errmessage{(Inkscape) Transparency is used (non-zero) for the text in Inkscape, but the package 'transparent.sty' is not loaded}%
    \renewcommand\transparent[1]{}%
  }%
  \providecommand\rotatebox[2]{#2}%
  \ifx\svgwidth\undefined%
    \setlength{\unitlength}{720bp}%
    \ifx\svgscale\undefined%
      \relax%
    \else%
      \setlength{\unitlength}{\unitlength * \real{\svgscale}}%
    \fi%
  \else%
    \setlength{\unitlength}{\svgwidth}%
  \fi%
  \global\let\svgwidth\undefined%
  \global\let\svgscale\undefined%
  \makeatother%
  \begin{picture}(1,0.48055556)%
    \put(0,0){\includegraphics[width=\unitlength]{model_S.pdf}}%
    \put(0.24440267,0.41129253){\color[rgb]{0,0,0}\makebox(0,0)[lb]{\smash{age in Gyr}}}%
    \put(0.14157955,0.07909474){\color[rgb]{0,0,0}\makebox(0,0)[lb]{\smash{median error APOGEE RC}}}%
    \put(0.22621089,0.26338539){\color[rgb]{0,0,0}\makebox(0,0)[lb]{\smash{age weighted model}}}%
  \end{picture}%
\endgroup%

%% file: gfx/chapter4/model_M.pdf_tex
\begingroup%
  \makeatletter%
  \providecommand\color[2][]{%
    \errmessage{(Inkscape) Color is used for the text in Inkscape, but the package 'color.sty' is not loaded}%
    \renewcommand\color[2][]{}%
  }%
  \providecommand\transparent[1]{%
    \errmessage{(Inkscape) Transparency is used (non-zero) for the text in Inkscape, but the package 'transparent.sty' is not loaded}%
    \renewcommand\transparent[1]{}%
  }%
  \providecommand\rotatebox[2]{#2}%
  \ifx\svgwidth\undefined%
    \setlength{\unitlength}{720bp}%
    \ifx\svgscale\undefined%
      \relax%
    \else%
      \setlength{\unitlength}{\unitlength * \real{\svgscale}}%
    \fi%
  \else%
    \setlength{\unitlength}{\svgwidth}%
  \fi%
  \global\let\svgwidth\undefined%
  \global\let\svgscale\undefined%
  \makeatother%
  \begin{picture}(1,0.49166667)%
    \put(0,0){\includegraphics[width=\unitlength]{model_M.pdf}}%
    \put(0.20854257,0.08007389){\color[rgb]{0,0,0}\makebox(0,0)[lb]{\smash{age in Gyr}}}%
    \put(0.14878433,0.4520148){\color[rgb]{0,0,0}\makebox(0,0)[lb]{\smash{median error APOGEE RC}}}%
    \put(0.75522775,0.45509741){\color[rgb]{0.80784314,0.74117647,0.39215686}\makebox(0,0)[lb]{\smash{age weighted model}}}%
  \end{picture}%
\endgroup%

%% file: Chapters/Chapter05.tex
\chapter{Summary}\label{ch:summary} 
The stellar \ac{imf} was inferred from the local chemodynamical evolution model of the \ac{mw} disc of \citet{JJ}. In a first attempt, volume-complete samples of Solar Neighbourhood stars from \ac{hip} and low-mass star counts from \citet{Just2015}, were used to constrain the \ac{imf} up to 6\,M$_\odot$.\\
In order to infer the high-mass index, the modular chemical evolution code  \emph{Chempy}, was developed. Its outcome was mapped onto \ac{apogee} \ac{rc} stellar elemental abundance distributions and their agreement was tested for different yield sets. The model favours a Salpeter-like high-mass slope and also determines infall and \ac{sn1a} parameters. The fiducial \ac{imf} values of the four-slope broken power-law are 
\begin{equation}
\begin{array}{rcl}
\alpha_0 &=& \text{1.26},\\
\alpha_1 &=& \text{1.49}\pm \text{0.08},\\
\alpha_2 &=& \text{3.02}\pm \text{0.06},\\
\alpha_2 &=& \text{2.28}\pm \text{0.09},\\
\mathrm{m}_0 &=& \text{0.5},\\
\mathrm{m}_1 &=& \text{1.39}\pm \text{0.05},\\
\mathrm{m}_2 &=& \text{6},\\
\end{array}
\end{equation} 
within the stellar mass regime, 0.08 < $\mathrm{m}/$M$_\odot$ < 100. The errors given are derived from the \ac{mcmc} simulation and are valid within model assumptions, but observational biases or model systematics are not included. The values without errors, were fixed empirically, using observational constraints which were not included in the simulation.\\

The results were obtained with Bayesian statistics, applied to the forward modelling technique. Focus was laid on synthesising realistic mock samples and on the likelihood construction, which is a statistical measure, allowing \ac{mcmc} algorithms to find model parameters, best representing the observational evidence. Vice versa the posterior \ac{pdf} can give insight on which data has the most discriminative power for the problem under investigation and also on the correlations between the different model parameters.\\
A desirable feature of the approach is that the likelihood of the abundance data is determined unsupervised. It is also much faster, than judging the matching of abundance patterns by eye. In the future it is intended to include two dimensional correlations of the data into the likelihood construction.\\
In conclusion, the presented technique is a valuable approach for the huge upcoming data sets which will increase the model complexity and, therefore, the demand for robust unsupervised parameter exploration.
\acresetall
\chapter{Outlook}\label{ch:outlook}
With the dawn of \ac{gaia} and the follow-up spectroscopic \ac{ges} survey, rich data sets are upcoming which will probe the chemical abundance and phase-space of the \ac{mw} in unprecedented detail. The analysis of these data and the incorporation of new-found structures will pose a huge challenge to Galaxy modelling. Techniques to lower the dimensionality of the data space, will become more important as shown by the first promising results from chemical tagging \citep{Hawkins2015}.\\
To account for the higher complexity, reflected by the data, the number of model parameters will increase. This will make parameter estimation techniques, which are able to incorporate lots of different data and at the same time capable of exploring a high-dimensional parameter space, a valuable asset.\\

It is planned to extend the local JJ-model to different Galactocentric radii. For the time being, the analysis of the parameter constraints, resulting from chemical abundances and possible extensions to \emph{Chempy}, are considered. To compare the discriminative power and the highest posterior probability of different survey data like \citet{Bensby2014}, \ac{apogee} or \ac{ges} will be an interesting endeavour and might shed light on survey biases and selection functions.\\
A study of the \ac{agb} mass range and its impact on the mass-sensitive carbon and nitrogen abundances could further constrain the \ac{imf}, independently from the approach in chapter\,\ref{ch:imf}. The inference of yield parameters should also be possible. It will be exciting to see the parameter constraints increasing, in parallel to the increase of observational data incorporated into the \ac{mw} model analysis. 

%% file: FrontBackmatter/Bibliography.tex
\manualmark
\markboth{\spacedlowsmallcaps{\bibname}}{\spacedlowsmallcaps{\bibname}} 
\refstepcounter{dummy}
\addtocontents{toc}{\protect\vspace{\beforebibskip}} 
\addcontentsline{toc}{chapter}{\tocEntry{\bibname}}
\bibliographystyle{jan_thesis}

\label{app:bibliography} 
\bibliography{Literature}

%% file: FrontBackmatter/Colophon.tex
\pagestyle{empty}

\hfill

\vfill

\pdfbookmark[0]{Colophon}{colophon}
\section*{Colophon}
This document was typeset in \LaTeX using a modified version of the typographical look-and-feel {\textit classicthesis} developed by Andr\'e Miede (GNU GPL).
 
\bigskip
\noindent Plots were created using {\textit matplotlib} together with {\textit inkscape}.\\